\documentclass[%
 reprint,
superscriptaddress,
 amsmath,amssymb,
 aps,
 prx,
floatfix,
longbibliography
]{revtex4-2}

\usepackage{graphicx}
\usepackage{xcolor}
\usepackage{dcolumn}
\usepackage{bm}
\usepackage[pdftex,colorlinks,citecolor=blue]{hyperref}
\usepackage{enumitem}
\setlist[enumerate]{
  label=(\roman*),
  leftmargin=20pt,     
  labelindent=0pt,
  itemsep=2pt,
  topsep=2pt
}
\usepackage[]{cleveref}
\usepackage{cancel}
\crefname{section}{\S\!}{\S\S\!}
\crefname{appendix}{\S\!}{\S\S\!}
\crefname{equation}{Eq.}{Eqs.}
\Crefname{equation}{Equation}{Equations}
\crefname{figure}{Fig.}{Figs.}
\Crefname{figure}{Figure}{Figures}
\usepackage{mdframed}

\newcommand{\nn}{\nonumber}
\newcommand\bh{\hat{\bm{b}}}
\newcommand\va{v_{\rm A}}
\newcommand\vs{v_{S}}
\newcommand\cs{c_{s}}
\newcommand{\gprl}[1]{\mathcal{K}_{#1}^{\|}}
\newcommand{\gprp}[1]{\bm{\mathcal{K}}_{#1}^{\perp}}
\newcommand{\gprps}[1]{\mathcal{K}_{#1}^{\perp}}
\newcommand{\zpm}{\bm{z}^{\pm}}
\newcommand{\zmp}{\bm{z}^{\mp}}
\newcommand{\zp}{\bm{z}^{+}}
\newcommand{\zm}{\bm{z}^{-}}
\newcommand{\fpm}{\bm{f}^{\pm}}
\newcommand{\fmp}{\bm{f}^{\mp}}

\newcommand{\ma}{\mathcal{M}_{\rm A}}

\newcommand{\chia}{\chi_{\rm A}}
\newcommand{\bkap}{\bm{\kappa}}
\newcommand{\kp}{\bm{k}_{\perp}}

\newcommand{\rsun}{R_{\odot}}
\newcommand{\Sm}{S_{\rho U}}
\newcommand{\Smv}{\bm{S}_{\rho U}}
\newcommand{\Su}{S_{ U}}

\newcommand{\Sth}{S_{\rm th}}
\newcommand{\Srho}{S_{\rho}}

\renewcommand{\d}{\delta}
\newcommand{\du}{\delta\bm{u}}
\newcommand{\db}{\delta\bm{B}}
\newcommand{\duprp}{\delta\bm{u}_{\perp}}
\newcommand{\duprl}{\delta {u}_{\|}}
\newcommand{\dbprp}{\delta \bm{B}_{\perp}}
\newcommand{\dbprl}{\delta {B}_{\|}}
\newcommand{\dVprlx}{\delta {\tilde{V}}_{\|}}

\newcommand{\zlpm}{z_{\|}^{\pm} }
\newcommand{\zlmp}{z_{\|}^{\mp} }

\newcommand{\tsmat}{\mathsf{S}}

\newcommand{\avgp}[1]{\langle{#1}\rangle_{\perp}}
\newcommand{\avg}[1]{\left\langle{#1}\right\rangle}
\newcommand{\avgi}[1]{\langle{#1}\rangle} 

\newcommand{\gradl}{\bh\cdot\!\nabla}  
\newcommand{\bhdg}{\nabla_{\|}}
\newcommand{\gradperp}{\nabla_{\!\perp}}
\newcommand{\rmd}{{\rm d}}

\newcommand{\grad}{\nabla}

\newcommand{\UperpII}{\bm{U}_{\perp 2}}
\newcommand{\Vpsi}{\bm{V}_{\!\psi}}
\newcommand{\Vpsit}{\widetilde{\bm{V}}_{\!\psi}}
\newcommand{\VU}{\bm{V}_{\!U}}

\newcommand{\Vrho}{\bm{V}_{\!\rho}}
\newcommand{\Vp}{\bm{V}_{\!p}}
\newcommand{\Vrhot}{\widetilde{\bm{V}}_{\!\rho}}
\newcommand{\Vpt}{\widetilde{\bm{V}}_{\!p}}
\newcommand{\taunl}{\tau_{\rm nl}}
\newcommand{\omeganl}{\omega_{\rm nl}}
\newcommand{\geff}{\bm{g}^{\perp}_{\rm eff}}

\newcommand{\zo}{\bm{Z}^{(0)}}
\newcommand{\dzp}{\delta \bm z^{+}}
\newcommand{\dzm}{\delta \bm z^{-}}
\newcommand{\dzpprp}{\delta \bm z^{+}_{\perp}}
\newcommand{\dzmprp}{\delta \bm z^{-}_{\perp}}
\newcommand{\dzpprl}{\delta z^{+}_{\parallel}}
\newcommand{\dzmprl}{\delta z^{-}_{\parallel}}
\newcommand{\imord}{\varsigma}
\newcommand{\ccomp}{c_{\rm comp}}
\newcommand{\Thetac}{\Theta_{\rm comp}}

\newcommand{\closesymbol}{\!}
\newcommand{\fourier}[1]{\tilde{#1}}  
\renewcommand{\fourier}[1]{#1}  

\newcommand{\rmi}{\text{i}}

\ifdefined\rmJ
\renewcommand{\rmJ}{\text{J}}
\else
\newcommand{\rmJ}{\text{J}}
\fi

\ifdefined\rmI
\renewcommand{\rmI}{\text{I}}
\else
\newcommand{\rmI}{\text{I}}
\fi

\newcommand{\besselarg}{a_\s}
\newcommand{\besselargint}{b_\s}

\ifdefined\partd
\renewcommand{\partd}[3][]{\frac{{\partial^{#1} #2}}{{\partial #3}^{#1}}}
\else
\newcommand{\partd}[3][]{\frac{{\partial^{#1} #2}}{{\partial #3}^{#1}}}
\fi

\renewcommand{\vec}[1]{\boldsymbol{#1}}
\ifdefined\vec  
\renewcommand{\vec}[1]{\boldsymbol{#1}}
\else
\newcommand{\vec}[1]{\boldsymbol{#1}}
\fi

\ifdefined\div
\renewcommand{\div}{{\vec{\nabla} \cdot}}
\else
\newcommand{\div}{{\vec{\nabla} \cdot}}
\fi

\newcommand{\gperp}{\vec{\nabla}_\perp}

\ifdefined\vr
\renewcommand{\vr}{{\vec{r}}}
\else
\newcommand{\vr}{{\vec{r}}}
\fi

\newcommand{\vk}{{\vec{k}}}
\newcommand{\vv}{{\vec{v}}}
\newcommand{\vkperp}{{\vk_{\closesymbol\perp}}}
\newcommand{\kperp}{k_\perp}
\newcommand{\kpar}{k_\parallel}
\newcommand{\vperp}{v_\perp}
\newcommand{\vvperp}{\vv_\perp}
\newcommand{\vpar}{v_\parallel}

\newcommand{\intv}[1][3]{{\int \rmd^{#1} \vec{v} \ }}

\newcommand{\s}{s}

\newcommand{\vths}[1][\s]{v_{\text{th}{#1}}}

\newcommand{\qs}[1][\s]{q_{#1}}

\newcommand{\ms}[1][\s]{m_{#1}}
\newcommand{\me}{\ms[e]}
\newcommand{\mi}{\ms[i]}

\newcommand{\rhos}{\rho_\s}

\newcommand{\dens}[1][\s]{n_{#1}}

\newcommand{\dns}[1][\s]{\delta n_{#1}}

\newcommand{\Ts}[1][\s]{T_{#1}}

\newcommand{\dTs}[1][\s]{\delta T_{#1}}

\newcommand{\ps}[1][\s]{p_{#1}}

\newcommand{\dps}[1][\s]{\delta p_{#1}}

\newcommand{\df}{\delta \closesymbol f}
\newcommand{\fs}{f_{\s}}

\newcommand{\Fs}{F_{\s}}

\newcommand{\dfs}{\delta \closesymbol f_{\s}}

\newcommand{\energy}{\varepsilon_\s}
\newcommand{\mus}{\mu_\s}

\newcommand{\pbra}[2]{\left\lbrace #1, #2 \right\rbrace}

\newcommand{\vRs}[1][\s]{{\vec{R}_{#1}}}

\newcommand{\hs}[1][\s]{h_{#1}}

\newcommand{\phipot}{\phi}  

\newcommand{\phipotkperp}{{\fourier{\phipot}}_{\vkperp}}
\newcommand{\dphipot}{\delta \phipot}  

\newcommand{\dphipotkperp}{{\fourier{\dphipot}}_{\vkperp}}

\newcommand{\dApar}{\delta \closesymbol  A_\parallel}

\newcommand{\dAparkperp}{\delta \closesymbol \fourier{A}_{\parallel{\vkperp}}}

\newcommand{\dBpar}{{\delta \closesymbol  B_\parallel}}

\newcommand{\dBparkperp}{{\delta \closesymbol  \fourier{B}_{\parallel\vkperp}}}

\newcommand{\ub}{\vec{b}}

\newcommand{\vdBperp}{{\delta \closesymbol  {\vec{B}}_{\closesymbol\perp}}}

\newcommand{\vdA}{\delta \closesymbol \vec{A}}

\newcommand{\vrhos}[1][\s]{\vec{\rho}_{#1}}

\newcommand{\vds}{\vec{v}_{\rmd \s}}

\newcommand{\vchi}{\vec{v}_{\chi}}

\newcommand{\avgRs}[1]{\left\langle #1 \right\rangle_{\vec{R}_\s}}

\newcommand{\avgRsinline}[1]{{\langle #1 \rangle_{\vec{R}_\s}}}
\newcommand{\avgr}[1]{\left\langle #1 \right\rangle_{\vec{r}}}

\newcommand{\fsa}[1]{\left< #1 \right>_{\psi}}

\newcommand{\Omegas}[1][\s]{\Omega_{#1}}

\renewcommand{\ub}{\hat{\vec{b}}}
\newcommand{\chipot}{\chi}
\newcommand{\curvature}{\vec{\kappa}}
\newcommand{\dupar}{\delta u_{\parallel}}
\newcommand{\dupars}{\delta u_{\parallel \s}}
\newcommand{\vduperp}{\delta \vec{u}_{\perp}}
\newcommand{\dT}{\delta T}
\newcommand{\nussp}{\nu_{\s \s'}}
\newcommand{\rmGamma}{\Gamma}
\newcommand{\drho}{\delta \rho}
\newcommand{\turbavg}[1]{\left< #1 \right>_{}}
\newcommand{\turbheating}{P_\s^{\text{turb}}}
\newcommand{\perpunitvectorbase}{\vec{e}}
\newcommand{\gprpfrac}[1]{\frac{\nabla_\perp #1}{#1}}

\begin{document}

\preprint{APS/123-QED}

\title{A Transport Theory of Turbulent Coronal Heating in General Geometry}

\author{Jonathan Squire}
\email{jonathan.squire@otago.ac.nz}
\affiliation{Department of Physics, University of Otago, Dunedin 9016, New Zealand}

\author{Benjamin D. G. Chandran}
\affiliation{Department of Physics and Astronomy, University of New Hampshire, Durham, New Hampshire 03824, USA}

\author{Toby Adkins}
\affiliation{Princeton Plasma Physics Laboratory, Princeton, New Jersey 08540, USA}

\author{William A. Clarke}
\affiliation{Rudolf Peierls Centre for Theoretical Physics, University of Oxford, Oxford OX1 3PU, UK}
\affiliation{University College, Oxford OX1 4BH, UK}
\affiliation{United Kingdom Atomic Energy Authority, Culham Science Centre, Abingdon OX14 4DB, UK}

\author{Romain Meyrand}
\affiliation{Department of Physics and Astronomy, University of New Hampshire, Durham, New Hampshire 03824, USA}
\affiliation{Department of Physics, University of Otago, Dunedin 9016, New Zealand}

\author{Matthew W. Kunz}
\affiliation{Department of Astrophysical Sciences, Princeton University, Princeton, New Jersey 08544, USA}
\affiliation{Princeton Plasma Physics Laboratory, Princeton, New Jersey 08540, USA}
\date{\today}%

\begin{abstract}
Magnetic geometry shapes how turbulence couples, transports, and dissipates energy in strongly magnetized plasmas. The solar corona, with its complex, structured maze of open and closed tubes and sharp transverse gradients, provides a prominent example; yet, most wave-turbulence models of coronal heating and solar-wind acceleration assume locally symmetric flux-tube geometries or add additional effects in ad hoc ways. Here we develop a geometry-complete multiscale transport theory for reduced-magnetodydrodynamic turbulence in an arbitrary background field, retaining squashing factors (magnetic shear), transverse gradients, curvature, and gravity at the same formal order as standard expansion-driven reflection. The theory couples fast, anisotropic fluctuations to slow evolution of the background through conservation laws, providing a unified description of wave propagation and reflection, turbulent heating, and cross-field transport of mass, momentum, and heat.
Applied to the corona, it yields a set of robust qualitative predictions. In smooth regions such as coronal-hole interiors, it recovers the familiar reflection-driven turbulent (RDT) cascade, a
baseline expectation that underlies some key successes of standard wave-turbulence models. In structured regions, however, additional geometry-driven channels can  dominate over RDT: distortion of the field-line mapping drives reflection even when parallel Alfvén-speed gradients are weak; curvature and non-radial geometry enable coupling to compressive responses and further heating channels; and waves  catalyze the relaxation of large-scale velocity shear  into heat. The same dynamics imply strong cross-field transport across open-closed interfaces, allowing continuous transverse exchange of mass, composition, momentum, and heat that could rival or exceed the field-parallel supply from the coronal base in streamer-adjacent open flux.
Taken together, these effects generically bias heating toward low altitudes in structured regions, providing a physical basis for coronal-hole--boundary corrections used in empirical wind-speed predictors. Likewise, as a closed set of conservation laws, the accompanying slow-timescale transport equations  could be evolved in time, providing a route to a global, geometry-aware model of the wave-driven corona and wind in which  all heating and cross-field transport emerge consistently at the order considered. More broadly, the framework provides an energy-consistent route to understanding the complex interplay of turbulence, geometry, and transport across a variety of astrophysical and terrestrial settings, for example in magnetospheres, accretion flows, jets, and magnetically confined fusion experiments. 
\end{abstract}

\maketitle


\section{Introduction}

\begin{figure}
\begin{center}
\includegraphics[width=1.0\columnwidth]{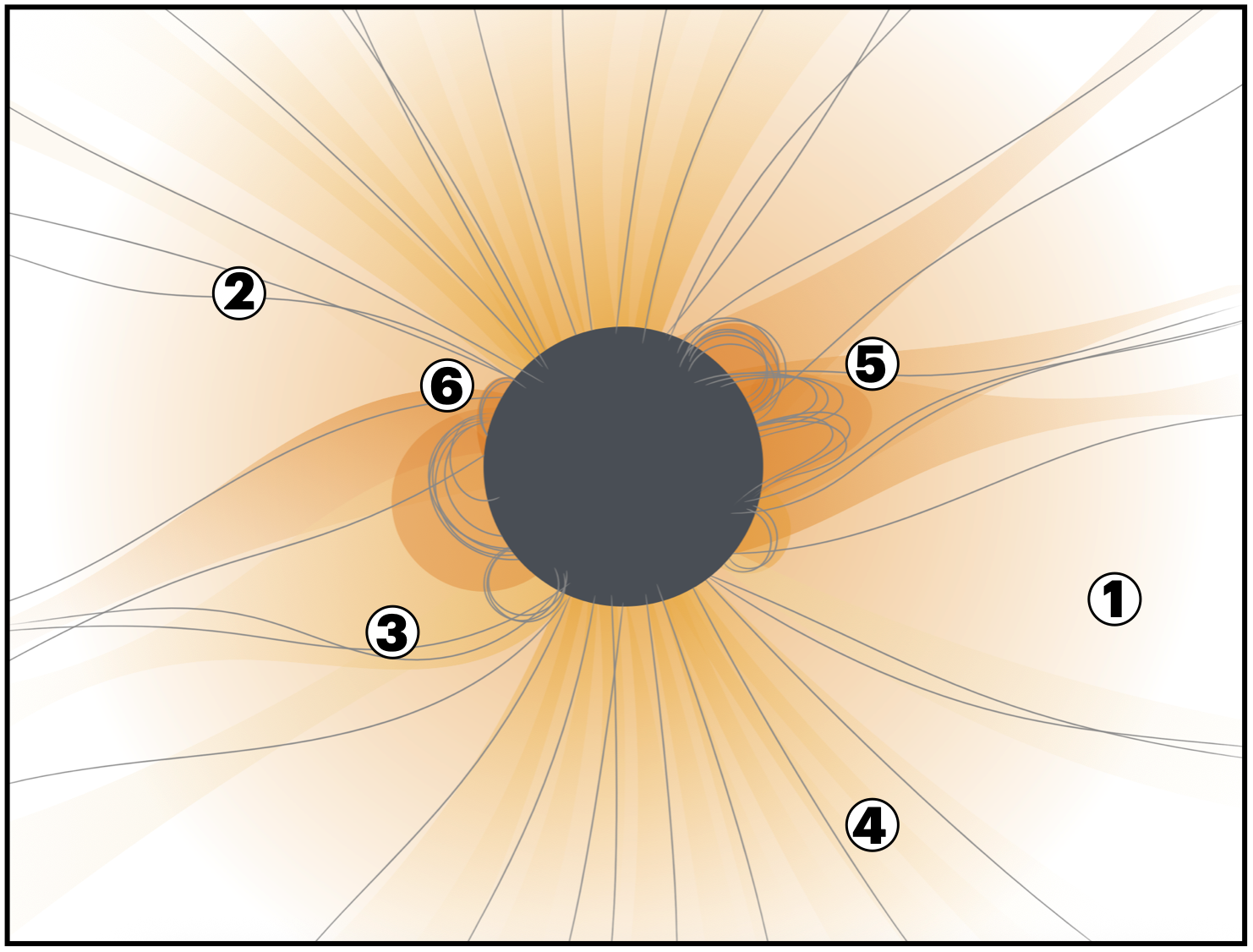}
\caption{
Schematic cartoon of the heating and transport pathways for the solar corona predicted by our multiscale RMHD theory and unified turbulence phenomenology. All effects are enabled by outward propagating Alfv\'enic fluctuations ($z^{+}$), assumed to be generated at the coronal base.
Numbers label distinct mechanisms:
(1) standard expansion-driven reflection-driven turbulent dissipation due to  Alfv\'en-speed gradients \cite{Velli1989}, dominant in nearly structure-free regions (e.g. coronal-hole centers).
(2) \emph{Q-reflection} caused by non-circular (squashed/distorted) flux-tube expansion, which can exceed the standard reflection wherever flux tubes deviate appreciably from circular expansion.
(3) curvature- and perpendicular-gravity-mediated generation of compressive fluctuations by $z^{+}$, yielding additional heating through three channels:
(3a) \emph{direct compressive feedback} (DCR), $z^{+}$-generated compressive perturbations feed back on $z^{+}$;
(3b) \emph{compressively catalyzed reflection} (CCR), $z^{+}$-generated compressive perturbations create counter-propagating ($z^{-}$) fluctuations that drive turbulence;
and (3c) \emph{Alfvén-catalyzed relaxation} (ACR), $z^{+}$-generated compressive perturbations enable the relaxation of cross-field inhomogeneity of the background.
(4) ACR-driven damping of cross-field shear/stream structure in the outflow, which turns on strongly as the wind becomes super-Alfvénic, thus  dissipating flow/stream structure into heat beyond the Alfvén radius.
(5) Cross-field mass and composition transport from dense, closed-field plasma into neighbouring open tubes, particularly effective near field cusps, and potentially helping to supply  slow-wind mass flux.
(6) Cross-field heat transport from hot closed regions  into surrounding open field lines.
Mechanisms (2), (3), and (6) preferentially energize structured regions at low altitudes, naturally linking enhanced low-coronal heating to structured backgrounds, and therefore slow-wind formation.
}\label{fig: overview}
\end{center}
\end{figure}

Magnetic-field geometry and inhomogeneity play a central role in how energy and momentum are transported in strongly magnetized plasmas, shaping turbulence and reconnection processes that drive the irreversible conversion of mechanical energy  into heat.  In many settings,  outcomes hinge on  global features such as curvature, stratification, and differential rotation: for example, pulsar properties depend sensitively on the global magnetospheric structure \cite{Spitkovsky2006}; 
galactic-disk evolution is likely influenced by magnetic-buoyancy instabilities controlled by stratification and field-line geometry \cite{Parker1966,Shu1974}; 
and the open-magnetic-flux geometry of stars sets wind torques \cite{Weber1967,Mestel1968}.  These examples motivate theoretical frameworks for studying astrophysical plasmas that treat geometry as a first-class dynamical ingredient rather than a perturbation around idealized configurations.

The solar corona and solar wind provide an unusually rich testbed for such  ideas: the wind is a canonical example of a magnetically guided outflow \cite{Parker1958}, while the corona exhibits ubiquitous wave activity, turbulence, and extreme spatial structuring over a vast range of scales.  Despite decades of work and its relative proximity to Earth, which have yielded increasingly stringent observational constraints, the physical origin of coronal heating and the mechanisms that set the fast/slow solar-wind dichotomy remain open \cite{Cranmer2019,Viall2020}.

A broad range of coronal-heating scenarios have been proposed and remain actively debated \cite{Klimchuk2015,DeMoortel2015,Reale2014,Cranmer2019}.  At a  coarse level, much of the community discussion is  organized around two limiting pictures: wave/turbulence-driven (WTD) dissipation, frequently emphasized in open-field regions  \cite{Cranmer2009}, and impulsive reconnection/loop-opening (RLO) energy release, commonly invoked in the lower corona \cite{Parker1988}.  This broad division is motivated in part by the dramatic changes in structural complexity between regions of the wind and corona: varying observations reveal order-unity density contrasts on tens-of-Mm scales \cite{DeForest2007}, fluctuation/turbulence amplitudes that vary strongly across neighbouring regions \cite{Tomczyk2007}, and narrow layers separating plasma with distinct composition and thermodynamics \cite{Chitta2020}.  
Regions with sharp variation in the magnetic field and its connectivity have been the focus of  RLO-based ideas, 
while the smoother field and vigorous fluctuations at higher altitudes  point to the dominance of WTD processes \cite{Chen2016}.


WTD models have been successful at reproducing key aspects of the solar wind and coronal structure, including aspects of the fast-/slow-wind dichotomy \cite{McComas2000,Cranmer2007,Cranmer2019}. 
A basic ingredient of most models is the partial \emph{reflection} of Alfv\'enic fluctuations by parallel gradients of the Alfv\'en speed, $\bh\cdot\nabla \va$, which converts outward-propagating waves into a counterpropagating component thereby driving a turbulent cascade \cite{Hollweg1986,Velli1989,Matthaeus1999}. 
Although the quantitative sufficiency of the reflection-driven cascade remains debated --- some direct simulations and  observational analyses suggest that, even in coronal holes, additional ingredients may be required \cite[e.g.,][]{Verdini2009,vanBallegooijen2016,Shoda2019,Chandran2019,Sioulas2025} --- it remains a natural baseline mechanism for open-field wave dissipation. 
Because $\bh\cdot\nabla \va$ is strongly influenced by flux-tube expansion, reflection-based WTD models naturally predict a dependence of wind properties on the magnetic expansion factor, a feature that is well supported observationally and used in current space-weather models \cite{Wang1990,Arge2000,MacNeice2018,Elliott2022}. 
At the same time, as noted above, this ``expansion/reflection'' ingredient represents only one aspect of coronal magnetic geometry, with the corona exhibiting many additional geometric features that are not treated systematically in standard flux-tube WTD formulations. These provide further channels for reflection, compressive-mode coupling, dissipation, and heating, raising the possibility that at least part of the low-coronal heating usually attributed to reconnection (RLO) instead reflects missing geometric physics in wave-driven models \cite{Downs2016,Morton2023}.

This motivates the core purpose of the present work.  We derive a new {multiscale reduced-MHD (RMHD) transport theory} and turbulence phenomenology for wave-driven processes in  magnetized plasmas with \emph{arbitrary magnetic geometry}: curved and non-radial fields, strong transverse structuring, and  varying field-line mapping (``$\mathcal{Q}$ factors'') are all incorporated within a single controlled expansion.  The result is a coupled set of equations for the fluctuations and the slow-timescale evolution of the background, with energetic consistency enforced by construction. This enables the systematic categorization of \emph{all}  geometry-mediated effects on wave/turbulence dissipation (within the ordering of the theory) --- reflection, mode coupling, heating, and cross-field transport channels --- without switching approximations between special cases. 
While various related geometric ingredients have been explored previously in particular settings --- e.g., resonant absorption \cite{Lee1986,Davila1987} or compressive coupling \cite{Magyar2019} due to perpendicular structure, curvature-induced coupling \cite{Southwood1985,Similon1989}, and phase mixing \cite{Heyvaerts1983} --- these effects have typically been treated in a piecemeal fashion rather than within a single energy-consistent transport framework \cite{VanDoorsselaere2020,Morton2023}. 

The multiscale strategy adopted here parallels modern gyrokinetic transport theory, where controlled asymptotics yield reduced fluctuation equations coupled self-consistently to transport-scale profile evolution \cite{Callen2010,Barnes2010,Abel2013}.  In fusion plasmas, this approach is not merely formal: large-scale gyrokinetic simulation and transport modeling have become central tools for predicting confinement and informing reactor design and operational optimization \cite{Jenko2000,Candy2009,Citrin2017}.  The present work brings a fluid-based version of the same multiscale transport viewpoint to heliophysics, further developing the theory into turbulence closures that make phenomenological predictions. With a general setting and the simplifications of a fluid model (compared to gyrokinetics), 
it also helps clarify aspects of how free energy is moved and shared between the background and fluctuations; these simplifications may also prove helpful in understanding turbulence in fusion devices.

Our resulting transport system is best viewed as a unification of numerous modeling ingredients, some of which have previously appeared in the coronal-heating and solar-wind literature in partially separated forms.   First, it contains the standard reflection-driven WTD backbone: outward Alfv\'enic fluctuations are partially reflected by large-scale {parallel} inhomogeneity, generating counterpropagating Elsasser fields that can sustain a turbulent cascade \cite{Hollweg1986,Matthaeus1999,Cranmer2019}.  Second, it provides a systematic route by which \emph{transverse} structuring drives WTD heating: the strong cross-field gradients and distorted field-line mapping that motivate Quasi-Separatrix Layer (QSL)/S-web perspectives --- including the ``squashing factor'' $\mathcal{Q}$ \cite{Titov2002,Antiochos2011} --- appear here  as explicit geometric drivers of reflection, coupling, and cross-field transport.  Third, the  equations contain large-scale shear/inhomogeneity driving terms similar in spirit to those used in ``Mixing Expansion Compression Shear'' (MECS)-style heliospheric closures \cite{Zank1996,Oughton2011,Breech2008,Usmanov2018,Zank2017}, but with coefficients and mechanisms fixed by the multiscale expansion rather than introduced phenomenologically.  Fourth, because the derivation retains fully general curved and non-radial geometry, it introduces novel heating mechanisms associated with field-line curvature and effective perpendicular forces (via solar gravity), opening channels for Alfv\'enic-compressive feedback and dissipation that are not represented in standard flux-tube formulations.  
Finally, a general prediction is that large perpendicular gradients drive \emph{cross-field transport} of density, composition, heat, and/or momentum \cite{Magyar2019}. Various interesting consequences arise; for instance, that some wind streams could be predominantly fuelled  via plasma transported via turbulence  from closed-field regions, as opposed to via the coronal base or loop opening. 

A key point is that all of these ingredients are obtained within one controlled approximation and are tied together by an exact global energy balance at the order retained, ensuring that disparate geometric effects are incorporated consistently rather than case-by-case. In \cref{fig: overview}, we provide an overview of various specific processes predicted to be  of relevance to the solar corona and wind.

The results of the theory have direct implications for the {global} organization of the solar wind into fast and slow streams.  A robust expectation from wind theory is that enhanced heating deposited  {below} the sonic point primarily increases the mass flux and tends to produce a {slower}, denser wind, whereas heating  higher up more efficiently increases the wind's terminal speed \cite{Leer1980,Hansteen1995,Cranmer2009,Chandran2021}.  Since many of the additional terms in our framework scale with transverse gradients, curvature, and non-radiality, it follows that structured regions (e.g., near coronal-hole boundaries and in connectivity-complex environments) are natural sites for stronger low-altitude wave-driven heating, while smooth, unstructured coronal-hole cores at larger 
altitudes mostly  reduce to the ``baseline'' expansion/reflection picture already included in standard WTD models \cite{Cranmer2007,vanderHolst2014,Mikic2018,Cranmer2019}.  This provides a concrete route to understand the physics behind empirical space-weather predictors such as  Wang--Sheeley--Arge (WSA) models \cite{Wang1990,Arge2000}, which supplement the correlation of expansion factor and wind speed with a further correlation of slower wind with angular distance to the nearest coronal-hole boundary \cite{Owens2005,MacNeice2018,Elliott2022}: in our framework, that distance acts as a proxy for the same geometric ingredients (squashing, transverse gradients, and field-line curvature/non-radiality) that all enhance low-altitude heating and thus produce slower wind.

For future interest, the slow-timescale transport equations also provide a natural strategy for global heliospheric modeling and, potentially, space-weather prediction.  In addition to furnishing local heating and transport rates, they contain a wave-driven Parker-like wind as part of a single coupled system: the large-scale outflow and thermodynamics can be evolved self-consistently with the fluctuation energetics, with heating and transport entering as explicit source terms. 
Structurally, this places the framework close to existing global wave-driven MHD models, which already evolve a large-scale outflow coupled to outward and reflected Alfv\'enic wave fields \cite{vanderHolst2014,Mikic2018,Sokolov2021,Parenti2022,Reville2022}, or to turbulence-transport closures that evolve turbulence energy, cross helicity, and correlation length in a Reynolds-averaged sense \cite{Zank2017,Usmanov2018,Chhiber2021}. The distinction here is how the system is derived: rather than appending a separately motivated turbulence or transport model to  MHD-like fluid equations, the fluctuation dynamics and the slow-timescale background evolution arise together from a single multiscale expansion. The resulting system shares much of the field-aligned wave-transport backbone of existing models, but additionally retains new geometry-mediated couplings --- perpendicular transport and new heating channels --- that are omitted in treatments derived from wave dynamics in locally one-dimensional flux tubes.  
Again, the same multiscale modeling ideas, with similar equations, have been used with substantial success for designing and understanding fusion devices \cite[e.g.,][]{Barnes2010,Citrin2017,Citrin2022}.

Finally, although our emphasis here is on the solar corona and wind, the formal structure we derive and discuss is broader: it provides a general description of low-frequency, strong-guide-field turbulence and wave-driven transport in arbitrary geometry.  It thus has clear application to other magnetized outflows and atmospheres --- for instance,  terrestrial, pulsar, or black-hole magnetospheres, accretion disks \cite{Kawazura2022}, or galactic winds and outflows.

\subsection{Outline of this paper}
Given the breadth of the multiscale framework and the number of distinct geometry-mediated effects it predicts, this paper is necessarily longer than is typical.  We have therefore endeavoured to make the key ideas and results of each part as self contained as possible, so that readers interested primarily in (for example) the phenomenology or empirical estimates need not follow every step of the derivation.  Two tables summarize notation and definitions for the main theory (\cref{tab: notation}) and phenomenology \cref{tab: notation_phenom})  to support this modular presentation.

The core theoretical development is outlined in \cref{sec:curved_RMHD}, where we present the multiscale transport theory itself.  For clarity, the detailed mathematics of the derivation is placed in  App.~\ref{app: derivation}, which is fully self contained and can be read without reference to the main text for readers uninterested in heliospherical/astrophysical applications.  The main text focuses on (i) stating the resulting equations and their structure, and (ii) providing physical interpretation of the couplings that emerge (e.g., how field geometry controls wave dynamics and couplings, and the limits that recover familiar wave-transport pictures).  The outcome is a coupled set of equations: reduced-MHD-like evolution equations for the fluctuations (\cref{sub: rmhd equations}), and transport-scale equations, driven by quadratic products of fluctuations,  for the slow evolution of the background (\cref{sub: transport in main text}). Some limitations of the theory are discussed in  \cref{subsec:limitations}.

In \cref{sec: phenomenology} we use the transport theory to develop a ``slaved-field'' phenomenology that estimates the size and qualitative impact of the new effects without requiring full numerical solutions of the fluctuation equations.  This section introduces the closure strategy, which is effectively  an extension of the standard reflection phenomenology of \citet{Dmitruk2002} to 
include the new geometric ingredients from the multiscale theory, treating backwards-propagating Alfv\'enic and compressive 
fluctuations on the same footing. Its predictions
are heating and transport rates that depend on the outward Alfv\'enic wave energy ($W^{+}_{\perp}$) and background geometrical coefficients.  

We then apply these ideas in \cref{sec: empirical} to obtain conservative empirical estimates in a minimal coronal model. The goal is not to provide a definitive calculation but instead  identify where and when the novel effects could plausibly matter.  Even under solar-minimum conditions with a simple prescribed magnetic field \cite{Banaszkiewicz1998}, the estimates suggest that multiple new mechanisms (summarised in \cref{fig: overview}) can be comparable to standard reflection-driven heating in different regions, motivating further observational,  theoretical, and numerical study.

Following the core theory in App.~\ref{app: derivation}, additional appendices cover various secondary issues and results. App.~\ref{app:parker_like_wind} discusses some subtleties in how the framework may be used as a global transport model for a Parker-like wind. App.~\ref{app: lowbeta_transonic} shows, via a different asymptotic expansion, that multiscale RMHD remains a correct model even for transonic fluctuations that 
are larger amplitude than formally allowed by the standard ordering. App.~\ref{app: instabilities} enumerates the wide range of standard instabilities contained within the multiscale RMHD system,  illustrating its potential utility  beyond the heliospheric context.  Finally, App.~\ref{app: gyrokinetics}, demonstrates how multiscale RMHD is recovered from  multiscale gyrokinetics \cite{Abel2013}, clarifying the relation of this work with fusion theories.

\begin{table}[!htbp] 
\caption{Common notations used in this work. Bold symbols denote vectors; subscripts $\perp$ and $\|$ indicate components perpendicular and parallel to $\bh$.}
\begin{ruledtabular}
\begin{tabular}{ll}
\textrm{Symbol} & \textrm{Definition} \\
\colrule
$\bm{r},\,R$ & Position vector, (spherical) radial coordinate \\
$\rho$ & Mass density \\
$\bm{u}$, $\bm{U}$ & Total velocity, mean flow; $\bm{U}=U\,\bh$ \\
$\bm{B}$, $B$ & Magnetic field, $B=|\bm{B}|$ \\
$p$, $T$ & Thermal pressure, temperature ($p=\rho {\rm R} T$) \\
$E_{\rm th}$ & Mean thermal energy density, $E_{\rm th}=p/(\gamma-1)$ \\
$s$ & Specific entropy, $s-s_0 = c_v \ln(p/\rho^\gamma)$ \\
$\sigma$ & Mass-weighted entropy, $\sigma=\rho s$ \\
$\Phi_{\rm grav}$ & Gravitational potential (assumed constant)\\
$\bm{\Omega}$ & Angular velocity of rotating frame \\
$\Phi_{\rm rot}$ & Effective centrifugal potential $\Phi_{\mathrm{rot}} =  -\left|\bm{\Omega}\times\bm r\right|^{2}/2$\\
$\Phi_{\rm tot}$ & $\Phi_{\rm tot}\equiv \Phi_{\rm grav} + \Phi_{\rm rot}$\\
$\bm{\Pi}$, $\eta$, & Viscous stress, resistivity \\
$\bm{q}$ & Heat flux \\
$S_{G}$ & External source of $G$ (heat, momentum, or mass)\\
\colrule
$\bh$ & Unit vector along mean field, $\bh=\bm{B}/B$ \\
$\ell$ & Arc-length coordinate along $\bh$; $\partial G/\partial \ell=\bh\cdot\nabla G$ \\
\colrule
$\gamma$ & Adiabatic index \\
$\rm R$ & Gas constant per unit mass, ${\rm R}=k_B/\bar{m}$ \\
$c_v$ & Specific heat at constant volume, $c_v = {\rm R}/(\gamma-1)$ \\
\colrule
$\va$ & Alfv\'en speed, $\va = B/\sqrt{4\pi\rho}$ \\
$\cs$ & Sound speed, $\cs^2 = \gamma p/\rho$ \\
$\vs$ & Slow-magnetosonic speed, $\vs^2 = \va^2/(1+\va^2/\cs^2)$ \\
$\beta$ & Plasma beta, $\beta = 8\pi p/B^2 = (2/\gamma)\,\cs^2/\va^2$ \\
$\ma$ & Alfv\'en Mach number $\ma\equiv U/\va$\\
\colrule
$\epsilon$ & Small expansion parameter, $\epsilon\sim k_\|/k_\perp\sim \delta \bm B/B$ \\
$l_{\|},l_{\perp}$ & Turbulence outer-scale correlation lengths \\
$k_\|,k_\perp$ & Turbulence wavenumbers ($k_{\|,\perp}\sim l_{\|,\perp}^{-1}$) \\
\colrule
$\gprl{G}$ & Parallel inverse  scale height $\gprl{G}\equiv\gradl \ln G$ \\
$\gprp{G}$ & Perpendicular inverse  scale length
$\gprp{G}\equiv\nabla_\perp\!\ln G$ \\
$\bm{\kappa}$, $\kappa$ & Field-line curvature, $\bm{\kappa}=\bh\!\cdot\!\nabla\bh$, $\kappa=|\bm{\kappa}|$ \\
$(\nabla\bh)_\perp$ & Perpendicular submatrix of $\nabla\bh$; Eq.~\eqref{eq: bh decomposition} \\
$\mathsf{I}_\perp$ & Identity tensor on the perpendicular subspace \\
$\tsmat$ & Symmetric traceless ``squeezing'' tensor in $(\nabla\bh)_\perp$ \\
$\mathsf{A}$ & Antisymmetric ``twist'' tensor in $(\nabla\bh)_\perp$ \\
$\bm{g}_{\rm eff}^{\perp,\|}$ & Effective gravities (including acceleration terms) \\
\colrule
$\langle\cdot\rangle$ & Turbulence average $\langle\cdot\rangle=\langle \avgp{\cdot}\rangle_{t}$ \\
$\avgp{\cdot}$, $\langle\cdot\rangle_t$ & Perp.~spatial and intermediate-time averages \\
$\delta g$ & First-order fluctuation of quantity $g$ ($\langle\delta g\rangle=0$) \\
$\duprp$, $\duprl$ & Perpendicular and parallel fluctuating velocities \\
$\dbprp$, $\dbprl$ & Perpendicular and parallel magnetic fluctuations \\
$\Phi$, $\Psi$ & Alfv\'enic potentials for $\duprp$, $\dbprp$ \\
$\bm{z}^\pm$ & Els\"asser fields, 
$\bm{z}^\pm = \duprp \mp \dbprp/\sqrt{4\pi\rho}$ \\
$\zlpm$ & Slow-mode Els\"asser variables,
$\zlpm = \duprl \mp \dVprlx$ \\
$\dVprlx$ & Slow-mode eigenfield, $\dVprlx = (\va^2/\vs)\,\dbprl/B$ \\
\colrule
$\mathcal{D}_{g}$ & Generic small-scale dissipation acting on  $g$  \\
$W_\perp^\pm$ & Alfv\'enic free energies, $W_\perp^\pm = \rho\langle {|\bm{z}^\pm|^2}\rangle/4$ \\
$W_\|^\pm$ & Slow-mode free energies, $W_\|^\pm = \rho\langle {|\zlpm|^2}\rangle/4$ \\
$W_s$ & Entropy-mode free energy \\
$W^{\rm tot}$ & Total fluctuating free energy \\
$W^{\perp}_u,\; W^{\perp}_B$ 
  & Alfvénic kinetic and magnetic energies    \\
$W^{\|}_u,\; W^{\|}_B$ 
  & Parallel kinetic and magnetic energies \\
$W^{\perp}_r,\; W^{\|}_r$
  & Residual energies, 
    $W^{\perp,\|}_r \equiv W^{\perp,\|}_u - W^{\perp,\|}_B$\\
    $p^{(2)}$, $p^{(2)}_{\rm tot}$ & Fluctuation contribution to thermal/total pressure \\
        \end{tabular}
\end{ruledtabular}\label{tab: notation}
\end{table}

\section{Multiscale MHD transport in general geometry}\label{sec:curved_RMHD}

This section provides a  brief overview of the core theory --- ``Multiscale Reduced MHD'' --- that provides the basis of our phenomenological results. 
The framework is based on ``multiscale gyrokinetics'' \cite{Callen2010,Barnes2010,Abel2013}, which has become a powerful workhorse in understanding magnetic-confinement fusion 
experiments. Our version, tailored to astrophysical situations, is in some ways  simpler than fusion methods, while being more complex in others: it is simpler because we start from  MHD, as opposed to the Vlasov--Maxwell set of equations, 
and are not concerned with finite-Larmor-radius effects; it is more complex because we allow background 
gradients  of all quantities in all directions, while  fusion-transport  theories restrict to closed flux surfaces, leading to  plasma 
quantities being constant along the field lines to leading order.  

The theory proceeds by postulating the 
 gyrokinetic/RMHD ordering \cite{Frieman1982,Schekochihin2009} and 
suitably defined averaging operators to expand the MHD equations. Each expansion order provides a different physical 
content: at first order, equilibrium relations and perpendicular pressure balance; at second order, equations for turbulent fluctuations, which are the generalization 
of standard RMHD to an arbitrary background geometry; and at third order, the so-called `transport equations', which describe the
effect of the turbulence on the slow evolution of the background density, momentum, magnetic field, and temperature.
Any energy lost/gained by the background on the slow transport timescale is gained/lost by the turbulent fluctuations through driving, dissipation, or fluxes, and we derive a global energy conservation law showing this explicitly. 

Taken together, these yield a  framework for understanding heating and transport induced
by waves launched from the solar surface into a background with any of (i) a general ({curved}, twisted, and sheared) mean magnetic field $\bm B=B\bh(\bm r)$, (ii) arbitrarily stratified density and temperature profiles, and (iii) a field-aligned trans-Alfv\'enic background flow $\bm U=U\bh$.
In the fluctuation equations, the theory captures WKB wave evolution and ``non-WKB''  reflection induced by flux-tube expansion, as studied in many previous works \cite{Velli1989,Cranmer2007,Chandran2009,Zank2017,Wang2022}, 
while also revealing a myriad of other, less-studied effects due to propagation through perpendicular gradients, curved magnetic fields, and gravitational fields.
In the transport equations, the theory captures the turbulent-dissipation-enhanced Parker-like wind acceleration along each 
field line, including modifications from wave-pressure gradients and other stresses \cite{Jacques1978,Cranmer2007} (some subtleties are discussed in App.~\ref{app:parker_like_wind}), as well as how perpendicular (cross-field) fluxes of mass, momentum, and energy  drive a slow reorganization of the background perpendicular structure.

The full derivation is long and technical, and is therefore given separately, in self-contained form, in App.~\ref{app: derivation}.
Here, we present the key results and equations, endeavouring to 
provide intuitive understanding without requiring the reader to follow the full derivation. 
We also  provide extended commentary on the physical content of different terms and effects, 
which will prove important for building a heating and transport phenomenology in \cref{sec: phenomenology}.

\subsection{Setup, notation, and splitting of mean and fluctuating components}
\label{subsec:fullMHD}

Our starting point is the set of  compressible magnetohydrodynamic (MHD) equations in 
a slowly and uniformly rotating frame of reference:
\begin{flalign}
&\frac{\partial \rho}{\partial t}  + \nabla\!\cdot\!\left(\rho\bm u\right)  =  0, \label{eq: rho}\\[0.25em]
&\rho\left(\frac{\partial \bm u}{\partial t}  + \bm u\cdot\nabla \bm u + 2\,\bm\Omega\times\bm u\right)
 = 
-\,\nabla \left(p + \frac{B^2}{8\pi} \right) \nn \\
&\qquad\qquad \qquad\qquad\quad+  \frac{\bm B\!\cdot\!\nabla\bm B}{4\pi}
 -  \rho\,\nabla \Phi_{\rm tot}
  +  \nabla\!\cdot\!\bm{\Pi}, 
 \label{eq: u}\\[0.35em]
&\frac{\partial \bm B}{\partial t} = \nabla\times\left(\bm u\times\bm B\right) 
 -  \nabla\times\left(\eta\,\nabla\times \bm B\right), \label{eq: B}\\[0.25em]
&\frac{\partial }{\partial t}\!\left(\frac{p}{\gamma-1}\right)
 +  \nabla\!\cdot\!\left(\frac{p}{\gamma-1}\,\bm u\right)
 = 
-\,p\,\nabla\cdot\bm u
 -  \nabla\cdot\bm q +  \Sth \nn\\
& \qquad\qquad\qquad \qquad\quad-  \bm{\Pi}:\nabla\bm u
 +  \frac{\eta}{4\pi}|\nabla\times\bm B|^{2}. \label{eq: p}
\end{flalign}
The notation is standard and listed in \cref{tab: notation}.  Via the definition of the  specific entropy $s=s_{0}+c_{v}\ln (p/\rho^{\gamma})$, where $c_{v}$ the heat capacity at constant volume and $s_{0}$ a reference entropy,
\cref{eq: p} can be written in the equivalent form
\begin{flalign}
&\rho T\left(\frac{\partial s}{\partial t}+\bm{u}\cdot\nabla s\right)
 = 
-\,\nabla\cdot \bm q + \Sth\nn\\
& \qquad\qquad\qquad \qquad\quad -  \bm\Pi:\nabla \bm u
 +  \frac{\eta}{4\pi}\left|\nabla\times \bm B\right|^{2} ,\label{eq: s}
\end{flalign}
which is used interchangeably with \eqref{eq: p}. The total effective potential $\Phi_{\rm tot}$ incorporates gravity and centrifugal effects (see \cref{eq: centrifugal}), and $\Sth$ is an external heat source or cooling function. Mass and momentum sources ($\Srho$ and $\Smv$, respectively) are omitted here for simplicity because they are not needed for later estimates, though they are included in the full derivation of App.~\ref{app: derivation}. The form of the dissipation, represented by the general viscous stress tensor $\bm{\Pi}$, the resistivity $\eta$, and the heat flux $\bm{q}$, is general and could incorporate kinetic effects if desired.
We use the sub- or super-scripts ${\perp}$ and $\|$ to 
denote quantities perpendicular and parallel to the mean magnetic field, respectively. Within our ordering, perpendicular fluctuations are
always Alfv\'enic and incompressible (see below), while  parallel fluctuations are compressive and polarized like  slow-magnetosonic modes (fast modes
are ordered out, as in standard RMHD).

\begin{figure}
\begin{center}
\includegraphics[width=1.0\columnwidth]{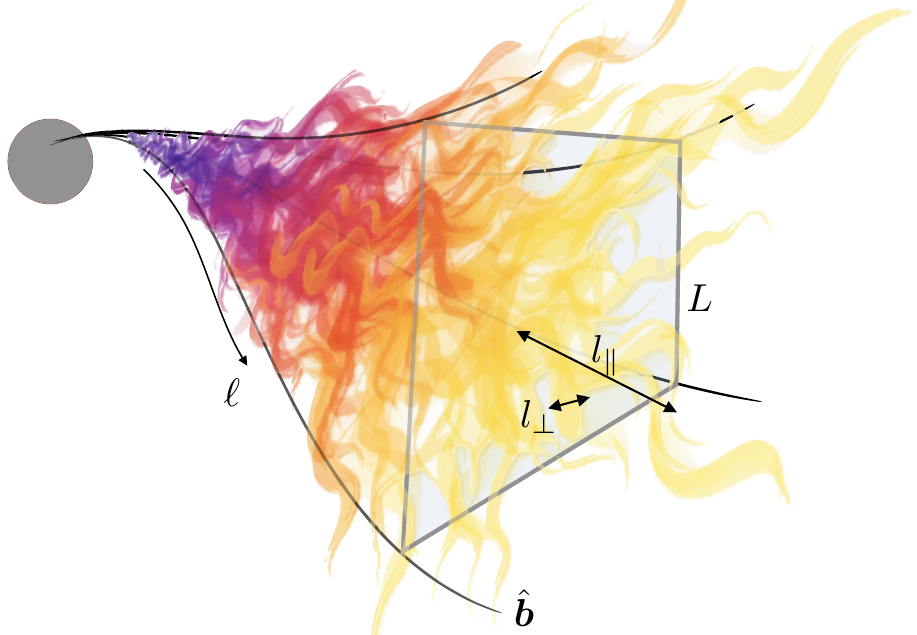}
\caption{Sketch of the expansion and averaging scheme. Fluctuations vary rapidly perpendicular to the background magnetic field $\bh$, while their variation in the parallel direction is on the same scale as the background. The turbulent average considers a perpendicular patch of size $L$ with $l_{\perp}\ll L \lesssim l_{\|}$. This implies that an average of a
perpendicular divergence is $\sim\! \epsilon$ smaller than the product of its parts, while the average itself commutes with the
parallel and large-scale gradient operators. The coordinate labelling distance along the field line is denoted by $\ell$}
\label{fig: average}
\end{center}
\end{figure}

We treat the turbulent fields as living on two widely separated sets of scales: fast, small-scale fluctuations, and a slowly varying, large-scale background. Formally, this separation is implemented by expanding every quantity \(g\) in powers of a small parameter \(\epsilon\) and, at each order, splitting it into its mean and fluctuating parts, viz.,
\begin{equation}
g = \sum_{n\ge0}\epsilon^n g^{(n)}, \: 
g^{(n)} = \langle g^{(n)} \rangle + \delta g^{(n)}, \: \langle \delta g^{(n)} \rangle = 0.
\end{equation}
Trading a slight notational ambiguity for simplicity, we  represent  background quantities $\langle {g^{(0)}}\rangle$ by their bare symbols $g$ (so, e.g., $\rho = \langle {\rho^{(0)}}\rangle$, $\bm B = \langle {\bm B^{(0)}}\rangle$), and use the unadorned symbol for first order quantities, $\delta g\equiv \d g^{(1)}$. 

The ``turbulent average'' \(\langle\cdot\rangle\) is defined as a perpendicular spatial average ($\avgi{\cdot}_\perp$) over a patch that is large compared to the fluctuation scale $\sim\!l_{\perp}$, composed with an intermediate-time average ($\avgi{\cdot}_t$) over a time that is long compared to the turbulent correlation time $\sim\!\omega$. By construction, this average removes all fast, small-scale structure and leaves only the slowly varying background, likewise having the property that it commutes with large-scale and slow-timescale derivatives. Within the ordering, any gradient of a mean or averaged quantity is \(\sim\!\epsilon\) smaller than a perpendicular gradient of a fluctuation, and  once a product of fluctuations such as \(\langle \delta \rho \,\duprp\rangle\) has been averaged, it behaves as a large-scale field. Note that the turbulent average does not operate in the direction parallel
to the background magnetic field, and there is no scale separation in this direction in keeping with the usual RMHD anisotropy $k_{\|}\sim \epsilon k_{\perp}$ (equivalently, $l_{\perp}\sim \epsilon l_{\|}$). This represents a  point of difference with fusion-transport theories, where the flux-surface average effectively operates as a
parallel average, annihilating parallel transport terms. 

Multiple-time-scale analysis  makes the separation between fast fluctuations and slow background evolution precise. We introduce two formal times, a fast time \(t\) associated with the wave/turbulent dynamics on the Alfv\'enic timescale ($\mathcal{O}(\epsilon)$ compared to perpendicular compressions) and a slow time \(\tau\) associated with transport, enacted by making the replacement
\begin{equation}
 \frac{\partial}{\partial t} \;\rightarrow\; \epsilon \frac{\partial}{\partial t} + \epsilon^3 \frac{\partial}{\partial \tau}.
\end{equation}
The \(\epsilon^3\) scaling for $\partial/\partial \tau$ reflects that transport-time dynamics are driven by inhomogeneous sources built from quadratic correlations of \(O(\epsilon)\) fields. This implies that the fast-time equations are solved on a fixed background; any averaged field depends only on $\tau$  and is therefore constant on the fast timescale. A quasi-stationary turbulent state --- an assumption of the theory from the outset --- then exists only if there are no secular (growing) terms in the fast-time solution. Multiple-time-scale analysis enforces this by requiring that the intermediate-time average of any fast time derivative vanishes. In practice, this means that fast derivatives of second-order mean quantities are traded for slow evolution of the zeroth-order background,  viz., for some field $g$, one identifies the fast \(\mathcal{O}(\epsilon)\)  change of  $\avgi{g^{(2)}}$ with the slow \(\mathcal{O}(\epsilon^3)\) change of the background $\avgi{g^{(0)}}$.

\subsubsection{The ordering parameter}

The expansion procedure yields equations describing how turbulence in any local patch depends on gradients of the slowly varying background. We thus introduce 
the following compact notation to capture the (inverse) scale length of a background $\mathcal{O}(\epsilon^{0})$ quantity $G$: 
\begin{equation}
\gprl{G} \equiv \bh\cdot\nabla\ln G, \quad
\gprp{G} \equiv \nabla_{\perp}\ln G.\label{eq: K defs main}
\end{equation}
The perpendicular  gradient operator $\nabla_\perp$ is defined by 
$\nabla_{\perp} G \equiv \nabla G- \bh\, \gradl G$
 (note that $\bh$ is the direction of the background field $\bm{B}=\langle \bm B^{(0)}\rangle$, and does not include contributions from perturbations).
The ordering parameter is then the usual anisotropic RMHD ordering parameter, with $\epsilon\ll1$ and 
\begin{align}
\epsilon &\sim\frac{\omega}{k_{\perp}\va} \sim \frac{k_{\|}}{k_{\perp}}\sim \frac{\delta \rho}{\rho}\sim \frac{\delta p}{p}\sim \frac{\delta s}{c_{v}}\sim\frac{|\delta \bm{B}|}{B}\sim \frac{|\delta \bm{u}|}{\va}\nn \\
&\sim  \frac{\kappa}{k_{\perp}}\sim \frac{\gprl{G}}{k_{\perp}} \sim  \frac{\gprps{G}}{k_{\perp}},
\label{eq: ordering}
\end{align}
where $\gprps{G}= |\gprp{G}|$ and $\bkap\equiv\bh\cdot\nabla\bh$ denotes the field-line curvature with $\kappa=|\bkap|$.
Under these assumptions, the parallel wavelength of fluctuations 
can be comparable to  background variation of any quantity ($\mathcal{K} \sim \kappa \sim k_{\|}$);  but --- in line with the
scale separation assumptions discussed above ---  fluctuations have smaller scales perpendicular to the magnetic field and small amplitude.
The equations we derive are thus reasonable for outer-scale fluctuations lower in the corona inside the
Alfv\'en point \citep{Cranmer2017}, becoming at best qualitatively valid at larger radii where $|\delta \bm{B}|\sim B$.

The relations \eqref{eq: ordering} are supplemented with the ordering $U\sim \va\sim\cs$ (respectively, the mean flow, Alfv\'en speed, and thermal sound speed), 
and the assumption that, to lowest order, the mean plasma flow aligns with the magnetic field $\bm{U}=U\bh$. The present theory could be generalized 
to non-field-aligned flows without field-aligned gradients (as for toroidal rotation in the fusion context), but given our intended application to 
the sub-Alfv\'en-radius corona, this is not our priority here. 
The rotation rate $\Omega$ is ordered comparable to the (inverse) transport timescale, $\Omega \sim \epsilon^{3}k_{\perp}\va$, in keeping with the idea that the rotation and field-reorganization times are similar in the solar context.


\paragraph*{A note on the subsonic ordering.}
Although \cref{eq: ordering} formally assumes subsonic fluctuations $|\d\bm u|/\cs\ll1$, this restriction can often be relaxed. This is relevant because many low-$\beta$ regions of the corona --- most notably coronal holes and related open-field structures at lower altitudes --- can have outward Alfv\'enic fluctuations with amplitudes $|\d \bm{u}|\sim \cs$ while still satisfying $|\delta \bm B|/B\ll1$ and $\delta \rho/\rho\ll1$. To justify the application to such regions, we show in App.~\ref{app: lowbeta_transonic} that there exists a novel distinct ordering, assuming low-$\beta$ and strongly dominant transonic outwards waves, that leads (at lowest order) to the \emph{same} fluctuation equations as the appropriate low-$\beta$ limit of the present system. Thus, while the strict derivation below uses the ordering \eqref{eq: ordering}, the resulting  equations  remain applicable more broadly than that formal ordering might at first suggest. Intuitively, this is expected because finite-amplitude Alfv\'enic fluctuations that maintain constant magnetic-field strength (spherical polarization) are nonlinear solutions on a constant background; they are therefore ignorant of the thermal sector and thus the sound speed \cite{Barnes1974,Hollweg1974}.

\subsection{The generalized RMHD equations} \label{sub: rmhd equations}

\subsubsection{Equilibrium}
At $\mathcal{O}(\epsilon)$ the average of the MHD system yields the following equilibrium relations: 
\begin{gather}
\rho\bm{U}\!\cdot\!\nabla\bm{U}= -\nabla \left(p+\frac{B^{2}}{8\pi}\right)+ \frac{\bm{B}\!\cdot\!\nabla\bm{B}}{4\pi} - \rho \nabla\Phi_{\rm grav},  \nn \\[0.5em]
\nabla\!\cdot(\rho \bm{U}) = 0,  \:\:
\rho T \bm{U}\!\cdot\!\nabla s = \Sth-\nabla\cdot \bm{q}.
\label{eq: equilibrium}
\end{gather}
Using $\nabla\cdot\bh=-B^{-1}\gradl{B} = -\gprl{B}$, which comes from  $\nabla\cdot(B\bh)=0$,  the continuity equation yields  the standard mass and flux conservation relation, $\gradl(\rho U/B)=0$, or \begin{equation}
\gprl{\rho}+\gprl{U}-\gprl{B}=0.\label{eq: mass conservation}
\end{equation}
Projection of the momentum equation perpendicular and parallel to $\bh$ yields 
\begin{equation}
\va^{2}(\bkap-\gprp{B})- \frac{{c}_{s}^{2}}{\gamma}\gprp{p} =  U^{2}\bkap + \nabla_{\perp}\Phi_{\rm grav}\equiv -\geff\label{eq: perp equil}
\end{equation}
and 
\begin{equation}
-\frac{{c}_{s}^{2}}{\gamma}\gprl{p}  = U^{2}\gprl{U}+\bh\cdot\nabla\Phi_{\rm grav}  \equiv -{g}^{\|}_{\rm eff},\label{eq: parallel equil}
\end{equation}
respectively.  We have written the right-hand sides in terms of the ``effective gravitational acceleration,'' whose perpendicular and parallel components are $\geff$ and  ${g}^{\|}_{\rm eff}$.
In the perpendicular direction, $\geff$ includes the local centrifugal acceleration due to the plasma's flow $\bm{U}=U\bh$ along  curved magnetic-field 
lines; in the parallel direction, ${g}^{\|}_{\rm eff}$ captures  the plasma's linear acceleration as well as the true gravity. 

The mean entropy equation is
\begin{align}
 \frac{p U}{\gamma-1}\gprl{s}=\Sth-\nabla\cdot\bm{q},
\label{eq:entropy-mean-main}
\end{align}
showing how $\gprl{s} = \gprl{p}-\gamma\gprl{\rho}$ quantifies the deviation from adiabatic evolution due to nonideal effects and sources (a mean heat flux $\bm{q}$ is included because it is more often relevant to the low-order equilibrium than viscous or resistive stresses).

\subsubsection{Constraints on the fluctuating variables} 
All vector perturbations are decomposed into perpendicular and parallel components, which are treated separately:
\begin{equation}
\duprp = (\mathsf{I} - \bh\bh)\cdot\d\bm{u},
\quad 
\duprl = \bh\cdot\d\bm{u},
\end{equation}
(where $\mathsf{I}$ is the identity) and similarly for $\bm B$.  Terms containing $\bh\cdot\nabla$ or $\nabla\cdot\bh$ acquire an extra $\epsilon$ relative to perpendicular derivatives; for instance, the latter two terms on the right-hand side of the equation
$$\nabla\cdot\delta \bm{u} = \nabla\cdot\duprp + \bh\cdot\nabla\duprl + \duprl\,\nabla\cdot\bh,$$ are one order in $\epsilon$ smaller than the first term. Consequently, at $\mathcal{O}(\epsilon)$,
the fluctuating parts of the continuity equation and $\nabla\cdot\bm B=0$ yield $\nabla\cdot\duprp = 0$ and $\nabla\cdot\dbprp = 0$. This
implies the equations can be written in potential form with
\begin{equation}
\duprp = \bh\times \nabla \Phi,\qquad \frac{\dbprp}{\sqrt{4\pi \rho}} = \bh\times \nabla\Psi,\label{eq: phi psi def}
\end{equation}
a clearly superior choice for numerical simulations, but one which hides the system's physical content. 
The fluctuating part of the momentum equation at $\mathcal{O}(\epsilon)$ yields perpendicular pressure balance
\begin{equation}
\frac{\delta p}{p}
=
-\gamma\frac{ \va^{2}}{\cs^{2}}
\frac{\delta B_{\parallel}}{B},
\label{eq:perp-press-balance}
\end{equation}
representing the slow-mode balance imposed by the time-scale ordering. 

\subsubsection{Consistency of ignoring the first-order background}
An average of the equations expanded to \(O(\epsilon^2)\) shows that all first-order mean fields are homogeneous and can be consistently absorbed into the lowest-order background. In the continuity, momentum, induction, and entropy equations, every quadratic fluctuation contribution at \(O(\epsilon^2)\) appears only as a total perpendicular divergence or curl, which becomes \(O(\epsilon^3)\) once the turbulent average is taken because large-scale gradients are \(O(\epsilon)\). As a result, the \(O(\epsilon^2)\) averaged system contains no fluctuation-driven sources for \(\langle \rho^{(1)} \rangle\), \(\langle \bm{u}^{(1)} \rangle\), \(\langle \bm{B}^{(1)} \rangle\), or \(\langle s^{(1)} \rangle\). Choosing these first-order means to vanish is therefore a consistent initialization: they remain zero on transport timescales, and the first nontrivial influence of the turbulence on the mean fields enters only at \(O(\epsilon^3)\).

\subsubsection{Alfvénic dynamics}

As in standard RMHD \cite{Schekochihin2009}, dynamical equations for the perpendicular (Alfvénic) fluctuations are obtained  by expanding the fluctuating part 
of the momentum and induction equations to $O(\epsilon^{2})$,  then applying the perpendicular projection operator. Written in terms of 
the   Els\"asser fields, $\zpm\equiv \delta\bm{u}_{\perp}\mp\delta\bm{B}_{\perp}/\sqrt{4\pi\rho}$, which are defined here so that $\zp$ propagates in the $+\bh$ direction, we find,
\begin{align}
&\frac{\partial \zpm}{\partial t} + (U\pm\va)(\bh\cdot\nabla + \bh\,\bkap\cdot)\zpm =
-\zmp\cdot\nabla\zpm - \frac{\nabla_{\perp} \tilde{p}}{\rho}\nn\\ 
&\quad- (U\mp\va)  \left[\zmp\cdot\nabla\bh -  \frac{1}{4} \gprl{\rho}(\zpm-\zmp)\right] \nn\\
 &\quad-2\bkap \left(U\duprl - \frac{B\dbprl}{4\pi\rho}\right) +\frac{ \d\rho}{\rho} \geff + \bm{\mathcal{{D}}}^{\perp}_{\pm},\label{eq: Alfvenic fluctuation equations main}
\end{align}
where $\bm{\mathcal{{D}}}^{\perp}_{\pm}$ represents the small-scale dissipative effects that act on $\zpm$.
The left-hand side comprises linear propagation effects at speed $U\pm \va$; the term $\bh\,\bkap\cdot\zpm$ is a geometric projection that enforces $\zpm\cdot\bh=0$ along curved field lines, which can be seen by dotting \eqref{eq: Alfvenic fluctuation equations main} with $\bh$, moving the $\bh$ through the $\bh\cdot \nabla$ operator (likewise, this term vanishes in the potential formulation; see  \cref{eq: potential form}).   The first  terms on the right-hand side, which take the same form regardless of geometry, are the familiar nonlinear advection and pressure, with $\tilde{p}$ chosen to enforce $\nabla\cdot\zpm=0$. The second line contains WKB growth and a generalized reflection term, which is discussed below (\S\ref{sub: RMHD structure}). The curvature- and gravity-driven couplings in the third line are novel to the present generalization --- they act as mutual sources/sinks between Alfvénic and compressive modes and vanish for a straight, homogeneous background for which  $\bkap=\gprp{G}=0$ for all background quantities $G$. 

\paragraph*{Relation to the classical flux-tube reduction.}  
Equations for fluctuations propagating in an inhomogeneous environment are well studied in the context of solar-wind modeling, with a long history  \citep[e.g.,][]{Whang1980,Velli1989,Zhou1990}. Standard derivations  \cite{Velli1993,Cranmer2005,Chandran2009,David2025} often simplify the system by introducing a spherical coordinate system $(R,\theta,\phi)$ whose polar axis coincides with the reference field line, then assuming all background quantities vary {only} along $R$.  The cross-sectional area $a(R)$ of that flux tube expands as $a\propto B^{-1}$, ensuring $\nabla\cdot\bm{B}=0$, and $B$, $\rho$ and $U$ can vary as arbitrary functions of $R$ subject to  mass conservation  \eqref{eq: mass conservation}. Assuming that the
fluctuations on top of this background, $\d\bm{u}$ and $\d\bm{B}$, are incompressible and perpendicular to $\bh$, one obtains (in the form given by \citet{David2025} equation 2.13 noting their $H_{\rho}^{-1}= -\gprl{\rho}$ and $H_{\rm A}^{-1}= \gprl{\va}$):
   \begin{align}
\frac{\partial \zpm}{\partial t} 
+ &(U \pm \va) \bh \cdot \nabla \zpm =-\zmp \cdot \nabla \zpm - \frac{\nabla_{\perp} \tilde{p}}{\rho} \nn\\ &
+ ({U \mp \va})\left( \frac{\gprl{\rho}}{4}\zpm+ \frac{\gprl{\va}}{2}\zmp \right) .
\label{eq: straight tube}\end{align}
Comparing to   \cref{eq: Alfvenic fluctuation equations main}, we see that our general version, which assumes nothing about the field-line geometry or symmetry, involves a similar set of terms other than the additional effects from curvature and gravity. We show below in \cref{sub: q decomposition} how the WKB and reflection terms in \cref{eq: Alfvenic fluctuation equations main} reduce to those in \eqref{eq: straight tube}.

\subsubsection{Compressive dynamics} 

Unlike the straight-field-line system \eqref{eq: straight tube}, Eqs.~\eqref{eq: Alfvenic fluctuation equations main} are not closed. We therefore need
equations for the compressive fluctuations, $\duprl$, $\dbprl$, and $\d s$. These are obtained from the momentum, induction, continuity, and entropy equations at $O(\epsilon^{2})$. For physical clarity in identifying slow-mode dynamics, we recast the compressive sector in the variables $\duprl$ and \begin{equation}
\dVprlx \equiv\dbprl \frac{\sqrt{1+\va^{2}/\cs^{2}}}{\sqrt{4\pi\rho}}=\frac{\dbprl}{B}\frac{\va^{2}}{\vs},
\end{equation}
 where $\vs=\va/\sqrt{1+\va^{2}/\cs^{2}}$ is the slow-magnetosonic speed. These combine into the slow-mode eigenfields $z_{\|}^{\pm}\equiv\duprl \mp \dVprlx$, which diagonalize the homogeneous linear dynamics and propagate along $\bh$ at $\vs$. The relation 
\begin{equation}
\frac{\d s}{c_{v}} = \frac{\d p}{p} - \gamma \frac{\d \rho }{\rho} = - \gamma \left(\frac{\vs}{\cs^{2}}\dVprlx +\frac{\d \rho }{\rho}\right)\label{eq:entropy relations}
\end{equation}
and perpendicular pressure balance \eqref{eq:perp-press-balance} close the system, allowing for equivalent formulations in terms of any two of the five variables $\d s$, $\d \rho$, $\d p$, $\dbprl$, or $\dVprlx$. 
A version that  clearly highlights the mode structure is
\begin{align}
&\frac{\rmd\duprl}{\rmd t} - \vs\bhdg\dVprlx =  \mathcal{D}^{\|}_{u}  \nonumber\\ 
&\qquad-\frac{\va}{2}\sum_{\pm} \zpm\cdot \left[\gprp{U} \ma \pm \gprp{B} - \bkap (\ma\pm1)\right]\nn\\& \qquad- U\duprl \gprl{U} + \vs\dVprlx (\gprl{\vs}+\gprl{\rho})+\frac{\d\rho}{\rho} g^{\|}_{\rm eff} ,\label{eq: parallel mom alfv}
\\[0.5em]
&\frac{\rmd\dVprlx}{\rmd t} - \vs \bhdg\duprl =   \mathcal{D}^{\|}_{\tilde{V}}
\nn\\ &\quad-\frac{\vs}{2}\sum_{\pm} \zpm\!\cdot\! \left[\gprp{B}\pm\gprp{U} \ma  \pm \bkap (\ma\pm1) - \frac{\gprp{p}}{\gamma}\right]\nn\\&\quad -\vs\left(\gprl{B} - \frac{\gprl{p}}{\gamma} \right)\!\duprl 
+  U\left(\gprl{\vs} - \frac{\vs^{2}}{\cs^{2}}\gprl{s}\right)\!\dVprlx, \label{eq: parallel induction alfv}
\\[0.5em]
&\frac{\rmd}{\rmd t} \frac{\d s}{c_{v}}=\mathcal{D}_{s}-\frac{1}{2}\sum_{\pm} \zpm\cdot\gprp{s}  -\left(\duprl  +  U \frac{\d p}{p}\right)\gprl{s},\label{eq: entropy}
\end{align}
where 
\begin{align}
  & \frac{\rmd}{\rmd t}  =\frac{\partial}{\partial t}+\duprp\cdot\nabla+U\bh\cdot\nabla,\nn\\
   & \bhdg  =\bh\cdot\nabla + \frac{\dbprp}{B}\cdot\nabla
\end{align}  are the convective derivative and the parallel derivative along the total (equilibrium plus perturbed) field line, respectively, and $\ma\equiv U/\va$ is the Alfv\'enic Mach number. The $\sum_{\pm}$ indicates the sum over $+$ and $-$ variables, and (as above) the $\mathcal{D}$ represent small-scale dissipative processes. Equivalent formulations in terms of $\dbprl/B$, $\d \rho/\rho$, and $\zlpm$ are given in App.~\ref{sub: compressive flucts app}.

\begin{figure*}
\begin{center}
\includegraphics[width=1.0\textwidth]{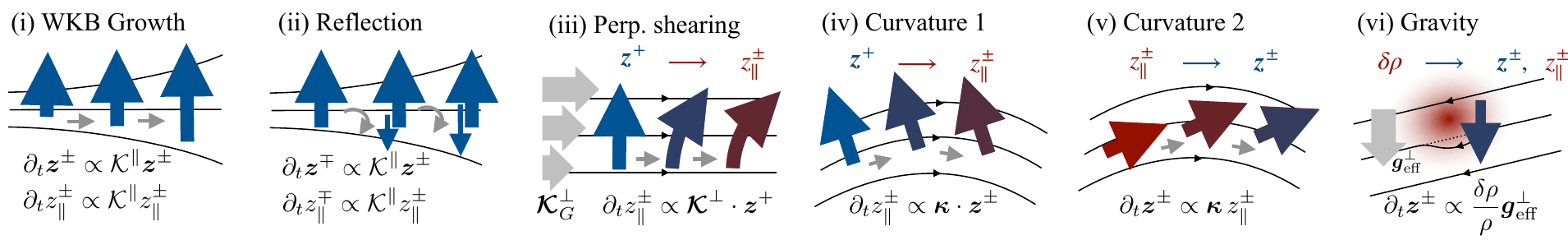}
\caption{The geometrical effects on propagating fluctuations contained in \eqref{eq: Alfvenic fluctuation equations main}, \eqref{eq: parallel mom alfv}, \eqref{eq: parallel induction alfv}, and \eqref{eq: entropy}. Blue and red lines illustrate outwards Alfv\'enic ($\zp$) and 
general compressive perturbations (represented as $\zlpm$), respectively, and thin grey arrows indicate heuristically the direction of wave propagation. Colors between the two extremes (e.g., purple) show
how geometrical effects can change the direction of a fluctuating field from 
perpendicular to parallel, and vice versa, as it propagates along field lines; such effects manifest mathematically as source terms in the relevant equations. Panels (i) and (ii) sketch how parallel gradients associated with field-line expansion drive WKB growth/decay of waves and reflection, respectively (while the diagram illustrates $\zpm$ in blue, these processes operate for slow modes also; see \cref{eq: slow mode zlpm}). Panel (iii) shows  how a perpendicular 
gradient in $\bm{U}$, $\rho$ or $|\bm{B}|$ (illustrated here by the grey arrows showing faster propagation at the top of the domain) shears  a perpendicular $\zpm$ fluctuation to create parallel component ($\zlpm$), as in the second lines of Eqs.~\eqref{eq: parallel mom alfv}--\eqref{eq: entropy}. Panels (iv) and (v) show how propagation 
in a curved field likewise causes Alfv\'enic-compressive coupling: because the background field changes direction along the propagation direction, an initially  perpendicular fluctuation yields a  parallel part as it propagates, and vice versa (mathematically, the $\bkap$-proportional terms in \eqref{eq: Alfvenic fluctuation equations main}, \eqref{eq: parallel mom alfv}, and \eqref{eq: parallel induction alfv}). Panel (vi) shows how, in the presence of gravity, a sinking overdensity will bend the field to generate Alfv\'enic fluctuations $\zpm$ (the $\geff$ proportional term in  \cref{eq: Alfvenic fluctuation equations main}). \ }
\label{fig: cartoons}
\end{center}
\end{figure*}

The left-hand sides of \cref{eq: parallel mom alfv,eq: parallel induction alfv} describe slow-wave propagation at $U\pm\vs$. When these equations are combined into the slow-mode eigenfields $\zlpm$ (\cref{eq: slow mode zlpm}), their parallel-gradient terms encode the WKB amplification and reflection of slow waves, in close analogy with the $\gprl{}$ terms in the Alfvénic sector.  The terms 
proportional to $\zpm$ in each equation encode the  generation of compressive fluctuations from transverse Alfvénic motions, a process related to resonant absorption \cite{Lee1986,Morton2023} or ``uniturbulence'' \cite{Magyar2019a,Magyar2019}.
We also see that parallel gravitational forces  ($g_{\rm eff}^{\|}$) drive $\duprl$, and thus slow waves, from entropy fluctuations (via $\d\rho$), while slow waves  likewise drive entropy fluctuations by tapping into parallel background entropy gradients. This implies that, except in specific circumstances (e.g., without additional heating/heat fluxes so that $\gprl{s}=0$), slow and entropy modes become linearly coupled even in the 
absence of $\zpm$.

\subsubsection{Structure of the generalized RMHD system}\label{sub: RMHD structure}

Before developing and applying a turbulent-heating phenomenology, it is helpful to interpret the physical roles of the $\nabla\bh$, curvature, and gradient terms, comparing couplings with their homogeneous counterparts and relating our system to the standard flux-tube form \eqref{eq: straight tube}.

The generalized RMHD system preserves the familiar terms from homogeneous RMHD, with slow 
waves and entropy modes nonlinearly advected only by Alfv\'enic fluctuations with no nonlinear feedback on the Alfv\'enic fluctuations \cite{Lithwick2001,Schekochihin2009}. However, background field geometry and stratification introduce new linear couplings. The system's key ingredients  can be summarised as:
\begin{description}
  \item[{Wave propagation}] each field is advected along $\bh$ at its characteristic speed: $(U\!\pm\!\va)$ for Alfvén modes,  $(U\!\pm\!v_{S})$ for slow modes, and $U$ for entropy modes.
  \item[{Parallel–gradient couplings}] factors such as $\gprl{B}$, $\gprl{U}$, and $\gprl{\rho}$ multiply the \emph{same-sign} field (e.g., $\zpm$ in $\partial\zpm/\partial t$) to yield WKB growth/decay, and the \emph{opposite-sign} field (e.g., $\zmp$ in $\partial\zpm/\partial t$) to induce non-WKB reflection. The latter converts outward- to inward-propagating energy. If desired, the WKB term can be eliminated by absorbing it into the
  field itself via an integrating factor, yielding the \emph{wave-action} form \cite{Heinemann1980,Chandran2009} (see \cref{subsec:RDT_basic}). (Eqs.~\eqref{eq: parallel mom alfv}--\eqref{eq: parallel induction alfv} are expressed in $\duprl$ and $\dVprlx$, hiding the separation of  WKB-amplification and reflection terms;  alternative forms isolating these terms are given in \cref{eq: slow mode zlpm,eq: f for slow modes} and yield the same effects for slow modes).
  \item[{Perpendicular/curvature couplings}] terms involving $\zpm\cdot\gprp{}$ or $\zpm\cdot\bkap$ let transverse Alfvénic motions generate parallel slow- or entropy-mode fluctuations.  The reverse transfer --- compressive feedback onto $\zpm$ --- does not occur via perpendicular gradients alone but is enabled in curved fields through $2\bkap(U\duprl-B\dbprl/4\pi\rho)$. 
  \item[{Buoyancy from stratification}] if field lines include a component perpendicular to gravity, or if $U^{2}\bkap\neq0$,  $\geff$ will be nonzero, and density perturbations   generate  Alfv\'enic $\zpm$ fluctuations. This is effectively a buoyancy feedback, the generalization of standard hydrodynamic gravity waves.
\end{description}
The physical reasons for these differing forms of feedback --- e.g., the importance of $\bkap$ and $\geff$ specifically for driving $\zpm$ --- can be
understood via geometrical arguments. These are illustrated graphically in \cref{fig: cartoons}.

As expected, for some parameter combinations, these additional couplings yield linear instabilities.
We show in App.~\ref{app: instabilities} how this recovers  well-known MHD instabilities,  as well as various generalizations, with 
little algebraic effort.

Note that our equations do not contain standard parametric decay, whereby a large-amplitude outwards propagating wave $\zp$ grows a linear 
instability that drives compressive fluctuations and a backwards propagating $\zm$. This coupling is related to the $\gprl{\rho}$- and $\gprl{B}$-dependent reflection terms 
in \cref{eq: Alfvenic fluctuation equations main} --- parallel variation of compressive quantities couples forward and backwards waves --- 
but the true instability would require higher-order terms in the expansion.
Similarly, the phenomenological model of \citet{vanBallegooijen2016}, where smaller-scale parallel variations from density   
fluctuations enhance reflection, could  be included via $\gprl{\rho}$; in the present model, true reflection of $\zp$ into $\zm$
via $\d\rho$ is formally $\mathcal{O}(\epsilon^{3})$.

\subsubsection{Geometric decomposition of Alfv\'enic reflection}\label{sub: q decomposition}

\begin{figure*}
\begin{center}
\includegraphics[width=1.0\textwidth]{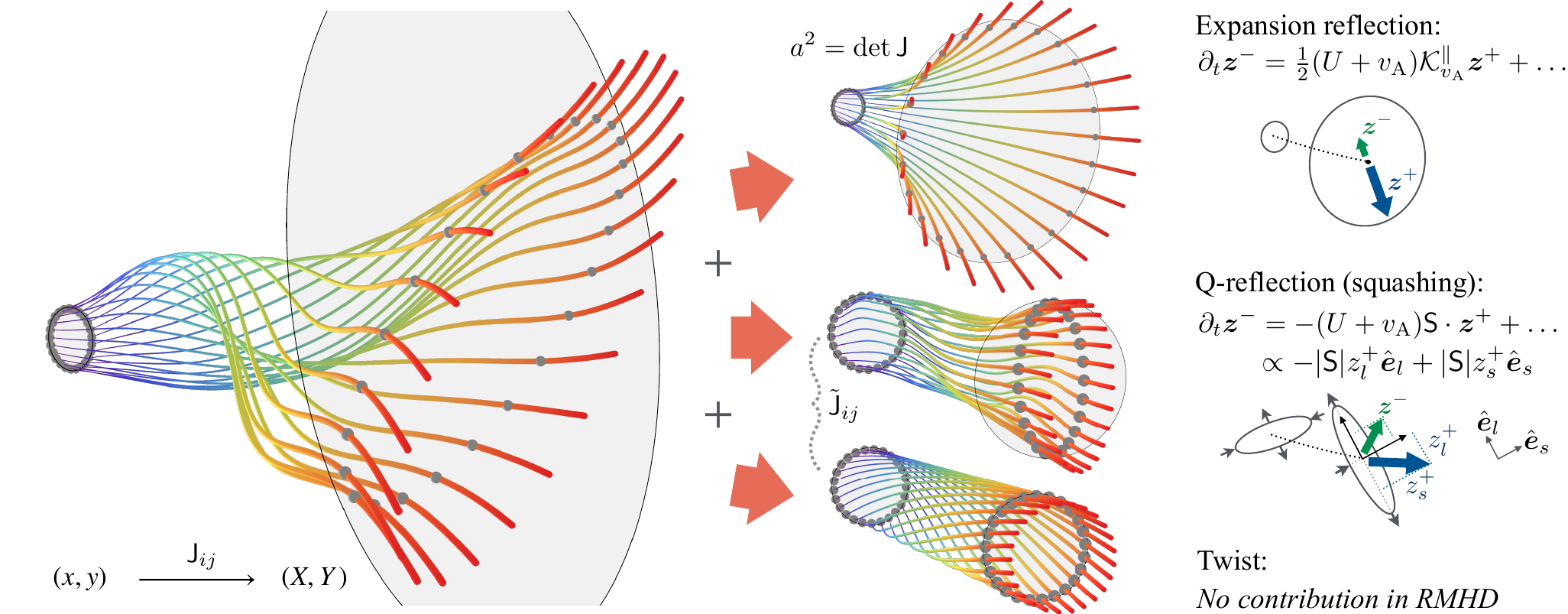}
\caption{Illustration of the field-line geometrical decomposition used to understand Alfv\'enic reflection (\cref{sub: q decomposition}). Any flux tube/field configuration, illustrated on the left and described by the field line mapping Jacobian $\mathsf{J}_{ij}$ (see \cref{{subsec:Q_reflection}}), can be decomposed into 
its curvature (not shown) and  perpendicular  $(\nabla\bh)_{\perp}$ tensor, the latter of which can reflect $\zp$ into $\zm$ waves. $(\nabla\bh)_{\perp}$ is further decomposed into its expansion ($\nabla\cdot\bh$), ``squashing'' ($\tsmat$), and ``twist'' ($\mathsf{A}$) parts (see \cref{eq: bh decomposition}), as illustrated in the middle group; these  
account for flux-tube area changes ($a^{2}={\rm det}\,\mathsf{J}$), volume-preserving ellipse transformations, and rotations ($\nabla\times \bh$), respectively. Expansion, the key component 
in ``standard'' phenomenological reflection-heating  theories, generally drives a $\zm$ antiparallel to $\zp$ (right side), while
``Q-reflection'' via squashing (a novel effect considered here), forces $\zm$ in the expansion/contraction directions of the ellipse with opposite signs, thus driving $\zm$ at a general angle compared to $\zp$. Twist has no contribution in the RMHD ordering because it forces a perpendicularly compressive $\zm$ from the $\zp$, which is rapidly eliminated by perpendicular pressure balance.     }
\label{fig: grad b}
\end{center}
\end{figure*}

The  term $\zmp\!\cdot\!\nabla\bh$ in \eqref{eq: Alfvenic fluctuation equations main} embodies the most general form of Alfvénic reflection. While 
this term has been written similarly in many previous works \cite{Zhou1990,Velli1993}, as far as we are aware its effects have only been considered
for a flux tube that possesses axisymmetry about $\bh$, when it reduces to a ``standard'' reflection form $\zmp \nabla\cdot\bh/2$ (see below). Here, we detail its more general effects, which have interesting implications for coronal heating.

 Because $\zpm\cdot\bh=0$,  only the {perpendicular} part of the matrix $\nabla\bh$ enters (likewise, since $\nabla\bh\cdot\bh=0$, $\zmp\!\cdot\!\nabla\bh$ contributes only to the Alfv\'enic sector).  We then define a locally field-aligned orthonormal basis $(\hat{\perpunitvectorbase}_{1},\hat{\perpunitvectorbase}_{2}, \bh)$ and write, without loss of generality,
\begin{equation}
(\nabla\bh)_{\perp}= \frac12(\nabla\cdot\bh)\,\mathsf I_\perp
		+\underbrace{\tsmat}_{\text{sym.\,traceless}}
                 +\underbrace{\mathsf{A}}_{\text{antisym.}}, \label{eq: bh decomposition}
\end{equation}
where $\mathsf I_\perp$ is the $2\times2$ identity in the perpendicular subspace,  the traceless symmetric matrix
$\tsmat$ describes {squashing} of the flux tube into an ellipse, and the (traceless) antisymmetric matrix $\mathsf{A}=\bh\!\cdot\!\nabla\times\bh\;
           (\hat{\perpunitvectorbase}_{1}\hat{\perpunitvectorbase}_{2}-\hat{\perpunitvectorbase}_{2}\hat{\perpunitvectorbase}_{1})/2,$ encodes the twist of the field line.
Their effects are illustrated graphically in \cref{fig: grad b} and can be understood as follows:
\begin{enumerate}
\item The isotropic piece, $(\nabla\!\cdot\!\bh)\mathsf I_\perp/2$, reproduces the familiar “magnetic-expansion’’ reflection: using $\nabla\cdot\bh=-\gprl{B}$  this term combines with  $\zmp \gprl{\rho}/4$  in \cref{eq: Alfvenic fluctuation equations main} to yield the reflection term
 $-(U\mp\va) \gprl{\va} \zmp/2$. This matches the standard form \eqref{eq: straight tube}, as appropriate under the assumption of a cylindrically symmetric flux tube with $\tsmat=\mathsf A=0$.
\item The antisymmetric ``twist'' part does not contribute at RMHD order. This can be proven by taking its curl, which shows that
$\zmp\!\cdot\!\mathsf{A}$ is a gradient of a scalar potential \footnote{The proof is straightforward using the potential form 
  for $\zpm$. Compute
$\nabla\times(\zp\cdot\mathsf{A}) = \nabla\times(\bh\times \nabla\zeta^{+}\cdot\mathsf{A})$, which gives $-\bh \,\mathsf{A}:\nabla\nabla\zeta^{+}$ to lowest order. 
The $\nabla\nabla$ operator is symmetric, showing that $\mathsf{A}:\nabla\nabla\zeta^{+}=0$, and thus that  $\zp\cdot\mathcal{A}$ is a purely compressive perpendicular  forcing on $\zm$. }. Any such gradient contribution is automatically  cancelled by the perpendicular pressure force that enforces $\nabla\cdot\boldsymbol{z}^{\pm}=0$ and so cannot drive significant reflection (equivalently, the term 
will not appear in the potential form of the equations \eqref{eq: potential form}).
\item The symmetric‐traceless tensor $\tsmat$ survives and represents reflection caused by area-preserving {elliptical deformations} of the flux surface.  By rotating into its local eigenbasis, $\hat{\perpunitvectorbase}_{l}$ and $\hat{\perpunitvectorbase}_{s}$, which are mutually perpendicular directions lying along the long and short axis of the ellipse, respectively, $\tsmat$ becomes $\tsmat=\mathrm{diag}(|\tsmat|,-|\tsmat|)$, where $|\tsmat|=({\tsmat_{ij}\tsmat_{ij}}/2)^{1/2}$ is the rate at which the long axis becomes longer (i.e., $\pm |\tsmat|$ are the eigenvalues of $\tsmat$).  Then 
$\zmp\cdot\tsmat= |\tsmat| z^{\mp}_{l}\,\hat{\perpunitvectorbase}_{l}- |\tsmat| z^{\mp}_{s}\,\hat{\perpunitvectorbase}_{s}$, i.e., the long-axis component of $\zpm$ reflects with  the   opposite sign to the short-axis component.  Although this will change the correlation of the forced $\zm$ with the driving $\zp$ compared to conventional RDT \cite{Perez2021,Meyrand2025,Bowen2025}, it is still expected to  inject energy into $\zmp$ at a rate comparable to the isotropic expansion term if $|\tsmat|$ and $\nabla\cdot\bh$ are comparable. 
\end{enumerate}

Collecting the non-vanishing pieces, the net Alfv\'enic reflection forcing  reads
\begin{equation}
\frac{\partial \zmp}{\partial t} =\dots + (U\pm\va)\left(\frac12\,\gprl{\va}\,\zpm-\zpm\cdot\tsmat\right)+\dots,
\label{eq:general_reflection_force}
\end{equation}
which will generally act as the primary driver of $\zm$ in the solar context (see \cref{sec: phenomenology}).  The first term is standard non-WKB reflection, while the second represents reflection by flux-surface squashing --- absent in straight-tube models but potentially dominant in regions where field lines become tangled.  A quantitative estimate of its contribution to heating, in particular the relation to the measured ``squashing factor'' $\mathcal{Q}$ from PFSS models, is deferred to \cref{sec: phenomenology}.

\subsubsection{Fluctuation energy in multiscale RMHD}

Dotting Eqs.~\eqref{eq: Alfvenic fluctuation equations main}, \eqref{eq: parallel mom alfv}, \eqref{eq: parallel induction alfv}, and \eqref{eq: entropy} with 
$ \zpm$, $\duprl$, $\dVprlx$, and $\d s$, respectively, then taking the turbulent average, one finds the   the generalized RMHD 
quadratic free-energy invariant. This is most naturally written in terms of the  Alfvénic, slow-mode, and entropy-mode  energies:
\begin{align}
&W^{\pm}_{\perp} \equiv \frac{1}{4}\rho \avg{|\zpm|^{2}},\qquad
W^{\pm}_{\|} \equiv \frac{1}{4}\rho \avg{|\zlpm|^{2}},\nn\\ 
&W_{s} \equiv \frac{1}{2}\frac{p}{\gamma(\gamma-1)}\avg{\frac{\d s^{2}}{c_{v}^{2}}},\label{eq: rmhd energies}
\end{align}
respectively.
Their sum
\begin{equation}
W^{\rm tot} \equiv W^{+}_{\perp} + W^{-}_{\perp} + W^{+}_{\|} + W^{-}_{\|} + W_{s}
\end{equation}
is the fluctuation's contribution to the second-order perturbation of the total MHD energy density,
\begin{equation}
E \equiv \frac12 \rho u^2 
+ \frac{B^2}{8\pi} + E_{\rm th} +  \rho\,\Phi_{\rm tot},
\end{equation}
where $E_{\rm th}={p}/(\gamma-1)$.
The compressive pieces combine so that \(W^{\|}_{+}+W^{\|}_{-}+W_{s}\) contains both the quadratic correction to \(E_{\rm th}\) and the kinetic and magnetic energies of slow-mode fluctuations. 

The evolution of \(W^{\rm tot}\) obeys a conservation law,
\begin{equation}
\nabla\cdot\left[\bh \sum_{\rm wave} (U + v_{\rm wave}) W_{\rm wave}\right]
= \mathcal{Y}^{\perp} + \mathcal{Y}^{\|} + D^{\rm tot},\label{eq: energy conservation}
\end{equation}
where \(v_{\rm wave} = \{\pm\va,\pm\vs,0\}\) for \(W_{\rm wave} = \{W^{\pm}_{\perp},W^{\pm}_{\|},W_{s}\}\), and \(D^{\rm tot}\) is the positive-definite function encoding the fluctuation's dissipation \eqref{eq:Dtot second order}. Note that the (fast) time derivative 
has disappeared from \eqref{eq: energy conservation} due to the application of the intermediate-time average as part of the turbulent average, encoding physically the idea that 
fluctuations  sit in quasi-steady balance between propagation, forcing, and dissipation.

The perpendicular source \(\mathcal{Y}^{\perp}\) can be written as a sum of five contributions,
\begin{equation}
\mathcal{Y}^{\perp} = \mathcal{Y}^{\perp}_{U} + \mathcal{Y}^{\perp}_{B} + \mathcal{Y}^{\perp}_{\rho} + \mathcal{Y}^{\perp}_{\rm th} - p \UperpII\cdot\gprp{p},
\end{equation}
where $\UperpII$ is a second-order perpendicular mean flow, which is set up by the fluctuations in order to maintain perpendicular force balance (its properties are discussed in App.~\ref{sub: perpendicular flow}, but it does not feature heavily in the core theory). The individual sources are
\begin{flalign}
&\mathcal{Y}^{\perp}_{U} \equiv - \rho U^{2}\,\VU\cdot(\bkap + \gprp{U}),\nn\\
&\mathcal{Y}^{\perp}_{B} \equiv\frac{B^{2}}{4\pi}\,\Vpsit\cdot(\bkap - \gprp{B}),\nn\\
&\mathcal{Y}^{\perp}_{g} \equiv \rho\,\Vrhot\cdot\geff,\nn\\
&\mathcal{Y}^{\perp}_{\rm th} \equiv - p \avg{\frac{\d \rho}{\rho}\duprp}\cdot\gprp{\rho}
- E_{\rm th} \avg{\frac{\d T}{T}\duprp}\cdot\gprp{T},
\label{eq: Y terms}
\end{flalign}
where 
\begin{align}
& 
\Vpsi \equiv \avg{\frac{\dbprl}{B}\duprp - \duprl\frac{\dbprp}{B}},\quad\Vrho \equiv \avg{\frac{\d\rho}{\rho}\duprp}, \nn\\ &
\VU \equiv \avg{\frac{\duprl\duprp}{U} - \frac{\dbprl\dbprp}{4\pi\rho\, U}},\quad\Vp \equiv \avg{\frac{\d p}{p}\duprp}\label{eq: V defs}
\end{align}
(defining $\Vp$ for later use), and
 \begin{equation}
\Vpsit = \Vpsi + \UperpII,\quad\Vrhot = \Vrho + \UperpII, \quad \Vpt = \Vp + \UperpII.
\end{equation}
The quantities $\VU$, $\Vpsit$, 
 $\Vrhot$, and $\Vpt$ will soon be identified as the turbulent transport velocities of $\bm{U}$, $\bm{B}$, $\rho$, and $p$, respectively. Note that  adding the contributions in $\mathcal{Y}^{\perp}$ cancels $\UperpII$  
 via the equilibrium condition \eqref{eq: perp equil}, explaining its lack of appearance in the fluctuation equations (this organization of terms will be clarified below).

 The terms \(\mathcal{Y}^{\perp}_{U}\), \(\mathcal{Y}^{\perp}_{B}\), and \(\mathcal{Y}^{\perp}_{g}\) describe exchange between the fluctuations and the free energy stored in, respectively, background kinetic energy, magnetic energy, and effective gravitational potential energy, while \(\mathcal{Y}^{\perp}_{\rm th}\) describes exchange with background density and temperature gradients (equivalently, pressure and entropy gradients). The parallel contribution \(\mathcal{Y}^{\|} \) encodes the wave-mean energy exchange,  via pressure work and Reynolds--Maxwell stresses, as the waves propagate along an inhomogeneous background. It  is given explicitly in \cref{eq: parallel energy nonconservation alfvenic,eq: parallel energy nonconservation comp} (split into its Alfv\'enic $\mathcal{Y}^{\|}_{\rm A}$ and compressive $\mathcal{Y}^{\|}_{C}$ contributions), generalizing  well-known wave ``work'' terms \cite{Chandran2015,Perez2021} to compressive fluctuations and a general background.
 
 In the homogeneous RMHD limit  \cite{Schekochihin2009}, all background gradients and curvature vanish, so \(\mathcal{Y}^{\perp} = \mathcal{Y}^{\|} = 0\) and each of the five energies \(W^{\pm}_{\perp}\), \(W^{\pm}_{\|}\), and \(W_{s}\) are individually conserved, simply moving along \(\bh\) at speed \(U+v_{\rm wave}\).

\subsection{Transport-timescale evolution of the slowly varying background}\label{sub: transport in main text}

The multiple-time-scale expansion and turbulent average ensure that any quadratic product appearing inside a total divergence or curl (e.g., \(\nabla\cdot(\duprp\d\rho)\) or \(\nabla\times(\du\times\db)\)) becomes a slowly varying, large-scale quantity once averaged. The result is that on the slow time \(\tau\), the background fields \(\rho\), \(\bm B\), \(U\), and \(E_{\rm th}=p/(\gamma-1)\) evolve under the influence of fluxes and sources produced by quadratic products of the small-scale fluctuations such as \(\avgi{\d\rho\duprp}\) or \(\avgi{\duprl\dbprp}\).

To obtain such evolution equations, we start from the conservative form of the MHD equations (Eqs.~\eqref{eq: conserved MHD rho}–\eqref{eq: conserved MHD E}), expand all fields \(g=\sum_n\epsilon^n g^{(n)}\), and then apply the turbulence average \(\langle\cdot\rangle\equiv\langle\avgp{\cdot}\rangle_t\) at \(\mathcal{O}(\epsilon^3)\). Because the average commutes with large-scale derivatives, and perpendicular divergences/curls of quadratic products gain an extra order of \(\epsilon\) (see App.~\ref{para: average and derivatives}), only a second-order expansion is needed for quantities inside a divergence/curl, dramatically simplifying the algebra. The procedure gives a closed set of transport equations for density, magnetic field, parallel momentum, and thermal energy in terms of the turbulent advection velocities \(\Vrho\), \(\Vpsi\), \(\VU\) (\cref{eq: V defs}) and the energy-exchange terms \(\mathcal{Y}^{\perp}_{U}\), \(\mathcal{Y}^{\perp}_{B}\), \(\mathcal{Y}^{\perp}_{\rho}\), \(\mathcal{Y}^{\perp}_{\rm th}\) (\cref{eq: Y terms}).

Viewed this way, the third-order transport system provides a non-trivial consistency check on the whole construction: it reproduces the fluctuating free-energy conservation law from the second-order system and fixes the dissipation rate \(D^{\rm tot}\) without additional assumptions. At the same time, the appearance of the same quadratic correlators in both the transport equations and the \(\mathcal{Y}^{\perp,\parallel}\) terms gives each forcing contribution a clear physical interpretation as the relaxation of a specific background gradient (in \(U\), \(B\), \(\rho\), or \(T\)) or as adiabatic exchange with the propagating wave field, rather than as an ad hoc source. Likewise, the method guarantees total energy conservation: any energy lost by the background is gained by the fluctuations, and vice versa \cite{Wang2022}.


\paragraph{Density.}

The continuity equation yields the density transport equation
\begin{equation}
\frac{\partial \rho}{\partial \tau}
=-\nabla\cdot\left(\avg{\d\rho\,\duprp}
   +\rho\UperpII\right)
= -\nabla\cdot(\rho\Vrhot).
\label{eq:rho-transport-main}
\end{equation}
Density is thus advected across the mean field by the effective perpendicular velocity \(\Vrhot\) \eqref{eq: V defs}, which combines the fluctuation-driven transport \(\Vrho\) with the self-induced mean flow \(\UperpII\).

\paragraph{Magnetic field.}

The induction equation in conservative form yields magnetic-field transport:
\begin{equation}
\frac{\partial \bm B}{\partial \tau}
= \nabla\times(\avg{\du\times\db} +
\UperpII\times\bm B) = \nabla\times(\Vpsit\times\bm B),\label{eq:B-transport-main}
\end{equation}
which shows that magnetic flux surfaces are frozen into, and advected perpendicularly by, the effective surface velocity \(\Vpsit\) \eqref{eq: V defs}.

A  nontrivial outcome of the induction equation at transport order is that there is no turbulent magnetic diffusion: the turbulent electromotive force $\bm{\mathcal{E}}= \avgi{\du\times\db}$ satisfies \(\bh\cdot\bm{\mathcal{E}}=0\), which implies it can be written as \(\bm{\mathcal{E}} = \Vpsi\times\bm{B}\) (as above). This means that the only effect of the fluctuations is to advect magnetic flux surfaces at the effective velocity \(\Vpsit\). This is a direct consequence of the incompressibility of RMHD Alfv\'enic fluctuations, which 
pushes \(\mathcal{E}_{\|}=0\) up one order, becoming $\mathcal{O}(\epsilon^{4})$ (see App.~\ref{sub: magnetic transport}). As a result, there is no anomalous ``turbulent resistivity’’ of the kind often invoked in stochastic flux-freezing or reconnection-diffusion arguments.
In multiscale gyrokinetic theory for fusion, this same effect manifests as 
the inability of the turbulence to modify the evolution of the safety factor \citep{Abel2013}.

\paragraph{Parallel momentum.}

Projecting the momentum equation along the mean field and averaging to transport order gives
\begin{align}
\frac{\partial(\rho U)}{\partial \tau}&
= -\nabla\cdot\left[\rho U(\VU+\Vrhot)\right]+\rho U\bkap\cdot(\VU+\Vrhot)\label{eq:Upar-transport-main}\\ 
&- \bh\cdot\nabla p_{\mathrm{tot}}^{(2)} + F^{\perp}_{\rm RM}
-2\nabla\cdot(\bh W_{r}^{\|})
-\rho\bh\cdot\nabla\Phi_{\rm rot}, \nn
\end{align}
where $p^{(2)}_{\rm tot}$ is the fluctuation's contribution to the total (thermal plus magnetic) pressure \eqref{eq: p2tot}, $F^{\perp}_{\rm RM} \equiv - W_{r}^{\perp}\gprl{B}+\rho\avg{\zp\cdot\tsmat\cdot\zm}$ is
the Reynolds/Maxwell force from the Alfv\'enic fluctuations due to the expanding/shearing field geometry, 
$W_{r}^{\perp}$ and $W_{r}^{\|}$ are the perpendicular and parallel residual energies (see \cref{tab: notation}), and 
$-\rho\bh\cdot\nabla\Phi_{\rm rot}$ is the centrifugal force due to the slowly rotating frame (the latter appearing here because the inverse rotation rate is 
ordered comparable to the transport timescale, as is appropriate in the solar context).
 
The first line of \eqref{eq:Upar-transport-main} captures the perpendicular transport of parallel momentum 
 by the turbulent flux \(\rho U(\VU+\Vrhot)\)  \eqref{eq: V defs}, which, as developed in \cref{subsec: stream_dissipation}, has interesting implications for heating the solar wind.
The remaining terms represent fluctuation-induced forces from wave pressure, Reynolds/Maxwell stresses ($F^{\perp}_{\rm RM}
-2\nabla\cdot(\bh W_{r}^{\|})$), and  rotation.

\paragraph{Thermal energy.}

On transport timescales, the thermal reservoir \(E_{\rm th}=p/(\gamma-1)\) is the ultimate sink for fluctuation energy: mechanical free energy in the waves is irreversibly converted into heat via viscous and resistive dissipation, \(\avgi{-\bm{\Pi}:\nabla\bm{u}}+\eta\avgi{|\nabla\times\bm{B}|^{2}}/4\pi\), 
while thermal fluctuation energy dissipates via heat fluxes. In formulating the thermal-energy equation, we therefore have a choice: we can either keep the dissipation function \({D^{\rm tot}}\) explicit, or eliminate it in favor of the quasi-stationarity condition for the fluctuations, which enforces that their local free-energy budget is balanced by the same transport-scale heating. These two viewpoints lead to equivalent, but physically complementary, forms for \(\partial E_{\rm th}/\partial \tau\), one emphasising the fluctuation's dissipation and the other emphasising the exchange with background gradients and parallel transport of wave energy. Explicitly:
\begin{subequations}
\begin{flalign}
& \frac{\partial E_{\rm th}}{\partial \tau}  =   -\nabla\cdot\left[ E_{\rm th} \left( \avg{\frac{\d p}{p}\duprp} + \gamma \UperpII  + \bh\avg{\frac{\d s}{c_{v}}\duprl}\right)\right]   \nn\\
&\qquad +D^{\rm tot}- \mathcal{Y}^{\perp}_{\rm th} - \mathcal{Y}^{\|}_{\rm th} + p\UperpII\!\cdot\!\gprp{p} - \nabla\cdot\avgi{\bm{q}^{(2)}}  \label{eq: thermal energy 1} \\[1em]
& =  -\nabla\cdot\left[ E_{\rm th} \left( \avg{\frac{\d p}{p}\duprp} + \gamma \UperpII  + \bh\avg{\frac{\d s}{c_{v}}\duprl}\right)\right] \nn\\
&\quad+\mathcal{Y}^{\perp}_{U}+\mathcal{Y}^{\perp}_{B}+\mathcal{Y}^{\perp}_{g}- \nabla\cdot\left[\bh \sum_{\rm wave} (U + v_{\rm wave})  W_{\rm wave} \right] \nn\\ &
\quad - p^{(2)}_{\rm tot}\nabla\cdot\bm{U} - U F^{\perp}_{\rm RM} \!- 2 W_{r}^{\|}\gradl U\!-\! \nabla\cdot\avgi{\bm{q}^{(2)}},  \label{eq: thermal energy 2} 
\end{flalign}\label{eq: thermal energy} 
\end{subequations}
where 
\begin{equation}
D^{\rm tot}
=\avg{-\bm\Pi:\nabla\bm u}
+\avg{\frac{\eta}{4\pi}|\nabla\times\bm B|^{2}}
+\avg{\frac{\d T}{T}\nabla\!\cdot\!\d\bm q}
\label{eq:total_dissipation}
\end{equation}
is the dissipation rate of fluctuations matching that in \cref{eq: energy conservation}, 
the various $\mathcal{Y}^{\perp}$ are defined above \eqref{eq: Y terms}, and $\mathcal{Y}_{\rm th}^{\|}$ is a parallel source (\cref{eq: Y prl th}).

 In the first form, \cref{eq: thermal energy 1}, \(D^{\rm tot}\) plays the role of the explicit heating rate from fluctuation dissipation (viscosity, resistivity, and small-scale heat flux), with the terms \(-\mathcal{Y}^{\perp}_{\rm th}-\mathcal{Y}^{\|}_{\rm th}\) subtracting the part of the energy that was extracted from  large-scale thermal gradients to drive those same fluctuations in the first place. In other words, fluctuations powered by \(E_{\rm th}\) and then dissipated do not change the net thermal energy budget when averaged; they mediate cross-field and parallel heat transport rather than providing a new source of heat. The divergence term in \eqref{eq: thermal energy 1} collects the turbulent heat fluxes, with a perpendicular flux \(E_{\rm th}\avgi{(\d p/p)\,\duprp}\) and a (likely subdominant) parallel flux along \(\bh\)  (\(\avgi{\d s/c_{v})\,\duprl}\)), together with the mean-flow advection \(\gamma E_{\rm th}\UperpII\). The final \(-\nabla\cdot\avgi{\bm{q}^{(2)}}\) term is the perturbation to the conductive heat flux, which is also likely to be unimportant.

In the second form, \cref{eq: thermal energy 2}, we have eliminated \(D^{\rm tot}\) using the quasi-stationary free-energy balance for the fluctuations. This makes explicit how the fluctuations tap into the background mechanical- and potential-energy reservoirs: the perpendicular source terms \(\mathcal{Y}^{\perp}_{U}\), \(\mathcal{Y}^{\perp}_{B}\), and \(\mathcal{Y}^{\perp}_{g}\) describe the fluctuation-mediated release of free energy stored in gradients of \(U\), \(B\), and the effective gravitational potential. The sum over waves gives the net change in parallel wave-energy flux across the patch, \(\nabla\cdot[\bh\sum_{\rm wave}(U+v_{\rm wave})W_{\rm wave}]\), accounting for any differences in the wave energy entering and leaving the averaging volume; this accounts for any dissipation of wave flux, as in, e.g., reflection-driven turbulence. The remaining terms (involving \(p^{(2)}_{\rm tot}\), \(F^{\perp}_{\rm RM}\), and \(W_{r}^{\|}\)) represent the work done by the fluctuating Reynolds and Maxwell stresses and by the fluctuating pressure on the mean flow, as seen in the momentum transport equation \eqref{eq:Upar-transport-main}.

Deriving \cref{eq: thermal energy} in this way automatically reproduces the fluctuating free-energy conservation law obtained from the second-order system, even though the calculation of the transport system never used the second-order equations explicitly. This is a stringent internal consistency check on the expansion and on the identification of all source terms. Finally, because we started from the conservative form of the MHD equations, total energy conservation is guaranteed: the transport-scale evolution of the mean mechanical and thermal energies sums to the third-order expansion of the conserved total energy, so any energy extracted from background gradients and waves in \cref{eq: thermal energy} is exactly balanced by a corresponding loss from the mechanical/potential reservoirs and/or the parallel wave-energy flux.

\paragraph{Cross-field fluxes.}

It is natural to measure transport relative to the motion of the magnetic surfaces, i.e., in the frame moving with \(\Vpsit\). This is standard in  fusion transport theory, where it is contained within the choice to use flux coordinates that evolve together with $\bm{B}$ (see App.~\ref{sub: third order summary}; \cite{Barnes2010,Abel2013}). To this end, we define the perpendicular cross-field fluxes \(\bm\Gamma_{G}\) for \(G=\{\rho,\rho U,E_{\rm th}\}\) by subtracting \(G\Vpsit\) from the stationary-frame fluxes. This gives
\begin{subequations}
\begin{align}
\bm\Gamma_{\rho}
&=\rho(\Vrho-\Vpsi),\label{eq: cross field fluxes rho}\\[0.5em]
\bm\Gamma_{\rho U}
&=\rho U(\VU+\Vrho-\Vpsi) \label{eq: cross field fluxes U},\\
\bm\Gamma_{\rm th}
&=E_{\rm th}\left(\avg{\frac{\d p}{p}\duprp}-\gamma\Vpsi\right).\label{eq: cross field fluxes th}
\end{align}\label{eq: cross field fluxes}\end{subequations}
The corresponding source terms in this frame (containing forces and heating) are collected in Eqs.~\eqref{eq:rho-flux-rel}–\eqref{eq:Eth-flux-rel} (note 
that the $\gamma \Vpsi$ in  $\bm\Gamma_{\rm th}$ is a choice, made in order to eliminate $\UperpII$ in the flux; this changes the heating terms in \eqref{eq: thermal energy 2} via a reorganization of the divergence).

For the applications in the following section, we use these transport equations in a simplified phenomenology, replacing the detailed quadratic correlators by simple closures  in order to estimate heating rates and cross-field fluxes \(\bm\Gamma_{\rho}\), \(\bm\Gamma_{\rho U}\), and \(\bm\Gamma_{\rm th}\)   in different regions of the corona and solar wind.

\subsection{Limitations of the multiscale expansion}\label{subsec:limitations}

Our multiscale expansion is derived within compressible MHD and a small-amplitude, long-wavelength, smooth-background ordering.  In applying it to the solar wind, the main limitations fall into four main categories: (i) kinetic effects, (ii) fluctuation amplitudes,   (iii) finite-Larmor-radius effects, and (iv) large-scale structure. Here we summarize where these assumptions are most likely to fail and why the reduced model  nonetheless remains useful.

\paragraph{(i) Kinetic physics.}
At large scales and low frequencies, collisionless plasmas admit fluid-like reductions.  The CGL system \cite{Chew1956} and Kulsrud's kinetic-MHD formulation \cite{Kulsrud1983} show that  long-wavelength dynamics are close to standard MHD, with the principal new ingredients being an anisotropic pressure and kinetic (Landau) damping of compressive fluctuations.  In this regime, shear-Alfv\'en fluctuations are  robust \cite{Foote1979} and  the {Alfv\'enic} sector satisfies almost the same nonlinear equations as in MHD \cite{Schekochihin2009,Kunz2015}.  By contrast, compressive/parallel perturbations (slow-/ion-acoustic-like and entropy/pressure-balanced components) can be  strongly influenced by Landau damping and kinetic effects, especially at higher $\beta$, which can alter their polarization and effective closure \citep[e.g.,][]{Snyder1997,Kunz2015}.  This will modify the details of curvature/gravity couplings that feed compressive fluctuations back onto the Alfv\'enic equations, as well as the compressive equations themselves.  On the other hand, our formulation is deliberately agnostic about the microscopic dissipation mechanism (all sinks enter through generic dissipation operators), so collisionless damping can be interpreted as providing an effective dissipation channel without changing the large-scale structure of the transport equations.  Moreover, there is both theoretical and numerical evidence that nonlinear turbulence can partially ``fluidize'' compressive dynamics by suppressing net phase mixing via stochastic echoes \cite{Schekochihin2016,Meyrand2019}.  Finally, pressure anisotropy in the solar wind is often bounded by mirror/firehose regulation \cite{Kunz2014,Chen2016,Verscharen2017} and pressure-anisotropy forces \cite{Squire2019,Squire2023,Majeski2024}, suggesting that isotropic-MHD dynamics can sometimes serve as a useful backbone with bounded corrections.

\paragraph{(ii) Fluctuation amplitudes at larger radii.}
The expansion formally assumes $\delta u_\perp/\cs\sim \epsilon\ll 1$ and $\delta u_\perp/\va\sim \delta B_\perp/B\sim \epsilon\ll 1$.  The former 
assumption (subsonic fluctuations) is likely broken in much of the open-field corona due to the low $\beta$; however, we show 
in App.~\ref{app: lowbeta_transonic} that the same equations result from a different (highly imbalanced, low-$\beta$) ordering with $\d u\sim \cs$, justifying
their use in such regions. The latter assumption (sub-Alfv\'enic fluctuations)
fails as the wind accelerates and normalized amplitudes $\d B_\perp/B$ approach unity near or above the Alfv\'en critical region \cite{Bale2019}.  These large-amplitude states in the solar wind often evolve toward nearly constant-$|\bm B|$, highly Alfv\'enic, spherically-polarized configurations and switchbacks \cite{Squire2020}; while RMHD appears to remain a surprisingly  reasonable approximation for such Alfv\'enic fluctuations \cite{Dmitruk2005,Meyrand2025,Abbas2026}, this certainly lies outside our ordering, motivating the development of complementary large-amplitude frameworks  \cite{Barnes1974,Mallet2021,Johnston2022}.

\paragraph{(iii) Finite-Larmor-radius (FLR) effects.}
Our reduced equations are intended for inertial-range dynamics with perpendicular scales assumed to lie well above ion kinetic scales.  While this is an excellent approximation for direct dynamical effects in the  wide inertial range of the solar wind \cite[e.g.,][]{Kiyani2015,Chen2016}, the ``helicity barrier'' effect \cite{Meyrand2021} shows that FLR effects can be of crucial significance by blocking small-scale dissipation of an imbalanced cascade with $z^{+}\gg z^{-}$. This can change 
the heating efficiency at a given amplitude (potentially invalidating aspects of our closures in \cref{sec: phenomenology} below), and is crucial for understanding the relative heating of different species \cite{Squire2022,Squire2023a,Adkins2025,Johnston2025,Zhang2025}. 

\paragraph{(iv) Large-scale structure and effects.}
Our ordering assumes that all equilibrium variation (e.g., gradients of $B$, $\rho$, $p$, and the field-line curvature) occurs on scales large compared to the perpendicular fluctuation scale, so that background-gradient operators are asymptotically smaller than $l_\perp^{-1}$.  This can break down in the presence of very sharp layers (e.g., narrow current layers), where equilibrium scales approach $l_{\perp}$.  In addition, we have assumed $\bm{U}$ is strictly parallel to $\bh$; this is reasonable close to the Sun but becomes increasingly inaccurate farther out as the Parker spiral develops a substantial transverse field component. Likewise, some other important effects in the outer heliosphere, such as pick up ions \cite{Zank1996},  are not included.

\paragraph{Outlook.}
Each of the above limitations can be plausibly addressed in future work. 
The success of fusion transport theories \cite{Abel2013} provides a path to address both (i) and (iii), at the cost of complexity, by including non-trivial parallel wind structure in multiscale gyrokinetics (or various simplifications, such as kinetic MHD or low-$\beta$ models \cite{Zocco2011,Schekochihin2019,Adkins2024}). Expansions at large amplitude (ii) are less certain but plausibly tractable in the limit of high imbalance $z^{+}\gg z^{-}$ \cite{Hollweg1974,Mallet2021}. Finally, the restriction to $\bm{U}\propto \bh$ is not strictly necessary \cite{Abel2013} so it may be straightforward to address aspects of (iv).

\subsection{Summary: various use cases for multiscale RMHD}\label{sub: sec 2 summary}

The multiscale RMHD framework describes both directions of the interaction between turbulence and large-scale structure: it provides a local theory of how magnetic geometry, stratification, and flow gradients shape the turbulence, while also quantifying how such turbulence feeds back on the background through transport, wave pressure, and heating. As such, there exist a variety of use cases for the system, which we summarize here 
before exploring their consequences for coronal heating in more detail below (\cref{sec: phenomenology,sec: empirical}).   

First, one may prescribe a particular geometry --- for example, tangled fields, velocity shear, stratified patches, or curved flux tubes --- to study its influence on fluctuation dynamics. This idea is  qualitatively  highlighted in \cref{fig: cartoons,fig: grad b}, which illustrate the ways in which geometrical effects drive novel couplings between types of fluctuations. It is important to realize that, although the multiscale RMHD  equations appear complex at first glance, they represent a significant simplification of the full MHD system --- for fluctuations, all gradient 
terms ($\gprp{p}$, $\gprl{\rho}$, $\bkap$ etc.) are functions only of the coordinate along the field line $\ell$ in a local domain (i.e., they are constant in the perpendicular directions). This allows complex geometrical effects to be studied straightforwardly with periodic boundary conditions in the perpendicular directions, or, by considering a local patch in the parallel direction, simplified yet further with constant coefficients in a fully periodic domain.
This approach could be used, for example, to 
predict local correlations between Alfv\'enic and compressive motions to compare with, and understand,  \emph{in-situ} solar-wind measurements. This complements standard MHD simulations which, for similar physical situations, are much more computationally demanding and difficult to interpret. Likewise, with suitable 
kinetic extensions (see \cref{subsec:limitations} above), the system would allow detailed study of how geometry controls heating partition 
between species and directions, an approach already pursued successfully for accretion disks in \citet{Kawazura2022} (whose system is a subset of multiscale RMHD). For such studies the relevant equations are \eqref{eq: Alfvenic fluctuation equations main} for $\zpm$ coupled to \eqref{eq: parallel mom alfv}, \eqref{eq: parallel induction alfv}, and \eqref{eq: entropy} for compressive fluctuations; various equivalent forms  given in Apps.~\ref{sec:alfvenic_fluctuations} and \ref{sub: compressive flucts app} may be more convenient for computational implementations.

Second, one may  prescribe a fixed background, either idealized or realistic, and use the transport equations  diagnostically to predict heating rates and cross-field fluxes. This approach  is pursued below in the coronal context using a simple phenomenological model (\cref{sec: phenomenology}), predicting how geometry controls wave-driven heating efficiency and altitude; it  is also common in fusion studies \citep[e.g.,][]{Citrin2022}. In the solar context, simulations with a wave source at
the  base would generalize local turbulent RMHD wave-heating models of open flux tubes or closed loops \cite[e.g.,][]{vanBallegooijen2011,Perez2013,vanBallegooijen2017} to arbitrary geometry, 
allowing the inclusion of more realistic physics for loops and slow-wind source regions. For such studies the relevant equations
are the same fluctuation equations, supplemented by the fluxes and sources of mass, magnetic flux, momentum, and heat given in \cref{eq:rho-transport-main,eq:B-transport-main,eq:Upar-transport-main,eq: thermal energy} with $\partial_{\tau}=0$, particularly the $\mathcal{Y}^{\perp}$ heating terms given in \cref{eq: Y terms}. 

Finally, one may also wish to solve the transport system self-consistently \emph{in time}, predicting the slow evolution 
of a complex background due to small-scale turbulence. 
In the heliospheric context, this  can yield
 a slow-time-dependent Parker-like wave-driven wind model that captures novel heating mechanisms and cross-field transport. With boundary data determined by  current magnetograms, the approach could yield a fully fledged space-weather model related to some current approaches, predicting wind speeds and properties as magnetic geometry varies \citep[e.g.,][]{Mikic2018,Parenti2022,Sachdeva2019}. This final use case involves some subtleties that require care, the details of which are discussed in App.~\ref{app:parker_like_wind}:
 in brief, the subsonic ordering, as well as  slow secular drifts not captured by the  multiscale expansion, mean that wave-driving terms are over-constrained 
 by the parallel equilibrium equations \eqref{eq: mass conservation}, \eqref{eq: parallel equil}, and \eqref{eq:entropy-mean-main}. Motivated 
by the widespread practice of including wave momentum and heating sources in Parker-like winds \citep[e.g.,][]{Jacques1978,Zank1996,Cranmer2007},  as well as the robustness of the fluctuation equations in the low-$\beta$, transonic regime where wave forces are large (see App.~\ref{app: lowbeta_transonic}), we propose to circumvent these issues by promoting $\mathcal{O}(\epsilon^{3})$ transport terms into the  $\mathcal{O}(\epsilon)$ parallel equilibria. 
This is 
effectively a ``resummation'' of the expansion,  adding together the first- and third-order mean equations, or equivalently, 
the system that results from considering the wave-driven mass fluxes, forces, and heating effects to be sources in the lowest-order constraints.

The full resummed system, as suitable for slow-timescale computational evolution,  is listed in \crefrange{eq:rho-transport-resummed}{eq:Eth-transport-resummed}. It recovers the familiar ingredients of standard wave-driven wind models: parallel momentum is driven by wave-pressure forces through $p_{\rm tot}^{(2)}$ and the wave Maxwell/Reynolds stress $F^{\perp}_{\rm RM}$ \cite{Perez2013}, while thermal energy changes arise from the dissipation of wave action (see \cref{eq: alfvenic heating expansion} below), together with any imposed cooling via \(\Sth\). In a simple limit (ignoring 
 fluctuations and setting 
 $\gamma=1$),  it  contains the \citet{Parker1958} isothermal wind (see App.~\ref{app:parker_like_wind}). What is new  is that these standard field-aligned effects appear alongside genuinely multidimensional transport and heating, including perpendicular transport of mass, momentum, magnetic flux, and  heat, as well as the fluctuation-mediated relaxation of perpendicular gradients in $U$, $B$, and gravity  (terms involving  $\mathcal{Y}^{\perp}$). The magnetic evolution
 moves around field lines at speed $\Vpsit$ without an enhanced turbulent resistivity, while the density evolution allows cross-field 
 mass transport, which breaks the standard flux-tube mass-conservation constraint $\rho U/B={\rm const.}$ ($\gprl{\rho}+\gprl{U}-\gprl{B}=0$).
The system would be solved as a collection of ``tubes,'' supplemented by parallel boundary conditions at the base, similar to current wave-driven heliospheric models \cite{vanderHolst2014,Sokolov2021,Parenti2022}. Flux-tube RMHD simulations or a closure (such as that presented shortly) are needed to specify how the 
fluctuation correlations depend on the background and its gradients along any given tube.

A constraint, which applies to any flux tube background (fixed or evolving), is that the transport time should be longer than the time needed for fluctuation information  to propagate the length 
of the flux tube, in order that the flux tube can be considered static for the fluctuations. Given that the wind propagates from the Sun to 1~AU in $2$--$5$ days, while the rotation rate and global reorganization time are closer to 25--30 days, this 
is likely reasonably  well satisfied in the heliospheric context.


\label{PHENOMENOLOGY STARTS HERE}

\section{Phenomenologies of coronal heating and transport}
\label{sec: phenomenology}

In this section, we develop the multiscale RMHD equations into phenomenologies
suitable for understanding the impact of general magnetic geometries and wave properties on heating of an open-flux atmosphere (e.g., the solar corona). 
Our goal is to develop simplified expressions --- broadly, those obtained via application of the standard assumption that
the effect of turbulence is to dissipate fluctuations at the rate $\taunl^{-1}=\omeganl$ --- suitable for constructing simple estimates of the relative size of different heating  and 
transport effects. 
Later, in \cref{sec: empirical}, we provide such estimates for a simplified coronal model. 

 We start by developing
the mathematical and physical framework to describe different heating processes, based on a simple generalization of standard theories \cite{Dmitruk2002} and the transport timescale evolution of 
thermal energy  \eqref{eq: thermal energy}. We then  break up the discussion into the relevant processes labeled in \cref{fig: overview}: in \cref{subsec:RDT_basic}, standard expansion-based reflection-driven turbulence (RDT), in a form that facilitates comparison with the following sections on novel effects; in \cref{subsec:Q_reflection}, ``Q-reflection'' in tangled fields;  in \cref{subsec:comp catalyzed}, 
``direct compressive feedback'' (DCF) and ``compressively catalyzed reflection'' (CCR); in 
\cref{subsec:comp_absorb}, ``Alfv\'en catalyzed relaxation'' (ACR) of background gradients; and in \cref{subsec: transport}, we estimate the turbulent fluxes of density, momentum, and heat in a convenient form.

Throughout this discussion, we treat each effect in isolation, thus implicitly assuming that effects do not interact with each other. This assumption is robust when 
one effect dominates over others, as is likely usually the case, but could 
be dubious in certain situations if effects cancel (particularly  ACR and DCF).
A reference for notations is provided in \cref{tab: notation_phenom}.

\subsection{Preliminaries}\label{subsec: preliminaries}

Our goal is to estimate heating rates and fluxes driven by Alfv\'enic fluctuations propagating outwards along
a flux tube of arbitrary geometry that follows a given field line, with arc-length coordinate $\ell$ so that $\gradl=\partial/\partial\ell$. All gradients and energies are thus
understood as transport-scale averages over the perpendicular cross-section, with spatial dependence along  \(\bh(\ell)\) (see \cref{fig: average}).
To understand heating, our starting point is  \cref{eq: thermal energy} in combination  with the assumption 
that the outwards Alfv\'enic energy dominates those of other fluctuations,
 $W_{\perp}^{+}\gg W_{\perp}^{-},\, W_{\|}^{\pm},\,W_{s}$. This results because $\zm$, slow, and entropy modes 
are each predominantly \emph{driven} by $\zp$ proportionally to terms that scale with parallel 
or perpendicular gradients (see \cref{fig: cartoons}) --- we thus term them ``slaved fields''. We will then estimate the magnitude of the slaved fields ($W_{\perp}^{-}$, $W_{\|}^{\pm}$, and $W_{s}$)
by balancing their forcing by $\zp$ with their turbulent dissipation rate. 

To formalize this idea, we carry out a subsidiary expansion of \eqref{eq: thermal energy} with $\{W_{\perp}^{-},\, W_{\|}^{\pm},\,W_{s}\}\sim \imord^{2 } W_{\perp}^{+}$, where $\imord \ll1$ is the ratio of slaved fluctuation amplitude to the primary $z^{+}$, keeping terms to $\mathcal{O}(\imord)$. Using that $p^{(2)}_{\rm tot} = W_{B}^{\perp} + \mathcal{O}(\imord^{2}) = (W_{\perp}^{+} -  W^{\perp}_{r})/2 + \mathcal{O}(\imord^{2})$, 
$W^{\perp}_{r} = \rho\avgi{\zp\cdot \zm}/2\sim \imord W_{\perp}^{+}$, and that $\mathcal{Y}^{\perp}_{U}\sim\mathcal{Y}^{\perp}_{B}\sim\mathcal{Y}^{\perp}_{g}\sim\imord W_{\perp}^{-}\va/L$ because $\duprp\approx \dbprp/\sqrt{4\pi\rho}\approx \zp/2$,
one finds
\begin{flalign}
&\frac{\partial E_{\rm th}}{\partial \tau} =  -\frac{\rho U \va}{U+\va}\gradl \left[\frac{(U+\va)^{2}}{\rho\, U \va} W_{\perp}^{+}\right]\nn\\
&\quad -\nabla\cdot\left[ E_{\rm th} \left(\frac{1}{2} \avg{\frac{\d p}{p}\zp} + \gamma \UperpII \right)\right] +\mathcal{Y}^{\perp}_{U}+\mathcal{Y}^{\perp}_{B}+\mathcal{Y}^{\perp}_{g}\nn\\
& \quad+ \ma W_{r}^{\perp} \gradl \va- \rho U  \avg{\zp\cdot\tsmat\cdot\zm}.\label{eq: alfvenic heating expansion}
 \end{flalign}
To obtain this equation, we have also used the identity $\nabla\cdot\bm{U}/2+ U \gprl{B} = U\gprl{\va}=\ma \gradl \va$ to combine instances of $W_{r}^{\perp}$ from $F^{\perp}_{\rm RM}$ and  $p^{(2)}_{\rm tot}$, and, in the first line, combined the $\mathcal{O}(\imord^{0})$ energy-flux and wave pressure terms, 
\begin{equation}
-\nabla\cdot\left[\bh (U+\va) W_{\perp}^{+}\right] -\frac{1}{2}W_{\perp}^{+}\nabla\cdot\bm{U},\end{equation}
into  the parallel gradient of the Heinemann--Olbert \emph{wave action} \cite{Heinemann1980} \begin{equation}
S^{*}_{\rm WA}\equiv \frac{(U+\va)^{2}}{\rho\, U \va} W_{\perp}^{+}.\label{eq: wave action def}
\end{equation}
 This form \eqref{eq: alfvenic heating expansion} accounts
for the fact that, in the absence of dissipation, $W_{\perp}^{+}$ itself need not be conserved as the wave propagates along a slowly varying background:
the wave can do work on the background via $W_{\perp}^{+}\nabla\cdot\bm{U}$, thereby changing $W_{\perp}^{+}$.
True heating from the loss of $\zp$ energy therefore arises only when $S^{*}_{\rm WA}$ decreases, which must be supplied by dissipation processes that couple $\zp$ to other fields (e.g., through $\zm\cdot\nabla$ or terms involving $\duprl$, $\dbprl$, or $\d\rho$). 
Mathematically, this manifests in \eqref{eq: alfvenic heating expansion} via the first term on the right-hand side being formally $\mathcal{O}(\imord^{0})$, with the
 consequence that \(\gradl S^{*}_{\rm WA}=0\) to lowest order in $\imord$. The coupling to the slaved fields provides 
 a next-order correction 
to the parallel gradient of \( S^{*}_{\rm WA}\), making this term the same size as those on the second and third lines.

Based on \cref{eq: alfvenic heating expansion}, we therefore identify three broad classes of processes related to heating:
\begin{enumerate}
\item \emph{Wave-action dissipation.} The first line of \cref{eq: alfvenic heating expansion} shows that  true heating from $\zp$ arises from a \emph{decrease} of the wave action $(U+\va)^{2}W_{\perp}^{+}/(\rho U\va)$ along the field. Wave-action changes can be enabled by dissipation associated with
either $\zm$ through $\zm\cdot\nabla\zp$ or, via the curvature or gravity, $\duprl$, $\dbprl$, or $\d\rho$ (see the final line of \cref{eq: Alfvenic fluctuation equations main}). The former ($\zm$) can be supplied 
via expansion, field shearing ($Q$-reflection), or compressive modes (compressive catalyzation).
The  residual-energy corrections (the final line of \eqref{eq: alfvenic heating expansion}), which account for the work done by the Reynolds/Maxwell stresses of the fluctuations, are considered as part of the same framework given their origin in reflection (see below).
\item \emph{Relaxation of large-scale gradients.} The terms $\mathcal{Y}^{\perp}_{U}+\mathcal{Y}^{\perp}_{B}+\mathcal{Y}^{\perp}_{g}$ correspond to a complementary channel in which Alfvénic fluctuations mediate the release of free energy stored in background gradients of $U$, $B$, and $\rho$ (through $\bm{g}_{\rm eff}$). In this case, the waves act as an intermediary that taps large-scale shear, magnetic gradients, and/or gravitational stratification and passes it to the small scales where it is dissipated into heat. This is related to ``resonant absorption'' processes studied in the solar-wind context \cite{Ionson1978,Lee1986,Davila1987} and would also capture heating from gradient-driven instabilities.
\item \emph{Heat transport.} The perpendicular flux term on the second line of \cref{eq: alfvenic heating expansion},
redistributes thermal energy across field lines rather than creating it. Locally this can appear as heating or cooling in a given patch, but in the global budget it is best interpreted as turbulent heat transport.
\end{enumerate}
The processes  treated below  in \S\S\ref{subsec:RDT_basic}--\ref{subsec:comp catalyzed}, \cref{subsec:comp_absorb}, and \cref{subsec: transport} fall into categories (i), (ii), and (iii), respectively. 

An important caveat of our method and the classification above is that in some situations, a process can cool rather than heat the plasma, meaning one process might partially compensate another. Our goal in treating them separately is thus to compare the {magnitude} of  different effects to motivate their interest for future study, as opposed to the more ambitious goal of  a true heating phenomenology.

\subsubsection{Amplitudes of slaved fluctuations}\label{subsec:amplitudes}
The slaved fields are taken to be in quasi-steady balance between linear driving and nonlinear cascade mediated by $\zp$. Because
all driving terms are proportional to $\zp$, whose fluctuations generically decrease in amplitude with scale as $z^{+}\propto k_{\perp}^{-1/4}$ or $z^{+}\propto k_{\perp}^{-1/3}$ \cite{Wicks2013,Schekochihin2020}, while the turbulent damping rate increases towards smaller scales as $\omeganl\sim k_{\perp} z^{+}$, 
we expect the forcing to be of relevance only at the outer scale; below the outer scale, the system will set up a nearly constant-flux cascade of the
slaved field that maintains quasi-steady state by dissipating its free energy into heat \cite{Dmitruk2002,Barnes2011,Adkins2022,Adkins2026}.  We thus estimate the turbulent dissipation rate via
the standard mixing-rate estimate, $\taunl^{-1}=\omeganl\propto \zp\cdot\nabla_{\perp}\sim z^{+}/l_{\perp}$, where $z^{+}$ and $l_{\perp}$
are the outer-scale amplitude and perpendicular scale of $\zp$, respectively. In 
all cases, we neglect all  terms in each equation other than $\zp$ driving and nonlinear damping; we will indicate the regime in which this closure is expected to hold, but leave a systematic treatment of other regimes for future work.
At the risk of offending certain readers, we retain numerical coefficients in these estimates, despite the obvious crudeness of equating $\zp\cdot\nabla$ with $z^{+}/l_{\perp}$ as an exact damping rate. We do so because in some instances the scaling exponents of the resulting power-law profiles depend on these coefficients, and they are therefore helpful for comparing different mechanisms. That said, the numerical coefficients in all closure 
estimates should be considered accurate only up to their order of magnitude. 

We now collect estimates for the slaved amplitudes for later use, introducing a consistent notation for the compressive fields whose driving is more intricate than that of $z^{-}$.  
\paragraph{Parallel gradients.} Parallel gradients drive $\zm$ fluctuations via the term $(U+\va)(\gprl{\va}\mathsf{I}_\perp/2+\tsmat)\cdot\zp$ (see \cref{eq: Alfvenic fluctuation equations main}), while $\zm$ is damped 
via $\zp\cdot\nabla\zm$. Balancing these terms per the scheme above gives a generalization of the outer-scale estimate of \citet{Dmitruk2002}: $z^{-}\sim l_{\perp} (U+\va)(\gprl{\va}/2+|\tsmat|)$. 
Details, along with an estimate of the approximation's validity range, are provided in \cref{subsec:RDT_basic} and \cref{subsec:Q_reflection}.

\paragraph{Perpendicular gradients.} $\duprl$, $\dVprlx$, and $\d\rho/\rho$ provide a convenient set of amplitudes from which to 
estimate heating, providing estimates for $\Vrho$, $\Vpsi$, and $\VU$. Using $\duprp\approx\zp/2$ and $\dbprp/\sqrt{4\pi\rho}\approx -\zp/2$, and keeping only  the relevant driving and nonlinear terms in  Eqs.~\eqref{eq: parallel mom alfv}, \eqref{eq: parallel induction alfv}, and \eqref{eq:density-evol} (for $\d\rho$) yields 
\begin{flalign}
& \zp\cdot\nabla \duprl +\frac{\vs}{\va} \zp\cdot\nabla\dVprlx\simeq {\va} \zp\cdot \bm{\mathcal{F}}_{u}, 
\nn\\
&\zp\cdot\nabla \dVprlx +\frac{\vs}{\va} \zp\cdot\nabla\duprl\simeq \vs \zp\cdot \bm{\mathcal{F}}_{B}, 
\nn\\
&\zp\cdot\nabla \frac{\d\rho}{\rho} -  \frac{\vs^{2}}{\cs^{2}}\zp\cdot\nabla\frac{\duprl}{\va}\simeq \zp\cdot \bm{\mathcal{F}}_{\rho},\label{eq: balance for compressive}
\end{flalign}
where 
\begin{align}
\bm{\mathcal{F}}_{u}  &\;\equiv\;
   -\gprp{U}\,\ma - \gprp{B}+\bkap(\ma + 1),\nn\\
\bm{\mathcal{F}}_{B}  &\;\equiv\;
   -\gprp{U}\,\ma - \gprp{B}
         -\bkap(\ma+1)+\frac{1}{\gamma}\gprp{p},\nn\\
\bm{\mathcal{F}}_{\rho}  &\;\equiv\;
  \frac{\vs^{2}}{\cs^{2}} \Bigl[ \gprp{U}\,\ma + \gprp{B}
         +\bkap(\ma+1)\Bigl]\nn\\[-0.7em]& \qquad\qquad+ \frac{1}{\gamma}\frac{\vs^{2}}{\va^{2}}\gprp{p}-\gprp{\rho}\label{eq: F definitions}
\end{align}	
are the forcing terms, and we have used the identity  $1+\cs^2/\va^2=\cs^2/\vs^2$ to get the expression for $\bm{\mathcal{F}}_\rho$. On each left-hand side of \cref{eq: balance for compressive}, the two nonlinear terms  correspond to 
advection by $\duprp$ and $\dbprp$, respectively, where the latter captures how $\zp$  distorts  the field lines, affecting slow-wave propagation or 
parallel compressions. For simplicity (see \footnote{Advection by the magnetic field in the compressive equations accounts for propagation of fluctuations along the true (wandering) field lines caused by $\zp$ fluctuations. Field-line tangling will generate small perpendicular scales and thus act as an effective cascade in both $k_{\perp}$ and $k_{\|}$ (the latter via 
reconnection of field lines enabled by dissipation \cite{Meyrand2019}). 
If such terms dominate, then the interpretation of Eqs.~\eqref{eq: balance for compressive} must be modified, with, for example, $\bm{\mathcal{F}}_{B}$ becoming 
more relevant to $\duprl$, and vice versa. 
In the slow-mode equations, the ratio of this $\dbprp$ advection to that from $\duprp$ is  $\simeq (\vs/\va)  \dVprlx/\duprl$ for $\duprl$, or $\simeq (\vs/\va) \duprl/\dVprlx$ for $\dVprlx$, and since $\vs/\va\sim \beta^{1/2}$ at $\beta\ll1$, the effect should be   subdominant in the $\beta\ll1$ corona for at least one of $\duprl$ and $\dVprlx$. Since
we generally care about whichever of $\duprl$ and $\dVprlx$ is larger, this justifies its neglect. In the  $\d\rho$ equation, 
perpendicular advection  yields a turbulent balance for $\duprl$ rather than $\d\rho$, with the interpretation that parallel compressions locally balance $\zp$ forcing, meaning \eqref{eq: compressive amplitudes} remains an appropriate estimate for $\d\rho$.}
 for discussion), we  consider only the former effect ($\duprp$ advection),
giving 
\begin{align}
&\frac{\duprl}{\va} \simeq l_{\perp} \frac{ {\zp\cdot\bm{\mathcal{F}}_{u} }}{z^{+}},\quad \frac{\d\rho}{\rho} \simeq  l_{\perp} \frac{ {\zp\cdot\bm{\mathcal{F}}_{\rho} }}{z^{+}}, \nn\\&
\frac{\dVprlx}{\vs} =\frac{\va^{2}}{\vs^{2}}\frac{\dbprl}{B}\simeq  l_{\perp} \frac{ {\zp\cdot\bm{\mathcal{F}}_{B} }}{z^{+}},\label{eq: compressive amplitudes}
\end{align}
where $\{\duprl,\dVprlx,\dbprl,\d\rho\}$ represent outer-scale amplitudes. In subsequent heating estimates we  typically require correlators of the form $\avgi{\zp\,\d g}$, with $\d g\in\{\duprl,\dVprlx,\d\rho\}$ obtained from balances of the form $\d g\propto \zp\cdot\bm{\mathcal{F}}_{G}$ for $G=\{u,B,\rho\}$. We therefore frequently encounter averages of the form $\avgi{\zp_i\zp_j}\,\mathcal{F}_{G,j}$, with indices understood to be perpendicular. Assuming $\zp$ is locally isotropic around $\bh$ and that its perpendicular components are uncorrelated, we take
$\avgi{\zp_i\zp_j}=\avgi{|\zp|^{2}}\delta_{ij}/2
=(z^{+})^{2}\delta_{ij}/2,$
and hence
\begin{equation}
\avgi{\zp\,(\zp\cdot\bm{\mathcal{F}}_{G})}
=\avgi{\zp_i\zp_j}\,\mathcal{F}_{G,j}
\simeq \frac{1}{2}(z^{+})^{2}\bm{\mathcal{F}}_{G},
\label{eq: direction of correlations}
\end{equation}
i.e., within this closure the vector correlator aligns with $\bm{\mathcal{F}}_{G}$.

Note that the $\bm{\mathcal{F}}$ forcing terms can be written in various ways using the equilibrium \eqref{eq: perp equil}, 
though some care is required because certain combinations nearly cancel in some regimes (for instance $\bkap= \gprp{B}$ in a 
force-free magnetic field). This is explored in \cref{sec: empirical}.

\paragraph*{Underlying assumptions.}
In the above derivation, we  neglected most effects in \crefrange{eq: parallel mom alfv}{eq: entropy} as being smaller than the turbulent damping, including: (i) slow-mode propagation, which couples $\duprl$ and $\dVprlx$; (ii) parallel compression ($\bhdg \duprl$), which couples $\duprl$ and $\d\rho$ (iii) slow-mode WKB/reflection effects, which  change the compressive amplitudes as they propagate; and (iv) coupling 
of density/entropy to slow modes via parallel buoyancy (the  $\d\rho$ term in \eqref{eq: parallel mom alfv}) or heating (the $\gprl{s}$ terms in \cref{eq: entropy}). 
Effects (i) and (ii) contain both  linear propagation/compression ($\vs\bh\cdot\nabla$), and the neglected nonlinear terms in \eqref{eq: balance for compressive}; the 
ratio of the former  to the nonlinear terms is $\chi_{\rm A}^{-1} = \va l_{\perp}/z^{+}l_{\|}$ and we expect $\chi_{\rm A}\simeq1$ 
for strong turbulence, so these terms are likely safely neglected (see \cite{Note2}). If they cannot be neglected --- i.e., 
if slow waves propagate before being damped --- it would be more appropriate to work from the equations for $\zlpm$ \eqref{eq: slow mode zlpm}.
Effects (iii) (WKB/reflection) and (iv) (slow/entropy coupling) are safely neglected if parallel correlation lengths are shorter than the parallel background variation:  their ratio with the 
propagation/compression effects (i)--(ii) is $\sim\! l_{\|}\gprl{g}$, where $g=\{B,p,\vs,s,\rho\}$,  and effects (i)--(ii) were argued to be themselves  
small. 
Finally, in taking $\zp\cdot\nabla$ to act as a turbulent dissipation with rate $\sim \omeganl$, we are implicitly assuming strong-turbulent mixing and thus $\chi_{\rm A}\gtrsim 1$; for $\chi_{\rm A}<1$, compressive parcels will be buffeted  weakly by high-frequency waves, which likely reduces the effectiveness of mixing, as in weak turbulence \cite{Schekochihin2020}.

Relaxing these assumptions would lead to a proliferation of different regimes --- while some of these are plausibly relevant for 
parameters in regions of the solar atmosphere, we leave a detailed characterization to future work.

\subsection{Expansion-driven RDT from parallel Alfv\'en-speed gradients}
\label{subsec:RDT_basic}

In the language developed above, ``standard'' reflection-driven turbulence (RDT) corresponds to
heating via dissipation of outward Alfv\'enic fluctuations generated through reflection from Alfv\'en-speed gradients: the wave action $S^{*}_{\rm WA}$ is no longer conserved because reflection from background $\va$ variation generates
a finite $\zm$, enabling a nonlinear cascade and dissipation of $W_{\perp}^{+}$.
This type of heating,  which is the only effect that survives in a straight flux tube with  no cross-field gradients, arises purely from the first class identified in \cref{subsec: preliminaries}:
a decrease $S^{*}_{\rm WA}$ along $\bh$. While the phenomenology presented 
here is effectively a review of the straight-tube theories of \citet{Chandran2009} and others \cite{Verdini2007,Dmitruk2002}, we present it in a form that is readily generalized and facilitates comparison with other  effects.
Likewise, within the framework of this paper, 
the theory can alternatively be thought of as characterizing how expansion specifically drives heating, even for a flux tube with more general geometry.

We start from the perpendicular Els\"asser
equations \eqref{eq: Alfvenic fluctuation equations main} in a straight flux tube with $\tsmat=0$, $\bkap=0$, and 
no perpendicular inhomogeneity, yielding the standard form \eqref{eq: straight tube} \cite{Chandran2009} (see also \cref{eq:general_reflection_force}). This straight-tube approximation should be valid when the slaved $z^-$ fluctuations created by $\gprl{\va}$ are large compared to the slaved fluctuations produced by $\tsmat$, $\bkap$, $\geff$, or perpendicular gradients. 
As in the wave-action discussion above, it is convenient to absorb the WKB evolution into
a rescaled field via the change of variables
\(\bm{f}^{\pm}\equiv (\ma^{1/2}\pm\ma^{-1/2})\zpm\), so that  $\avgi{|\bm{f}^{+}|^{2}}/4=S^{*}_{\rm WA}$ \cite{Heinemann1980}.
\Cref{eq: straight tube} then becomes the \emph{wave-action} system
\begin{equation}
\frac{\partial \bm{f}^{\pm}}{\partial t}
+ (U \pm \va)\!\left(\bh\cdot\nabla \bm{f}^{\pm}
-\frac{\gprl{\va}}{2}\,\bm{f}^{\mp}\right)
= -\bm{z}^{\mp}\cdot\nabla\bm{f}^{\pm}
-\nabla \tilde{p}_{*},
\label{eq:wave_action_straight}
\end{equation}
where the role of $\nabla\tilde{p}_{*}$ is to ensure that  $\nabla\cdot\bm{f}^{\pm}=0.$
Absent nonlinearity or reflection, $\bm{f}^{+}$ maintains constant amplitude as it propagates in the flux tube, 
making the form \eqref{eq:wave_action_straight} particularly convenient.

We  wish to understand how $\bm{f}^{+}$ dissipates due to the 
self-generated $\bm{f}^{-}$ via the third term ${\propto}\gprl{\va}$ in \eqref{eq:wave_action_straight}.
Representing the outer-scale rms amplitude of $\fpm $ as $f^{\pm}$ to build a simple phenomenology, \cref{eq:wave_action_straight} can be partitioned into three physically distinct pieces at the outer correlation scales \(l_{\|}\) and \(l_{\perp}\):
\begin{equation}
\left(\frac{\partial}{\partial t}+U\bh\cdot\nabla\right)f^{\pm}
\!\!
\underbrace{\pm\frac{\va}{l_{\|}}f^{\pm}}_{\text{propagation}}
\!\!-\!\!
\underbrace{\mathcal{R}^{\pm}_{\rm A}f^{\mp}}_{\text{reflection}}
\!=
-\underbrace{\omeganl^{\pm}f^{\pm}}_{\text{cascade}},
\label{eq:cartoon}
\end{equation}
where
\begin{equation}
\mathcal{R}_{\rm A}^{\pm}=\frac{(U\pm\va)\gprl{\va}}{2},\quad
\omeganl^{\pm}\sim\frac{z^{\mp}}{l_{\perp}} \label{eq: reflection defs}
\end{equation}
are the reflection and nonlinear rates. 
We also define the non–dimensional ratios
\(
\chi_{\mathrm{A}}^{\pm}\equiv\omeganl^{\pm}/({\va}/{l_{\|}})\) and 
\(\chi_{\exp}\equiv\omeganl^{\pm}/\mathcal{R}_{\rm A}^{\pm}\) \cite{Goldreich1995,Meyrand2025}.

\paragraph{Standard strong-turbulence phenomenology.}
\label{subsubsec:breakdown}

Let us specialize to the sub-Alfv\'enic corona, $\ma\ll1$, which simplifies the relation between $\zpm$ and $\fpm$ to $z^{\pm}\approx  \ma^{1/2}f^{\pm}$ (this removes  algebraic complications associated with $R_{\rm A}$, where $\ma=1$, but is not necessary \cite{Chandran2009}).
To derive a heating rate, the phenomenology assumes
\begin{enumerate}
\item \(z^{-}\ll z^{+}\), with  \(z^{-}\) set by the {instantaneous} balance between reflection and nonlinear damping (slaving):
      \begin{equation}
 \mathcal{R}_{\rm A}^{-}f^{+}\sim\omeganl^{-}f^{-}\implies
      z^{-}\simeq\mathcal{R}_{\rm A}^{-} z^{+}\frac{l_{\perp}}{z^{+}}
      \propto \frac{\va l_{\perp}|\gprl{\va}|}{2}.\label{eq: standard phen zm}
\end{equation}
\item In the \(z^{+}\) equation, reflection is negligible and the nonlinearity causes damping, giving
 \(
 \va\,\bh\cdot\nabla f^{+}\sim f^{+}z^{-}/l_{\perp}
      \).
Combining this with \Cref{eq: standard phen zm}  yields \(\bh\cdot\nabla f^{+}= \pm f^{+}\gprl{\va}/2\), which integrates to
       \begin{equation}
      f^{+}(\ell)\propto\va^{\pm 1/2} \implies z^{+}\propto \ma^{1/2} \va^{\pm1/2}.\label{eq: standard phen zp}
\end{equation}
      The sign of the exponent of $\va$ is positive (negative) where \(\va\) decreases (increases) with $\ell$, ensuring it always represents damping.
\end{enumerate}
The turbulent heating rate from the dissipation of wave action in  \cref{eq: alfvenic heating expansion} is therefore \cite{Dmitruk2002}
\begin{equation}
Q_{\rm A}^{\rm exp} = -\frac{\rho U\va}{U+\va}\bh\cdot\nabla S^{*}_{\rm WA} \simeq  \frac{z^{-}}{l_{\perp}} W_{\perp}^{+}
      \simeq W^{+}_{\perp}\va\frac{|\gprl{\va}| }{2},
\label{eq:Q_RDT}
\end{equation}
where $W_{\perp}^{+} = \rho(z^{+})^{2}/4$ (i.e., equating the outer-scale amplitude with $\avgi{|\zp|^{2}}^{1/2}$).
Interestingly, $Q_{\rm A}$  is independent of the nonlinear scale \(l_{\perp}\): a smaller \(l_{\perp}\) increases the damping rate but reduces the driven \(z^{-}\) amplitude in the same proportion.
The heating rate from $\zm$ dissipation is likewise
\begin{equation}
Q_{\rm A}^{-} \simeq \frac{z^{+}}{l_{\perp}} W_{\perp}^{-} \simeq \frac{z^{-}}{z^{+}}Q_{\rm A}^{\rm exp} \ll Q_{\rm A}^{\rm exp}.
\end{equation}
Because the phenomenology's assumptions break down as $z^{-}$ approaches $z^{+}$ (see below), 
the heating from $z^{-}$ dissipation is always subdominant within its range of validity (though this is commonly included in phenomenological modeling \cite[e.g.,][]{Cranmer2007,Usmanov2014,Reville2020}).

Note that we have deliberately ignored the $W^{\perp}_{r}$  contribution in  \cref{eq: alfvenic heating expansion}, which arose from the expansion-induced Reynolds/Maxwell stress $F^{\perp}_{\rm RM}$. 
The contribution is large in magnitude when  $\zm$ maintains the same phase as its $\zp$ forcing ($\avgi{\zp\cdot\zm}\sim -z^{+}z^{-}$), {viz.,} when $\zm$
is not strongly scrambled in phase by turbulent dissipation. Generically $\gprl{\va}W_{\perp}^{r}$ should be positive, since $\partial\zm/\partial t\propto +\gprl{\va}\zp$ and $W_{\perp}^{r}=\avgi{\zm\cdot\zp}/2$, signaling an additional effective heating 
when the Maxwell dominates the Reynolds stress. However, this effect should 
not be treated as heating in the usual sense --- its appearance in the energy equation is due to the additional force on
the wind, thereby changing its adiabatic cooling ---  so care is required in interpreting its effect \cite{Perez2021,Meyrand2025}. The same effect from squashing, via $\avgi{\zp\cdot\tsmat\cdot\zm}$, involves the 
same physics and will likewise be ignored below.

\paragraph*{Underlying assumptions.}
The above derivation rests on the following assumptions:
\begin{enumerate}
\item[\textit{a}.] a single outer perpendicular scale  $l_{\perp}$ for both \(z^{+}\) and \(z^{-}\);
\item[\textit{b}.] anomalous coherence, whereby $\zm$ remains nearly coherent with $\zp$ despite propagating in the opposite direction (this allows the neglect of ${\va}/{l_{\|}}$ in the $z^{-}$ balance);
\item[\textit{c}.] a strong cascade with \(\chi_{\mathrm{A}}\gtrsim1\), so that $z^{\pm}$ fluctuations cascade to dissipate at rate $z^{\mp}/l_{\perp}$;
\item[\textit{d}.] rapid adjustment of \(z^{-}\) compared with the radial scale of \(\va\), requiring $\va/\omeganl^{-}\ll (\gprl{\va})^{-1}$ (the left- and right-hand sides being the damping distance  for $z^{-}$ and the  lengthscale of $\va$ variation, respectively);
\item[\textit{e}.] negligible reflection in the \(z^{+}\) equation, because $z^{+}\gg z^{-}$.
\end{enumerate}
Conditions \textit{a} through \textit{c} are discussed extensively in \citet{Meyrand2025}, while a weak  (\(\chi_{\mathrm{A}}\lesssim1\)) phenomenology to address \textit{c} is developed in \citet{Chandran2019}. Conveniently it transpires that conditions  \textit{d} and \textit{e} are both equivalent to \(\chi_{\exp}\gtrsim1\), which marks the domain where classical RDT applies in any case.

\paragraph{Radial evolution of the turbulence.}
\label{subsubsec:radial_breakdown}

Considering the balance $\mathcal{R}_{\rm A}^{-}z^{+}\!\sim\!\omeganl^{-}z^{-}$, a physically enlightening way to write \cref{eq: standard phen zm} is \cite{Meyrand2025,Abbas2026} 
\begin{equation}
\frac{z^{-}}{z^{+}}\!\simeq\!\chi_{\exp}^{-1}.
\end{equation}
Hence the imbalance (normalized cross-helicity),
\[
\sigma_{c}\equiv\frac{W_{\perp}^{+}-W_{\perp}^{-}}{W_{\perp}^{+}+W_{\perp}^{-}}
           \;=\;\frac{1-\chi_{\exp}^{-2}}{1+\chi_{\exp}^{-2}}
           \;\approx\;1-\frac{2}{\chi_{\exp}^{2}}+\ldots,
\]
is controlled directly by $\chi_{\exp}$  \citep{Meyrand2025}: larger amplitudes and/or weaker expansion ($\chi_{\exp}\!\gg\!1$) drive highly imbalanced turbulence ($\sigma_{c}\approx 1$), whereas smaller amplitudes and/or stronger expansion ($\chi_{\exp}\!\rightarrow\!1$) drive the system towards balanced turbulence. 

Moreover, because $l_{\perp}$ grows rapidly with $\ell$, approximately as 
the flux tube width $\propto\!B^{-1/2}$, we  expect that  in nearly all situations, $\omeganl\sim z^{+}/l_{\perp}$, and therefore $\chi_{\exp}$ falls rapidly with radius (this is effectively always true beyond the Alfv\'en point \cite{Meyrand2025,Abbas2026}). The outcome is a decrease in imbalance with $\ell$, as observed in the solar wind.
Moreover, any process that modifies $\chi_{\exp}$ --- e.g.,\ differing initial amplitudes, expansion histories, or flux-tube widths --- will set distinct radial profiles of $\sigma_{c}$.  This offers some explanation for the observed spread of cross helicities between fast and slow solar wind streams \cite{Meyrand2025}: more overall 
expansion, associated with slower wind \cite{Wang1990}, will cause lower $\chi_{\exp}$ at any given radius, thereby leading to lower imbalance, as observed 
\cite{Bruno2013}.

\paragraph{Breakdown of the phenomenology.}

Likewise, if \(\chi_{\exp}\) drops below unity --- either because the tube’s geometric expansion is extremely large or once \(z^{+}\) decays --- the entire phenomenology breaks down as the turbulence approaches $\sigma_{c}\approx 0$.  In this regime, the turbulent damping is too slow to keep up with the growth of $z^{-}$, so our assumed balance fails. Simulations in the super-Alfv\'enic regime show that turbulent heating shuts off as it becomes balanced, with the turbulence ``freezing'' into magnetic structures with large negative $W_{r}^{\perp} $\citep{Meyrand2025,Abbas2026}. While this has not been well explored in sub-Alfv\'enic flows, at the least, we can expect qualitatively different, nearly balanced fluctuations once $\chi_{\exp}<1$.

A second potential shutdown occurs if the  \(l_{\perp}\) grows sufficiently to make the cascade \emph{Alfvénically weak}, \(\chi_{\mathrm{A}}\lesssim 1\)  (\(\omeganl\lesssim {\va}/{l_{\|}}\)).  
Whether significant heating persists in this limit is unsettled: analytic arguments and some simulations suggest it can \citep{Chandran2019}, whereas idealized (expanding box) simulations show heating shutting off \citep{Meyrand2025}.  
If the latter is correct (heating shuts off), dissipation halts where \(\chi_{\mathrm{A}}\simeq1\); if the former is, the ultimate cutoff would occur once \(\chi_{\exp}\simeq\chi_{\mathrm{A}}^{-1}>1\), which results from applying assumption \textit{d} above to weak turbulence scalings.


\begin{table}
\caption{Useful definitions for the transport, phenomenology, and empirical theories.}
\begin{ruledtabular}
\begin{tabular}{ll}
\textrm{Symbol} & \textrm{Definition} \\
\colrule
$\mathcal{Y}^{\perp}$, $\mathcal{Y}^{\|}$ & Fluctuation--mean energy exchange; \cref{eq: Y terms}.\\
$\UperpII$ & Second-order mean perpendicular flow\\ 
$\Vrho,\Vpsi,\VU,\Vp$ & Turbulent transport velocities; \cref{eq: V defs}\\
$\Vrhot,\Vpsit,\Vpt$ & $\Vrhot\equiv \Vrho+\UperpII$, $\Vpsit\equiv \Vpsi+\UperpII$, $\Vpt\equiv \Vp+\UperpII$\\ 
$\bm{\Gamma}_{\rho},\bm{\Gamma}_{\rho U},\bm{\Gamma}_{\rm th}$ & Perp.~mass/momentum/heat flux; \cref{eq: cross field fluxes}\\
$D^{\rm tot}$ & Total fluctuation dissipation rate; \cref{eq:Dtot second order}\\ 
$F^{\perp}_{\rm RM}$ & Alfv\'enic Reynolds/Maxwell force; \cref{eq: RM perp def}\\
\colrule
 $S^{*}_{\rm WA}$ &  Alfv\'enic wave action, \cref{eq: wave action def} \\
$\bm{f}^{\pm}$ & WKB rescaled $\zpm$, $\bm{f}^{\pm} \equiv (\ma^{1/2}\pm\ma^{-1/2})\zpm$\\
$\bm{\mathcal{F}}_{u},\bm{\mathcal{F}}_{B},\bm{\mathcal{F}}_{\rho}$ & Slaved forcing vectors; \cref{eq: F definitions}\\ 
$\omeganl$, $\taunl$ & Nonlinear rate/time, $\taunl^{-1}\sim\omeganl\sim z^{+}/l_{\perp}$\\
$\mathcal{R}_{\rm A}^{\pm}$ & Expansion reflection rate $\mathcal{R}_{\rm A}^{\pm}={(U\pm\va)\gprl{\va}}/{2}$\\
$\mathcal{R}_{\mathcal Q}^{\pm}$ & Squashing reflection rate $\mathcal{R}_{\mathcal Q}^{\pm}=(U\pm\va)|\tsmat|$\\
$\chi_{\rm A}$ & Critical balance parameter, $\chi_{\rm A}=\omeganl/k_{\|}\va$\\
$\chi_{\exp}$ & Reflection nonlinear ratio, $\chi_{\exp}=\omeganl/\mathcal{R}_{\rm A}$\\ 
$\chi_{\exp,\mathcal Q}$ & Squashing nonlinear ratio, $\chi_{\exp,\mathcal Q}=\omeganl/\mathcal{R}_{\mathcal Q}$\\
\colrule
$\eta_{\rm turb}$ & Turbulent diffusivity $\eta_{\rm turb}=z^{+}l_{\perp}/4$\\ 
RDT & Reflection driven turbulence \cref{subsec:RDT_basic}\\ 
$Q_{\rm A}^{\rm exp}$ & Expansion RDT heating rate; \cref{eq:Q_RDT}\\
$K^{\rm damp}$ & Inverse heating length; \cref{eq: final QRDT full,eq: Qccr K defns,eq:K_ACR}\\ 
$a$ & Expansion factor of tube; \S\ref{subsec:Q_reflection}\\
$\mathcal{Q}$ & Squashing factor of field-line mapping; \S\ref{subsec:Q_reflection}\\
$\mathsf{J}$ & Field-line mapping Jacobian; \cref{eq:J_evolution}\\ 
DCF & Direct compressive feedback; \S\ref{sub: DCF}\\
CCR & Compressively catalyzed reflection; \S\ref{sub: CCR}\\
ACR & Alfv\'en-catalyzed relaxation; \S\ref{subsec:comp_absorb}\\
\colrule
$\ell$ & Field-line parallel coordinate \\
$R$ & Spherical radius\\
$R_{\rm A}$ & Alfv\'en radius where $\ma=1$ \\
$\mathcal{F}^{\rm crit}$ & Forcing (gradients) needed  to dominate RDT\\
$\Thetac$ & Compressive mediator; $\kappa$ or $|\geff|/\va^{2}$, Eq.~\eqref{eq:Kcomp_generic}\\
$\ccomp$ & Coefficient in $K^{\rm damp}$ for DCF/CCR/ACR\\
$L_{\perp,\rho}$, $L_{\perp,p}$ & Perpendicular gradient scales, $L_{\perp,G}=|\gprp{G}|^{-1}$\\
\end{tabular}
\end{ruledtabular}
\label{tab: notation_phenom}
\end{table}

\subsection{Q-reflection: RDT in tangled fields}
\label{subsec:Q_reflection}

We now extend standard expansion-driven RDT to build a \emph{geometry-agnostic} phenomenology of reflection that remains valid when the guide field
is arbitrarily distorted  via expansion, twist, and shear (see \S\ref{sub: q decomposition}).  This amounts to retaining the effects of  $\tsmat$ in \cref{eq: Alfvenic fluctuation equations main} (see \cref{eq:general_reflection_force}), but still neglecting the curvature- and gravity-mediated feedback from compressive 
fluctuations (recall that field-line twist does not contribute in the RMHD ordering). Importantly, in many physically relevant situations, standard reflection,
\( \mathcal{R}_{\rm A}^{\pm}\propto\gprl{\va}\,z^{\mp} \), will  plausibly become sub-dominant to \(\tsmat\cdot \zpm\), implying that reflection by
\emph{squashing} of flux surfaces dominates over reflection arising from gradients in the Alfv\'en speed.   
As argued in \S\ref{sub: q decomposition}, 
the field-line twist
\(\mathsf{A}\) does not contribute to reflection at RMHD order, so its only contribution is to 
change the field-line foot point mapping. 

\paragraph{Squashing factor and its link to \(\tsmat\).}
The ``{squashing factor}'' $\mathcal{Q}$ \citep{Titov2002} has been extensively studied 
in past works and computed from solar magnetic-field data, particularly for
its application to ``S-web'' models of slow-wind sources \citep{Antiochos2011}. 
We thus start by relating $\mathcal{Q}$ to $\tsmat$ in order to link $\mathcal{Q}$ to the turbulent heating phenomenology discussed below.

$\mathcal{Q}$ is defined via the Jacobian \(\mathsf{J}_{ij}\equiv\partial X_{i}/\partial x_{j}\) that maps one
perpendicular plane \((x,y)\) to another \((X,Y)\) by following field lines. 
This mapping is shown graphically in \cref{fig: grad b} as the coordinate transformation between the black circles that mark the intersection of field lines with the illustrated perpendicular planes. 
The determinant of   \(\mathsf{J}_{ij}\) gives the area change associated with the mapping, and is thus 
the expansion factor $a^{2}$, while
\begin{equation}
\mathcal{Q}\equiv\frac{\mathsf{J}_{ij}\mathsf{J}_{ij}}{\det\mathsf{J}} = \frac{\mathsf{J}_{ij}\mathsf{J}_{ij}}{a^2}
\end{equation}
measures the magnitude of $\mathsf{J}$ absent expansion, and thus how much neighboring lines ``tangle''.

Because the transverse coordinate of a field line satisfies \(\rmd\bm{x}/\rmd \ell=\bh(\bm{x})\), a first-order Taylor expansion of \(\bm{x}+\delta\bm{x}\) shows that $\mathsf{J}$ satisfies
\begin{equation}
\frac{\rmd\mathsf{J}}{\rmd \ell}= \nabla\bh \cdot \mathsf{J}.
\label{eq:J_evolution}
\end{equation}
Rescaling $\mathsf{J}$ by $a$ to define \(\mathsf{J}=a\,\tilde{\mathsf{J}}\) with \(\det\tilde{\mathsf{J}}=1\), \cref{eq:J_evolution} becomes (see \S\ref{sub: q decomposition})
\begin{equation}
\frac{\rmd\ln a}{\rmd\ell}= \frac{1}{2}\nabla\cdot\bh,
\qquad
\frac{\rmd\tilde{\mathsf{J}}}{\rmd\ell}= (\tsmat+\mathsf{A})\cdot\tilde{\mathsf{J}},\label{eq: Jtilde eq}
\end{equation}
showing how, as expected, the trace of \(\nabla\bh\) governs expansion (encoded in $a$), while its traceless part,
\(\tsmat+\mathsf{A}\), governs shape distortion (encoded in $\tilde{\mathsf{J}}$).  
Differentiating \(\mathcal{Q}=\tilde{\mathsf{J}}_{ij}\tilde{\mathsf{J}}_{ij}\) then yields
\begin{equation}
\frac{\rmd\mathcal{Q}}{\rmd\ell}     =2\,\mathrm{Tr}\!\left(\tilde{\mathsf{J}}^{T}\tsmat\tilde{\mathsf{J}}\right),
\label{eq:dQds_short}
\end{equation}
since the symmetric matrix product has eliminated the antisymmetric part \(\mathsf{A}\). Because $\tilde{\mathsf{J}}$ evolves without knowing about the expansion,  Eq.~\eqref{eq:dQds_short} shows
 explicitly how the symmetric-traceless part of $\nabla\bh$ (encoded in \(\tsmat\)) is the primary driver of $\mathcal{Q}$ evolution, mirroring its exclusive role in Alfv\'en-wave reflection physics (see \S\ref{sub: q decomposition}).

\paragraph{Reflection from squashed field lines.}
In the presence of squashing, we retain the effect of $\tsmat$ in \cref{eq:wave_action_straight}, which becomes
\begin{align}
\frac{\partial \bm{f}^{\pm}}{\partial t} 
+ (U \pm \va)&\left(\bh\cdot\nabla \bm{f}^{\pm}
-\frac{\gprl{\va}}{2}\,\bm{f}^{\mp} + \tsmat\cdot \bm{f}^{\mp}\right)
\nn\\ &= -\bm{z}^{\mp}\cdot\nabla\bm{f}^{\pm}
-\nabla \tilde{p}_{*}. 
\label{eq:wave_action_with_S}
\end{align}
Analogous to \cref{subsec:RDT_basic}, we define a squashing-driven
reflection rate  
\( \mathcal{R}_{\mathcal{Q}}^{\pm}\equiv\left(U\pm\va\right)|\tsmat| \),
where \( \pm |\tsmat|=({\tsmat_{ij}\tsmat_{ij}}/2)^{1/2}\) are the eigenvalues of $\tsmat$, which characterize the local deformation rate.
Recall that the only difference between 
this and standard reflection is that the components of $\zm$ along the long and short directions
of the ellipse pick up opposite signs, thus rotating $\zm$ compared to its source $\zp$ (see \cref{fig: grad b} and \S\ref{sub: q decomposition}). On the level of this phenomenology, its effects are thus identical to expansion reflection;
balancing against the cascade gives an analogous $\chi_{\exp}$ parameter
\begin{equation}
\chi_{\exp,\mathcal{Q}}\;\equiv\;\frac{\omeganl^{-}}{\mathcal{R}_{\mathcal{Q}}^{-}}
                  \;=\;\frac{z^{+}/l_{\perp}}
                           {\left(U-\va\right)|\tsmat|},
\end{equation}
which also has the property that  \(z^{+}/z^{-}\simeq\chi_{\exp,\mathcal{Q}}\) when
\( \chi_{\exp,\mathcal{Q}}\gtrsim1\) (see below).

\paragraph{Heating rate and evolution  of \(\zpm \).}
Repeating the  
arguments of §\ref{subsec:RDT_basic}, we start by 
applying the same wave-action transformation to \cref{eq: Alfvenic fluctuation equations main}, which
simply transforms $(U\mp\va)\zmp\cdot\tsmat$ into $(U\pm\va)\fmp\cdot\tsmat$. 
The balance between driving and dissipation for $f^{-}$ thus gives 
\begin{equation}
z^{-}\sim \va l_{\perp}\left(\frac{|\gprl{\va}|}{2} + |\tsmat|\right)
\end{equation}
(cf. \cref{eq: standard phen zm}).
The nonlinear dissipation of $f^{+}$ along $\ell$,  $\va\,\bh\cdot\nabla f^{+}\sim f^{+}z^{-}/l_{\perp}$
can be integrated with the aid of
\( \rmd\ln\mathcal{Q}^{1/2}/\rmd\ell= \gprl{\mathcal{Q}}/2\approx |\tsmat| \)
(the approximation is exact if the principal axes of
\(\tsmat\) do not rotate along the field line; see below). This  yields
\begin{equation}
z^{+}\;\propto\;\ma^{1/2}\va^{\pm1/2}(\ell)\,\mathcal{Q}^{-1/2}(\ell),
\end{equation}
where we have assumed for simplicity that $\mathcal{Q}$ increases with $\ell$.
The heating rate becomes  (cf.~\cref{eq:Q_RDT}), 
\begin{align}
Q_{\rm A}^{\rm full} & \simeq \frac{1}{8}\rho (z^{+})^{2}\va(|\gprl{\va}| + |\gprl{\mathcal{Q}}|)\label{eq: final QRDT full}\\
& = W_{\perp}^{+}\va K^{\rm damp}_{\rm A};\quad K^{\rm damp}_{\rm A} \equiv \frac{1}{2}(|\gprl{\va}| +|\gprl{\mathcal{Q}}|),\nn
\end{align}
where the second expression identifies the Alfv\'enic heating scale $K^{\rm damp}_{\rm A}$ for helpful comparison with other 
mechanisms below. We see that 
 gradients of \(\mathcal{Q}\) act analogously to
those of $\va$ in governing reflection; however, in complex-field regions they can be orders of
magnitude larger, with solar extrapolations often finding
\(\mathcal{Q}\sim10^{5}\) between the low corona and the height of the
sonic point \citep{Titov2002,Antiochos2011,Cranmer2017}, while $\va$ likely only decreases by a factor of several across the same range.

Just as $\chi_{\exp}\lesssim 1$ marks the boundary where standard  RDT theory ceases to apply, so 
too does $\chi_{\exp,\mathcal{Q}}\lesssim 1$ for squashed fields. Thus, for very large squashing rates
our phenomenology ceases to apply, presumably with the result that Alfv\'en waves are ``stuck'' and energy 
cannot be transported to larger altitudes.

\paragraph{A subtlety: rotation of the squashing plane.}
In our phenomenology, $\tsmat$ appears in two distinct ways: directly 
through the reflection source $\tsmat\cdot\bm{z}^{+}$ in \cref{eq:wave_action_with_S}, and indirectly via the relation 
$\gprl{\mathcal{Q}}/2\approx|\tsmat|$, which relates the local reflection 
rate into gradients of $\mathcal{Q}$. These are strictly equivalent only 
when the principal axes of $\tsmat$ maintain the same orientation with $\ell$.

For the former, the reflection magnitude 
$|\tsmat\cdot\bm{z}^{+}|=|\tsmat||\bm{z}^{+}|$ is insensitive to the 
orientation of $\bm{z}^{+}$, since $\tsmat$ has eigenvalues $\pm|\tsmat|$ 
(only the direction of the reflected $\bm{z}^{-}$ depends on alignment 
with the principal axes $\hat{\perpunitvectorbase}_{l,s}$ of $\tsmat$). 
Likewise, $\mathsf{A}$ does not appear in 
\cref{eq:wave_action_with_S}, so reflection itself is unaffected 
by field-line twist.
For the latter, the relation of $\tsmat$ with $\mathcal{Q}$ is more delicate, because 
$\gprl{\mathcal{Q}}=2\,\mathrm{Tr}(\tilde{\mathsf{J}}^{T}\tsmat
\tilde{\mathsf{J}})/\mathcal{Q}$ depends on the alignment of 
$\tilde{\mathsf{J}}$ with the axes of $\tsmat$. To understand this, we write 
$\tsmat=|\tsmat|\,\mathsf{R}\,\mathrm{diag}(1,-1)\,\mathsf{R}^{T}$, where 
$\mathsf{R}=\exp(\phi\,\mathsf{R}_{0})$,  $\mathsf{R}_{0}\equiv 
\hat{\perpunitvectorbase}_{1}\hat{\perpunitvectorbase}_{2}
-\hat{\perpunitvectorbase}_{2}\hat{\perpunitvectorbase}_{1}$ is the 2D 
rotation generator, and $\phi(\ell)$ is the angle between the fixed perpendicular basis ($\hat{\perpunitvectorbase}_{1},\hat{\perpunitvectorbase}_{2}$) and the 
eigenbasis of $\tsmat$ ($\hat{\perpunitvectorbase}_{l},\hat{\perpunitvectorbase}_{s}$). We then transform $\tilde{\mathsf J}$ into the  
principal-axis frame via $\hat{\mathsf{J}}\equiv\mathsf{R}^{T}\tilde{\mathsf{J}}$ and \cref{eq: Jtilde eq} becomes
\begin{equation}
\frac{d\hat{\mathsf{J}}}{d\ell} 
= \left[\,|\tsmat|\,\mathrm{diag}(1,-1) 
+ \left(\frac{1}{2}\bh\cdot\nabla\times\bh-\dot{\phi}\right)\,\mathsf{R}_{0}\,\right]\hat{\mathsf{J}},
\label{eq:Jhat_rotated}
\end{equation}
where $\bh\cdot\nabla\times\bh/2$ arises as the (single) 
component of $\mathsf{A}$ and $\dot{\phi}\equiv d\phi/d\ell$ is the 
rate at which the eigenframe of $\tsmat$ rotates along the field line. 
Both contributions can act in the same way, rotating 
$\hat{\mathsf{J}}$ away from alignment with $\mathrm{diag}(1,-1)$, which 
suppresses $\mathcal{Q}$ growth. Only when $|\bh\cdot\nabla\times\bh/2-\dot{\phi}|
\ll|\tsmat|$ does $\hat{\mathsf{J}}$ remain nearly diagonal, giving 
$\mathcal{Q}(\ell)\approx \exp\!\big(2\int_{0}^{\ell}|\tsmat|(\ell')
\,d\ell'\big)$.

When this condition fails, the correspondence 
$\gprl{\mathcal{Q}}/2\approx|\tsmat|$ breaks down: the reflection rate 
$|\tsmat|$ remains a robust driver of $\bm{z}^{-}$ (independent of 
$\mathsf{A}$), but $\gprl{\mathcal{Q}}/2$ falls below $|\tsmat|$, so 
using gradients of $\mathcal{Q}$ from solar field-line extrapolations 
as a proxy for the local reflection rate will likely \emph{underestimate} the squashing-driven heating. 
Further work is needed to characterize this regime quantitatively.

\paragraph{Summary.} The framework above reduces to classical RDT when pure
expansion dominates and generalizes it when
\(|\tsmat|\) becomes larger than $\gprl{\va}$.  It therefore supplies a single phenomenology
capable of handling both smooth and  tangled magnetic geometries.
It ignores, however,  the effect of perpendicular gradients or field-line curvature, to which 
we now attend.

\subsection{Compressive feedback on \texorpdfstring{$\zp$}{z$^{+}$}}\label{subsec:comp catalyzed}

The previous two sections considered how $\zm$, generated directly from $\zp$ via expansion ($\gprl{\va}$) or squashing ($\tsmat$),
turbulently heat the plasma by damping the  wave action (i.e., $\zp$). 
Two other processes enable changes to the wave-action  at the same formal order in our equations, but rely on $\zp$ driving compressive fluctuations via perpendicular gradients. These  always occur concurrently, and lead to the same scaling within the (drastic) approximations of our slaved phenomenology. Their mechanisms are as follows:
\begin{enumerate}
\item \emph{Direct compressive feedback (DCF).} Compressive fluctuations ($\duprl$, $\dbprl$, and $\d\rho$) are driven by $\zp$ through perpendicular gradients, curvature, and/or effective gravity (see \cref{fig: cartoons}). These compressive fluctuations then couple directly to $\zp$, to change the wave action, via the curvature and/or $\geff$ (but \emph{not} perpendicular gradients; see the final line of \eqref{eq: Alfvenic fluctuation equations main}). 
\item \emph{Compressively catalyzed reflection (CCR).} The compressive fluctuations  driven  in (i) also drive $\zm$ fluctuations via the curvature and/or effective gravity, hence ``catalyzing'' reflection of $\zp$. These $\zm$ fluctuations turbulently dissipate wave action via $\zm\cdot\nabla\zp$.
\end{enumerate}

These processes could in principle occur in isolation from the parallel reflection mechanisms (\cref{subsec:RDT_basic} and \cref{subsec:Q_reflection}) in a flux tube without parallel Alfv\'en-speed gradients or squashing (such that the second line of \eqref{eq: Alfvenic fluctuation equations main} was zero). More generally, they 
will occur in tandem with parallel-reflection mechanisms  in a flux tube of general geometry, and we 
will compare their magnitudes to understand their potential role in coronal structure.  Because they 
both involve the continual production and dissipation of compressive fluctuations, they 
necessarily also occur concurrently with Alfv\'en-catalyzed relaxation, which is treated separately below (\cref{subsec:comp_absorb}) because of its different energy source/sink (large-scale 
gradients versus wave-action dissipation).

\subsubsection{Direct compressive feedback}\label{sub: DCF}

The relevant terms in the $\zp$ equation \cref{eq: Alfvenic fluctuation equations main} are 
\begin{equation}
\frac{\partial}{\partial t} \zp|_{\rm DCF} = \dots  -\,2\bkap\!\left(U\,\duprl - \frac{B\,\dbprl}{4\pi\rho}\right)
 + \frac{\d\rho}{\rho}\,\bm g_{\rm eff}^{\perp},
\end{equation}
where we keep only the explicit curvature and effective-gravity couplings to the compressive fields.  Their contribution to the evolution of the Alfv\'enic energy (or wave action) is
\begin{flalign}
&-Q_{\rm DCF}=\frac{\partial}{\partial t} \left. W_{\perp}^{+}\right|_{\rm DCF}
= \frac{\rho}{2}\avgi{\zp\cdot\partial_{t}\zp|_{\rm DCF}}\nn\\
&\quad= -\frac{\rho}{2}\avg{\zp\!\cdot\!\Bigl[ 2\bkap\!\left(U\,\duprl - \frac{B\,\dbprl}{4\pi\rho}\right)
 - \frac{\d\rho}{\rho}\,\bm g_{\rm eff}^{\perp}\Bigr]\!}.
\label{eq:ccr_Wp_general}
\end{flalign}
Inserting  the slaved-regime amplitudes from  \cref{subsec:amplitudes} and assumptions about the isotropy of $\zp$ around $\bh$ \eqref{eq: direction of correlations}, we obtain
\begin{align}
Q_{\rm DCF}
\simeq  \rho\frac{z^{+}l_{\perp}}{2}\!\! \left[ \bkap\cdot \left(\va U \bm{\mathcal{F}}_{u}- {\vs^{2}} \bm{\mathcal{F}}_{B} \right)-  \frac{1}{2}\geff\cdot \bm{\mathcal{F}}_{\rho}\right],\label{eq: Qccr 1 first}
\end{align}
where  $\bm{\mathcal{F}}_{u}$, $\bm{\mathcal{F}}_{B}$, and $\bm{\mathcal{F}}_{\rho}$ are defined in \cref{eq: F definitions}.
 Using $\chi_{\rm A} = z^{+} l_{\|}/(\va l_{\perp})$, which is expected to be $\chi_{\rm A}\simeq 1$ for strong turbulence, we can rewrite \eqref{eq: Qccr 1 first} in the useful form:
 \begin{align}
Q_{\rm DCF} \simeq W_{\perp}^{+} \va (K^{\rm damp}_{{\rm DCF},u}+K^{\rm damp}_{{\rm DCF},B}+K^{\rm damp}_{{\rm DCF},\rho}),\label{eq: Qccr 2 first}
\end{align}
with
 \begin{align}
K^{\rm damp}_{{\rm DCF},u} &= 2 \chi_{\rm A}^{-1}\frac{U}{\va} l_{\|} \bm{\mathcal{F}}_{u}\cdot\bkap, \nn\\
K^{\rm damp}_{{\rm DCF},B}&=-2 \chi_{\rm A}^{-1}\frac{\vs^{2}}{\va^{2}} l_{\|} \bm{\mathcal{F}}_{B}\cdot\bkap, \nn\\
K^{\rm damp}_{{\rm DCF},\rho} &= -\chi_{\rm A}^{-1} l_{\|}\bm{\mathcal{F}}_{\rho} \cdot\frac{\geff}{\va^{2}}.\label{eq: Qccr K defns}
\end{align}
The  $K^{\rm damp}$ are recognized as inverse heating lengthscales,  allowing direct comparison to the expansion/squashing induced heating rate
\eqref{eq: final QRDT full}. This heating is thus governed by the ratio of the parallel correlation length $l_{\|}$ to the perpendicular lengthscales that govern compressive driving (${\mathcal{F}}^{-1}_{G}$), in addition to another lengthscale ($\kappa^{-1}$ or $\va^{2}/|\geff|$). The relative size of these terms, or even their sign, 
is not at all obvious, given the complexity of the $\mathcal{F}_{G}$ terms, and will likely vary dramatically between 
different regions; we  explore this further below.

Physically, the picture is that $\zp$ does work against curvature and $\geff$ to excite compressive motions; those motions are then rapidly damped by the cascade at rate $\omeganl$, but in the process feed back on the $\zp$ that created them. We note that nothing in the structure of \eqref{eq:ccr_Wp_general} {guarantees} that these terms heat, rather than cool, the plasma, although  if the background is (nonlinearly) stable then heating should be expected. For 
example, the term multiplying $\bkap$ in $\bm{\mathcal{F}}_{u}$ ($\bm{\mathcal{F}}_{B}$) is positive (negative), such that the signs in \eqref{eq: Qccr K defns}  ensure waves propagating through a curved field yield $Q_{\rm DCF}>0$ from both $\duprl$ and $\dbprl$; likewise, in a stably stratified system with $\gprp{\rho}$
aligned with $\geff$ (light fluid ``on top'' of heavy fluid), the negative sign of $\gprp{\rho}$ in $\bm{\mathcal{F}}_{\rho}$ yields net heating. 
In the opposite case, where the background gradients support a free-energy reservoir (but not necessarily a linear instability)  the same couplings can in principle {feed} energy back into the Alfvénic component, making $Q_{\rm DCF}<0$ and thereby growing Alfv\'enic fluctuations as they 
propagate through the region. We provide further commentary on this subtlety below (\cref{eq: ACR DCF comparison}).

\subsubsection{Compressively catalyzed reflection (via \texorpdfstring{$\zm$}{z$^{-}$})}\label{sub: CCR}

In addition to feeding back directly on $\zp$, the compressive fluctuations described in \cref{subsec:comp catalyzed}  generate  a counter-propagating $\zm$, which can then damp $\zp$ through the usual nonlinear term $\zm\cdot\nabla\zp$.  Isolating curvature, effective gravity, and the $\zp$ nonlinearity in  \cref{eq: Alfvenic fluctuation equations main} for $\zm$, and also neglecting the time-derivative because the compressive drive is assumed to balance quasi-steadily with the present turbulent dissipation, we obtain
\begin{equation}
\zp\cdot\nabla\zm
\simeq
-2\bkap\!\left(U\,\duprl - \frac{B\,\dbprl}{4\pi\rho}\right)
+ \frac{\d\rho}{\rho}\,\bm g_{\rm eff}^{\perp},
\end{equation}
or
\begin{equation}
z^{-} \sim \frac{l_{\perp}}{z^{+}}
\left|
2\bkap\!\left(U\,\duprl - \frac{B\,\dbprl}{4\pi\rho}\right)
-\frac{\d\rho}{\rho}\,\bm g_{\rm eff}^{\perp}
\right|
\end{equation}
under the usual strong-turbulence assumptions.
This $\zm$ in turn damps $\zp$ through
\begin{equation}
\frac{\partial}{\partial t} \zp|_{{\rm CCR}} =  -\zm\cdot\nabla\zp \sim - \frac{z^{-}}{l_{\perp}}\zp
\end{equation}
so the associated heating rate is
\begin{align}
&Q_{\rm CCR}
\simeq
-\frac{\rho}{2}\avg{\zp\cdot \frac{\partial}{\partial t}\zp|_{{\rm CCR}}} 
\simeq W^{+}_{\perp}\frac{ z^{-}}{l_{\perp}}
\nn\\ 
&\:\:\simeq \rho\frac{z^{+} l_{\perp}}{2}
\left| \bkap\left(\va U {\mathcal{F}}_{u}- {\vs^{2}} {\mathcal{F}}_{B}\right)-  \frac{1}{2}\geff  {\mathcal{F}}_{\rho}\right|,
\end{align}
where $ {\mathcal{F}}_{G}= |\bm{\mathcal{F}}_{G}|$.
We see that the magnitude of $Q_{\rm CCR}$ is
the same as $Q_{\rm DCF}$ \eqref{eq: Qccr 1 first}, other than the dot products in the latter; it likewise also heats according to analogous scalings as those given in \cref{eq: Qccr 2 first}, with \begin{align}
K^{\rm damp}_{{\rm CCR},u} &= 2 \chi_{\rm A}^{-1}\frac{U}{\va} l_{\|} {\mathcal{F}}_{u}\,\kappa, \nn\\
K^{\rm damp}_{{\rm CCR},B}&=2 \chi_{\rm A}^{-1}\frac{\vs^{2}}{\va^{2}} l_{\|} {\mathcal{F}}_{B}\,\kappa, \nn\\
K^{\rm damp}_{{\rm CCR},\rho} &= \chi_{\rm A}^{-1} l_{\|}{\mathcal{F}}_{\rho} \frac{|\geff|}{\va^{2}}.\label{eq: Qccr2 K defns}
\end{align} The important difference, which justifies the removal of the signs in \eqref{eq: Qccr2 K defns}, is that  CCR should  always 
\emph{dissipate} wave action to cause {heating}, because it involves nonlinear/turbulent feedback on $z^+$ via $\zm\cdot\nabla\zp$ (DCF, in contrast, acts via the linear compressive feedback terms).
In the likely more common case where the waves liberate free energy via DCF, the heating rates of the two effects should be additive, yielding
\begin{equation}
Q_{\rm CFeed} \simeq Q_{\rm DCF} + Q_{\rm CCR}\simeq 2Q_{\rm DCF},
\end{equation}
with $Q_{\rm DCF}$ given by \cref{eq: Qccr 2 first}.

Finally, we note that the ratio of $z^{-}$ produced via CCR to that produced by direct reflection is simply the same as the 
ratios of their heating rates.

\subsection{Alfv\'en-catalyzed relaxation (ACR)}
\label{subsec:comp_absorb}

Alfv\'enic fluctuations can enable the {relaxation} of large-scale perpendicular gradients of the background flow, magnetic field, and effective potential energy, with the liberated free energy ultimately dissipated into heat by the turbulent cascade.
In our transport system, this channel appears through the perpendicular source terms
$\mathcal{Y}^{\perp}_{U}$, $\mathcal{Y}^{\perp}_{B}$, and $\mathcal{Y}^{\perp}_{g}$ in \cref{eq: Y terms}, which represent fluctuation-mediated exchange with background kinetic, magnetic, and effective-potential energy reservoirs, respectively.
We term this process \emph{Alfv\'en-catalyzed relaxation} (ACR) to emphasize that $\zp$ waves act as the catalyst that allows perpendicular gradients to drive compressive fluctuations, which are then dissipated into heat. It is similar to  relaxation enabled by turbulence induced by an instability, as studied regularly in fusion contexts \cite{Jenko2002,Barnes2010,Garbet2010} (see also \cref{subsec: instabilities} below).

In contrast to compressive feedback on $\zp$ (\cref{subsec:comp catalyzed}), which heats by {damping $\zp$ energy} through compressive feedback, 
the Alfv\'enic wave action is not damped at all in ACR.  In practice, both operate concurrently, with the separation made to highlight their physical differences: the same compressive fluctuations that dissipate (ACR) also feed back on $\zp$ and can change $S^{*}_{\rm WA}$, so the corresponding heating/cooling rates should add at the level of our phenomenology. 

To quantify ACR we must work from the transport-scale thermal-energy equation \eqref{eq: thermal energy}, rather than purely from the fluctuating equations directly (i.e., the forcing terms $\bm{\mathcal{F}}_{G}$ used in CCR). This is because $\mathcal{Y}^{\perp}_{\rm th}$, which is contained implicitly  in the dissipation rate of fluctuations, represents the extraction of large-scale {thermal} free energy by fluctuations that is subsequently dissipated back into thermal energy. Thus, a fraction of the turbulent heating implied by the slaved closure (determined by $\bm{\mathcal{F}}_{G}$)  does not contribute to true net heating, because the driver of the fluctuations is the thermal energy itself (see discussion beneath \cref{eq:total_dissipation}). 
This does not imply that fluctuations driven by thermal gradients are irrelevant --- they still drive transport, which  will be estimated below in \cref{subsec: transport}.


\paragraph{A convenient frame: transport relative to flux surfaces.}
Because cross-field transport is most naturally measured relative to the motion of magnetic flux surfaces, we work in the frame moving with the effective surface velocity $\Vpsit$.
This choice removes most explicit dependence on the self-induced mean flow $\UperpII$, which is difficult to close phenomenologically and enters via $\Vpsit$ and $\Vrhot$ in \cref{eq: thermal energy}.
In this frame, the contribution of perpendicular gradient relaxation to the local heating rate can be read from the $\mathcal{Y}^{\perp}$ terms (cf.\ \cref{eq:Eth-flux-rel}) as
\begin{flalign}
&Q_{\rm ACR}
\equiv \mathcal{Y}^{\perp}_{U}+\mathcal{Y}^{\perp}_{B}+\mathcal{Y}^{\perp}_{g}
- \,\nabla\!\cdot\!(p\Vpsit) \label{eq:Q_ACR_start}\\
&\quad= - p\,\nabla\!\cdot\!\Vpsit-\,\rho U^{2}\,\VU\!\cdot\!(\bkap+\gprp{U})
+ \rho(\Vrho-\Vpsi)\!\cdot\!\geff ,
\nn
\end{flalign}
where the final term on the first line arises in order to eliminate $\UperpII$ from the flux and $\mathcal{Y}^{\perp}$ terms (see \cref{sub: third order summary}). The terms on the second line, which we use below, represent (from left to right) compressional work associated with flux-surface convergence/divergence, turbulent transport of parallel momentum, and work done moving mass across $\geff$, respectively.

\paragraph{Closure for turbulent advection velocities.}
We now estimate $\Vrho$, $\Vpsi$, and $\VU$ in terms of the outer-scale Alfv\'enic fluctuations, adopting the same strong-turbulence closure as above, with $\duprp\simeq \zp/2$ and $\dbprp/\sqrt{4\pi\rho}\simeq-\zp/2$.
From the definitions \eqref{eq: V defs} we obtain
\begin{align}
\Vpsi
&= \avg{\frac{\dbprl}{B}\,\duprp - \duprl\frac{\dbprp}{B}}
\simeq \frac{1}{2}\avg{\zp\!\left(\frac{\duprl}{\va}+\frac{\dbprl}{B}\right)\!},
\nn\\
\VU
&= \avg{\frac{\duprl\duprp}{U}-\frac{\dbprl\dbprp}{4\pi\rho\,U}}
\simeq \frac{\va}{U}\,\Vpsi
= \ma^{-1}\Vpsi,
\nn\\
\Vrho
&=\avg{\frac{\d\rho}{\rho}\,\duprp}
\simeq \frac{1}{2}\avg{\zp\,\frac{\d\rho}{\rho}} .
\label{eq:V_def_ACR}
\end{align}
To estimate the compressive fields appearing in these correlators, we use the same slaving and isotropy assumptions as in \cref{subsec:amplitudes} (\cref{eq: compressive amplitudes}), yielding
\begin{align}
\Vpsi
&\simeq \eta_{\rm turb}
\left(\bm{\mathcal{F}}_{u}+\frac{\vs^{2}}{\va^{2}}\bm{\mathcal{F}}_{B}\right),\quad \Vrho
\simeq \eta_{\rm turb}\bm{\mathcal{F}}_{\rho}, 
\label{eq:Vpsi_est_ACR}
\end{align}
where  $\eta_{\rm turb}\equiv z^{+}l_{\perp}/4$ has been defined as the $z^{+}$-induced turbulent diffusion coefficient.

\paragraph{Heating rate.}
Substituting \cref{eq:Vpsi_est_ACR} into \cref{eq:Q_ACR_start} yields a heating rate that can be written in a form
similar to our earlier damping-length notation.
We define the compressional heating rate $Q_{{\rm ACR},\psi}
\equiv
- p\,\nabla\!\cdot\!\Vpsit$, but will not attempt to estimate this as it is the only part that depends on  $\UperpII$ through $\Vpsit=\Vpsi+\UperpII$. Its physical content is the same as if $\Vpsit$ were a normal large-scale flow in the plasma (as opposed to, in part, turbulent transport), and in  the presence of continual and significant compression/expansion of magnetic-field lines,  it could cause substantial heating/cooling.  
The total heating rate can then be expressed in terms of $W_{\perp}^{+}=\rho(z^{+})^{2}/4$ and the turbulence-strength parameter
$\chi_{\rm A}=(z^{+}l_{\|})/(\va l_{\perp})$ as (cf. \cref{eq: final QRDT full,eq: Qccr 2 first}),
\begin{align}
Q_{\rm ACR}
&\simeq Q_{{\rm ACR},\psi}
+ W_{\perp}^{+}\va\!\left(
K^{\rm damp}_{{\rm ACR},U}
+K^{\rm damp}_{{\rm ACR},g}
\right),
\label{eq:Q_ACR_final_form}
\end{align}
with
\begin{align}
K^{\rm damp}_{{\rm ACR},U}
&\equiv
- \chi_{\rm A}^{-1}\frac{U}{\va}
l_{\|}\,
\biggl(\bm{\mathcal{F}}_{u}+\frac{\vs^{2}}{\va^{2}}\bm{\mathcal{F}}_{B}\biggr)\!\cdot\!
\bigl(\bkap+\gprp{U}\bigr),
\nn\\
K^{\rm damp}_{{\rm ACR},g}
&\equiv
\chi_{\rm A}^{-1}\,
l_{\|}\,
\biggl(\bm{\mathcal{F}}_{\rho}-\bm{\mathcal{F}}_{u}-\frac{\vs^{2}}{\va^{2}}\bm{\mathcal{F}}_{B}\biggr)\cdot
\frac{\geff}{\va^{2}}.
\label{eq:K_ACR}
\end{align}

\subsubsection{Cancellations and the comparison to DCF}\label{eq: ACR DCF comparison}

The ACR heating form \eqref{eq:K_ACR}  is  similar to that of direct compressive feedback (DCF)
\eqref{eq: Qccr K defns}, but differs in detail. One difference, of significant interest in the solar-wind context, is that ACR can  liberate free energy stored  in flow gradients (not only curvature- and gravity-related free energy), as signalled via the appearance of
 $\gprp{U}$ in \cref{eq:K_ACR}. The sign of $\gprp{U}$ in $\bm{\mathcal{F}}_{u}$ and $\bm{\mathcal{F}}_{B}$ is negative,  such that waves passing through regions with perpendicular flow gradients naturally liberate the free energy contained in those gradients into thermal energy; this allows, for example, the dissipation of solar-wind stream structure to heat the plasma (see \cref{subsec: stream_dissipation}). 

A second difference, which is subtle and requires further study, is that some  of the ACR heating terms  in $K^{\rm damp}_{{\rm ACR},U}$ and $K^{\rm damp}_{{\rm ACR},g}$  are  reversed compared to their counterparts in DCF.
This implies that, for some effects, heating/cooling via ACR offsets cooling/heating via DCF, such 
that free-energy of $\zp$ that is gained/lost as  it propagates is lost/gained from the \emph{background}, without changing the thermal energy.  It is helpful to consider the $\bm{\mathcal{F}}_{\rho}\cdot\geff$ term as the simplest example, which 
is positive in \eqref{eq:K_ACR} but negative in \eqref{eq: Qccr K defns}. Since $\bm{\mathcal{F}}_{\rho}=-\gprp{\rho}+\dots$, for a stably stratified 
atmosphere ($\nabla_{\perp}{\rho}$ aligned with $\geff$, with denser plasma on the ``bottom''), we see that 
$K^{\rm damp}_{{\rm ACR},g}$ is negative, while $K^{\rm damp}_{{\rm DCF},\rho}$ is positive. Physically, this corresponds to the waves turbulently flattening a density gradient and thereby \emph{increasing} the background free energy, with 
the waves damped out (via DCF) to compensate. In an unstably stratified atmosphere, 
the opposite occurs: flattening the density gradient via diffusion releases free energy, but this all goes into growing $\zp$ as
it propagates. In both cases, the DCF/ACR combination provides no net heating.

This example might serve to make a reader suspicious of our approach, or  of the DCF/ACR split in the first place (we recall that 
this split was motivated by the fact that DCF changes the wave action, while ACR does not). To this we can offer two responses. First, CCR --- which relies on true 
turbulent dissipation of $\zp$ via $\zm$ --- will act in concert and  always {heat} the plasma, thus leading (according to our closure) to net heating for both signs of stratification at the same rate at which DCF/ACR move free energy into or out of $\zp$.
Second, this type of cancellation occurs only in very specific cases;  the effect of other terms (e.g., curvature) is less 
clear,  complicated by the fact that  we
worked in the magnetic field frame for ACR, but the stationary frame for DCF (this was by necessity  to avoid computing $\UperpII$). 
Numerical simulations are clearly required to understand this ACR/DCF interplay and more complex geometrical effects.

\subsection{Transport}\label{subsec: transport}

In addition to the conversion of Alfv\'enic wave action or large-scale gradients into heat (\S\ref{subsec:RDT_basic}--\ref{subsec:comp_absorb}), the same multiscale dynamics drives cross-field transport of mass, momentum, and thermal energy.  Here, we derive these fluxes in terms of the 
large-scale gradients using the same slaved-field phenomenology as above. This yields diffusion-like equations for the large-scale fields, revealing a number of 
interestingly non-trivial effects.

Our starting point is the generic moving-flux-surface conservation law (see App.~\ref{sub: third order summary})
\begin{equation}
\left.\frac{\partial G}{\partial\tau}\right|_{\psi}+\nabla\cdot\bm{\Gamma}_{G}=S_{G},
\end{equation}
where $G\in \{\rho,\rho U,E_{\rm th}\}$ and $\left.\partial/\partial \tau\right|_{\psi}$ denotes the transport-time derivative taken in the frame moving with the magnetic field lines, so that $\bm{\Gamma}_{G}$ represents a {cross-field} transport relative to that motion.  For the conserved quantities of interest, the fluxes were given in  \cref{sub: transport in main text}:
\begin{subequations}
\begin{gather}
\bm\Gamma_{\rho}
=\rho(\Vrho-\Vpsi),\label{eq: cross field fluxes 2 rho}\\
\bm\Gamma_{\rho U}
=\rho U(\VU+\Vrho-\Vpsi),\label{eq: cross field fluxes 2 U}\\
\bm\Gamma_{\rm th}
=E_{\rm th}\left(\avg{\frac{\d p}{p}\duprp}-\gamma\Vpsi\right),\label{eq: cross field fluxes 2 th}
\end{gather}\label{eq: cross field fluxes 2}\end{subequations}
so that each transport law has the schematic form $\left.\partial/\partial \tau\right|_{\psi} G=-\nabla\cdot(G\,\bm V_{G})+\cdots$ with an associated effective transport velocity $\bm V_{G}\equiv \bm\Gamma_{G}/G$.

In order to interpret physically the appearance of terms in $\bm V_{G}$,
a useful reference point is ordinary diffusion: if $\bm\Gamma_{G}= -D \nabla_{\perp} G=-D G\gprp{G} $, then $\left.\partial/\partial \tau\right|_{\psi} G=\nabla\cdot(D\nabla_{\perp} G)$ tends to \emph{flatten} $G$ (provide transport down its gradient).  More generally, writing $\bm\Gamma_{G}=G\,\bm V_{G}$ clarifies that (i) a contribution $\bm V_{G}\propto- \nabla_{\perp}G\propto -\gprp{G}$ produces standard diffusion of $G$, while (ii) any part of $\bm V_{G}$ that does \emph{not} depend on $\nabla G$ acts as a {pinch/advection}, driving more complex dynamics (e.g., up-gradient accumulation).  In particular, if
$\bm V_{G}$ contains a term $\propto\! +\gprp{H}=\nabla_{\perp}{H}/H$ (for some other large-scale quantity $H$), then for $G>0$ this advects $G$ \emph{up} the $H$ gradient, i.e., toward regions of larger $H$ (a ``pinch'' toward maxima of $H$); likewise, a term $\propto -\nabla_{\perp}{H}$ drives $G$ toward minima of $H$.  Whether this up-gradient advection produces a net local gain of $G$ is determined by the sign of $-\nabla\cdot(G\bm V_{G})$: locally, $\nabla\cdot\bm\Gamma_{G}>0$ corresponds to a net \emph{loss} of $G$ from the patch, while $\nabla\cdot\bm\Gamma_{G}<0$ corresponds to a net \emph{gain}.  In what follows we will therefore distinguish diffusion (terms $\propto-\nabla_{\perp}{G}$ that flatten $G$) from pinch terms, which involve gradients of other fields, e.g., $\gprp{B}$ or $\gprp{U}$, and can drive systematic up-gradient transport and set nontrivial stationary profiles. Note that such ``pinch'' terms do not imply a violation of the second law of thermodynamics: they reflect the fact that the {invariant} mixed by the turbulence is not $G$ itself but some $H$-weighted combination  of variables \cite{Boxer2010}.

\paragraph{Approximate transport velocities.}
We estimate the transport coefficients using the same strong-damping ordering as in \S\ref{subsec: preliminaries} and the same velocity definitions as in \cref{eq:V_def_ACR}.  It is helpful to  recall
$\eta_{\rm turb}\equiv{z^{+}l_\perp}/{4}$, leading to
\begin{align}
&\Vpsi\simeq \eta_{\rm turb}\left(\bm{\mathcal{F}}_{u}+\frac{\vs^{2}}{\va^{2}}\bm{\mathcal{F}}_{B}\right),\nn\\&
\Vrho\simeq\eta_{\rm turb}\,\bm{\mathcal{F}}_{\rho},\qquad
\VU\simeq \ma^{-1}\Vpsi,
\label{eq: V_transport_estimates}
\end{align}
together with the pressure-velocity correlator entering \cref{eq: cross field fluxes 2},
\begin{equation}
\avg{\frac{\d p}{p}\duprp\!}
\;\equiv\;\Vp
\simeq
-\frac{\gamma}{2}\frac{\va^{2}}{\cs^{2}}\avg{\frac{\dbprl}{B}\zp\!}
\simeq
-\eta_{\rm turb}\gamma \frac{\vs^{2}}{\cs^{2}}\bm{\mathcal{F}}_{B},
\label{eq: Vp_estimate}
\end{equation}
where we have used the same isotropy assumption \eqref{eq: direction of correlations} for $\zp$. 

Substituting the definitions of $\bm{\mathcal{F}}_{u}$ and $\bm{\mathcal{F}}_{B}$ yields an explicit form for the flux-surface advection velocity,
\begin{align}
\frac{\Vpsi}{\eta_{\rm turb}}
\simeq &
-\left(1+\frac{\vs^{2}}{\va^{2}}\right)\!\left(\gprpfrac{B}+\ma\,\gprpfrac{U}\right)
\nn\\ &\quad+\frac{\vs^{2}}{\cs^{2}}(\ma+1)\bkap
+\frac{1}{\gamma}\frac{\vs^{2}}{\va^{2}}\gprpfrac{p}.
\label{eq: Vpsi_general}
\end{align}
Using the perpendicular equilibrium force balance \eqref{eq: perp equil}, \cref{eq: Vpsi_general} can be manipulated into a range of forms, but 
none are obviously more enlightening than others. Since $\VU\simeq \ma^{-1}\Vpsi$, we will not list it again here. 

Despite the complexity of $\Vpsi$,  the thermal transport velocities  simplify considerably in their cross-field form: Using \eqref{eq: V_transport_estimates}--\eqref{eq: Vp_estimate} and the identity
\(\vs^2/\va^2+\vs^2/\cs^2=1\), the curvature and pressure-gradient
pieces cancel from the cross-field density and thermal-energy fluxes, giving (without further approximation)
\begin{align}
\frac{\Vrho-\Vpsi}{\eta_{\rm turb}}
&\simeq
-\gprpfrac{\rho}
+2\left(\gprpfrac{B}+\ma\gprpfrac{U}\right),
\label{eq: Vrho_minus_Vpsi}
\\
\frac{\Vp-\gamma\Vpsi}{\eta_{\rm turb}}
&\simeq
-\gprpfrac{p}
+2\gamma\left(\gprpfrac{B}+\ma\gprpfrac{U}\right).
\label{eq: Vp_minus_gammaVpsi}
\end{align}

\paragraph{Magnetic-surface advection.}
Because $\Vpsi$ is, by construction, the transport velocity of magnetic field lines (or flux surfaces) themselves, \cref{eq: Vpsi_general} shows that the turbulence drives a familiar ``diffusive'' tendency for $B$ through the $ -\nabla_{\perp} B/B$ contribution, but with systematic drifts associated with $\nabla_{\perp}{U}$, $\nabla_{\perp}{p}$, curvature $\bkap$,   or equivalently $\bm g_{\rm eff}^{\perp}$ after using the equilibrium.  An interesting regime for the solar wind is found by assuming $\bkap$ and $\bm g_{\rm eff}^{\perp}$  are negligible and eliminating $\nabla_{\perp}{p}$ in favor of $\nabla_{\perp}{B}$; this gives \begin{equation}
\frac{ \Vpsi}{\eta_{\rm turb}}\simeq -2\gprpfrac{B}-\ma\left(1+\frac{\vs^{2}}{\va^{2}}\right) \gprpfrac{U},\label{eq:flux-surface-V}
\end{equation}
highlighting how standard diffusion ($\propto\!-\nabla_{\perp}{B}$) combines with an advective bias proportional to $-\ma\,\gprp{U}$, so that flux surfaces tend to be transported toward minima of $U$. Note that the sign convention of $U$ is fixed because we have taken both $U$
and $\zp$ to move in the same direction (outwards).

\paragraph{Density and pressure: diffusion plus a turbulent pinch.}
From \cref{eq: cross field fluxes 2 rho,eq: Vrho_minus_Vpsi} we find
\begin{equation}
\frac{\bm\Gamma_{\rho}}{\eta_{\rm turb}}
\simeq
-\nabla_{\perp}\rho + 2\rho\left(\gprpfrac{ B}+\ma\gprpfrac{ U}\right),
\label{eq: Gamma_rho_pinch}
\end{equation}
showing how the density  transport involves a genuinely diffusive part
$\bm\Gamma_{\rho,{\rm diff}}=-\eta_{\rm turb}\nabla_{\perp}\rho$, which leads to the density evolution $\partial \rho/\partial {\tau}|_{\psi} = \nabla_{\perp}\!\cdot(\eta_{\rm turb} \nabla_{\perp}\rho)+\dots$, thus smoothing density
gradients. The remaining terms are more interesting, acting as a {pinch} that transports particles up gradients of $B$
and, for outward-propagating $\zp$,   up gradients of $U$.  Similarly, using \cref{eq: cross field fluxes th} and
\cref{eq: Vp_minus_gammaVpsi} gives
\begin{equation}
\frac{\bm\Gamma_{\rm th}}{\eta_{\rm turb}}
\simeq
-\nabla_{\perp}E_{\rm th} + 2  \gamma E_{\rm th} \left(\gprpfrac{ B}+\ma\gprpfrac{ U}\right),
\label{eq: Gamma_th_pinch}
\end{equation}
showing that pressure (and hence thermal energy) is likewise subject to  diffusion, plus a pinch toward stronger $B$ and larger $U$.

If \(\ma\) is treated as approximately constant across the transport layer,
the stationary profiles implied by $\bm\Gamma_{\rho}=0$ and $\bm\Gamma_{\rm th}=0$ are 
\begin{equation}
\rho \propto B^{2}\,U^{2\ma},
\quad
p \propto B^{2\gamma}\,U^{2\gamma\ma},\label{eq: stationary pinch profiles}
\end{equation}
and, by extension, a temperature that is also peaked towards stronger $B$ and $U$,
$T\propto B^{2(\gamma-1)}\,U^{2(\gamma-1)\ma}$.
In the limit $U=0$, these reduce to the well-known ``turbulent-pinch'' effect scalings,
$\rho\propto B^{2}$ and $p\propto B^{2\gamma}$, which were previously derived from interchange/mixing and quasilinear
treatments for dipole-like confinement and magnetospheric plasmas, and observed experimentally in a levitated-dipole device
\cite{Hasegawa1987,Kouznetsov2007,Kobayashi2010,Boxer2010}. The same physics has
 emerged essentially \emph{directly} from the multiscale RMHD transport theory,  without assuming a particular  geometry
or mode structure. The method simultaneously generalizes the pinch to situations with  field-aligned flows,
showing how density, pressure, and temperature will be drawn towards faster-moving regions of the plasma. The physics has
interesting implications for the sourcing of slow wind from boundaries between open- and closed-field regions in the corona, which 
will be explored below. 

\paragraph{Momentum transport.}
Momentum transport, like heat transport,
involves additional forces --- most importantly the wave-pressure term
$\bh\cdot\nabla p_{\mathrm{tot}}^{(2)}\approx \bh\cdot\nabla W^{+}_{\perp}/2$ and $F^{\perp}_{\rm RM}$; see \cref{eq:Upar-transport-main} --- 
but turbulent momentum fluxes could play an interesting role by redistributing parallel flows across the magnetic  field (e.g., between solar-wind streams).
Unfortunately, the momentum flux $\bm\Gamma_{\rho U}=\rho U\,(\VU+\Vrho-\Vpsi)$ cannot be significantly simplified in general, due
to the complexity of $\VU=\ma^{-1}\Vpsi$ (\cref{eq: Vpsi_general}). 
For intuitive understanding, it is therefore  useful to recast the transport in terms of $U$ alone, since $U$ is the quantity
most directly associated with streams and is easier to interpret than $\rho U$. The price is that converting from a conserved
variable to a primitive one introduces additional terms that depend explicitly on $\partial_{\tau}\rho$ and on perpendicular gradients
of $\rho$ and $U$.

To obtain the transport equation for $U$ we use the identity
$\rho\,\partial U /\partial \tau|_{\psi}=\partial(\rho U)/\partial \tau|_{\psi} -U\partial \rho/\partial \tau|_{\psi}$,
and  the density transport $\partial\rho/\partial \tau|_{\psi}=-\nabla\cdot(\rho \Vrhot)$ and momentum transport equation \eqref{eq:Upar-transport-main}.
This gives, after straightforward rearrangement,
\begin{align}
\left.\frac{\partial U}{\partial \tau}\right|_{\psi}
&= -\nabla\cdot(U\,\VU)
-\Vrho\cdot(\grad_\perp{U}-U\bkap)
\nn\\
&-U\VU\cdot\left(\gprpfrac{\rho}-\bkap\right)+\nabla_\perp{U}\cdot\Vpsi +U\bkap\cdot\UperpII\nn\\&
 -\frac{1}{2\rho}\bh\cdot\nabla W_{\perp}^{+}
+\dots ,
\label{eq:U-transport-primitive}
\end{align}
where $\dots$ denotes the remaining force terms from \cref{eq:Upar-transport-main}, which are smaller in $W_{\perp}^{-}/W_{\perp}^{+}\ll1$ and $W_{\|}^{\pm}/W_{\perp}^{+}\ll1$ (except the centrifugal force $-\bh\cdot\nabla\Phi_{\rm rot}$).  The first term $-\nabla\cdot(U\VU)$ is the ``flux-like'' piece responsible for
diffusion/pinch of the flow speed itself, whereas the next two terms
have arisen from  
the conversion between $\rho U$ and $U$ (combined with the curvature contributions); they represent the momentum change from advection of velocity by the density-carrying drift, or vice versa. 

It is helpful to consider the nearly radial, far-outflow regime in which
$\bm g_{\rm eff}^{\perp}=\bkap=0$. Rearranging \eqref{eq: Vpsi_general} using the equilibrium \eqref{eq: perp equil} for $\nabla_{\perp}p$ gives
\begin{equation}
\frac{\VU}{\eta_{\rm turb}}
\;\simeq\;
-\left(1+\frac{\vs^{2}}{\va^{2}}\right)\gprpfrac{U}
-\frac{2}{\ma}\gprpfrac{B},\label{eq: Vu straight field}
\end{equation}
so that the ``flux'' term in \cref{eq:U-transport-primitive} indeed corresponds to (i) a standard diffusive smoothing of $U$ through the first term
$\propto \!-\nabla_{\perp }U$, albeit with a slightly enhanced diffusion coefficient $\eta_{\rm turb}(1+\vs^{2}/\va^{2})$ and (ii) a pinch/advection of $U$ toward weaker $B$ through the second term $\propto \!-\nabla_{\perp} B$.
This picture, which is complementary to that for the magnetic advection (\cref{eq:flux-surface-V}) and predicts a similar effect,  shows that solar-wind streams should evolve to have faster $U$ in regions of weaker $B$. This is indeed observed \cite{Bale2021}.  

A caveat is that the flux-based picture is valid only so long as the non-flux terms in \eqref{eq:U-transport-primitive} are smaller than the flux term $\nabla\cdot(U\VU)$. In turn, this condition is satisfied when $\rho$ and $U$ are nearly homogeneous, varying only a small amount in magnitude compared to 
their size. To see this, consider the background $G=G_{0}+\Delta G(\bm{x})$ (for $G=\{U,\rho\}$) with $\Delta G\ll G_{0}$ and $G_{0}$ slowly varying such that the contribution to $\nabla_\perp{G}/G$ from $\Delta G$ dominates that from $G_{0}$; then a $\Delta U$ that varies on scale $K$ causes the flux term to be $\sim K^{2}\Delta U$, while the other terms are  $\sim K^{2} (\Delta G)^{2}/G_{0}$, which is $\sim \Delta G/G_{0}$ smaller. 
This type of quasi-homogeneous background could be relevant, for example, to streams, but likely not to open-closed boundaries. A final subtlety to note is that while the $2/\ma$ coefficient in \eqref{eq: Vu straight field} becomes large for small $\ma$, the underlying momentum flux remains finite because $\rho U \VU = \rho \va \Vpsi$.

\paragraph{Remarks on the interpretation of transport physics.}
Our emphasis in this section has been on interpreting the turbulent fluxes $\bm{\Gamma}_{G}$ as \emph{outputs} of an assumed background, but it is worth stressing that the transport equations constitute a closed, nonlinear dynamical system, given a closure for the fluctuations' correlators.  In that context, our phenomenological approach has effectively assumed a separation of tasks: we estimate heating and fluxes for a prescribed background and then infer instantaneous consequences.  A fully self-consistent model would instead \emph{evolve} the background and the turbulence simultaneously, since changes in $\rho$, $B$, $U$, and $T$ feed back on the fluctuations and therefore on the dissipation and transport coefficients. 
More robust predictions will therefore come from either (i) joint time-dependent modeling of transport and fluctuation energetics (e.g., a global corona), a task we discuss  in  \cref{app:parker_like_wind}, or (ii) stability analyses of the coupled transport-closure system around a background state, to determine when the fluxes act diffusively versus when they lead to runaway reorganization. We leave these tasks for future work.

\subsection{Instabilities}\label{subsec: instabilities}

The transport theory above treats the external, solar-forced $\zp$ as the primary driver of fluctuations, with slaved fields and fluxes determined by the balance between their $\zp$ forcing and nonlinear damping at the mixing rate $\taunl^{-1}\sim\omeganl\sim z^{+}/l_\perp$.  A natural question is whether and when this picture can be superseded by {gradient-driven} instabilities that tap free energy in the mean gradients to generate fluctuations and transport. Related processes are well studied fusion gyrokinetics, where instabilities are generally agreed to be the primary driver of turbulent transport  \cite[e.g.,][]{Barnes2011,Garbet2010,Adkins2022,Nies2026,Adkins2026}. While the full range of instabilities supported by the multi-scale RMHD system is  complex and a detailed analysis belongs elsewhere (we treat several in App.~\ref{app: instabilities}), 
it is helpful to provide some simple dimensional estimates as a  minimal way to assess this competition between local-instability and wave-driven heating/transport in the solar context.

The relevant comparison is of a linear growth rate $\gamma$ to the turbulent decorrelation rate imposed by $\zp$ {at the same perpendicular scale}.  If the instability at perpendicular scale $k_{\perp}\sim l_{\perp}^{-1}$ has
\begin{equation}
\gamma \;\lesssim\; \omeganl(l_{\perp}) \sim \frac{z^{+}(l_\perp)}{l_\perp},
\label{eq:instability_swamping}
\end{equation}
then coherent growth will be sheared/decorrelated by the incoherent waves propagating through the domain.  Conversely,  with $\gamma\gtrsim \taunl^{-1}$, instability could develop and dominate the fluctuation energy, transport, and heating.

Without committing to any specific mode, a useful dimensional estimate for the \emph{fastest} growth rate supported by Eqs.~\eqref{eq: Alfvenic fluctuation equations main} and \eqref{eq: parallel mom alfv}--\eqref{eq: entropy} is  
\begin{equation}
\gamma \sim V\,|\gprp{ {G}}|\implies \frac{\gamma}{\omeganl}
\;\sim\;
\frac{V}{z^{+}}\;l_\perp\,|\gprp{ {G}}|,
\label{eq:gamma_dimensional}
\end{equation}
where ${G}$ denotes whichever combination of $p$, $\rho$, $B$, or geometric factors ($\bkap$, $\geff$) supplies the free energy for the instability, and $V$ is a characteristic speed associated with the instability's drive.   For example, buoyancy- or interchange-type instabilities in a stratified medium, which are likely 
to be of most relevance for coronal quasi-equilibria, can be cast in this class with $V\sim \cs$: with a gravitational force (buoyancy) one has $|\bm g_{\rm eff}^{\perp}|\sim \cs^{2}|\gprp{ p}|$, so that $\gamma^{2}\sim |\geff||\gprp{\rho}|\sim\cs^{2}|\gprp{\rho}|\,|\gprp{ p}|$, while magnetic interchange instabilities replace part of the gradient dependence with the magnetic curvature $\bkap$ (see Apps.~\ref{appsec: parker instability} and \ref{appsec: interchange instability}). For shear-driven modes, for example the magnetorotational instability, one should use $V\sim U$ in \eqref{eq:gamma_dimensional} instead (see Apps.~\ref{sec: magnetorotational}).

Two caveats are worth emphasizing.  First,  the RMHD ordering precludes dependence of $\gamma$ on  $|k_{\perp}|$ directly (the only appearance  $k_{\perp}$ makes in the equations is in the nonlinear terms), although instabilities generically depend on the direction of $\bm{k}_{\perp}$ compared to background gradients (see \cref{eq: full_disp_rel}). This implies that instabilities  grow at similar rates up 
to the largest perpendicular scales available in the system (i.e., those comparable to the gradient lengthscales). Given that we have assumed 
scale separation between $l_{\perp}$ and these global scales, the relevant $\omeganl$  at these larger scales would be  smaller than $z^{+}/l_\perp$ evaluated at the $\zp$ outer scale, implying instability-driven dynamics could drive larger-scale rearrangements even if \cref{eq:instability_swamping} holds at the scales where $\zp$ resides (which could, of course, also put them outside of the assumed ordering).  Second, not all instabilities obey the dimensional estimate \cref{eq:gamma_dimensional}: for example, resistive/tearing-type modes depend explicitly on dissipation parameters, as do double-diffusive instabilities, while some  shear-driven instabilities such as Kelvin--Helmholtz, fall outside the RMHD ordering assumptions.

Overall, given that $z^{+}$ and $\cs$ are likely similar in much of the corona/solar wind, while $l_{\perp}|\gprp{ {G}}|<1$, the estimate \eqref{eq:gamma_dimensional} suggests that outward-propagating Alfv\'enic fluctuations should typically dominate over instabilities for small-scale dynamics. However,  a quasi-global-scale instability that drives rearrangements on scales larger than the pre-existing $\zp$ could change this picture in specific regions.  

\label{EMPIRICAL ESTIMATES SECTION}
\section{Empirical estimates of coronal heating and transport}\label{sec: empirical}

This section is dedicated to estimating the relevance of the effects described above to our 
best-studied astrophysical example in which they may be important: coronal heating and the 
acceleration of the solar wind.
We have shown that the wave-driven heating rate depends not only on the parallel Alfvén-speed
gradient that drives classical reflection, but also on
perpendicular gradients of most background quantities, the squashing factor $\mathcal{Q}$, field-line angle (via $\geff$),
and field-line curvature. Broadly speaking, these terms  become important whenever the
background varies across the mean magnetic field.
The aim of this section is therefore twofold:
\begin{enumerate}
\item[(i)] to assemble simple profiles of
\(U\), \(B\), \(\rho\), and \(\va\)
that reproduce the canonical “fast’’ and “slow’’ solar-wind branches, allowing simple estimates of expansion-based heating, and
\item[(ii)] to use those profiles as a \emph{minimum‐complexity}
reference against which  novel effects can be
compared directly with the standard phenomenologies.
\end{enumerate}
We note that our transport equations (\S\ref{sub: transport in main text}), which allow one to evolve mean  wind-speed and temperature profiles based on the effect of the small-scale turbulence, in principle
provide a complete framework for the development of a fully self-consistent coronal and wind model based on the wave-driven heating and transport phenomenologies (see App.~\ref{app:parker_like_wind}).
However, their solution involves similar complexity to standard 2D or 3D phenomenological wind models \cite{vanderHolst2014,Lionello2014}, which  lies beyond the scope of the current work. We thus undertake the less ambitious task of providing order-of-magnitude estimates and comparing our
results to standard expansion-based reflection. A quick reference for notation is provided in \cref{tab: notation_phenom}.

\paragraph*{A minimalist approach.} 
To justify this approach, we note that the lower solar corona is observationally highly structured across neighbouring flux tubes on transverse scales from a few Mm to tens of Mm \cite{Cranmer2019}.  Polar coronal holes contain plume/interplume structure with order-unity density contrasts \cite{DeForest2007,Raymond2014}.  In addition, coronal-wave amplitudes might vary  across adjacent structures \cite{Tomczyk2007}, and high-resolution spectroscopy/rasters resolve sharp cross-field boundaries separating plasma with distinct thermodynamic properties \cite{Chitta2020}.  These observations motivate the consideration of large transverse gradients in the operators that enter our heating/transport expressions.

However, accurately capturing such  fine structure  to evaluate heating rates would
require a full 2-D or 3-D  model on which to compute relevant terms. This could be obtained from, for example, global MHD-like simulations \citep{Lionello2009,vanderHolst2014,Mikic2018},
or empirical constructions in different regions, but is necessarily highly uncertain and complex. Moreover, 
 the underlying field geometry, and thus  transverse gradients and curvature,
change dramatically over the solar cycle.
At solar maximum the dipole‐like approximation breaks down,
active‐region streamers proliferate, and high-$\mathcal{Q}$ corridors
can open and close over short timescales.

Given these uncertainties, we adopt a deliberately conservative benchmark:
the analytic field of \citet{Banaszkiewicz1998}
together with analytic wind-speed fits to the
numerical ZEPHYR solutions of \citet{Cranmer2007}.
The model  is appropriate only for solar-minimum conditions,
thereby providing a  {lower bound} on
heating effects associated with structured backgrounds, and we do not directly use its lateral structure, which 
can become exaggerated by 1D solution models \cite{Lionello2014} (see below).
If the novel effects already rival or exceed the
classical expansion term in this smoothed environment,
they must be at least as important in the real,
transversely structured corona. Likewise, we do not include any detailed
model of closed-field geometry, although the theory should apply equally well and be of similar interest to 
heating in such regions \cite{vanBallegooijen2011,Downs2016,vanBallegooijen2017}.
This exercise acts to demonstrate that, outside of smooth coronal-hole centers, the new effects are plausibly dominant at low altitudes and/or in the presence of streams;
but we emphasize that all details of the estimates we show are highly uncertain with strong time variability, serving only as order-of-magnitude estimates and motivation for future work.

\subsection{Model corona and solar wind}\label{sub: model corona and solar wind}

\begin{figure*}
\begin{center}
\includegraphics[height=0.6\columnwidth]{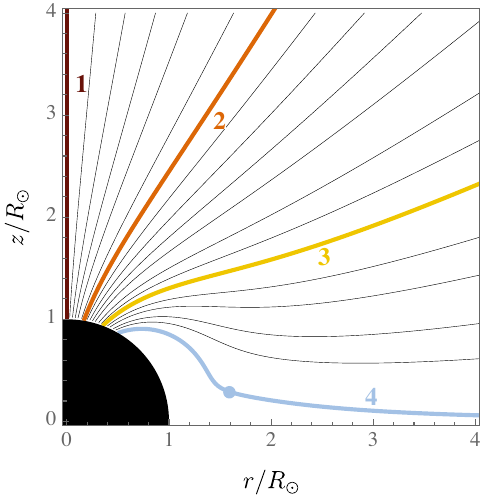}~~~
\includegraphics[height=0.6\columnwidth]{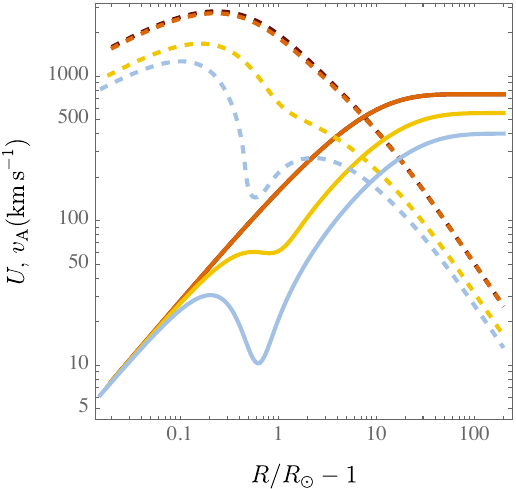}~~~
\includegraphics[height=0.6\columnwidth]{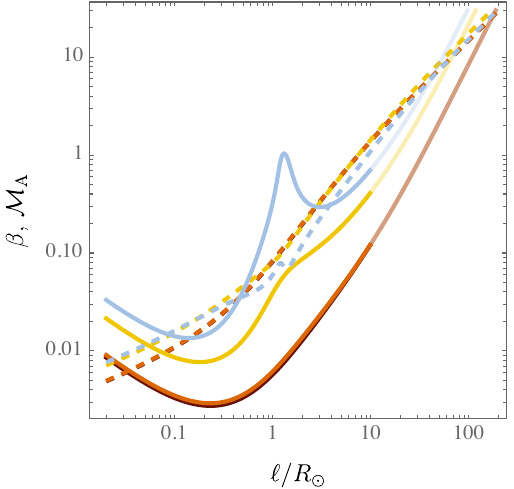}
\caption{The  coronal-wind model used in this section with  representative open flux tubes.
Left: meridional view of the corresponding magnetic geometry of \citet{Banaszkiewicz1998} (thin black curves), with the four selected flux tubes (field lines) highlighted and numbered; these span a range from polar/coronal-hole-like flux tubes (1--2) to a streamer-belt-adjacent, near-ecliptic flux tube (4). 
Middle: background outflow speed $U$ (solid) and Alfv\'en speed $v_{\rm A}$ (dashed) along four representative flux tubes of the analytic minimum-corona model, as functions of heliocentric height $R/R_{\odot}-1$ (for comparison to \citet{Cranmer2007} figure 11; note that tubes 1 and 2 mostly overlap). Right: plasma $\beta = 8\pi p /B^{2}$ (solid) and Alfv\'enic Mach number $\ma=U/\va$ (dashed) along the four flux tubes, as functions of field-line coordinate $\ell$; the lighter lines in $\beta$ beyond $\ell\approx 10\rsun$ indicate that the isothermal assumption $T=10^{6}{\rm K}$ is incorrect (see e.g., \citet{Cranmer2007} figure 9).
Together, these profiles provide a minimally structured baseline against which we test how geometric effects (squashing, curvature, and effective gravity) can modify reflection, heating, and transport relative to standard expansion-driven models. Note, however, that we do not directly compute any transverse gradients from the model for the estimates presented below.}
\label{fig: model}
\end{center}
\end{figure*}

Figure \ref{fig: model} illustrates the minimal‐complexity background atmosphere we adopt to estimate  the relevance of the heating phenomenologies described above to various regions.   As justified above, these are based on the \citet{Banaszkiewicz1998} solar-minimum-like analytic magnetic-field, with the wind speed fit from the one-dimensional flux‐tube profiles of \citet{Cranmer2007}. The latter was chosen as a standard empirical-model reference 
 of a wave/turbulence-driven corona, based on the same magnetic-field model, but the details of the wind model are not crucial to our main conclusions.

The left panel of \cref{fig: model} shows the magnetic field lines of the analytic dipole-quadrupole model (parameters $K{=}1.0$, $a_1{=}1.538$, $Q{=}1.5$; \cite{Banaszkiewicz1998}), which provides a convenient $B(r,z)$ for both coronal‐hole centers and the helmet-streamer cusp ($r$ and $z$ here are cylindrical coordinates).  Along each open field line we prescribe a solar-wind outflow speed $U$ (solid curves, middle panel) using the following simple empirical fit  to \citet{Cranmer2007}'s figure 11:
\begin{align}
U(\tilde{R})\;=&\;v_\infty
\left(1-e^{-(\tilde{R}^\alpha-R_1^\alpha)/R_{\rm out}^{\alpha}}\right)
\left(1-e^{-\tilde{R}^2/w_{\rm cusp}^2}\right)\nn\\
&+v_\infty^{{\rm max}}\,\left(1-e^{-\tilde{R}^\alpha/(R_{\rm out}^{\max})^{\alpha}}\right)
e^{-\tilde{R}^2/w_{\rm cusp}^2}.\label{eq: model wind}
\end{align}
Here $\tilde{R}(\ell)\equiv R(\ell)-R_{\odot}$ is the  distance from the solar surface, where $\ell$ is the distance measured along a  field line, and  $v_\infty$ varies between the limiting ZEPHYR speeds $v_\infty=v_\infty^{{\rm max}}=750\ \mathrm{km\,s^{-1}}$ (polar fast wind) and $v_\infty^{{\rm min}}=400\ \mathrm{km\,s^{-1}}$ (streamer-edge slow wind) reported by \citet{Cranmer2007}.  The shape parameters  
$R_1$, $w_{\rm cusp}$ and $R_{\rm out}$ are taken to be linear functions of $v_\infty$ as it is varied in the range $\{v_\infty^{{\rm min}},v_\infty^{{\rm max}}\}$ between the values $R_1/\rsun=\{0.5,0\}$, $w_{\rm cusp}/\rsun=\{0.35,1\}$ and $R_{\rm out}/\rsun=\{14,6\}$ (the ``max'' superscript indicates the value at $v_\infty=v_\infty^{{\rm max}}$ so that $R_{\rm out}^{\rm max}/\rsun=6$).
The Gaussian with width $w_{\rm cusp}^2$ is needed to reproduce the double-humped $U(\ell)$ wind profile found for high-expansion flux tubes, which 
results from the existence of more than one critical point \citep{Vasquez2003} (note that the Gaussian contributions in \cref{eq: model wind} cancel at $v_\infty=v_\infty^{{\rm max}}$ as  fast streams have a standard structure).  The dashed curves in the middle panel of \cref{fig: model} show the corresponding Alfvén speed $\va(\ell)$ obtained by (i) setting the 1 AU proton density $n_{p}$ to vary linearly from $n_p{=}2.5\ \mathrm{cm^{-3}}$ (fast wind) to $7\ \mathrm{cm^{-3}}$ (slow wind) in accord with \citet{McComas2000}, and (ii) tracing the mass flux back to the low corona under the  flux-conservation constraint $\rho U/B=\text{const}$.  The latitudinal dependence of $v_\infty$ at 1 AU is taken to be  
$v_\infty=750-350\,\exp[-(\theta_{0}/30^\circ)^4]\;\mathrm{km\,s^{-1}}$, where $\theta_{0}$ is the angle from the ecliptic, providing an approximate match to  the Ulysses latitude survey.
Finally, in order to estimate $\beta$, we supplement the model by 
assuming  fixed temperature $T=10^{6}{\rm K}$;  while inappropriate at larger radii as the wind cools,  this will only be used for near-sun computations where the wind is approximately isothermal \cite{Cranmer2007,Chandran2021}. The right panel of \cref{fig: model} shows $\beta$ and $\ma=U/\va$ along each field line. We see that $\beta$ approaches ${\sim}1$ beyond $\ell\sim \rsun$ on field-line 4; this is because $B$ decreases rapidly near the magnetic-field cusp, and is at least somewhat consistent with measurements \cite{Gary2001,Mancuso2003}.

This model yields an internally consistent set of $U$, $\rho$, $\va$,  $B$, and $p$ as a function of $\ell$. 
It thus lets us compute field-aligned gradients, expansion factors, dimensionless parameters ($\ma$ and $\beta$), and canonical reflection-driven heating rates without ambiguity, and thereby estimate the size of the physical 
ingredients  involved in the novel terms (e.g., perpendicular scale lengths)  that would be needed to have them dominate. 
While a reader may be forgiven for thinking that our field and flow \eqref{eq: model wind} are unnecessarily complex for such a task, 
the estimation is difficult to achieve  without some moderately realistic geometry, because quantities such as $U$, $\rho$, and $\va$ vary by orders of magnitude in  $R$ across short distances, but must 
remain consistent with each other (e.g., via mass and flux conservation) and with constraints from 1AU and other observations. As an example, the parallel Alfv\'en speed gradient involves both density and magnetic field gradients, which tend 
to cancel out in the low corona. This yields a maximum in $\va$, often decreasing the very gradients that drive reflection-based heating \cite{vanBallegooijen2016}.

\paragraph*{Fluctuation model and heating reference.}
In what follows we compare most of the novel heating and transport estimates to a single ``standard'' coronal-heating rate, because all candidate rates vary by orders of magnitude along a realistic flux tube.  We take this reference to be the familiar expansion/reflection-driven turbulent heating rate $Q_{\rm A}^{\exp}$ (our \cref{eq:Q_RDT}): variants of this are widely used in wave/turbulence-driven wind models \cite[e.g.,][]{vanderHolst2014,Mikic2018}, thus providing a natural normalization to understand where novel effects could be of interest.  Although the ZEPHYR solutions \cite{Cranmer2007} were obtained by evolving the wind with a different frequency-dependent reflection and phenomenological cascade model, we now argue that $Q_{\rm A}^{\exp}$ is also a reasonable approximation to the {Alfv\'enic} part of ZEPHYR's heating model, which dominates for the altitudes of focus here (see figure 9 of \cite{Cranmer2007}).  Specifically, ZEPHYR computes reflection heating by solving linear non-WKB transport equations for $z^{\pm}$ and defining a spectrum-averaged reflection coefficient $\mathcal{R}\sim z^{-}/z^{+}$ \citep{Barkhudarov1991,Cranmer2007}.  Their turbulent heating rate (their equation 47) reduces, for strong imbalance ($z^{+}\gg z^{-}$) and an efficient cascade ($\mathcal{E}_{\rm turb}\!\to\!1$ when $t_{\rm eddy}\ll t_{\rm ref}$; their eqs.~48--49), to the schematic form $Q_{\rm A}^{\rm Cran}\sim W^{+}(z^{+}/l_{\perp})\,\mathcal{R}$.  In the relevant short-wavelength  limit $|k_{\parallel}/\gprl{\va}|\gg 1$ (where  $|\gprl{\va}| = H_{\rm A}^{-1}$ in their notation), these wave equations yield the standard scaling $\mathcal{R}\sim 1/(|k_{\parallel}H_A|)$ up to order-unity factors \citep{Barkhudarov1991,Velli1993,Meyrand2025}.  Translating to our notation hence gives  $Q_{\rm A}^{\rm Cran}\sim W^{+}\va \chi_{\rm A}|\gprl{ \va}|$, where $\chi_{\rm A}\equiv z^{+}l_{\parallel}/(\va l_{\perp})$.  Thus, when $\chi_{\rm A}\sim 1$ (strong, critically balanced turbulence), $Q_{\rm A}^{\rm Cran}$ matches $Q_{\rm A}^{\exp}$ at the level needed for an order-of-magnitude reference. This  provides further justification for our use of $Q_{\rm A}^{\exp}$ to assess where  new channels are plausibly important, since it is $Q_{\rm A}^{\rm Cran}$ that drives the wind structure fitted by \cref{eq: model wind}.

To evaluate heating/transport rates we must specify combinations of $l_{\parallel}$, $l_{\perp}$, and $z^{+}$ (or equivalently $W^{+}$ and $\chi_{\rm A}$).  We therefore parameterize the fluctuation geometry using $(l_{\parallel},\chi_{\rm A})$, taking $l_{\parallel}$ from the observed dominant outer-scale period $T\simeq 2\,{\rm min}$ of near-Sun fluctuations \citep{Huang2023}, which fixes the base parallel scale as $l_{\parallel,0}\simeq Tv_{\rm A,0}\simeq 0.17\,\rsun\,(v_{\rm A,0}/1000\,{\rm km\,s^{-1}})$. We then propagate $l_{\|}$ along the tube using the WKB/eddy-stretching estimate $l_{\parallel}(\va+U)=\text{const}$.  For simplicity we adopt $\chi_{\rm A}= z^{+}l_{\|}/\va l_{\perp}=1$ (propagation critical balance \cite{Schekochihin2020}). Although turbulence may become weak ($\chia<1$) at larger altitudes because $l_{\perp}$  increases due to flux-tube expansion \cite{Chandran2019}, a reduction in  $\chi_{\rm A}$ generically \emph{increases} the relative effectiveness of compressive channels  at fixed $l_{\|}$ (see, e.g., \cref{eq: Qccr K defns}), so this assumption is deliberately conservative.  With this choice, $W^{+}$ itself cancels out of most heating \emph{ratios} against $Q_{\rm A}^{\exp}$ (the only exception is the cumulative mass/composition loading estimate in \cref{subsec:rho_transport_closed}, which requires an explicit model for $W^{+}$).  

These assumptions are intentionally minimal and are meant as a proof-of-concept normalization. A fully self-consistent global model would evolve the background  and $W^{+}_{\perp}$ together, a task that we defer to future work.

\begin{figure}
\begin{center}
\includegraphics[width=1\columnwidth]{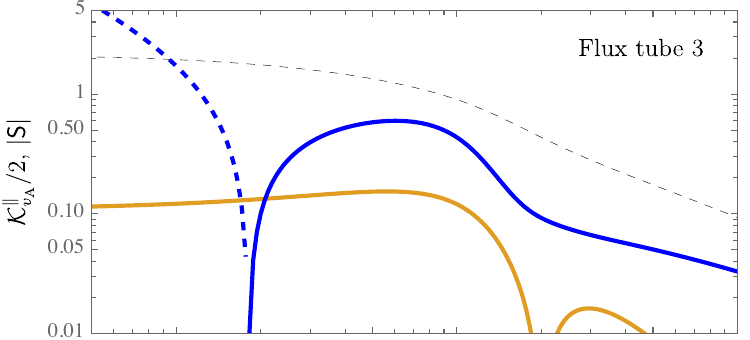}\\[-0.cm]
\includegraphics[width=1\columnwidth]{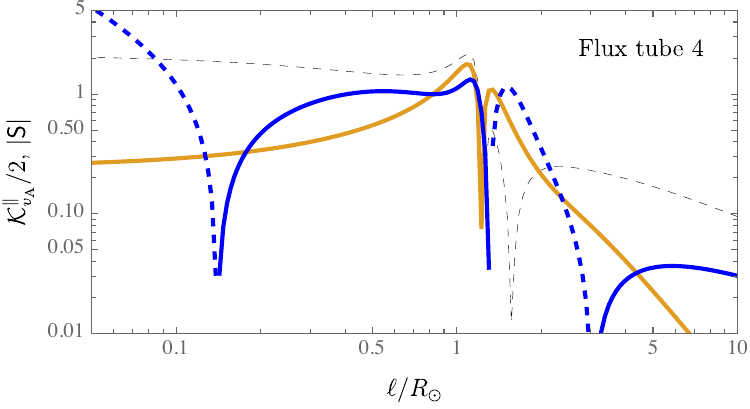}
\caption{
Comparison of squashing-driven reflection to standard Alfv\'en-speed-gradient reflection along two representative open flux tubes in the model corona of \cref{fig: model} (Flux tube~3: upper panel; Flux tube~4, near-ecliptic/streamer-belt-adjacent: lower panel). 
The standard local reflection coefficient $|\gprl{\va}|/2$ in blue (dashed regions show where $\gprl{\va}>0$) is contrasted with the local squashing rate $|\tsmat|$ that governs Q-reflection heating in orange.
The thin dashed line shows the tube expansion rate $\nabla\cdot\bh$ for reference.
Because $|\gprl{\va}|$ vanishes at extrema of $\va(\ell)$, $Q$ reflection can provide reflection and heating where expansion-driven reflection is locally suppressed. Likewise, despite the simplicity of the analytic model, $|S|$ on  the near-ecliptic tube becomes comparable to $|\gprl{\va}|/2$ over $\ell\simeq 0.8$--$3\,R_{\odot}$,  consistent with enhanced low-coronal heating in more structured regions and hence with slower-wind tendencies when heating is deposited at lower altitudes \cite{Hansteen1995}.}
\label{fig: Q estimates}
\end{center}
\end{figure}
\subsection{$Q$ reflection from squashed flux tubes}
\label{subsec:Q_reflection_compare}

To assess whether squashing-driven reflection can plausibly compete with (or dominate over) standard expansion-driven reflection in the low corona, we compare the local deformation rate $|\tsmat|$ to the usual RDT reflection coefficient $|\gprl{\va}|/2$ along representative field lines.  In the phenomenology of \S\ref{subsec:Q_reflection}, these enter on essentially the same footing: the driven inward fluctuations scale as
\(
z^{-}\sim \va l_{\perp}\big(\gprl{\va}/2+|\tsmat|\big)
\),
so the relative importance of heating from $Q$ reflection is controlled by the ratio $|\tsmat| / (|\gprl{\va}|/2)\approx |\gprl{\mathcal{Q}}|/|\gprl{\va}|$.
The comparison serves to demonstrate that, even for simple smooth  magnetic fields (in our case, analytically prescribed), reflection heating can be 
underestimated by expansion alone.

\paragraph{Computation from the magnetic field.}
Along each model field line we compute the perpendicular gradient of the unit-vector field $\bh(\bm{r})$ and extract the tensor that controls shape deformation in the perpendicular plane.  Concretely, we form the projected gradient
\(
(\nabla\bh)_{\perp} \equiv (\mathsf{I}_{\perp}-\bh\bh)\cdot\nabla\bh\cdot(\mathsf{I}_{\perp}-\bh\bh),
\)
evaluate it along the field-line trajectory, and represent it in the orthonormal perpendicular basis $\{\hat{\perpunitvectorbase}_{1},\hat{\perpunitvectorbase}_{2}\}$, where $\hat{\perpunitvectorbase}_{1}$ is perpendicular to $\bh$ in the $r,z$ plane and $\hat{\perpunitvectorbase}_{2}$ is the azimuthal unit vector.   This yields a $2\times2$ matrix that we decompose into its isotropic, symmetric-traceless, and antisymmetric parts,
\begin{equation}
(\nabla\bh)_{\perp}= \frac12(\nabla\cdot\bh)\mathsf I_\perp
		+{\tsmat}
                 +{\mathsf{A}}, \label{eq: bh decomposition 2}
\end{equation}
where $\mathsf I_\perp$ is the $2\times2$ identity in the perpendicular subspace
 and $\nabla\!\cdot\!\bh = {\rm Tr}[(\nabla\bh)_{\perp}]$ is the field-line expansion shown with thin dotted lines in \cref{fig: Q estimates}.  The quantity relevant for $Q$ reflection is the magnitude of the symmetric-traceless part,
$|\tsmat|=({\tsmat_{ij}\tsmat_{ij}}/2)^{1/2}$,
which is basis-invariant and equals the magnitude of the eigenvalues of $\tsmat$ (i.e. the local squashing/stretching rate).  As discussed in \S\ref{subsec:Q_reflection}, $\mathsf{A}$ does not contribute to reflection at RMHD order, so plotting $|\tsmat|$ provides a direct local diagnostic of the strength of squashing-driven reflection.  Combining the
field expansion $\nabla\cdot\bh$ with the density gradient $\gprl{\rho}$, we then compare $|\tsmat(\ell)|$ to the standard reflection coefficient $|\gprl{\va}(\ell)|/2$ along the same tube.

\paragraph{Results and interpretation.}
\Cref{fig: Q estimates} shows this comparison for open field lines 3 and 4 (yellow and blue in \cref{fig: model}), at latitudes of around $30^{\circ}$ and $0^{\circ}$, respectively.  Two qualitative points are immediate.

First, the standard reflection coefficient $|\gprl{\va}|/2$ is necessarily small in the neighbourhood of any extremum of $\va(\ell)$: the blue curve drops to zero at the maximum of $\va$ (e.g., Line 3 at $\ell\approx 0.2\rsun$), a consequence of the 
cancellation of $B$ and $\rho$  gradients in $\va\propto B/\sqrt{\rho}$.  This effect is exacerbated for flux tube 4  near the ecliptic  (bottom panel), which exhibits multiple $\va$ extrema and hence multiple intervals in which $|\gprl{\va}|$ becomes small.  Since $|\tsmat|$ does not vanish at $\va$ extrema, $Q$ reflection provides an additional channel that naturally ``fills in'' these gaps, smoothing the effective reflection/heating profile through $\va$-extremum regions. This could complement 
nonlocal effects that have been discussed previously for such regions \cite{Cranmer2005,vanBallegooijen2016}.

Second, even for this smooth analytic field, $Q$ reflection can be comparable to (or exceed) standard reflection over an extended low-coronal range on the near-ecliptic line: in the lower panel, $|\tsmat| \simeq |\gprl{\va}|/2$ for $\ell\simeq 0.8$--$3\,\rsun$ and $|\tsmat| \gtrsim |\nabla\cdot\bh|$ for $\ell\simeq 1.2$--$2\,\rsun$.  This implies that wave energy would be reflected and dissipated more efficiently at lower altitudes than would be inferred from expansion-driven reflection alone, consistent with the idea that increased geometrical complexity near the streamer belt preferentially enhances low-coronal heating. An enhanced wave dissipation at lower altitudes is well known to push wind solutions toward slower, denser outflows \cite{Hansteen1995}, so the trend seen here is in the correct direction for existing slow-wind phenomenology.  

Given that realistic coronal fields are far more structured than the present analytic model, the $|\tsmat|$ values inferred here are very conservative, likely even for solar minimum. Indeed,  for the near-ecliptic case shown, the total squashing remains modest, $\mathcal{Q}\approx 27$ along the entire tube, whereas extrapolations and models of the minimum corona often contain narrow corridors and quasi-separatrix layers with orders-of-magnitude larger $\mathcal{Q}$ \cite{Titov2002,Antiochos2011}, implying much larger local $|\tsmat|$. We therefore view \cref{fig: Q estimates} as a conservative demonstration that squashing-driven reflection is plausibly important across a broad range of regions outside the smoothest fields in coronal-hole centers, motivating further study.


\subsection{Compressive heating mediated by  curvature or gravity: comparison with reflection}
\label{sec: empirical compressive}

Our aim in this subsection is not to predict an absolute compressive heating rate, which would require details of perpendicular gradients in regions where our model is a poor representation of the real corona (including near the closed-open boundary of flux tube 4). Rather we ask a simpler question: {given a minimally structured background field}, what level of perpendicular structuring would be required for the compressive channels to compete with standard reflection heating?  Concretely, we compare the compressive heating (damping lengths) implied by the slaved DCF/CCR/ACR phenomenologies to the usual reflection scale $\gprl{\va}/2$, using only quantities that can be computed directly from field-aligned quantities in  the background model (field-line geometry, $\bkap$, and $\geff$). 


\paragraph{A unified scaling for DCF/CCR/ACR.}
Although direct compressive feedback (DCF), compressively catalyzed reflection (CCR), and Alfv\'en-catalyzed relaxation (ACR) differ in the detailed combinations of forcing vectors that appear in their $K^{\rm damp}$ (cf.\ \cref{eq: Qccr K defns,eq: Qccr2 K defns,eq:K_ACR}), they share the same basic structure in the strong-turbulence slaved regime:
\begin{equation}
K^{\rm damp}_{\rm comp}
\;\sim\;
\chi_{\rm A}^{-1}\ccomp\,l_{\|}
\mathcal{F}
\Thetac
\label{eq:Kcomp_generic}
\end{equation}
where $\mathcal{F}$ represents one of the forcing magnitudes built from perpendicular gradients (the various $\mathcal{F}_{u,B,\rho}$ in \cref{eq: F definitions}, with units of inverse length), and $\Thetac(\ell)=\kappa$ or 
$\Thetac(\ell) = {|\geff|}/{\va^{2}}$
is the  geometric \emph{compressive mediator} supplied by field-line curvature or effective gravity.
The dimensionless 
prefactor $\ccomp$ is $1$ for the $\geff$-mediated mechanisms (e.g., $K^{\rm damp}_{{\rm CCR},\rho}$ in \cref{eq: Qccr2 K defns}), while for the $\kappa$-mediated mechanisms (e.g., $K^{\rm damp}_{{\rm CCR},u}$ or $K^{\rm damp}_{{\rm CCR},B}$)  $\ccomp=2U/\va$ or $\ccomp=2\vs^{2}/\va^{2}$.
The notable exception to the  form \eqref{eq:Kcomp_generic} is the explicit $\gprp{U}$ factor in $K^{\rm damp}_{{\rm ACR},U}$, which is most interesting for stream structure at larger altitudes and will be treated separately below in \cref{subsec: stream_dissipation}.

\begin{figure}
\begin{center}
\includegraphics[width=1\columnwidth]{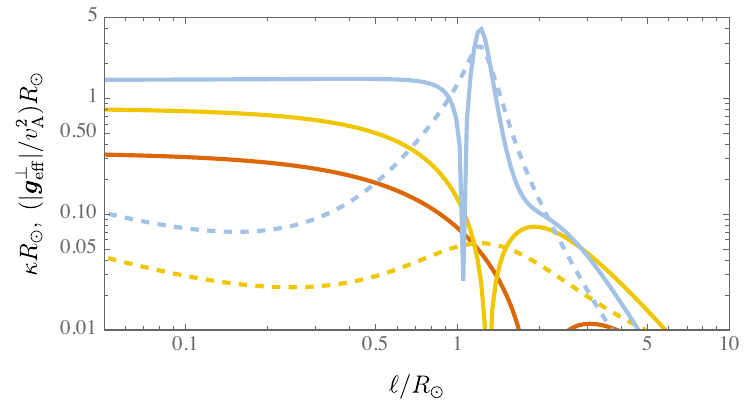}
\caption{
The ``compressive mediators'' $\Thetac$ that enable compressive coupling (see \cref{eq:Kcomp_generic}), namely the field-line curvature $\kappa$ (solid) and the normalized effective perpendicular gravity  $|\geff|/\va^{2}$ (dashed), plotted in units of $1/\rsun$.  The DCF and CCR compressive heating channels, as well as a portion of ACR, depend on the product of the 
compressive mediator with a forcing function $\bm{\mathcal{F}}$ (see text and \cref{fig: Lcrit}).}
\label{fig: Fcrit}
\end{center}
\end{figure}

\begin{figure}
\begin{center}
\includegraphics[width=1\columnwidth]{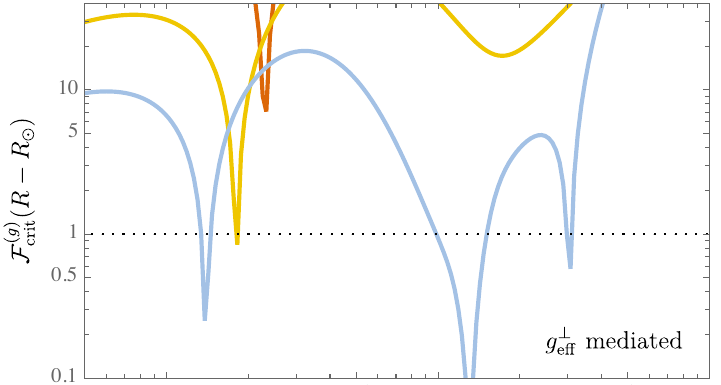}\\\includegraphics[width=1\columnwidth]{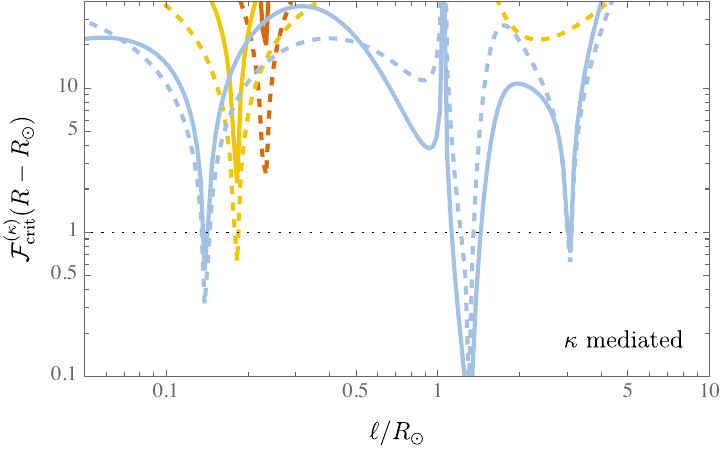}
\caption{
The critical perpendicular structure scale for compressive channels to dominate over standard reflection along flux tubes 2--4 of the model (colors as in \cref{fig: model}).  
The curves show $\mathcal{F}_{\rm crit} (R-R_{\odot})$, where $\mathcal{F}_{\rm crit}$ is defined by \cref{eq:Fcrit_def}; perpendicular structure with $\mathcal{F}_{u,B,\rho}\gtrsim \mathcal{F}_{\rm crit}$ is sufficient (within the slaved strong-turbulence closure) for the DCF/CCR/ACR compressive pathways to dominate over reflection-driven damping at that location. The top panel shows the $\geff$-mediated channels with $\ccomp=1$ (see \cref{eq: Qccr K defns,eq: Qccr2 K defns,eq:K_ACR}), and the bottom panel shows $\kappa$-mediated channels with $\ccomp=2(\vs/\va)^{2}$ (solid lines) or $\ccomp = 2\ma$ (dashed lines).  
We  normalize by height above the surface, $R-\rsun$, so that the dotted lines show  $\mathcal{F}_{\rm crit} = (R-\rsun)^{-1}$ (i.e., structure comparable to global scales),  highlighting that the required structure is not necessarily extreme in the low corona; for example, flux tube 4 in  the $\geff$-mediated heating (light blue), shows that density or magnetic structure with $\gprps{\rho,B}\gtrsim 5/\rsun$  over the range $\ell/R_{\odot}\sim 0.6$--$3$ would already make compressive heating dominate RDT  in this simple  background field.}
\label{fig: Lcrit}
\end{center}
\end{figure}

\paragraph{Measurement from the coronal model.}
The compressive mediators, $\kappa$ and $|\geff|/\va^{2}$, can be computed directly from the background coronal model.  The curvature $\bkap=\bh \cdot\nabla\bh$ is obtained from the analytic magnetic field by differentiating the unit vector $\bh$ explicitly.  The effective gravity is computed from the true solar gravitational field written in terms of the escape speed,  $|\nabla\Phi|=v_{\rm esc}^{2}/(2R)$ and therefore
$|\nabla\Phi|/\va^{2}\approx 0.2/R$ for $\va\sim 10^{3}~{\rm km\,s^{-1}}$ in the low corona.
We then project this acceleration perpendicular to the guide field, $\nabla_{\perp}\Phi=(\mathsf{I}_{\perp}-\bh\bh)\cdot\!\nabla\Phi$, and include the centrifugal contribution from field-line curvature, $U^{2}\bkap$, so that $\geff=-\nabla_{\perp}\Phi - U^{2}\bkap$.  The centrifugal term  is negligible in our model: close to the Sun $\ma\ll 1$, while at larger distances $\kappa$ is small. \Cref{fig: Fcrit} plots $\kappa(\ell)$ and $|\geff(\ell)|/\va^{2}(\ell)$, in   units of $\rsun$.


\paragraph{Critical lengthscales of background quantities.}
To quantify the critical perpendicular lengthscales needed for compressive heating to dominate, 
we compare   $K^{\rm damp}_{\rm comp}$ in \eqref{eq:Kcomp_generic} with the damping scale of standard RDT, $K^{\rm damp}_{\rm A,exp} = \gprl{\va}/2$. This defines the \emph{critical compressive forcing magnitude} for which $K^{\rm damp}_{\rm comp}>K^{\rm damp}_{\rm A,exp}$:
\begin{equation}
\mathcal{F}_{\rm crit}(\ell)
\;\equiv\;
\frac{1}{\ccomp}\,
\frac{\chi_{\rm A}\,}{l_{\|}\Thetac(\ell)} \frac{|\gprl{\va}|}{2}.
\label{eq:Fcrit_def}
\end{equation}
In a  perpendicularly  structured  background, an $\mathcal{F}\in \{\mathcal{F}_{u},\mathcal{F}_{B},\mathcal{F}_{\rho}\}$ that satisfies  $\mathcal{F}\gtrsim \mathcal{F}_{\rm crit}$ implies $K^{\rm damp}_{\rm comp}>K^{\rm damp}_{\rm A,exp}$, such that compressive heating  dominates RDT-driven heating at that location.
 \Cref{fig: Lcrit} plots $\mathcal{F}_{\rm crit}$ for the field lines 1--3 for $\Thetac=|\geff|/\va^{2}$  and $\Thetac=\kappa$ in the top and bottom panels, respectively.
Noting that the $\mathcal{F}$ expressions \eqref{eq: F definitions} contain various inverse length scales (some multiplied by dimensionless factors), we 
normalize by the (inverse) distance to the solar surface $(R-R_{\odot})^{-1}$ to  provide an immediate geometric reference.

\paragraph{Implications.}
Even for our minimally structured analytic field, the resulting critical scales $\mathcal{F}_{\rm crit}$ are not obviously extreme.  For the $\geff$-mediated channels with $\ccomp=1$ (see \cref{eq: Qccr2 K defns}), the upper panel of \cref{fig: Lcrit} indicates that over the range  $\ell/R_{\odot}\sim 0.6$--$3$ along the streamer-belt-adjacent flux tube 4 (blue line), 
density structuring ($\gprp{\rho}$ in $\bm{\mathcal{F}}_{\rho}$) on perpendicular scales $\lesssim \!0.2\rsun$ ($\gprp{\rho}\gtrsim 5$) would be sufficient for compressive damping to dominate over  reflection; such perpendicular structure scales are plausible given known coronal structuring \cite{Klimchuk2006}.  For curvature-mediated channels, we have $\ccomp=2\vs^{2}/\va^{2}$ for  $\dbprl$ channels (via $\mathcal{\bm{F}}_{B}$; solid lines) and $\ccomp=2\ma$ for $\duprl$ channels (via $\mathcal{{F}}_{u}$; dashed lines). We see that $\dbprl$ channels are potentially of similar importance to $\geff$-mediated heating on flux tube 4, due to the modest $\beta$ and curvature (see \cref{fig: model}); for example, noting the $\gprp{p}/\gamma$ in $\bm{\mathcal{{F}}}_{B}$, we see that transverse pressure structure with scales
 $\lesssim \!0.1-0.5\rsun$ would drive competitive heating rates across the range  $\ell/R_{\odot}\sim 1.2$--$3$. Although at first glance similar, the $\duprl$ channels (dashed lines) are likely less important: the $\bkap-\gprp{B}$ term in $\bm{\mathcal{F}}_{u}$ vanishes for a force-free field and the other terms pick up another $\ma$ factor, which is small in these near-Sun regions.

As with the $Q$-reflection comparison above, our estimates are deliberately conservative as the \citet{Banaszkiewicz1998} field, which determines $\bkap$ and $\geff$, is rather smooth and large scale.  The purpose of \cref{fig: Fcrit,fig: Lcrit} is therefore to provide a simple, geometry-based demonstration that compressive channels can plausibly  competite with reflection in structured regions, and to motivate more detailed modeling that includes thermodynamics and realistic cross-field structure. For example,  
these additional heating sources could potentially be included straightforwardly as
additional effects in current structured wind models (\citet{Evans2012} apply a similar approach heuristically, with promising results),
allowing more accurate determination of the dynamical importance for coronal heating and wind acceleration. Like for $Q$ reflection, we are 
drawn to observe that such effects will lead to enhanced heating 
in more structured regions, generically at lower altitudes. This would drive slower, denser outflows \cite{Hansteen1995}, 
in agreement with well-known observational associations \cite{Abbo2016}.

\begin{figure}
\begin{center}
\includegraphics[width=1.0\columnwidth]{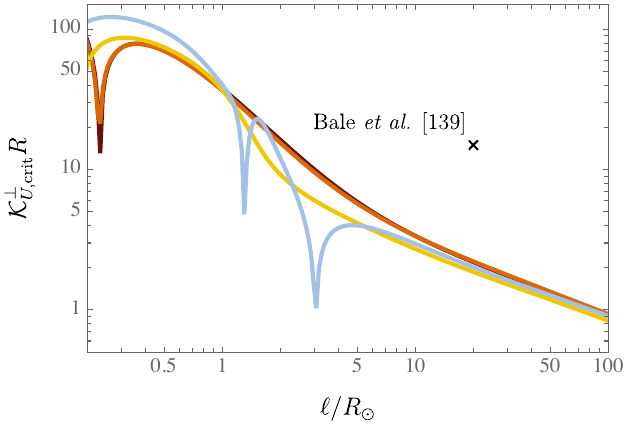}
\caption{Estimate of the perpendicular stream lengthscale required for dissipation of flow gradients (ACR) to compete with standard reflection-driven heating.  Plotted is the critical perpendicular gradient $\gprps{U,{\rm crit}}$, normalised by the heliocentric distance $R$, along field lines 1--4 of the model (color scheme as in previous figures).  By construction, transverse flow structure with $ \gprps{U}\gtrsim \gprps{U,{\rm crit}}$ is sufficient for outward-propagating Alfv\'enic turbulence to dissipate the free energy in \(\gprp{U}\) at a rate comparable to the usual reflection channel via the \(\gprp{U}\) contribution to ACR in $K^{\rm damp}_{{\rm ACR},U}$  \eqref{eq:K_ACR}.  The required transverse gradients become less restrictive with increasing $\ell$, reflecting the growing importance of this mechanism as the wind becomes trans-Alfv\'enic. The $\times$ shows an observational estimate from near-Sun PSP-observed streams \cite{Bale2021}, while at large radii even the global wind structure ($\gprp{U}\sim R^{-1}$) could be sufficient to enable the mechanism to dominate heating. }
\label{fig: stream dissipation}
\end{center}
\end{figure}

\subsection{Heating via the dissipation of flow gradients}
\label{subsec: stream_dissipation}

The ACR heating rate (\cref{eq:Q_ACR_final_form}--\ref{eq:K_ACR}) contains one qualitatively distinct contribution that we omitted in the treatment of curvature- or gravity-mediated channels  above (\cref{sec: empirical  compressive}): the  $\gprp{U}$ in $K^{\rm damp}_{{\rm ACR},U}$ enables $\zp$ to tap perpendicular gradients of the {mean solar-wind speed} $U$ and convert the associated kinetic free energy into heat.  Because such $\gprp{U}$ structure is most naturally associated with solar-wind streams and becomes more dynamically relevant as the wind accelerates (i.e., away from the highly structured lower corona), we treat this “stream-dissipation” channel separately here. 

\paragraph{Straight-field estimate.}
To isolate the basic scaling, we consider a nearly radial, far-outflow regime in which $\bkap=\bm g^{\perp}_{\rm eff}=0$, to neglect the curvature/gravity couplings in \cref{eq:K_ACR} and simplify the $\bm{\mathcal{F}}$ expressions.
The combination of forcing terms that sets $K^{\rm damp}_{{\rm ACR},U}$ is
$\bm{\mathcal{F}}_{u}+(\vs/\va)^{2}\bm{\mathcal{F}}_{B}= \ma\VU/\eta_{\rm turb}$, which in this limit
simplifies to the form already given in \cref{eq: Vu straight field}, 
so that \cref{eq:K_ACR} reduces to
\begin{align}
&K^{\rm damp}_{{\rm ACR},U}\frac{\chi_{\rm A}}{l_{\|}}
\simeq
-\ma
\left(\bm{\mathcal{F}}_{u}+\frac{\vs}{\va}\bm{\mathcal{F}}_{B}\right)\!\cdot\!\gprp{U}
\nn\\
&\quad\simeq
\ma^{2}\!\left(1+\frac{\vs^{2}}{\va^{2}}\right)\!|\gprp{U}|^{2}+ 2\ma\gprp{B}\!\cdot\!\gprp{U}.
\label{eq:K_ACR_U_straight}
\end{align}
Thus, even without curvature or gravity, ACR provides an effective damping of kinetic free energy stored in cross-field shear.  The $\ma$-dependence is the key new feature: the stream-dissipation channel is  weak deep in the corona (small $\ma$) but naturally strengthens as the wind approaches the Alfv\'en point $R_{\rm A}$ where $\ma=1$.

\paragraph{Critical shear and a transverse scale.}
As in previous sections, we compare this channel to standard reflection heating by equating $K^{\rm damp}_{{\rm ACR},U}$ to the reflection damping length $\gprl{\va}/2$.
For the purposes of a simple estimate we neglect the mixed term
$2\ma\,\gprp{B}\!\cdot\!\gprp{U}$;
this yields the critical shear, analogous to $\gprps{\rm crit}$ or $\mathcal{F}_{\rm crit}$ above,
\begin{equation}
\gprps{U,{\rm crit}}
\;\equiv\;
\ma^{-1} \left(1+\frac{\vs^{2}}{\va^{2}}\right)^{-1/2}
\left(\frac{\chi_{\rm A}}{l_{\|}}\frac{|\gprl{\va}|}{2}\right)^{1/2}.
\label{eq:Ucrit_def}
\end{equation}
To plot this, as in \cref{sec: empirical  compressive}, we take $l_{\|}\sim\va\tau$ at the base with $\tau\simeq2$ min and use $\chi_{\rm A}=1$ for simplicity, also neglecting the $1+\vs^{2}/\va^{2}$ factor, which would provide a factor $\simeq\!2$ enhancement in heating at higher $\beta$ but requires a more accurate temperature model to evaluate. The resulting $\gprps{U,{\rm crit}}$ is shown in \cref{fig: stream dissipation}.  The interpretation is that if a particular stream satisfies $\gprps{U}\gtrsim \gprps{U,{\rm crit}}$ at a given $\ell$, then wave-catalyzed dissipation of flow gradients will exceed reflection heating within our closure.

Note that, although it is omitted from \cref{fig: stream dissipation}, the mixed contribution
$2\ma\,\gprp{B}\!\cdot\!\gprp{U}$ in \cref{eq:K_ACR_U_straight} could be  important at lower altitudes because it scales as $\propto\ma$ rather than $\propto\ma^{2}$. 
Its comparative magnitude and sign depend on the relative orientation of the $B$ and $U$ gradients and therefore the detailed stream structure, suggesting interesting possibilities such as misaligned gradients \emph{suppressing} the dissipation of streams at low altitudes; we defer such refinements to future work.

\paragraph{Implications for stream evolution.}
Because of the explicit $\ma^{2}$ scaling in \cref{eq:K_ACR_U_straight}, this channel naturally “turns on’’ as the wind accelerates, suggesting a simple route by which stream structure can be dissipated into heat beginning around the Alfv\'en point (where $\ma\sim1$) and continuing outward.
Indeed, in situ measurements show pronounced, patchy transverse structure in the young solar wind, including Alfv\'enic switchback/microstream intervals with order-unity changes in $B$ and significant changes in $\ma$ and $U$ over modest angular extents \cite{Bale2021,Fargette2021}.  Using the transverse velocity contrast $\Delta U/U$ across a structure of perpendicular wavenumber $K_{\perp}$ to estimate $\gprps{U}\sim(\Delta U/U)\,K_{\perp}$, the results of  \citet{Bale2021} conservatively suggest $\Delta U/U\simeq 1/5$ on scales of around $5^{\circ}$ longitude ($K_{\perp}\simeq 70/R$), giving 
$\gprps{U} \simeq15/R$ at $R\simeq 20 \rsun$. As shown on \cref{fig: stream dissipation}, this is well above $\gprps{U{\rm crit}}$,  suggesting that wave-catalyzed dissipation of cross-field shear should dominate significantly over reflection-driven heating in comparably structured flows. This offers a possible  explanation for the recent results of \citet{Sioulas2025}, who show that heating rates predicted by RDT for $R>R_{\rm A}$ are less than those inferred directly from the plasma's thermodynamics. 
Likewise, this heating will necessarily erode the stream structure itself (as predicted
by our transport equations; see \cref{subsec: transport}) and  it is indeed observed that lateral stream structure  disappears as 
the wind propagates well beyond $R_{\rm A}$ \cite{Horbury2023}.

At larger radii, we see that even \emph{global} wind structure, with $\gprps{U}\sim R^{-1}$ should 
be sufficient to enable heating that dominates that from reflection, although our model here does not include a Parker spiral, which will grow in importance with $R$.

\paragraph{Relation to shear-driven heating models.}
A long-standing picture for solar-wind heating and stream evolution invokes dissipation triggered by velocity shear instabilities, treating stream interfaces or global structure as an \emph{injection} mechanism for fluctuations \cite{Roberts1987,Roberts1992}.  ACR heating is thus a qualitatively distinct mechanism and the classical Kelvin-Helmholtz (KH) instability often invoked for shear injection lies outside the multiscale RMHD ordering because it requires \(k_{\|}\!\sim\!k_{\perp}\) (indeed,  a pure shear, $\gprp{U}$, is linearly stable in our equations; see App.~\ref{app: instabilities}).
In the heliospheric turbulence-transport literature, shear driving is commonly modelled as a source term that amplifies the fluctuation energy at a rate \(\gamma_{\rm sh}\sim C_{\rm sh}U/r\) \cite{Zank1996,Matthaeus1999,Breech2008,Oughton2011,Zank2017}, where \(C_{\rm sh}\equiv (\Delta U/U)(r/\Delta r)\) encodes a transverse velocity jump \(\Delta U\) across a layer of width \(\Delta r\), so that \(\gamma_{\rm sh}=\Delta U/\Delta r\). In this phenomenology, $C_{\rm sh}$ is an effective efficiency factor that encapsulates how strongly the mean shear adds energy to fluctuations; this could involve direct driving by KH instability, or a shear enhancement of existing 
fluctuations (\cite{Roberts1992}; a situation similar in spirit to ACR, although, as shown below, with different predicted scalings).

A minimal mixing-length heating estimate for shear injection is then \(Q_{\rm sh}\sim \rho\,\delta u^{2}\gamma_{\rm sh}\), with \(\delta u\sim l_{\perp,{\rm sh}}\gamma_{\rm sh}\) set by the scale of the resulting turbulence \(l_{\perp,{\rm sh}}\) (in the simplest picture, \(l_{\perp,{\rm sh}}\sim \Delta r\)). This gives \(Q_{\rm sh}\sim \rho\,l_{\perp,{\rm sh}}^{2}(C_{\rm sh}U/r)^{3}\). By contrast, our ACR shear channel corresponds to fluctuation-catalyzed dissipation of {existing} flow gradients rather than fluctuation injection followed by dissipation; using \(\gprp{U}\sim (\Delta U/U)/\Delta r = C_{\rm sh}/r\) and  the same approximations as in Fig.~\ref{fig: stream dissipation} yields \(Q_{{\rm ACR},U}\sim \rho\,z^{+}l_{\perp}(C_{\rm sh}U/r)^{2}\). Their ratio is therefore
\(Q_{\rm sh}/Q_{{\rm ACR},U}\sim (\Delta U/z^{+})(l_{\perp,{\rm sh}}/\Delta r)(l_{\perp,{\rm sh}}/l_{\perp})\),
so shear-driven injection can dominate for large velocity contrasts and/or when shear injects at large \(l_{\perp,{\rm sh}}\).

However, the physical interpretation of the shear source amplification rate  $\gamma_{\rm sh}$ is not unique: it either presupposes some generic mechanism by which the mean shear continually transfers energy into fluctuations (without specifying the detailed route \cite{Roberts1992}), or else it is meant to represent injection by a specific linear instability such as KH. In a sense, ACR provides a specific physical 
 mechanism for realising the former case in the presence of pre-existing Alfv\'enic fluctuations $z^{+}$ (its different scaling
 arises because the $z^{+}$ provides the turbulent damping, as opposed to the shear-produced fluctuations).    In the latter case of a linear instability, a magnetized plasma imposes an additional and physically important constraint: in the simplest incompressible setting with a field parallel to the shear, instability requires sufficiently super-Alfv\'enic jumps, \(\Delta U \gtrsim 2\va\) (up to order-unity factors depending on field obliquity and compressibility \cite{Chandrasekhar1961,Miura1982}). In many solar-wind contexts this favors regions where magnetic tension is weakest (e.g. reduced \(\va\),  particularly large \(\Delta U\), and/or $\ma\gg1$), whereas elsewhere the shear may remain  stable and the injection picture is correspondingly less direct. In such regimes the ACR channel provides a complementary route: it does not rely on  instability, but instead predicts a {continual} dissipation of \(\gprp{U}\) mediated by outward-propagating $\zp$, with an efficiency that strengthens as \(\ma\) increases (cf.\ Fig.~\ref{fig: stream dissipation}). This offers an ``Alfv\'enic'' mechanism for relaxing stream gradients or large-scale wind structure  when shear-driven  injection is weak or absent, particularly at small $\Delta U$.

\begin{figure}
\begin{center}
\includegraphics[width=1.0\columnwidth]{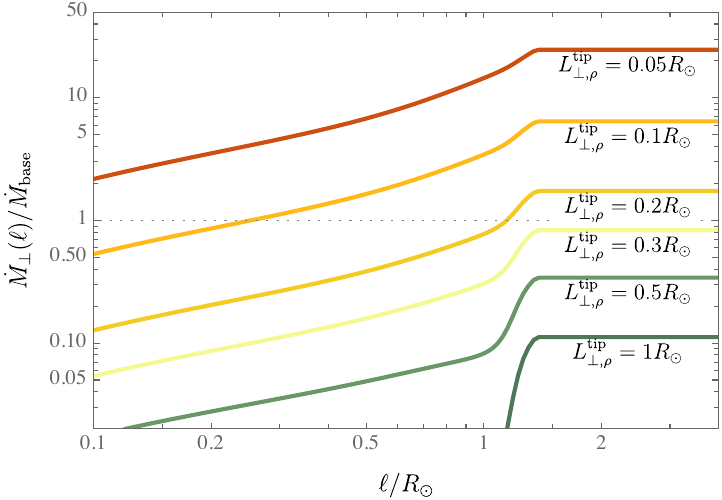}
\caption{Cumulative mass (and composition) loaded onto an open flux tube by cross-field transport ($\bm\Gamma_{\rho}$) from a neighboring closed-field reservoir.  We plot the cumulative added mass flux $\dot{M}_{\perp}(\ell)$ along flux tube~4, normalized to the “base” mass flux $\dot{M}_{\rm base}\equiv \rho U L_{\perp,\rho}^2$ of a tube of the same width.  Transport is assumed to operate only up to the ``tip'' of the streamer/closed-field region (marked at $\ell_{\rm tip}\simeq 1.35,\rsun$ in Fig.~\ref{fig: model}), after which $\dot{M}_{\perp}$ saturates --- if a curve crosses unity (dotted line) before $\ell_{\rm tip}$, it implies that the mass loading from cross-field transport would exceed that assumed to come from the base.  The different curves correspond to different assumed boundary widths at the closed-field tip, $L^{\rm tip}_{\perp,\rho}$, with $\gprps{\rho}\sim L_{\perp,\rho}^{-1}$ mapped along the tube by flux expansion.  Even relatively broad boundaries ($L^{\rm tip}_{\perp,\rho}\sim 0.2$–$0.3,\rsun$) can supply an $\mathcal{O}(1)$ fraction of the slow-wind mass flux, implying that continual turbulent transport across open/closed interfaces could substantially mass- and composition-load open-field plasma.}
\label{fig: density transport}
\end{center}
\end{figure}
\subsection{Density and composition transport from closed-field regions}
\label{subsec:rho_transport_closed}

An interesting consequence of the transport system (\S\ref{subsec: transport}), which seems to have been under-appreciated in past literature (but see  \citet{Magyar2019}),  is that Alfv\'enic fluctuations generically drive a \emph{cross-field} mass flux whenever they propagate through regions with perpendicular density gradients.  In our closure, this flux is mediated by the same slaved compressive responses that underlie \S\ref{subsec:comp_absorb}--\S\ref{subsec: transport}: the fluctuations generate an effective turbulent diffusivity/pinch, and the corresponding cross-field density flux,
\(\bm\Gamma_{\rho}=\rho(\Vrho-\Vpsi)\) \eqref{eq: cross field fluxes 2 rho},
acts to mix plasma across flux surfaces.  In the corona this opens up a continuous pathway for plasma (and therefore its compositional signatures) to leak from dense closed-field regions into neighboring open flux, supplementing whatever mass is supplied along the open tube from the low atmosphere or reconnection.

To illustrate the potential magnitude of this effect, we estimate the cumulative mass loading of an open field line by transport from a neighboring closed region.  We focus on open flux tube 4 of the minimal coronal model (Fig.~\ref{fig: model}), and imagine that for \(\ell<\ell_{\rm tip}\) it is adjacent to a pseudo-/helmet-streamer-like closed region that is denser and slower than the open tube, with cross-field transport ceasing once the closed flux terminates around the streamer tip at $\ell\simeq \ell_{\rm tip}$.  We take \(\ell_{\rm tip}=1.35\,\rsun\), indicated by the dot in Fig.~\ref{fig: model}.

\paragraph{A simple flux-tube width model.}
We do not know the perpendicular width over which the closed-to-open transition occurs, so we parameterize it by a flux-tube width \(L_{\perp,\rho}\).  Specifically, we prescribe the width at the streamer-tip location,
\(L_{\perp,\rho}^{\rm tip}\equiv L_{\perp,\rho}(\ell_{\rm tip})\),
and extrapolate it along the open line using magnetic-flux conservation, i.e.,
$L_{\perp,\rho}(\ell)\;\simeq\;L_{\perp,\rho}^{\rm tip}[{B(\ell_{\rm tip})}/{B(\ell)}]^{1/2}$.
This ansatz is not self-consistent once significant mass is added --- the tube will gain mass flux compared to that added at the base --- but it provides a transparent way to connect \(\gprp{\rho}\) and \(\gprp{U}\) to a single geometric scale, and is sufficient for an order-of-magnitude assessment.

For comparison we define a “base” mass flux along the open tube associated with the same width,
\[
\dot M_{\rm base}\;\equiv\;\rho(\ell)\,U(\ell)\,L_{\perp,\rho}^{2}(\ell),
\]
which is constant with \(\ell\) by construction in our simple coronal model. This provides a useful 
normalization, allowing direct comparison of the transport-supplied mass with the mass flux associated with $\simeq\!400{\rm kms}^{-1}$ slow wind  (the absolute normalization of the density itself cancels out of this plotted ratio, because we assume the density in the closed region also scales with $\rho(\ell)$).

\paragraph{Cross-field density flux and cumulative mass loading.}
Using the slaved closure, the cross-field density flux is \eqref{eq: Vrho_minus_Vpsi},
\begin{equation}
\bm\Gamma_{\rho}
\simeq
\rho\,\eta_{\rm turb}
\Bigl[-\gprp{\rho}+2\bigl(\gprp{B}+\ma\,\gprp{U}\bigr)\Bigr],
\end{equation}
with \(\eta_{\rm turb}\equiv z^{+}l_{\perp}/4\).  It is useful to rewrite the prefactor as
\(\rho\,\eta_{\rm turb}=\chi_{\rm A}^{-1}W_{\perp}^{+}l_{\|}/\va\),
as for the heating rates discussed above, in order to use the same estimates for $l_{\|}$ and assume $\chi_{\rm A}\simeq1$.
We then estimate the \emph{cumulative} mass flux loaded into the open tube by integrating the cross-field flux over the boundary length encountered along the tube of width $L_{\perp,\rho}(\ell)$.  This gives
\begin{flalign}
&\dot M_{\perp}(\ell)
\;\sim\;
\int_{0}^{\min(\ell,\ell_{\rm tip})}
\!\!\mathrm{d}\ell'\;
|\bm\Gamma_{\rho}(\ell')|
L_{\perp,\rho}(\ell')
\label{eq:Mdot_add_def}\\
&\;\simeq
\int
\mathrm{d}\ell'\;
\frac{W_{\perp}^{+}\,l_{\|}}{\va\,\chi_{\rm A}}
\left|
-\gprp{\rho}
+2\bigl(\gprp{B}+\ma\,\gprp{U}\bigr)
\right|
L_{\perp,\rho}.\nn
\end{flalign}
To close the gradients, we take the neighboring closed region to be denser and slower than the open tube across the same interface, so that the gradients oppose each other, setting \(\gprp{\rho}\sim -L_{\perp,\rho}^{-1}\) and \(\gprp{U}\sim +L_{\perp,\rho}^{-1}\) (with this sign convention \(\dot M_{\perp}>0\) injects plasma from closed to open).  The \(\gprp{B}\) contribution is taken directly from the background coronal model; the \citet{Banaszkiewicz1998} configuration is force free over the region we consider, so   \(\gprp{B}=\bkap\), providing an intuition for how this contribution changes sign near the streamer tip.

We set the base fluctuation amplitude to \(z^{+}\simeq20~\mathrm{km\,s^{-1}}\), a conservative estimate of low-coronal Alfv\'enic motions inferred from non-thermal line widths and transverse-wave measurements \cite{Tomczyk2007,McIntosh2011}, and propagate \(z^{+}(\ell)\) using the same reflection/WKB estimate as in \cref{eq: standard phen zp}.  This is not fully self-consistent since the transport and heating we compute will change $z^{+}$ and it applies most readily to the open tube rather than the closed region, but it captures some leading geometrical and damping trends required for an empirical estimate.

\paragraph{Results: Transport can supply an {order-unity} fraction of the slow-wind mass flux.}
Figure~\ref{fig: density transport} shows \(\dot M_{\perp}(\ell)/\dot M_{\rm base}\), i.e.\ the cumulative mass flux after transport loading, normalized by the base flux for the same \(L_{\perp,\rho}\).  Different curves correspond to different interface widths at the streamer tip, \(L_{\perp,\rho}^{\rm tip}/\rsun= \{0.05,0.1,0.2,0.3,0.5\}\).  Two qualitative points stand out.

First, even relatively broad cross-field transition layers can drive substantial mass loading: for \(L_{\perp,\rho}^{\rm tip}\simeq 0.3\,\rsun\) (which, for reference,  gives $L_{\perp,\rho}\simeq 0.05\rsun \approx 35{\rm Mm}$ at $\ell=0.1\rsun$), the integrated transport contribution becomes comparable to the  mass flux of the open tube in this model.  In other words, a continuous, turbulence-mediated leakage of plasma from closed regions can in principle supply an \emph{order-unity} fraction of the mass flux of the slow wind, without requiring discrete eruptive events.

Second, the “bump” in \(\dot M(\ell)\) near \(\ell\simeq\ell_{\rm tip}\) is associated with the cusp-like geometry of the streamer-tip region, where the field becomes weaker towards the closed-field region thus flipping the direction of $\gprp{B}$ compared to $\gprp{\rho}$.  The effect is a manifestation of the ``pinch'' towards regions of stronger field predicted by our transport theory and supported by experiment (see \cref{subsec: transport} \cite{Boxer2010}); its effect relative to the density-gradient-induced transport increases with $L_{\perp,\rho}^{\rm tip}$ because it is a consequence of the magnetic geometry only.  This feature --- enhanced closed to open-field transport near  pseudo-/helmet-streamer tips --- should be generic because such  configurations involve field-strength minima near the cusp, thus driving plasma into surrounding regions with stronger field. 
\paragraph{Implications for slow-wind structure and composition.}
Because \(\bm\Gamma_{\rho}\) is a flux of plasma mass, it will carry whatever composition and charge-state signatures characterize the neighboring closed region.  Thus, even if the open tube is fed from the base by relatively “fast-wind-like” plasma, continuous cross-field transport provides a natural route for mixing in the enhanced FIP bias, charge states, and other closed-field signatures often associated with slow wind \cite{Geiss1995,Abbo2016,Verscharen2019}.  This mechanism is complementary to interchange reconnection and streamer-top release of blobs/outflows: those processes can inject plasma intermittently, whereas the transport channel described here is continuous and directly tied to the presence of pre-existing turbulence and cross-field gradients.  

Finally, while we have focused on streamer-adjacent open flux for concreteness, the same mechanism can operate wherever strong cross-field density contrasts coexist with driven  Alfv\'enic turbulence, particularly near  active-region boundaries. The same cross-field fluxes should also mediate mixing {between neighboring closed loops/strands}, implying that density and compositional contrasts below a diffusion length are erased over a diffusion time, setting a minimum transverse scale for long-lived structuring at a given turbulence level. 
This connects naturally to the observation that coronal EUV emission frequently appears organized into finite-width threads/strands, although it remains observationally difficult to separate real substructure and instrumental/line-of-sight effects \citep[e.g.,][]{Aschwanden2005,DeForest2007,Reale2014,Morton2023a}.

\begin{figure}
\begin{center}
\includegraphics[width=1.0\columnwidth]{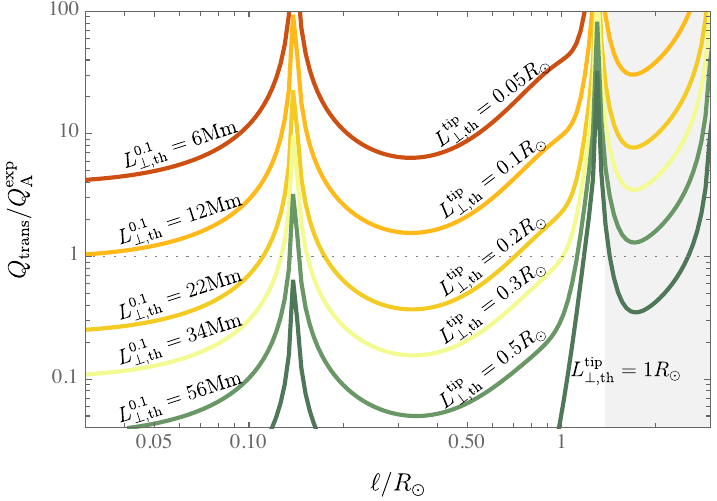}
\caption{
Estimate of the effective heating due to cross-field heat transport ($\bm{\Gamma}_{\rm th}$) from a hotter  closed-field region into a neighboring open flux tube (flux tube 4 in \cref{fig: model}). We plot $Q_{\rm trans}/Q_{\rm A}^{\exp}$, where $Q_{\rm trans}$ is the local heating/cooling implied by the divergence of the cross-field thermal-energy flux and $Q_{\rm A}^{\exp}$ is the expansion-driven RDT reference rate \eqref{eq:Q_RDT}. Curves correspond to different assumed boundary widths at the cusp, $L_{\perp,{\rm th}}^{\rm tip}$, mapped along the tube by flux expansion (as in \cref{fig: density transport}). The closed-field tip  $\ell_{\rm tip}\simeq1.35\,R_{\odot}$ --- beyond which there is no more heating because there is no neighboring closed region --- is marked by the gray shading, and the horizontal dotted line marks where $Q_{\rm trans}$ will dominate over reflection heating. The implied effective heating can be substantial close to the Sun, even for a relatively smooth open/closed boundary layer.}
\label{fig: heat transport}
\end{center}
\end{figure}

\subsection{Heat transport from closed-field regions}
\label{subsec: heat transport}

Just as the density flux enables cross-field mass exchange, the thermal-energy flux in \cref{eq: cross field fluxes 2 th} implies a \emph{cross-field heat transport} whenever an open-flux tube lies adjacent to a hotter closed-field region (helmet/pseudo-streamers, active-region loops, etc.).  In this case, $\nabla\!\cdot\bm\Gamma_{\rm th}$ acts as an effective local heating term for the open region (and likewise, a cooling in the closed region) providing continuous effective heating of
the open-field plasma without requiring eruptive release.

\paragraph{A simple boundary geometry and the required temperature contrast.}
For concreteness we adopt the same geometry as in \cref{subsec:rho_transport_closed}: open field line~4 is imagined to run adjacent to a closed-field region up to $\ell_{\rm tip}$ (the dot in Fig.~\ref{fig: model}), beyond which the adjacency ends.  The magnitude of the temperature contrast across helmet/pseudo-streamer boundaries is not tightly constrained in general, but there is some observational evidence that they are hotter than neighboring coronal-hole/open-flux regions  \citep[e.g.,][]{Ichimoto1996}.  Since the transport we estimate below is  proportional to the transverse temperature and pressure gradients, the key requirement for this mechanism to matter is simply an $\mathcal{O}(1)$ contrast across a finite boundary width.  The same logic will apply to active-region boundaries, where both temperature contrasts and structure scale are typically more extreme.

\paragraph{Thermal-energy flux and a temperature-only heat flux.}
The cross-field term in the thermal-energy equation \eqref{eq: thermal energy} is the divergence of
$\bm\Gamma_{\rm th}=E_{\rm th}\left(\Vp-\gamma\Vpsi\right)$ (\cref{eq: cross field fluxes 2 th}), where $\Vp\equiv\avg{(\d p/p)\,\duprp}$.
Under the slaved closure (see \cref{eq: Vp_minus_gammaVpsi}),
\begin{equation}
\bm\Gamma_{\rm th}
\simeq
-{\eta_{\rm turb}} E_{\rm th}\left[\gprp{p}
-2\gamma(\gprp{B}+\ma\,\gprp{U})\right],
\label{eq:heatflux_Eth_form}
\end{equation}
so $\bm\Gamma_{\rm th}$ contains both genuine heat transport and the thermal-energy component associated with mass transport (since $E_{\rm th}\propto\rho T$).
If the goal is to isolate {heating from temperature exchange} across the boundary, it is convenient to form a temperature-only flux by subtracting the thermal-energy flux implied by $\bm\Gamma_{\rho}$:
\begin{align}
\bm{V}_{T}&\equiv \frac{\bm\Gamma_{T}}{T}
\equiv\;
\frac{\bm\Gamma_{\rm th}}{E_{\rm th}}-\frac{\bm\Gamma_{\rho}}{\rho}
\nn\\&\simeq
-\eta_{\rm turb}\left[
\gprp{T}-2(\gamma-1)(\gprp{B}+\ma\,\gprp{U})
\right].
\label{eq:Gamma_T_def}
\end{align}
In this form, $-\eta_{\rm turb}\,E_{\rm th}\,\gprp{T}=-\eta_{\rm turb}E_{\rm th}\nabla_{\perp} T/T$ is the diffusive-like heat flux that results from temperature gradients, while the remaining terms constitute a pinch that transports temperature toward larger $B$ and (for outward-propagating $\zp$) toward larger $U$.

\paragraph{A local heating estimate and comparison to RDT heating.}
To connect with the earlier heating phenomenology, we write $\eta_{\rm turb}$ in terms of $W_{\perp}^{+}$ and $\chi_{\rm A}$ as above
and estimate the local heat input into the open region as the divergence of \eqref{eq:Gamma_T_def},
$Q_{\rm trans}\sim -\nabla\!\cdot\!\left(E_{\rm th}\bm V_{T}\right)$.  Using $E_{\rm th}=\rho\va^{2}\beta/[2(\gamma-1)]$ and estimating $\nabla_{\perp}\!\cdot(\cdots)$ by a transverse scale $L_{\perp,{\rm th}}(\ell)^{-1}$ allows straightforward comparison to the expansion-driven RDT heating rate $Q_{\rm A}^{\exp}$ (\cref{eq:Q_RDT}):
\begin{align}
\frac{Q_{\rm trans}}{Q_{\rm A}^{\exp}}&
\simeq 
\frac{\beta}{2(\gamma-1)}\chi_{\rm A}^{-1}l_{\|} \times \nn\\ &
\frac{\gprp{p}\cdot\left[
\gprp{T}-2(\gamma-1)(\gprp{B}+\ma\,\gprp{U})
\right]}{|\gprl{\va}|/2}.
\label{eq:Qtrans_over_QRDT}
\end{align}
This form is schematically identical 
to \cref{eq:Kcomp_generic} for ACR, DCF, and CCR, with  the compressive mediator replaced by  the gradient combination in $\bm\Gamma_{T}$ and $c_{\rm comp }\propto \beta$.

To evaluate \cref{eq:Qtrans_over_QRDT} empirically we make the same simplifying assumptions as in \cref{subsec:rho_transport_closed}: we parameterize the transverse gradients by a boundary width at the streamer tip ($\ell\approx 1.35 \rsun$), $L_{\perp,{\rm th}}^{\rm tip}$, and propagate it along the tube by magnetic-flux expansion, $L_{\perp,{\rm th}}(\ell)\propto B(\ell)^{-1/2}$.  We take the gradients $\gprp{p}$, $\gprp{T}$, and $\gprp{U}$ to be $\sim L_{\perp,{\rm th}}(\ell)^{-1}$ and aligned such that the closed region is hotter, denser, and slower than the open tube.  The remaining geometric factor $\gprp{B}$ is computed directly from the model (and changes sign along the field line).  We use the same outer-scale inputs as before ($l_{\|}\sim \va\tau$ with $\tau\simeq 2$~min and $\chi_{\rm A}=1$).

\paragraph{Empirical result and interpretation.}
The resulting estimates are shown in \cref{fig: heat transport}, where we also list the implied boundary widths around $\ell\simeq 0.1\,\rsun$ to  illustrate that  large low-coronal effective heating rates may not require unrealistically sharp gradients. We see that  the implied $Q_{\rm trans}$ dominates over RDT across the full length of the tube for $L_{\perp,{\rm th}}^{\rm tip}\gtrsim 0.1\rsun$, and is  important   near the tip for all choices of $L_{\perp,{\rm th}}^{\rm tip}$ in \cref{fig: heat transport}, due to the higher $\beta$ and cusp geometry. 
The latter effect arises for the same reason as in the density transport estimates (\cref{fig: density transport}): the pinch term in \cref{eq:Gamma_T_def} tends to transport heat and temperature toward stronger $B$ (see \cref{eq: stationary pinch profiles}), so near cusp-like regions where $B$ becomes small the net effect is generically to drive heat \emph{away} from the cusp and into neighboring stronger-field regions.  This produces the large localized heating enhancement near $\ell\sim \ell_{\rm tip}$ even for large $L_{\perp,{\rm th}}^{\rm tip}$, suggesting that open flux adjacent to streamer cusps can be a preferred site for transport-driven effective heating.

\paragraph{Implications and broader contexts.}
If closed-field regions are indeed maintained at higher temperature than neighboring open flux (whether by enhanced wave dissipation or other mechanisms), then heat transport provides a continuous way to heat the open-field boundary layer and, by extension, to help establish the conditions associated with slow wind: strong heating in structured regions low in the atmosphere (inside the sonic point) will promote mass loading without requiring proportionate acceleration, adding to the 
Q-reflection and compressive effects already discussed above \S\ref{subsec:Q_reflection_compare}--\ref{sec: empirical compressive}.  The same transport channel should be even more effective near active-region boundaries, where temperature contrasts and fine-scale structuring are larger, and where open/closed complexity (including narrow corridors) is commonplace.  Quantifying those regimes, however, requires a more realistic global background, so we again defer detailed analysis to future work.

\section{Summary and conclusions}\label{sec:summary}

The purpose of this paper is to provide a coherent, \emph{geometry-complete} framework for turbulence- or wave-driven heating and transport in strongly magnetized astrophysical plasmas, with the solar corona and solar wind as the primary application. The central idea is to retain arbitrary smooth geometrical features as {first-class dynamical ingredients}: rather than considering curvature, magnetic shear, stratification, or cross-field gradients as separate effects, to be treated individually in an {ad hoc} manner, we derive a single multiscale reduced-MHD (RMHD) transport system in which all such effects enter on the same footing and can thus be organized systematically. This strategy is motivated by the multiscale gyrokinetic transport theory developed for the study of turbulence in magnetic-confinement-fusion devices \citep{Callen2010,Barnes2010,Abel2013}, which couples reduced fluctuation dynamics to slow profile evolution; this theory  is increasingly being used to inform predictive modeling and configuration design in
fusion plasmas \citep[e.g.,][]{RobergClark2023,Rodriguez2024,DiSiena2026}. Key differences between that fusion-focused theory and the astro-/helio-focused one presented here are our allowance for arbitrary smooth parallel structure and open field lines, the inclusion of gravity, and the use of a fluid (as opposed to kinetic) starting point, motivated by the large-scale, rather than gyroscale, dynamics of interest.

The resulting multiscale RMHD system contains the full set of ingredients normally retained in wave/turbulence-driven (WTD) theories of the corona and solar wind --- field-aligned propagation on an inhomogeneous background, WKB evolution, non-WKB reflection, and wave-pressure (ponderomotive) forces --- while also exposing a wide family of geometry-mediated couplings omitted in standard formulations.  This breadth leads to a natural two-part structure for both the paper and this summary.  First, we present the core multiscale theory as a general framework for low-frequency, strong-guide-field turbulence, heating, and transport in arbitrary smooth geometry. This framework can be applied well beyond the heliosphere --- for example, as a local model of systems where turbulence is driven by instabilities rather than by boundary-launched waves \cite{Kawazura2022}, or even as a simplified  way to interpret and understand physical effects contained within fusion gyrokinetics (see App.~\ref{app: gyrokinetics}). Second, because our primary  interest is Alfv\'en-wave-driven coronal heating and solar-wind acceleration, we use the theory to construct a unified, general-geometry phenomenology and conservative empirical estimates for the heliosphere. These clarify how outward Alfv\'enic fluctuations can dissipate into heat or drive transport  in a structured corona, with various novel pathways highlighted schematically in \cref{fig: overview}.

\subsection{Multiscale RMHD transport theory (key results)}

The key results of the general multiscale theory are summarized here. 
The multiscale theory is derived in self-contained form in App.~\ref{app: derivation}, and its relation to the fusion gyrokinetic theory of \citet{Abel2013} is detailed in App.~\ref{app: gyrokinetics}.

\paragraph*{Geometry-complete multiscale RMHD transport system at $O(\epsilon^3)$.---} We derive a multiscale expansion of the MHD equations in the standard anisotropic RMHD (gyrokinetic) ordering, separating fast, small-scale fluctuations from slow, large-scale transport evolution. The result is a closed, coupled system consisting of: (i) generalized RMHD equations for Alfv\'enic and compressive/entropy fluctuations on an inhomogeneous background, and (ii) transport-time evolution equations for $\rho$, $\bm B$, $U$, and thermal energy driven by quadratic correlators of the waves/fluctuations. The framework is \emph{geometry-complete at $\mathcal{O}(\epsilon^3)$}, in the sense that it retains all couplings induced by general background perpendicular and parallel gradients, field-line curvature ($\bm\kappa$), gravitational forces $\bm{g}_{\rm eff}$, and the magnetic deformation tensor $\nabla\bh$ (including expansion and magnetic shear or flux-tube ``squashing''), with a background field-parallel flow $U$.

\paragraph*{Energetic consistency and a closed energy theorem at transport order.---}
The multiscale construction is energetically consistent by design: any energy lost/gained by the background on the slow transport timescale is gained/lost by the turbulent fluctuations through driving, dissipation, or fluxes \cite{Abel2013,Wang2022}. We derive a global energy conservation law showing this explicitly (App.~\ref{subsub: local energy conservation}). Although this means that the equations involve terms that are expected to be negligible in the solar context (e.g., quadratic correlators of compressive fluctuations), it provides a controlled bookkeeping of wave-mean exchange, ensuring that geometry-mediated reflection, coupling, heating, pressure work,  and transport are neither double-counted nor omitted.

\paragraph*{A taxonomy of geometry-mediated couplings, including perpendicular-gradient-driven compressive responses.---}
The generalized fluctuation system makes explicit how classes of background structure drive dynamical coupling between fluctuations, as summarized in \cref{fig: cartoons}. Parallel gradients separate cleanly into WKB amplification/decay and non-WKB reflection of both Alfv\'en and slow waves, the former (WKB physics) conveniently expressible in wave-action variables (\cref{eq:wave_action_straight,eq: f for slow modes}). Perpendicular gradients, curvature, and gravity  drive conversion from Alfv\'enic to compressive/entropy fluctuations, providing a controlled route to generate compressive fluctuations from transverse motions; this is closely analogous in spirit to resonant-absorption  and ``uniturbulent'' processes in transversely inhomogeneous plasmas \cite{Lee1986,Goossens2011,Magyar2019,Morton2023}. Curvature and perpendicular gravity also enable the reverse feedback from compressive fluctuations onto Alfv\'enic fluctuations \cite{Similon1989}, opening additional channels for dissipation of wave action that are absent in straight-field RMHD.

\paragraph*{Transport equations for global, 3D structured wave-driven winds, and multiple routes to heating.---}
The transport system is constructed as a series of conservation laws for mass, flux, momentum, and energy, making it suitable (given closures or solutions for fluctuation correlators) for slow-timescale evolution of a fully three-dimensional structured atmosphere/outflow. Along each field line it can be reduced to Parker-like wind equations (see App.~\ref{app:parker_like_wind}), including wave-pressure (ponderomotive) forces and other $O(\epsilon^3)$ stresses, while simultaneously retaining turbulence-induced field-perpendicular fluxes of mass, momentum, heat, and magnetic surfaces. A key conceptual outcome is that heating arises not only from dissipation of wave action (a historical focus of solar-wind modeling), but also because waves enable the dissipation/relaxation of {large-scale free energy} stored in mean gradients (flow shear, magnetic/thermal gradients, effective potential energy). These effects are retained within the same energetic framework.

\paragraph*{Magnetic-field evolution: no turbulent resistivity at transport order.---}
An interesting specific outcome of the transport-time induction equation is that magnetic flux surfaces remain flux-frozen at the order retained, viz., the large-scale $\bm{B}$ evolution can be expressed purely  in terms of a turbulence-induced effective perpendicular velocity, because the parallel component of the turbulent EMF vanishes at $\mathcal{O}(\epsilon^3)$: $\mathcal{E}_\parallel\equiv\bh\cdot\avgi{\delta\bm u\times\delta\bm B} = \mathcal{O}(\epsilon^4)$ (see App.~\ref{sub: magnetic transport}). This arises as a nontrivial consequence of RMHD incompressibility and enforced scale separation, and is the RMHD analogue of the result in fusion multiscale gyrokinetics that turbulence cannot accelerate the evolution of the safety factor $q$ \cite{Abel2013}. It rules out the mean-field-dynamo ``$\beta$-effect'' for RMHD-like anisotropic fluctuations, as often discussed in the context of dynamo theory \cite{Rincon2019}. It also breaks a key assumption behind stochastic-flux-freezing and ``reconnection-diffusion''  \cite{Eyink2011,Lazarian2020}, implying that scale-separated RMHD turbulence cannot enhance large-scale reconnection via an $\mathcal{E}_\parallel$ at transport order.


\subsection{Coronal heating and solar-wind applications}

We now summarize some interesting implications for coronal heating and the acceleration of the solar wind.  To extract transparent trends and order-of-magnitude scalings from the multiscale RMHD system without a full numerical solution of the coupled fluctuation-transport system, we use a unified ``slaved-field'' phenomenology (\cref{subsec:amplitudes}) in which compressive and reflected waves are assumed to arise from the balance of forcing and turbulent dissipation provided by outward Alfv\'enic fluctuations. This  is the straightforward generalization of standard reflection-driven turbulence closures \cite{Dmitruk2002} to arbitrary geometry.  The accompanying empirical estimates are compared to reflection-driven heating estimates on a highly simplified model corona (\cref{fig: model}) to highlight  where structured geometry  can plausibly compete with, or dominate, standard expansion/reflection-driven turbulence. These results are intended as demonstrations of potential importance or novel effects, not as calibrated predictions; robust quantitative results will require jointly evolving turbulence on a more realistic background.

\paragraph*{Generic conclusion: structured regions preferentially energize low altitudes (\cref{fig: overview}).---}
A robust expectation from wind theory is that heating deposited below the sonic point increases mass flux and tends to produce slower, denser wind, whereas heating higher up more efficiently increases terminal speed \cite{Leer1980,Hansteen1995,Withbroe1988}. Because many new heating sources in the geometry-complete system scale with transverse gradients, curvature, field-line non-radiality, and mapping deformation, structured regions (coronal-hole boundaries, streamer-adjacent corridors, cusp environments) are natural sites for enhanced low-altitude wave-driven heating and transport. This could provide a unifying physical theory for why ``structure'' correlates empirically with slow-wind-like conditions, although significant further
work is needed to test the picture quantitatively.

On the other hand, the geometry-complete analysis does {not} \emph{replace} the standard reflection-driven-turbulence (RDT) picture \cite{Dmitruk2002}. 
In the limit of a relatively smooth, weakly structured open flux tube --- the idealized situation most closely associated with coronal-hole centres and fast-wind streams --- new geometric terms become small and the theory reduces to the usual expansion/reflection problem \cite{Velli1989,Chandran2009}. 
This provides an important consistency check: novel mechanisms identified here do not spoil the baseline RDT phenomenology in the smooth-field limit, which has seen important successes in reproducing key features of heliospheric structure \cite[e.g.,][]{Cranmer2007,Mikic2018,Cranmer2019}  (though some studies suggest that additional physics may be needed to capture heating rates quantitatively \cite{vanBallegooijen2016,Shoda2018,Chandran2019,Shoda2019,Sioulas2025}). 
One straightforward outcome of the RDT phenomenology remaining relevant is the prediction that the turbulence imbalance (normalized cross helicity) will usually decrease with heliocentric radius (see \cref{subsec:RDT_basic}), and, insofar as larger expansion is associated with slower wind, will tend to be lower in slow-wind streams than in fast-wind streams, as generally observed \cite{Bruno2013,Abbo2016}.

\paragraph*{$Q$-reflection: a new reflection channel from flux-tube squashing.---}
In addition to the standard RDT from non-WKB reflection driven by field-aligned Alfv\'en-speed gradients, $\propto\bh\cdot\nabla \va$ \cite{Velli1989}, the symmetric-traceless magnetic deformation tensor $\mathsf S$ drives wave reflection by area-preserving squashing of flux tubes (see \cref{fig: grad b} and \cref{subsec:Q_reflection}). This links turbulent dissipation directly to the field-line mapping Jacobian that underlies the ``squashing factor'' $\mathcal{Q}$ of  \citet{Titov2002} and subsequent S-web theories \cite{Antiochos2011}, with $\mathsf S$ playing the role of an effective ``$\bh \cdot\nabla \ln \mathcal{Q}$''. The effect can remain strong where $\bh\cdot\nabla \va$ is small, and is already competitive with expansion-driven reflection even in a highly smooth solar-minimum model (\cref{fig: Q estimates}).  Since measured/inferred $\mathcal{Q}$ values in the real corona can be orders of magnitude larger than in such minimal fields, the same mechanism naturally implies wave-driven dissipation concentrated preferentially at low altitudes in globally structured regions. The effect could also enhance heating at low altitudes in coronal holes if local plume structure creates  complex field structure \cite{Bora2025}, potentially helping to explain low-altitude heating deficits 
seen in simulations that include only reflection \cite{vanBallegooijen2016,Chandran2019}.

\paragraph*{Curvature/gravity mediated compressive pathways: DCF and CCR.---}
Curved or non-radial magnetic fields enable outward Alfv\'enic fluctuations to self-generate compressive responses that can change wave action through two novel effects (i) \emph{direct compressive feedback} (DCF; \cref{sub: DCF}) and (ii) \emph{compressively catalyzed reflection} (CCR; \cref{sub: CCR}). The former (DCF) arises from the direct feedback of slaved compressive fluctuations on  outward-propagating Alfv\'en waves (see \cref{fig: cartoons}), while the 
latter (CCR) involves slaved compressive fluctuations driving inward-propagating Alfv\'enic fluctuations that then turbulently dissipate the outwards waves.   These channels can  provide additional heating routes that are absent in straight-field RMHD but can be competitive with expansion-driven reflection  for plausible levels of perpendicular structuring (\cref{fig: Lcrit}), again favoring stronger wave-driven heating at low altitudes in regions of greater magnetic complexity.

\paragraph*{Conversion of background free energy into heat: ACR.---}
In addition to dissipation of wave action, the transport system predicts that outwards Alfv\'enic fluctuations  drive the dissipation of large-scale free energy stored in background cross-field gradients, converting it into heat through \emph{Alfv\'en-catalyzed relaxation} (ACR; \cref{subsec:comp_absorb}).  The key conceptual point is that waves can heat not only by cascading their own energy, but also by enabling the turbulent relaxation of an otherwise stable perpendicular equilibrium (similar to transport). As well as acting on curved or non-radial fields with a similar strength to DCF and CCR, this mechanism provides a route to dissipate cross-field velocity-shear (stream) structure  as the wind  approaches trans- or super-Alfv\'enic speeds. Our slaved estimates (\cref{subsec: stream_dissipation}) and PSP measurements \cite{Bale2021} suggest the effect could   dominate reflection-driven heating near and beyond the Alfv\'en point where $U\approx \va$ (see \cref{fig: stream dissipation}). The mechanism involves some similar features to empirical shear-transport turbulence-source  prescriptions from past literature and transport modeling \cite{Roberts1987,Zank1996,Breech2008}, but differs in its qualitative interpretation and quantitative predictions.

\paragraph*{Cross-field mass and composition transport: fueling slow wind from closed-field regions.---}
The transport equations imply a generic cross-field plasma mass flux whenever Alfv\'enic turbulence propagates through regions with perpendicular density gradients (\cref{subsec: transport}). In the corona, this opens a continuous pathway for mass (and compositional signatures) to leak from denser closed-field reservoirs into neighboring open flux \cite{Magyar2019}, complementing intermittent interchange-reconnection or blob-release scenarios \cite{Sheeley1997,Fisk2003}. Conservative empirical estimates suggest that even relatively broad open/closed boundary layers can supply an $\mathcal{O}(1)$ fraction of the slow-wind mass flux via continual transport (\cref{fig: density transport}). This provides a natural continuous mechanism to import closed-field compositional signatures (first-ionization-potential bias and charge states) into open-field wind \cite{Geiss1995,Abbo2016,Verscharen2019}. Cusp-like geometries generically drive an enhancement near streamer tips because there is a  predicted ``density pinch'' toward regions with stronger magnetic fields, $\rho\propto B^{2}$. This pinch has been long-discussed and observed experimentally in levitated-dipole experiments \cite{Hasegawa1987,Kouznetsov2007,Kobayashi2010,Boxer2010}; our theory 
recovers these results and predicts how the effect persists in general geometry.

\paragraph*{Cross-field heat transport from closed regions.---}
Heat and momentum are transported across the field by essentially the same multiscale physics as mass (\cref{subsec: transport}):  Alfv\'enic fluctuations propagating through cross-field temperature  structure drive a cross-field thermal-energy flux from hotter to cooler regions.  In the corona this provides a continuous, non-eruptive way to heat open-field boundary layers and streamer-adjacent corridors from closed-field reservoirs, complementing other mechanisms.  Conservative empirical estimates suggest that the implied {effective} low-altitude heating can be large compared to standard expansion-driven RDT, even for relatively broad open/closed transition layers (\cref{fig: heat transport}).  As for density, cusp-like geometries leave a distinctive imprint: the same pinch physics that draws plasma mass toward stronger magnetic fields implies that  temperature transport is biased toward stronger-field regions (i.e., it moves heat away from  cusps); this produces a localized effective heating burst in our estimates near the streamer-tip region (\cref{fig: heat transport}).

\subsection{Outlook and future extensions}

\paragraph*{Global wind modeling and space-weather applications.---}
In operational space-weather forecasting,  solar-wind conditions in the inner heliosphere (within ${\sim}0.1$AU) are still commonly constructed using fast empirical or semi-empirical coronal mappings based on large-scale magnetic structure, rather than being predicted from  first-principles heating physics \cite{Arge2000,Odstrcil2003,Pomoell2018}.  The original Wang--Sheeley (WS) approach parameterized the  wind speed primarily by the flux-tube expansion factor $f_s$, often in a simple form such as $U_\infty \simeq U_0 + U_1 f_s^{-\alpha}$ \cite{Wang1990}, where $U_0$, $U_1$, and $\alpha$ are empirical parameters and $U_\infty$ is a speed far from the solar surface.  Wave/turbulence-driven (WTD) theories provide a physics-based rationale for why such a proxy can work: flux-tube expansion controls field-strength variation along the tube (and hence gradients in $\va$), which control Alfv\'enic wave reflection and turbulent dissipation \cite{Velli1989,Cranmer2007,Chandran2009,Cranmer2017}.  However, WS has largely been supplanted by Wang--Sheeley--Arge (WSA) models that augment expansion with an additional dependence on the proximity of a field line to coronal-hole boundaries (or related open-closed boundary metrics) with representative form
\begin{equation}
U_\infty \simeq U_0 + \frac{U_1}{(1+f_s)^\alpha}\,\Bigl[1-e^{-(\theta_b/\theta_0)^\beta}\Bigr],
\end{equation}
where $\theta_b$ measures distance to the nearest coronal-hole boundary and $\theta_0$ and $\beta$ are further  parameters \cite{Arge2000,Arge2004}. The additional dependencies reflect the empirical fact that expansion alone is not sufficient to predict wind speed near structured regions \cite{Arge2000,Dakeyo2024}; indeed, it has been argued that the boundary-distance ``corrections'' via $\theta_{b}$ can be more important than the original expansion term via $f_{s}$  \cite{Riley2015,MacNeice2018,Dakeyo2024}.  

Our geometry-complete multiscale RMHD theory and slaved phenomenology provide a natural physics basis for why a $\theta_b$-type dependence should exist: near open-closed boundaries and other structured regions, complex geometrical features, for which  $\theta_b$ is a crude proxy, 
activate additional  heating and transport channels beyond standard $\bh\!\cdot\!\nabla \va$ reflection. These additional channels generally heat and add mass at lower altitudes (within the sonic point) thereby driving denser, slower wind than would otherwise occur.  This offers a concrete WTD-based interpretation  for why boundary-adjacent flux tubes tend to produce slower wind than would be expected from expansion alone. 
Of course, the true predictions of the multiscale framework are too complex to be treated via a single parameter $\theta_{b}$, but the trend is promising for future study.

Of similar interest is the potential to explain the diversity of slow-wind types, often separated into Alfv\'enic and non-Alfv\'enic wind \cite{DAmicis2021}. The generalized RDT phenomenologies of \cref{subsec:RDT_basic,subsec:Q_reflection} predict that the Alfv\'enic imbalance depends on the competition between nonlinear turnover and expansion/squashing, and is therefore controlled by the corresponding ratios of these rates, $\chi_{\rm exp}$ and $\chi_{\rm exp,\mathcal{Q}}$ (see \cref{tab: notation_phenom}). As $\chi_{\rm exp}$ or $\chi_{\rm exp,\mathcal{Q}}$ approach unity, reflection becomes strong enough that the fluctuation field evolves toward a more balanced state ($z^{-}\sim z^{+}$) and the simple  phenomenology breaks down. This suggests that weakly Alfv\'enic or non-Alfv\'enic slow wind may be favored in streams for which the surviving outward fluctuation amplitude has been sufficiently reduced by larger altitudes that $\chi_{\rm exp}\!\sim\!1$; for example through enhanced low-altitude dissipation in regions of large $\mathcal{Q}$ gradients, or through cross-field mass loading from neighboring closed flux that damps $z^{+}$ (see \cref{eq:perp-pm-app}). By contrast, Alfv\'enic slow wind may correspond to streams that lack sufficient high-altitude heating to become fast, but still retain sufficiently large outward amplitudes that $\chi_{\rm exp}\!>\!1$ so that the turbulence is imbalanced. This interpretation appears qualitatively consistent with current observations, which suggest that Alfv\'enic slow wind carries larger fluctuation amplitudes and wave-energy fluxes than  non-Alfv\'enic wind \cite{DAmicis2021,Rivera2025,DAmicis2025}, although clearly further theoretical and observational work is needed.

More broadly, the transport-time equations contain the ingredients needed for a dynamic, geometry-aware 3D heliosphere model, in which heating, wave-pressure forces, and cross-field transport emerge together from the one multiscale expansion (see App.~\ref{app:parker_like_wind}). Given a turbulence closure or a direct numerical solution for the fluctuation correlators, as well as a realistic time-dependent magnetic boundary condition, they therefore define a self-consistent evolution of the fluctuation energetics and  slowly varying 3D background, providing a formalism to extend physics-based heliosphere models that explicitly evolve averaged outwards-wave dynamics or turbulence closures in open and/or closed regions \cite{vanderHolst2014,Sokolov2021,Parenti2022,Gombosi2018,Usmanov2018,Downs2016}. This retains new geometry-mediated channels such as perpendicular transport, which are mostly omitted in incompressible or field-aligned treatments of fluctuations, as well as ensuring conservation of 
total energy, momentum, and mass by construction as the background evolves. Likewise, the dependence of the heating and transport terms on magnetic and plasma geometry is fixed by the multiscale expansion, leaving only a small set of turbulence parameters to be constrained against observations, rather than supplied through separately motivated heating prescriptions \citep[e.g.,][]{Evans2012} or empirical relations \cite{Arge2000,Pomoell2018}. Extensions to include Parker-spiral geometry, larger amplitudes,  and other outer-heliosphere effects (see below) would allow the model to address not only inner-heliospheric wind launching but also the subsequent wind-speed evolution and damping of stream structure farther out.

\paragraph*{Further extensions and broader applications.---}
The multiscale expansion adopted here is deliberately idealized --- small-amplitude, long-wavelength, fluid (MHD) fluctuations on a smooth background. This has necessarily led to the neglect of a number of important effects (see detailed discussion in \cref{subsec:limitations}), but the simplification is also part of the motivation: retaining a clean asymptotic structure is what allows the coupled fluctuation-transport system to remain energetically conservative at transport order, 
while the fluid model renders many effects physically transparent by diluting the equations' complexity. The most obvious next steps are therefore to work towards incorporating missing physics, including kinetics, finite-Larmor-radius effects (in order to capture the helicity barrier \cite{Meyrand2021}), large-amplitude Alfv\'enic fluctuations with $|\dbprp|/B\sim 1$, and Parker-spiral geometry.  A kinetic model, either via full gyrokinetics \cite{Abel2013} or  kinetic MHD \cite{Kulsrud1983}, would be more physically complete for the weakly collisional corona  and would  allow the same framework to make more specific predictions for heating partition and wind type. For example, channels that dissipate Alfv\'enic wave action through a perpendicular cascade (e.g., reflection) are expected to bias toward perpendicular ion and/or electron heating \cite{Adkins2025}, whereas compressive dissipation (e.g., ACR) provides a natural route to predominantly parallel heating \cite{Schekochihin2009,Howes2010,Kawazura2020,Howes2024}.  These
could lead to interesting correlations between geometric structure, wind type, and nonthermal features of the background plasma such as temperature anisotropies and  interspecies differences.
Another direction of research is to extend the  small-amplitude ordering, which is often not formally satisfied in the solar wind: beyond  the Alfv\'en point $|\dbprp|/B\sim 1$ but $|\bm{B}|$ remains nearly constant (spherically polarized), while  coronal fluctuations are generally transonic  ($z^{+}\sim c_s$). We argue in App.~\ref{app: lowbeta_transonic} that the impact of the latter on our predictions is minimal, but  it will be important to understand more concretely how the governing equations break down at large amplitudes and, one hopes,  develop complementary multiscale schemes that capture additional effects (for example, the expansion-induced modification to spherically polarized Alfv\'en-wave structure  \cite{Barnes1974,Mallet2021}).

More broadly, while the heliospheric applications emphasize boundary-launched Alfv\'enic driving, the core multiscale system is equally applicable to turbulence driven by instabilities and  other magnetized environments where geometry and transport are central.  Disk dynamics  and jet outflows are natural targets because the same separation between fast turbulence and slow transport underlies accretion-disk modeling; indeed, the RMHD model used by \citet{Kawazura2022} and \citet{Kawazura2024} to study turbulent heating partition in magnetorotational turbulence corresponds to a subset of the present equations (see App.~\ref{sec: magnetorotational}). This suggests a clear path to extend those successes to more general geometry and other instabilities.  Magnetospheric plasmas (either terrestrial or pulsar) provide another compelling arena, where strong inhomogeneity, curvature, and interchange-like mixing make cross-field pinches and transport ideas directly relevant (indeed, the turbulent-pinch scalings recovered here were motivated originally by magnetospheric physics \cite{Hasegawa1987}).  Similar multiscale ideas may also help in understanding angular-momentum transport in stellar interiors, where global geometry couples internal-wave fluxes to slow rotational evolution and magnetic stresses/instabilities  \cite{Talon2008,Aerts2019,Fuller2019}.

\acknowledgements

The authors thank K.~Abbas, V.~Davis, L.~Huckle, Z.~Johnston, A.~A.~Schekochihin, and M.~Zhang for helpful discussions.
JS and TA acknowledge the support of the Royal Society Te Ap\=arangi, through Marsden-Fund grant MFP-UOO2221. The work of T.A. was also supported in part by the Laboratory Directed Research and Development (LDRD) Program at the Princeton Plasma Physics Laboratory for the U.S. Department of Energy under Contract No. DE-AC02-09CH11466. The United States Government retains a non-exclusive, paid-up, irrevocable, world-wide license to publish or reproduce the published form of this manuscript, or allow others to do so, for United States Government purposes.
BC and RM were supported in part by the
National Aeronautics and Space Administration (grant numbers 80NSSC24K0171 and NNN06AA01C) and the
U.S. Department of Energy (grant number DE-SC0026201).
WC was supported by an Oxford--Radcliffe Scholarship in Theoretical Physics from the University College, Oxford, the UKAEA, a James Fairfax--Oxford Australia Scholarship, and the Simons Foundation via a Simons Investigator Award to A.~A.~Schekochihin .
MWK was supported by the National Aeronautics and Space Administration (NASA) under Grant No.~80NSSC24K0171 issued
through the Heliophysics, Theory, Modeling and Simulation Program.

\appendix

\label{APPENDIX STARTS HERE}

\section{Derivation of Multiscale Reduced MHD}\label{app: derivation}

\subsection{Governing MHD equations}
Our starting point is the MHD equations in an external gravitational potential and uniformly rotating frame:
\begin{align}
&\frac{\partial \rho}{\partial t}  + \nabla\!\cdot\!\left(\rho\bm u\right)  =  \Srho, \label{eq:cont}\\[0.25em]
&\rho\left(\frac{\partial \bm u }{\partial t}  + \bm u\cdot\nabla \bm u + 2\bm\Omega\times\bm u\right)
 = 
-\,\nabla \left(p + \frac{B^2}{8\pi} \right) \nn \\
&\qquad+  \frac{\bm B\!\cdot\!\nabla\bm B}{4\pi}
 -  \rho\nabla \Phi_{\rm tot}  + \Smv-\bm{u}\Srho
 +  \nabla\!\cdot\!\bm{\Pi}, 
 \label{eq:mom}\\[0.35em]
&\frac{\partial \bm B}{\partial t}  = \nabla\times\left(\bm u\times\bm B\right) 
 -  \nabla\times\left(\eta\,\nabla\times \bm B\right), \label{eq:ind}\\[0.25em]
&\frac{\partial E_{\rm th}}{\partial t} 
 +  \nabla\!\cdot\!\left(E_{\rm th}\,\bm u\right)
 = 
-\,p\,\nabla\!\cdot\!\bm u
 -  \nabla\!\cdot\!\bm q+  \Sth\nn\\
&\qquad -  \bm{\Pi}:\nabla\bm u
 +  \frac{\eta}{4\pi}\,|\nabla\times\bm B|^{2}
  . \label{eq:intE}
\end{align}
Here 
\(\rho\) is the mass density, \(\bm u\) is the plasma velocity, \(p\) is the  thermal pressure, $E_{\rm th}=p/(\gamma-1)$ is the thermal energy, and \(\gamma\) is the adiabatic index.  The magnetic field is
\(\bm B\) with  \(B^2 \equiv \bm B\!\cdot\!\bm B\) and $\bh=\bm{B}/B$.
Momentum dissipation (viscosity) is represented by the general stress tensor \(\bm{\Pi}\), resistivity by a magnetic diffusivity \(\eta\),  and thermal physics by a heat flux \(\bm q\).    We will often 
list the dissipative (viscous and resistive) terms as $\mathcal{D}_{u}$ and $\mathcal{D}_{B}$ as a convenient shorthand
to represent their effect in dissipating fluctuations, and our theory does not rely on many specific features 
of the form of this dissipation.  External  heating or cooling sources (e.g., radiative cooling) are collected in \(\Sth\),  any external forces in the momentum source $\Smv$, and any external mass source in $\Srho$ (the physical motivation behind retaining $\Su$ and $\Srho$ for solar-coronal applications is discussed in App.~\ref{app:parker_like_wind}).
The angular velocity of the uniformly rotating frame is \(\bm{\Omega}\), and the external gravitational 
potential is $\Phi_{\mathrm{grav}}$. Gravity and the centrifugal contribution then combine to yield an effective potential \(\Phi_{\rm tot}=\Phi_{\mathrm{grav}}+\Phi_{\mathrm{rot}}\) with
\begin{equation}
\Phi_{\mathrm{rot}}(\bm r)  =  -\frac{1}{2}\,\left|\bm{\Omega}\times\bm r\right|^{2}, \label{eq: centrifugal}
\end{equation}
where $\bm{r}$ is the position vector.

We also define the Alfv\'en speed,  sound speed, and slow-magnetosonic speed,
$$\va \equiv \frac{B}{\sqrt{4\pi\rho}}, \quad
\cs^2 \equiv \frac{\gamma p}{\rho},\quad v_{S}^{2}\equiv \frac{\va^{2}}{1+\va^{2}/\cs^{2}},$$  respectively, as well as $\beta\equiv 8\pi p/B^{2} = (2/\gamma)\, \cs^{2}/\va^{2}$. For most calculations, rather than the pressure or internal energy we  will  use the specific entropy per mass,
\begin{equation}
s-s_{0} \equiv c_{v}\ln\frac{p}{\rho^\gamma},\label{eq:entropy def}
\end{equation}
where $c_{v}$ is the heat capacity 
at constant volume, $s_{0}$ is a reference entropy, and $\gamma$ is the usual adiabatic constant. Combining \cref{eq:cont} and \cref{eq:intE} yields
\begin{align}
\rho T \left(\frac{\partial}{\partial t}+\bm{u}\cdot\nabla \right)s
 = &
-\,\nabla\cdot \bm q  +\Sth- \gamma c_{v}T \Srho \nn\\ &-  \bm\Pi:\nabla \bm u
 +  \frac{\eta}{4\pi}\,\left|\nabla\times \bm B\right|^{2},\label{eq: entropy app}
\end{align}
where $p = \rho {\rm R} T$ and ${\rm R}=k_{B}/\bar{m}$ is the specific gas constant per unit mass such that $c_{v}={\rm R}/ (\gamma-1)$ ($\bar{m}$ is the mean particle mass).

The parallel and perpendicular correlation lengths of the turbulent fluctuations are denoted 
by $l_{\|}$ and $l_{\perp}$ respectively, with inverses $k_{\|}$ and $k_{\perp}$.

\subsection{Notation, ordering, and geometrical identities}\label{sub: notation}
We employ an asymptotic expansion that separates rapidly varying turbulent fields from a slowly varying background along open magnetic field lines. An average
\(\langle\dots\rangle\) over intermediate perpendicular scales and times annihilates fluctuations and preserves means. However,  unlike fusion transport theory, which is based on the same ordering
scheme in gyrokinetics from the Vlasov equation \citep{Frieman1982,Schekochihin2009},
no flux-surface average is introduced, and  parallel energy and momentum fluxes remain explicit. This 
allows the theory to capture heating and momentum transfer from externally driven waves passing through the local domain.

This subsection provides the definitions and notation required to proceed with the derivation, particularly 
focusing on the properties of the average that allow the mean-fluctuation split. An illustration of the geometrical setup is shown in \cref{fig: average}.

\paragraph{Ordering and mean–fluctuation expansion.}
We expand every field in powers of some small parameter \(\epsilon\),
\[
g  =  \sum_{n\ge0}\epsilon^n\,g^{(n)}, \qquad
g^{(n)}  =  \langle g^{(n)} \rangle  +  \delta g^{(n)} ,
\]
where \(\langle \cdot \rangle\) is the \emph{turbulence average} (defined below as a composition of a perpendicular spatial average and an intermediate time average). By definition,
\[
\langle \delta g^{(n)} \rangle  =  0 ,
\]
and, in general, zeroth-order mean quantities are represented by an unadorned symbol: 
\[
\bm{B}\equiv \langle\bm{B}^{(0)} \rangle,\: \bm{U}\equiv \langle\bm{u}^{(0)} \rangle,\:
{p}\equiv \langle{p}^{(0)} \rangle,\: {\rho}\equiv \langle{\rho}^{(0)} \rangle,\:{s}\equiv \langle{s}^{(0)}\rangle,
\]
with the mean magnetic-field direction $\bh$ thus defined from $\bm{B}$. 
Note that this does risk ambiguity in the representation of the full, unexpanded quantities (e.g., Eqs.~\eqref{eq:cont}--\eqref{eq: entropy app})
but the distinction should be clear from the context.
Likewise, for ease of notation, $\delta g\equiv \d g^{(1)}$ for any quantity $g$.
 
 \paragraph{Projection and parallel/perpendicular split.}
Define the projection operators
 parallel and perpendicular to the mean-field direction $\bh$:
\begin{equation}
a_{\|} \equiv \bh\cdot \bm a, \quad
\mathcal{P}_{\perp}[\bm a]=\bm a_{\perp} \equiv \bm a - \bh\, a_{\|}.\label{eqapp: projectors}
\end{equation}
Under our ordering below, these split the perturbations into Alfv\'enic (perpendicular) fluctuations and compressive (parallel) perturbations, such that the expansion proceeds using
$\d\bm{u} = \duprp + \duprl \bh$ and $\d\bm{B} = \dbprp + \dbprl\bh$.
We likewise define the gradient operator,
\[
\nabla_{\perp} g \equiv \nabla g- \bh\,\gradl g,
\]
where $g$ is some scalar. The symbol $\nabla_{\|}$ will be used below to denote the (nonlinear) gradient along the perturbed field line (note the notational difference compared to some past related works \cite{Schekochihin2009,Kunz2015}).

To capture the local variation of slowly varying background quantities we define the compact notation, 
\begin{equation}
\gprl{G} \equiv \gradl\ln G, \quad
\gprp{G} \equiv \nabla_{\perp}\ln G, \quad
\gprps{G} \equiv |\gprp{G}|,\label{eq: K defs}
\end{equation}
which will be used to emphasize that  background gradients should be considered as local parameters in the fluctuation equations.
Unlike in the fusion context, where the 
nested flux surfaces guarantee that all background gradients are in the direction perpendicular to those flux surfaces (usually denoted ${\rm d}\ln p/{\rm d}\psi$ etc. \cite{Barnes2010}), our expansion
allows  the background quantities to both vary in the different directions (e.g., $\gprp{\rho}$ 
and $\gprp{B}$ need not be parallel) and have nonzero field-aligned gradients (e.g., $\gprl{p}\neq0$). Note that corrections to the $\gprl{G}$ and $\gprp{G} $ quantities due to the fluctuating fields occurs at a higher order than will be considered here.

The entropy's notation differs slightly: because $s$  already contains a logarithm, it is more natural to define  $\gprl{s}\equiv c_{v}^{-1}\gradl s$ and $\gprp{s}\equiv c_{v}^{-1}\nabla_{\perp} s$.

 \paragraph{Expansion parameter}
 
To specify $\epsilon$, gradients of any mean quantity are ordered with  parallel derivatives of fluctuations, while  
perpendicular derivatives of fluctuations are $\epsilon^{-1}$ larger (see \cref{fig: average}).  Parallel/compressive fluctuations are the same formal order as perpendicular ones.  This ordering retains non-WKB wave reflection physics in the fluctuation equations. It is equivalent to multiscale gyrokinetics \citep{Callen2010,Barnes2010,Abel2013} and identical to the derivation of standard RMHD with slow modes in \citet{Schekochihin2009}, other than the addition of slow variation of the background. Explicitly, we take:
\begin{align}
\epsilon  &\sim 
\frac{\delta \rho}{\rho}\sim\frac{\delta p}{p}\sim\frac{\delta s}{s}\sim
\frac{|\delta \bm B|}{B}\sim\frac{|\delta \bm u|}{\va}\sim
\frac{k_{\|}}{k_{\perp}}\sim\frac{\kappa}{k_{\perp}}
\nonumber\\[-0.2em]
&\sim 
\frac{\gprl{\va}}{k_{\perp}}\sim\frac{\gprl{\rho}}{k_{\perp}}\sim
\frac{\gprps{\rho}}{k_{\perp}}\sim\frac{\gprps{U}}{k_{\perp}}\sim
\frac{\gprps{B}}{k_{\perp}} \sim \frac{\omega}{k_{\perp}\va}.
\label{eq:ordering}
\end{align}
Here \(\omega^{-1}\) characterises the wave/turbulent period of fluctuations (it will soon be identified with the Alfv\'en frequency), and \( \bm\kappa \equiv \bh\cdot\nabla \bh\) is the field-line curvature with \(\kappa=|\bm\kappa|\).
In keeping with an ordering appropriate for the Sun, we will order $\nabla\Phi\sim \epsilon k_{\perp}\va^{2}$, but relegate rotation to the transport timescale, $\Omega\sim \epsilon^{3}k_{\perp}\va$.

The mean flow is assumed to be field-aligned at zeroth order, \(\bm U = U\,\bh\) with \(U\sim \va\), which differs from fusion momentum transport 
theories involving net toroidal rotation. For consistency, the mean external force $\Smv$ is assumed to align with the background-field direction  $\Smv=\Sm\bh$. We will show below it is consistent to neglect a first-order mean perpendicular flow; a second-order mean \(\bm U_{\perp 2}\) must be retained because it can be generated by fluctuations.
We take $\Srho$, $\Sm$, and $\Sth$ to have no perturbed part, in keeping with the idea that any slowly varying 
sources could be collected fully into a lowest-order part (transport-order corrections to  sources, for instance a perpendicular force, could be added trivially if desired).

\paragraph{Perpendicular spatial average \(\avgp{(\cdot)}\).}\label{app para: perp average}
The  perpendicular spatial average mentioned above is defined over a patch of size \(L\) (area $L^{2}$) satisfying
\(l_{\perp}\ll L \lesssim l_{\|}\), so that the averaging scale is comparable to the variation scale of background quantities (see \cref{fig: average}),
\[
\avgp{g}  \equiv  \frac{1}{L^{2}}\int \mathrm{d}^2\bm{r}_{\perp}\,g.
\]
By construction, the spatial average of a fluctuation vanishes, \(\avgp{\delta g}=0\),
while the spatial average of a mean equals the mean itself because its scale of variation is on scales larger than $L$.

\paragraph{Intermediate-time average \(\langle\,\cdot\,\rangle_{t}\).}
The intermediate time average is defined as a standard average over the intermediate time window \(T\), which  satisfies
\(\omega^{-1}\ll T \ll \tau\), where
 \(\tau\) is the slow transport time. 
To separate fast and slow temporal dynamics, the two timescales $t$ and $\tau $ are treated formally via multiple-time-scale analysis. Noting that the fast time scale is already $\mathcal{O}(\epsilon)$ (see \eqref{eq:ordering}), we   replace \(\partial/\partial t \rightarrow \epsilon \partial/\partial t  + \epsilon^{3}\,\partial/\partial \tau \), where  the \(\epsilon^{3}\) scaling for $\partial/\partial \tau$ arises 
from the fact that transport-time-scale dynamics are driven by inhomogeneous sources built from quadratic correlations of \(O(\epsilon)\) fields.

A quasi-periodic ``fluctuating'' fast-time solution exists only if {secular} terms disappear, which is enforced in multiple-time-scale analysis by ensuring that  the time average of the 
 fast-time-scale equations vanishes.
Practically, this condition transfers the nonzero average of the quadratic sources into the slow evolution of the background.
By  demanding the disappearance of fast secular growth terms, multiple-time-scale analysis eliminates fast-time-scale derivatives of  second-order mean quantities in favor of the slow
evolution of mean backgrounds on the transport timescale  \(\tau\). As a less formal, heuristic description, we note that a change in some second-order mean  represents an \(O(\epsilon^{2})\) addition 
to the background; a change in $\langle {{g}^{(2)}}\rangle_{\perp}$ at a rate of   \(O(\epsilon)\) is therefore physically 
equivalent to a change in $g^{(0)}$ at a rate of  \(O(\epsilon^{3})\),   effectively providing the identification
$\langle \partial\langle {g^{(2)}} \rangle_{\perp} /\partial t\rangle_{t}  \rightarrow  \partial g^{(0)}/\partial \tau$.

\paragraph{Turbulent average and total derivatives.}\label{para: average and derivatives}

Combining the perpendicular and time averages discussed above, the \emph{turbulence average} is the composition
\[
\langle g \rangle  \equiv  \langle\,\avgp{g}\,\rangle_t .
\]
To the order retained in our calculations, this  commutes with large-scale spatial derivatives 
and  the slow-time-scale derivative \(\partial/\partial \tau \).

A property used extensively in our derivation is that total derivatives of turbulent averages of quadratic products of fluctuations gain an additional order in $\epsilon$ when the gradient is commuted through the average. So, for example, $\avg{\nabla\cdot(\duprp \d \rho)} = \nabla\cdot\avg{\duprp \d \rho} + \mathcal{O}(\epsilon^{4})$.
Intuitively, this is straightforward to understand: the average makes the product a large-scale quantity, which thus gains 
an order in $\epsilon$ upon taking its derivative. More formally, we 
can understand this as a consequence of the average shifting the derivative to the patch's boundary. Consider, for simplicity, a square patch with normal $\hat{\bm{n}}$ for a case that
is homogeneous in one perpendicular direction (denoted $y$) with gradients in the other ($x$). We
 examine $\avg{\nabla\cdot(\duprp\d\rho)}$, which is 
\begin{equation}
\left< \frac{1}{L^{2}}\int dx\,dy\, \nabla\cdot(\duprp\d\rho) \right>_{t}= \frac{1}{L^{2}}\int_{{C}} \hat{\bm{n}} \cdot \langle \duprp\d\rho \rangle_{t},
\end{equation}
where ${C}$ denotes the line integral around the patch. Given the homogeneity in $y$, combined with the time average and short correlation length ($l_{\perp}\ll L $), the segments of  ${C}$ at different $y$ cancel, 
giving 
\begin{equation}
\frac{ \langle \langle \d u_{x}\d\rho\rangle_{t} \rangle_{y}|_{x+L}-\langle \langle \d u_{x}\d\rho\rangle_{t} \rangle_{y}|_{x}}{L},\label{eq: stokes theorem thing}
\end{equation}
where $\langle \cdot\rangle_{y}=L^{-1}\int_{y}^{y+L}dy\,$. The $y,t$ average eliminates $l_{\perp}$-scale variation of $\d u_{x}\d\rho$ (i.e., $\langle \langle \d u_{x}\d\rho\rangle_{t} \rangle_{y}$ is a large-scale quantity), while the $x$ difference in \eqref{eq: stokes theorem thing} yields an $x$ derivative, viz., $\partial\langle \langle \d u_{x}\d\rho\rangle_{t} \rangle_{y}/\partial x\approx \partial\avg{\d u_{x}\d\rho}/\partial x$ at $\mathcal{O}(\epsilon^{3})$. 
An almost identical procedure applies to the curl of a vector quantity for treating the  induction equation: derivatives in the perpendicular 
direction reduce to boundary terms under the average, with these then becoming a large-scale perpendicular derivative similar to \cref{eq: stokes theorem thing}, while parallel derivatives commute trivially with $\avgp{\cdot}$.

Practically, this means that in order to compute fluxes in the \emph{third-order} equations, 
we need only keep total derivatives of \emph{second-order} products.

\paragraph{Geometrical identities for \(\bh\).}
The following identities of $\bh$, which are derived from $\nabla\cdot\bm{B}=0$ and $\bh\cdot\bh=1$, will be used extensively:
\[
\nabla\cdot \bh = -\,\gprl{B}, \qquad
\nabla \bh \cdot \bh = \bm 0\implies
 \bkap\cdot\bh=0 .
\]
The latter identity means that in a locally field-aligned orthonormal basis $(\bh,\hat{\perpunitvectorbase}_{1},\hat{\perpunitvectorbase}_{2})$, the $\nabla\bh$ tensor takes the form
\begin{equation}
\nabla\bh = \begin{pmatrix} 0 & \bm{0} \\
\bm{\kappa} & (\nabla \bh)_{\perp}
\end{pmatrix}.\label{eq: grad bh}
\end{equation}
On the perpendicular subspace we define,
\[
(\nabla \bh)_{\perp}
= \frac{1}{2}(\nabla\cdot \bh)\, \mathsf{I}_{\perp}
+ \mathsf{S}
+ \mathsf{A}
\]
where \(\mathsf{I}_{\perp}\) is the $2\times 2$ identity and \(\mathsf{S}\) and \(\mathsf{A}\) is the symmetric traceless “squeezing”  and antisymmetric ``twist'' tensors, respectively (their detailed construction and role are developed in the main text).

\subsection{First order: equilibria and $O(\epsilon)$ constraints on the fluctuations}\label{subapp: first order}

 We now apply the expansion from \S\ref{sub: notation} to the governing equations \eqref{eq:cont}--\eqref{eq:intE},
using the turbulence average $\langle\cdot\rangle$ to separate each 
equation into its mean and fluctuating parts. Noting that the turbulence average annihilates the 
fast time derivative, this yields,
at first order,  mass/flux conservation and equilibrium conditions on the mean quantities, as well 
as several important constraints on the fluctuations.

\paragraph{Continuity.}
Averaging the continuity equation at $O(\epsilon)$ gives the standard mass and flux conservation constraint
\begin{equation}
\nabla\cdot(\rho\, U\bh) = U\gradl \rho + \rho \gradl U + \rho U \nabla\cdot\bh =  S_{\rho},
\label{eq:first-cont-mean}
\end{equation}
which  yields
\begin{equation}
\gprl{\rho}+\gprl{U}-\gprl{B}=(\rho U)^{-1}S_{\rho}, 
\label{eq:mass-flux-invariant}
\end{equation}
or equivalently, that $ \rho U/B$ is constant along a field line in the absence of a mass source \cite{Leer1980}. 

The fluctuating part of the $O(\epsilon)$ continuity equation is 
\begin{align}
\rho\,\nabla\cdot\delta\bm u  =  0\quad\implies\quad
\nabla\cdot\delta\bm u_{\perp}  =  0,
\label{eq:div-up}
\end{align}
since $\bh\cdot\nabla\,\delta u_{\parallel}$ and $\delta u_{\parallel}\,\nabla\cdot\bh$
are $O(\epsilon^{2})$.

\paragraph{Momentum.}
Averaging the momentum equation at $O(\epsilon)$ gives
\begin{align}
\rho\,\bm{U}\cdot\nabla\bm{U}
=& 
-\,\nabla\left(p+\frac{B^2}{8\pi}\right)
 + \frac{\bm B\cdot\nabla\bm B}{4\pi}\nn\\&
 - \rho\,\nabla\Phi_{\mathrm{grav}}+\bh(\Sm - U\Srho),
\label{eq:first-mom-mean}
\end{align}
where  $\bm U=U\,\bh$.
It is helpful to decompose \eqref{eq:first-mom-mean} into its perpendicular and parallel parts giving, 
\begin{equation}
\va^{2}(\bkap-\gprp{B})- \frac{{c}_{s}^{2}}{\gamma}\gprp{p} = U^{2}\bkap + \nabla_{\perp}\Phi_{\rm grav}\label{eqapp: perp equil}
\end{equation}
and 
\begin{equation}
- \frac{{c}_{s}^{2}}{\gamma}\gprl{p} + \Sm-U\Srho=  \gradl\Phi_{\rm grav}+U^{2}\gprl{U} ,\label{eqapp: parallel equil}
\end{equation}
respectively.  The right-hand sides of \cref{eqapp: perp equil,eqapp: parallel equil} can be considered
as an effective gravity $\bm{g}_{\rm eff}$, with the flow contribution representing the effective force from centrifugal or linear acceleration, respectively:
\begin{align}
&\geff=-U^{2}\bkap - \nabla_{\perp}\Phi_{\rm grav},\nn\\
&{g}^{\|}_{\rm eff}=-U^{2}\gprl{U}-\gradl\Phi_{\rm grav};\label{eq: effective gravity}
\end{align}
see main text for further discussion. 

At $O(\epsilon)$, the fluctuating part of the  momentum equation gives
\begin{equation}
-\nabla_{\perp}
\left(\delta p + \frac{B\,\delta B_{\parallel}}{4\pi}\right)
= 0
\implies
\frac{\delta p}{p}
=
-\gamma\frac{ \va^{2}}{\cs^{2}}
\frac{\delta B_{\parallel}}{B},
\label{eq:perp-pressure-balance}
\end{equation}
i.e., perpendicular pressure balance at leading order, as in gyrokinetics or straight-field RMHD \citep{Schekochihin2009}.

\paragraph{Entropy}
At $O(\epsilon)$ the mean (entropy) equation is
\begin{align}
\rho T(\bm U\cdot\nabla s)& = \Sth- \gamma c_{v}T \Srho-\nabla\cdot\bm{q}\nn\\ \implies \gprl{s}&= \frac{\gamma-1}{p U}(\Sth- \gamma c_{v}T \Srho-\nabla\cdot\bm{q}),
\label{eq:entropy-mean}
\end{align}
showing that  $\gprl{s}$ quantifies the total deviation from adiabatic evolution along a 
field line due to nonideal effects and heating/mass sources.
We include the mean heat flux here at lowest order explicitly (unlike the viscous and resistive effects) because it is more often relevant to the low-order equilibrium; for instance, in modeling a kinetic plasma where 
a heat flux strongly modifies the background.

Fluctuations do not contribute to the entropy equation at $O(\epsilon)$, but \cref{eq:entropy def} yields a useful 
relation for future reference:
\begin{equation}
\frac{\d s}{c_{v}} = \frac{\d p}{p} - \gamma \frac{\d \rho}{\rho}.\label{eq: s p and rho}
\end{equation}
Likewise, equilibrium gradients are related via $\gprp{s} = \gprp{p}-\gamma \gprp{\rho}$ and $\gprl{s} = \gprl{p}-\gamma \gprl{\rho}$.

\paragraph{Induction and solenoidality.}
With $\bm U\parallel\bm B$, the mean induction equation contributes no new
constraint at this order. Solenoidality at $O(\epsilon)$ gives
\begin{align}
\nabla\cdot\delta\bm B_{\perp} \;=\; 0,
\label{eq:div-bp}
\end{align}
and, together with \eqref{eq:div-up}, allows the standard potential representations
$\delta\bm u_{\perp}=\bh\times\nabla\Phi$ and
$\delta\bm B_{\perp}/\sqrt{4\pi\rho}=\bh\times\nabla\Psi$.

\paragraph{Summary of first-order constraints.}
The lowest-order background obeys mass-flux conservation \eqref{eq:mass-flux-invariant} and the
equilibrium force balance \eqref{eqapp: perp equil}–\eqref{eqapp: parallel equil}.

Fluctuations are  incompressible in the perpendicular direction
\eqref{eq:div-up}–\eqref{eq:div-bp} and satisfy  perpendicular pressure balance
\eqref{eq:perp-pressure-balance}. Via \eqref{eq: s p and rho}, one has the 
choice to express thermal fluctuations in terms of  two of  $\d p/p $, $\d s$, $\d\rho/\rho $, or $\d T/T=\d p/p-\d\rho/\rho$. 
We will use various options below to optimize compactness and physical clarity.

\subsection{Second order: the generalized RMHD equations}\label{subapp: second order}

At second-order, we will obtain the dynamical equations for Alfv\'enic and compressive fluctuations, 
{viz.,} the generalized RMHD equations. Before 
doing this, however, it is necessary to justify the procedure
by deriving the averaged equations  at second order (\S\ref{subsub: aver at second order}). We will 
find that 
the $O(\epsilon^{2})$ averaged  equations are homogeneous, meaning that all  first-order mean quantities $\langle \rho^{(1)}\rangle,\,\langle \bm u^{(1)}\rangle$ etc.~can be set to zero and will remain zero. This simplifies
the system significantly. It occurs because all  quadratic fluctuation averages reduce to perpendicular
divergences or curls at this order, and are therefore $O(\epsilon^3)$ under our averaging scheme. Rotation is ordered $O(\epsilon^3)$ and therefore also absent, while we assume the sources have only  $O(\epsilon)$ parts.

\subsubsection{Averaged equations at second order}\label{subsub: aver at second order}
Throughout this subsection we will use the shorthand notation  $g^{(1)}$ for the {first-order mean}
(i.e.\ $g^{(1)}\equiv\langle g^{(1)}\rangle$ for $g=\rho,\,\bm{u}$ etc.).
\paragraph{Continuity.}
At second  order, the average of the continuity equation is
\begin{align}
\nabla\cdot\!(\rho^{(1)}\,\bm U + \rho\,\bm u^{(1)})
 + 
\cancel{\,\big\langle \nabla\cdot\!\left(\duprp\,\d\rho\right)\big\rangle\,}
 =  0.
\label{eq:cont-second-mean}
\end{align}
The quadratic term is a total perpendicular divergence and therefore contributes only at $O(\epsilon^3)$ (see \S\ref{app para: perp average}). It is thus  dropped here.

\paragraph{Momentum.}
Because dissipation through $\bm\Pi$ is assumed to act more strongly on the small scales than large, and therefore only on  fluctuating quantities, we 
start by dropping the dissipation in the averaged momentum equation at $O(\epsilon^2)$, giving
\begin{align}
\rho\bm U&\cdot\nabla \bm u^{(1)}
 + 
\rho\bm u^{(1)}\cdot\nabla \bm U + \rho^{(1)}\bm U\cdot\nabla \bm U 
= -\rho^{(1)}\,\nabla\Phi_{\mathrm{grav}} \nn\\
& -\,\nabla\!\left(p^{(1)} + \frac{B\,B^{(1)}}{4\pi}\right)
 + \frac{
\bm B^{(1)}\!\cdot\nabla\bm B + \bm B\!\cdot\nabla\bm B^{(1)}}{4\pi}\nn\\
&- u_{\|}^{(1)}\Srho- \cancel{\rho\big\langle \duprp\cdot\nabla \duprp\big\rangle\,} + 
\cancel{\frac{\big\langle \dbprp\cdot\nabla \dbprp\big\rangle}{4\pi}}.
\label{eq:mom-second-mean-hom}
\end{align}
Using the first-order constraints $\nabla\cdot\duprp=\nabla\cdot\dbprp = 0$, the fluctuation average becomes
$\langle \duprp\cdot\nabla \duprp\rangle=\langle \nabla\cdot(\duprp\,\duprp)\rangle$
and similarly for $\dbprp$. Since these are total divergences, they are
$O(\epsilon^3)$ under averaging and thus removed as in \cref{eq:cont-second-mean}.

\paragraph{Induction.}
Averaging the induction equation at $O(\epsilon^2)$ yields, ignoring the effect of resistive 
dissipation on the averaged field,  
\begin{align}
\nabla\times\!(\bm u^{(1)}\times\bm B + \bm U\times\bm B^{(1)})
+
\cancel{\big\langle \nabla\times(\d\bm u\times\d\bm B)\big\rangle}
= 0.
\label{eq:ind-second-mean}
\end{align}
Like the divergence terms above, the curl of the fluctuations is a boundary term under the perpendicular spatial average and is thus $O(\epsilon^3)$; it drops here.

\paragraph{Entropy.}
The viscous and resistive heating terms in the entropy equation enter at $O(\epsilon^3)$ because they balance the dissipation of fluctuation energy, which appears from $\d \bm{u}\cdot \partial_{t}\d \bm{u}$ and similar terms (see below). We likewise assume there is no second-order contribution to $\Sth$ (this can be absorbed into the first-order part). 
The averaged entropy equation is then,
\begin{align}
\bm u^{(1)}\cdot\nabla s
& + 
\bm U\cdot\nabla s^{(1)}
 + 
\cancel{\,\big\langle \duprp\cdot\nabla \d s\big\rangle\,}
= -\frac{\nabla\cdot\langle {\bm q}^{(1)}\rangle}{\rho T} \nn\\
&\qquad- \frac{p^{(1)}}{p}\frac{\Sth-\nabla\cdot\bm{q}}{\rho T} + \frac{\rho^{(1)}}{\rho}\frac{\gamma c_{v}\Srho}{\rho}
\label{eq:ent-second-mean}
\end{align}
with the quadratic average moving to $\mathcal{O}(\epsilon^{3})$ because it is a total divergence.
Taking the first-order averaged heat flux $\langle {\bm q}^{(1)}\rangle$ to be a function of first-order mean quantities, we see that this equation is also homogeneous. 

\paragraph{Consequence.}
Equations \eqref{eq:cont-second-mean}, \eqref{eq:mom-second-mean-hom}, \eqref{eq:ind-second-mean}, and
\eqref{eq:ent-second-mean} are {homogeneous} in $\rho^{(1)},\,\bm u^{(1)},\,\bm B^{(1)}$ and $s^{(1)}$.
Thus there is no forcing of first-order means by fluctuations at $O(\epsilon^2)$: if these are zero initially --- effectively a choice of the  initial conditions --- 
they remain zero on transport timescales, giving
\begin{equation}
\rho^{(1)}=s^{(1)}=p^{(1)}=0,\quad \bm u^{(1)} = \bm B^{(1)}=0.
\end{equation}
Fluctuation effects first enter the mean evolution at $O(\epsilon^3)$
via divergences/curls of quadratic fluxes and the averaged dissipation (heating).

Note that this differs from the gyrokinetic system in standard treatments \cite{Abel2013}, where 
neoclassical effects --- arising from the competition between particle drifts and collisions --- arise at this order and
 cause important corrections to the distribution function. Their absence here is 
 a result of starting from collisional MHD, effectively taking the dynamics of all species to be identical (see App.~\ref{app: gyrokinetics}).  Such 
 effects will also exist in the solar coronal and other astrophysical plasmas, so may be interesting to explore in future work.

\subsubsection{Alfv\'enic fluctuations (perpendicular dynamics)}
\label{sec:alfvenic_fluctuations}
The fluctuating part of the expansion at  second order   splits cleanly into dynamical equations
for the Alfv\'enic (field-perpendicular) dynamics and compressive (parallel) dynamics. As 
discussed in the main text (see \cref{fig: cartoons}), curvature and perpendicular gradients couple 
these linearly, while the nonlinearity is unchanged compared to straight-field RMHD.

\paragraph{Setup}The derivation of the Alfv\'enic equations starts with application of the perpendicular projector \cref{eqapp: projectors} to the momentum  and  induction equations.
It is convenient to record the following $\perp$–$\|$ decompositions (valid to leading order in $\epsilon$). For any background
$\bm G=G\,\bh$ with $G= U$ or $B$ and any fluctuation $\bm{g} = \d\bm u$ or $\d\bm B$,
\begin{subequations}\label{eq:PP-splits}
\begin{align}
\d\bm g\cdot\nabla\bm G
&= 
\big(\, G\,\d\bm g_\perp\cdot\nabla\bh  +  \d g_\|\,G\,\bkap \,\big)
\nn\\ &\quad+ 
\bh\,\big(\, G\,\d\bm g_\perp\cdot\gprp{G}  +  G\,\d g_\|\,\gprl{G} \,\big),\\
\bm G\cdot\nabla\,\d\bm g
&= 
\big(\, G\,\bh\cdot\nabla\d\bm g_\perp  +  G\,\bh\,\bkap\cdot\d\bm g_\perp  +  G\,\d g_\|\,\bkap \,\big)
\nn\\ &\quad+ 
\bh\,\big(\, G\,\bh\cdot\nabla\d g_\|  -  G\,\bkap\cdot\d\bm g_\perp \,\big),\label{eq: perp parallel expansion 2}\\
\d\bm g\cdot\nabla\,\d\bm g
&= 
\big(\d\bm g_\perp\cdot\nabla\d\bm g_\perp \big)
 + 
\bh\,\big(\d\bm g_\perp\cdot\nabla\d g_\| \big).
\end{align}\label{eq: perp parallel expansion}\end{subequations}
The $\bh$ term in the perpendicular part of \eqref{eq: perp parallel expansion 2} is needed 
to ensure this part of $\bm G\cdot\nabla\,\d\bm g$ remains locally perpendicular to $\bh$; the negative contribution lies along $\bh$. 
This term, which accounts for the parallel transport of vectors along a curved $\bh$, is purely geometrical and disappears in the potential formulation of the equations (\cref{eq: potential form} below).

It is convenient to define the total time and parallel derivatives,
\begin{align}
\frac{\rmd}{\rmd t}\;&\equiv\; \frac{\partial}{\partial t} + \bm U\cdot\nabla + \duprp\cdot\nabla, \nn\\
\bhdg \;&\equiv\; \bh\cdot\nabla + \frac{\dbprp}{B}\cdot\nabla,\label{eq: total time deriv}
\end{align}
as well as  the ``Alfv\'enic'' versions 
\begin{equation}
\frac{\rmd^{\rm A}}{\rmd t}\equiv \frac{\rmd}{\rmd t} + \bm U\,\bkap\cdot,\quad \bhdg^{\rm A} \equiv\; \bhdg + \bh\,\bkap\cdot.\label{eq: alfvenic total time}
\end{equation}

\paragraph{Dissipation}
We will  use the notation $\mathcal{D}^{\perp,\|}_{g}$  to reference the small-scale dissipation of any 
quantity in the 
perpendicular/parallel direction  (e.g., $\bm{\mathcal{D}}^{\perp}_{u}$ and $\mathcal{D}^{\|}_{u}$   for the velocity dissipations). These arise at $\mathcal{O}(\epsilon^{2})$ from viscosity, resistivity, or heat fluxes, as required to dissipate fluctuation energy at the same rate it is produced. However, it is not worthwhile
to work out the form of each $\mathcal{D}$ in detail in terms of $\nabla\cdot\bm{\Pi}$, $\eta$, and $\nabla\cdot\bm{q}$, because (i) the collisional forms are often inappropriate anyway, particularly in the solar wind, and (ii) artificial hyper-diffusion, or non-physical values of dissipation coefficients, are almost always used in turbulence simulations. The only required properties of $\mathcal{D}$ are
(i) that their action on mean quantities is negligible (see \cref{subsub: aver at second order}) and (ii) that they provide a positive-definite sink for the energy associated with the 
fluctuation in question.

\paragraph{Velocity and magnetic-field perturbations }
We expand the momentum \eqref{eq:mom} and induction \eqref{eq:ind} equations then apply the perpendicular projection with \eqref{eq: perp parallel expansion}, yielding
\begin{flalign}
\frac{\rmd^{\rm A}}{\rmd t}\duprp&
 +  U\,\duprp\cdot\nabla\bh
= 
\frac{B}{4\pi\rho}(
\bhdg^{\rm A} \dbprp
 +  \dbprp\cdot\nabla\bh
)
\nn\\
&-2\,\bkap\!\left(U\duprl  -  \frac{B\dbprl}{4\pi\rho}\right)
 + \frac{\d\rho}{\rho}\,\bm g_{\rm eff}^{\perp} -\frac{ \nabla_{\perp}\tilde{p}}{\rho} \nn\\ &
 -\duprp\frac{\Srho}{\rho}+ \bm{\mathcal{D}}^{\perp}_{u}
\label{eq:perp-mom}
\end{flalign}
and
\begin{align}
\frac{\rmd^{\rm A}}{\rmd t}\dbprp
 & +  B\duprp\cdot\nabla\bh
= 
B\bhdg^{\rm A} \duprp
 +  U\dbprp\cdot\nabla\bh \nn\\
&-U\dbprp(\gprl{U}-\gprl{B})+ \bm{\mathcal{D}}^{\perp}_{B}.
\label{eq:perp-ind}
\end{align}
We have used $\nabla\cdot\bm{U}=U(\gprl{U}-\gprl{B})$ in \cref{eq:perp-ind}, while in \cref{eq:perp-mom}, $\nabla_{\perp}\tilde{p} = \nabla_{\perp}\d (p+B^{2}/8\pi)^{(2)}$ is fixed by the requirement that $\nabla\cdot\duprp = 0$ from the first-order equations.
Here and below, where we list $\Srho$ in the fluctuation equations, it should be considered as convenient shorthand for the gradient combination $\rho U(\gprl{\rho}+\gprl{U}-\gprl{B})$, in keeping with the idea that the fluctuations' evolution can be computed for any arbitrarily varying background flux tube.

\paragraph{Els\"asser form}
Define
\begin{equation}
\zpm  \equiv  \duprp \mp \frac{\dbprp}{\sqrt{4\pi\rho}}.
\end{equation}
Pulling the density gradients through the Alfv\'enic total time and parallel derivatives, then adding and subtracting \eqref{eq:perp-mom} and \eqref{eq:perp-ind} yields:
\begin{align}
\frac{\partial \zpm}{\partial t}
& +  (U\pm\va)\,\big(\bh\cdot\nabla + \bh\,\bkap\cdot\big)\,\zpm
 \nn\\ &+(U\mp\va)  \left[\zmp\cdot\nabla\bh -  \frac{1}{4} \gprl{\rho}(\zpm-\zmp)\right]
   \nn\\
=& - \zmp\cdot\nabla \zpm
 - \rho^{-1}\nabla_{\perp} \tilde p  - \zpm \frac{\Srho}{\rho}\nn\\
& -\,2\bkap\!\left(U\,\duprl - \frac{B\,\dbprl}{4\pi\rho}\right)
+ \frac{\d\rho}{\rho}\,\bm g_{\rm eff}^{\perp}+ \bm{\mathcal{D}}^{\perp}_{\pm}.
\label{eq:perp-pm-app}
\end{align}
For $\Srho=0$, this becomes \cref{eq: Alfvenic fluctuation equations main}  in the main text, where we discuss extensively its physical 
content and how it reduces to the now-standard 
straight-field form of  \citet{Chandran2009} (see also \citet{Velli1993}). The more general form \eqref{eq:perp-pm-app} demonstrates that 
adding mass to the system via $\Srho$ causes $\zpm$ to damp in time.

\paragraph{Potential form}
Because $\nabla\cdot\duprp=\nabla\cdot\dbprp=\nabla\cdot\zpm = 0$, these equations can be written in terms of the Alfv\'enic potentials \citep{Schekochihin2009}:
\begin{equation}
\duprp = \bh\times \nabla \Phi,\qquad \frac{\dbprp}{\sqrt{4\pi \rho}} = \bh\times \nabla\Psi.\label{eq: phi psi defn}
\end{equation}
This is achieved by taking $\bh\cdot\nabla \times $\eqref{eq:perp-mom} and uncurling \eqref{eq:perp-ind}, making use of the identities $\bh\cdot\nabla \times (\duprp\cdot\nabla\bh) = - (\nabla\bh)_{\perp}:\nabla_{\perp}\nabla_{\perp}\Phi + \nabla\cdot\bh\, \nabla_{\perp}^{2} \Phi$ and $\bh\cdot\nabla \times (\bh\cdot\nabla\duprp ) = (\nabla\bh)_{\perp}:\nabla_{\perp}\nabla_{\perp}\Phi + \bh\cdot\nabla \nabla_{\perp}^{2} \Phi$ (valid only to leading order in $\epsilon$). This yields
 \begin{subequations}
\begin{align}
 \frac{\rmd}{\rmd t}\nabla_{\perp}^{2}\Phi&  = \va \bhdg \nabla_{\perp}^{2}\Psi + U \gprl{B} \nabla_{\perp}^{2}\Phi - \va \gprl{\va}  \nabla_{\perp}^{2}\Psi\nn\\
 & +2\,\bh\times \bkap\cdot\left(U^{2}\nabla\frac{\duprl}{U} - \va^{2}\nabla\frac{\dbprl}{B}\right) \nn\\ &-\bh\times \bm g_{\rm eff}^{\perp} \cdot \nabla\frac{\d \rho}{\rho} + \mathcal{D}^{\perp}_{\Phi}-\frac{\Srho}{\rho}\nabla_{\perp}^{2 }\Phi,\label{eq: potential form phi}\\
  \frac{\rmd}{\rmd t}\Psi =& \va \bh\cdot\nabla\Phi  + \va\gprl{B} \Phi - U \gprl{\va}\Psi+ \mathcal{D}^{\perp}_{\Psi}-\frac{\Srho}{\rho}\Psi.\label{eq: potential form psi}
\end{align}\label{eq: potential form}\end{subequations}
This form is perhaps less physically intuitive than those above but provides the clearly superior choice for computational studies. A specific case with $U=0$ in a straight flux tube ($\bkap=\bm g_{\rm eff}^{\perp}=0$) was derived and solved by \citet{vanBallegooijen2011} to study turbulent heating of the low corona. Another, with $U=0$, $\gprl{\rho}=0$, and low $\beta$, was derived by \citet{Strauss1976}  for studying fluctuations in tokamaks. One 
can also combine Eqs.~\eqref{eq: potential form} into a set for the potentials of $\zpm$, $\zeta^{\pm}= \Phi \mp \Psi$; this form yields terms  $\propto(\nabla\bh)_{\perp ij}\zeta^{\pm}_{,ij}$ that cause reflection from expansion and/or squashing, as in \cref{eq:perp-pm-app}. 

\subsubsection{Compressive  fluctuations (parallel dynamics)}\label{sub: compressive flucts app}

\paragraph{Entropy.}
The fluctuating entropy equation becomes
\begin{align}
\frac{\rmd}{\rmd t}\,\frac{\d s}{c_{v}}
=&-\,\duprp\cdot\gprp{s} -\duprl\,\gprl{s} -  U \frac{\d p}{p}\gprl{s}\nn\\&+ \gamma  \frac{\d T}{T}\frac{\Srho}{\rho}+ \mathcal{D}_{s},
\label{eq:s-fluct1}
\end{align}
where $\d T/T=\d p/p - \d\rho/\rho$.
Using the $O(\epsilon)$ perpendicular pressure balance \eqref{eq:perp-pressure-balance}, 
this can be rewritten as
\begin{align}
\gamma \frac{\rmd}{\rmd t}\!\left(\frac{\va^2}{\cs^2}\frac{\dbprl}{B} + \frac{\d\rho}{\rho}\right)
=&\;\duprp\cdot\gprp{s} + \left(\duprl+U\frac{\d p}{p}\right)\gprl{s}\nn\\ &-\gamma \frac{\d T}{T}\frac{\Srho}{\rho}-\mathcal{D}_{s},
\label{eq:s-combo}
\end{align}
which we will make use of soon.

\paragraph{Continuity.}
The fluctuating part of the continuity equation at $\mathcal{O}(\epsilon^{2})$ becomes,
\begin{align}
\frac{\rmd}{\rmd t}\!\left(\frac{\d\rho}{\rho}\right)
&=  -\,\duprp\cdot\gprp{\rho}
 -  \duprl\,\gprl{\rho} - \frac{\d\rho}{\rho}\frac{\Srho}{\rho}
 -  (\nabla\!\cdot\d\bm u)^{(2)},
\label{eq:cont-2nd-fluct}
\end{align}
where $(\nabla\!\cdot\d\bm u)^{(2)} = \nabla\cdot\duprp^{(2)} + \nabla\cdot(\bh\,\duprl)$.

\paragraph{Parallel induction equation.}
Starting from the parallel fluctuating part of the induction equation with  the identities \eqref{eq: perp parallel expansion}, then using 
\cref{eq:cont-2nd-fluct} to eliminate $(\nabla\!\cdot\d\bm u)^{(2)}$, we obtain
\begin{align}
\frac{\rmd}{\rmd t}\,& \dbprl
-  B\,\frac{\rmd}{\rmd t}\!\left(\frac{\d\rho}{\rho}\right)
 -  B\,\bhdg\,\duprl -   B\frac{\d\rho}{\rho}\frac{\Srho}{\rho}
 \nn\\
 &= U\,\dbprp\!\cdot\!\big(\bkap + \gprp{U}\big)
 -  B\duprp\!\cdot\!\big(\bkap + \gprp{B} - \gprp{\rho}\big) \nn\\
&\quad-B\,\left(\gprl{B}-{\gprl{\rho}}\right)\,\duprl
 + U\,\gprl{B}\,\dbprl + \mathcal{D}^{\|}_{B}.
\label{eq:par-ind-raw}
\end{align}
Eliminating $\rmd/\rmd t (\d\rho/\rho)$ via \eqref{eq:s-combo} and using pressure balance \eqref{eq:perp-pressure-balance}  to write $\d p/p$ in terms of $\dbprl/B$ gives
\begin{align}
\Bigg(1&+\frac{\va^2}{\cs^2}\Bigg)\frac{\rmd}{\rmd t}
\frac{\dbprl}{B}
- 
\bhdg\duprl
={\mathcal{D}}^{\|}_{B} \label{eq:par-ind-comb}\\
&+  {U}\frac{\dbprp}{B}\!\cdot\!\big(\bkap + \gprp{U}\big)
 -  \duprp\!\cdot\!\Big[\bkap + \gprp{B} - \frac{1}{\gamma}\,\gprp{p}\Big]\nonumber\\
 &+  \Bigg(\frac{\gprl{p}}{\gamma}-\gprl{B}\Bigg)\duprl + U\frac{\va^{2}}{\cs^{2}} \left(\gprl{\beta} - \gprl{s} + \gamma \frac{\Srho}{\rho U}\right)\frac{\dbprl}{B} \nn
\end{align}
and
\begin{align}
&\Bigg(1+\frac{\cs^2}{\va^2}\Bigg)\,
\frac{\rmd}{\rmd t}\!\left(\frac{\d\rho}{\rho}\right)
+\bhdg\duprl = 
-  {U}\frac{\dbprp}{B}\!\cdot\!\big(\bkap + \gprp{U}\big) \nn\\
&\quad +  \duprp\!\cdot\!\left[\bkap + \gprp{B} + \frac{1}{\gamma}\frac{\cs^{2}}{ \va^{2}}\,\gprp{p}-\left( 1+\frac{\cs^{2}}{\va^{2}}\right)\gprp{\rho} \right]\nonumber\\
&\quad+  \!\left(\frac{1}{\gamma}\frac{\cs^{2}}{ \va^{2}}\gprl{s}\!-\!\gprl{\rho}\!+\!\gprl{B}\!\right)\duprl + U\!\left(\!\gprl{\beta} \!-\! \gprl{s} \!+ \!\gamma \frac{\Srho}{\rho U}\!\right)\!\!\frac{\dbprl}{B}\nn\\ &
\quad- \left(1+\frac{\cs^{2}}{\va^{2}}\right)\frac{\d\rho}{\rho}\frac{\Srho}{\rho}+{\mathcal{D}}^{\|}_{\rho},
\label{eq:density-evol}
\end{align}
where the  ${\mathcal{D}}^{\|}_{B}$ in \eqref{eq:par-ind-comb} has changed its normalization from  \eqref{eq:par-ind-raw}, but will not prove necessary 
to keep track of these details. The dissipation ${\mathcal{D}}^{\|}_{\rho}$ in \cref{eq:density-evol}  arises from the density's coupling to $\dbprl$ and $\d s$ fluctuations.

\paragraph{Parallel momentum equation.}
Applying the parallel projection to the fluctuating momentum equation at $\mathcal{O}(\epsilon^{2})$, and using pressure balance \eqref{eq:perp-pressure-balance} and the identities \eqref{eq: perp parallel expansion}, yields
\begin{align}
&\frac{\rmd\,\duprl}{\rmd t}
 -  \va^{2}\,\bhdg\frac{\dbprl}{B}
=  \mathcal{D}^{\|}_{u}+  \frac{\d\rho}{\rho}g^{\|}_{\rm eff} \nn\\
&\quad
+  \va^{2}\frac{\dbprp}{B}\!\cdot\!\big(\gprp{B}-\bkap\big)
 -  U\,\duprp\!\cdot\!\big(\gprp{U}-\bkap\big)
\nn\\
&\quad+2\va^{2}
\gprl{B} \frac{\dbprl}{B}-U\,\duprl\left(\gprl{U} +\frac{\Srho}{\rho U}\right).
\label{eq:par-mom}
\end{align}

\paragraph{Slow-mode variables.}
Introduce the slow-mode eigenmode variables $\zlpm = \duprl \mp  \dVprlx$ with 
$ \dVprlx \;\equiv\; ({\va^{2}}/{\vs})({\dbprl}/{B}).$
Rearranging Eqs.~\eqref{eq:par-ind-comb} and \eqref{eq:par-mom}
yields
\begin{align}
&\frac{\rmd\,\duprl}{\rmd t} 
 -  \vs\,\bhdg\dVprlx
= \mathcal{D}^{\|}_{u} +  \frac{\d\rho}{\rho}g^{\|}_{\rm eff}
 \nn\\
 &\quad+\va^{2}\frac{\dbprp}{B}\!\cdot\!\big(\gprp{B}-\bkap\big)
-U\,\duprp\!\cdot\!\big(\gprp{U}-\bkap\big)
\nn\\ &\quad -U\,\duprl \left(\gprl{U} +\frac{\Srho}{\rho U}\right)
 +   \vs\dVprlx(\gprl{\vs}+\gprl{\rho}),
\label{eq:par-mom-slow}
\\[0.25em]
&\frac{\rmd\, \dVprlx}{\rmd t}
 -  \vs\,\bhdg\duprl
=  \mathcal{D}^{\|}_{V}\label{eq:par-ind-slow}
 \\
&\quad
-\vs\,\duprp\!\cdot\!\Big(\bkap + \gprp{B} - \frac{1}{\gamma}\,\gprp{p}\Big) +{U \vs}\frac{\dbprp}{B}\!\cdot\!\big(\bkap+\gprp{U}\big)\nn\\
&\quad-\vs\!\left(\!\gprl{B} \!-\! \frac{\gprl{p}}{\gamma} \!\right)\!\duprl \!+\!  U\left[\gprl{\vs} \!-\! \frac{\vs^{2}}{\cs^{2}}\!\left(\!\gprl{s}\!+\! \gamma \!\frac{\Srho}{\rho U}\!\right)\right]\!\dVprlx,\nn
\end{align}
where $\mathcal{D}^{\|}_{V}$ involves a mix of resistive and heat-flux dissipation, since $\dVprlx$ involves magnetic and pressure perturbations. Rewriting \eqref{eq:par-mom-slow}-\eqref{eq:par-ind-slow} in terms of $\zpm$ rather than $\duprp$ and $\dbprp$ yields Eqs.~\eqref{eq: parallel mom alfv}-\eqref{eq: parallel induction alfv} in the main text.

\paragraph{Slow-mode Els\"asser form}
Setting $\Srho=\Sm=0$ for simplicity, such that $g^{\|}_{\rm eff} = p\gprl{p}/\rho$ and $\gprl{\rho}+\gprl{U}-\gprl{B}=0$, then writing $\d\rho$ in terms of $\dVprlx$ and $\d s$ and combining Eqs.~\eqref{eq:par-mom-slow}-\eqref{eq:par-ind-slow} gives the eigenmode form,
\begin{align}
&\frac{\rmd\zlpm}{\rmd t} \pm \vs \bhdg\zlpm = \mathcal{D}^{\|}_{\pm}\nn\\
&\quad
- \frac{1}{2} (U\mp\vs)\left[\zlpm(\gprl{U} - \gprl{\vs})+ \zlmp(\gprl{U} + \gprl{\vs})\right]\nn\\
&\quad  - \frac{1}{2}U \gprl{s} \left[ \frac{\vs^{2}}{\cs^{2}}(\zlpm - \zlmp) \mp \frac{\vs}{\gamma} \zlmp \right]- \frac{T \d s}{\gamma(\gamma-1)}\gprl{p}
\nn\\
&\quad-\frac{\duprp}{U}\cdot\left[ U^{2}(\gprp{U}-\bkap) \mp U\vs \Big(\bkap + \gprp{B} - \frac{1}{\gamma}\,\gprp{p}\Big)\right]\nn\\
&\quad+\frac{\dbprp}{B}\cdot\left[ \va^{2}(\gprp{B}-\bkap) \mp U \vs(\bkap+\gprp{U}) \right] .\label{eq: slow mode zlpm}
\end{align}
Terms proportional to $\zlpm$ capture the slow mode's WKB growth/decay, while those proportional to $\zlmp$ and $\d s$ are 
a generalized slow-mode reflection. In an isentropic background $\gprl{s}=0$, the former --- WKB physics via the term $ (U\mp\vs)\zlpm(\gprl{U} - \gprl{\vs})/2$ --- is the exact generalization (with $\va\rightarrow\vs$) of the same term for Alfv\'en waves (see \cref{eq:perp-pm-app}):
$$\frac14 (U\mp\va) \gprl{\rho} = -\frac12(U\mp\va)(\gprl{U}-\gprl{\va}). $$
We can thus  form ``slow-mode wave-action variables'' --- 
the slow-mode equivalent of the (Alfv\'enic) \citet{Heinemann1980} variables  --- by defining
\({f}^{\pm}_{S}\equiv \mathsf{F}_{S}^{\pm}(R)\zlpm\), with  $\mathsf{F}_{S}^{\pm}(R) = \mathcal{M}_{S}^{1/2}\pm\mathcal{M}_{S}^{-1/2}$, 
where $\mathcal{M}_{S} \equiv U/\vs$ (cf.~$\bm{f}^{\pm}$ in \cref{subsec:RDT_basic}). 
Writing \cref{eq: slow mode zlpm} in terms of ${f}^{\pm}_{S}$ gives, 
\begin{align}
&\frac{\rmd  f_{S}^{\pm}}{\rmd t} \pm \vs \bhdg f_{S}^{\pm} + \frac{1}{2}f_{S}^{\mp}(U\pm\vs)(\gprl{U} + \gprl{\vs})\nn\\
&\quad+ \frac{\mathsf{F}^{\pm}_{S}T \gprl{p}}{\gamma(\gamma-1)} \d s + \frac{U}{2}\left[\frac{\vs^{2}}{\cs^{2}} f_{S}^{\pm} -\frac{\mathsf{F}^{\pm}_{S}}{\mathsf{F}^{\mp}_{S}}\left(\frac{\vs^{2}}{\cs^{2}}\mp \frac{2\vs}{\gamma U}\right)\!f_{S}^{\mp} \right]\gprl{s} \nn\\
&\quad= \{\duprp,\,\dbprp\: \text{terms}\} +  \mathsf{F}^{\pm}_{S}\mathcal{D}^{\|}_{\pm},\label{eq: f for slow modes}
\end{align}
where the $\duprp$, $\dbprp $ terms on the final line are just those from \eqref{eq: slow mode zlpm} multiplied by $\mathsf{F}^{\pm}_{S}$.
When $\gprl{s}=0$, the substitution has
eliminated the WKB growth term  in \cref{eq:wave_action_straight}, thereby isolating reflection and entropy coupling. If $\gprl{s}\neq0$, signifying continual heating/cooling of the background, 
we see an additional growth of slow waves arising from exchange with the background thermal energy because pressure fluctuations drive entropy fluctuations (specifically, the contribution can be traced to the $\d p/p$ term in \cref{eq:s-fluct1}).
This similarity to the Alfv\'enic sector does not carry over to reflection, which scales with $\gprl{U}+\gprl{\vs}$ for slow modes  (when $\gprl{s}=0$) and $\gprl{\va}$ for Alfv\'en waves. 

The form \eqref{eq: f for slow modes} is not useful for our purposes, because  slow modes do not interact nonlinearly, and because of the relevant timescales in the solar wind --- we argue in \cref{subsec:comp_absorb} that $\duprl$ and $\dbprl$ are turbulently damped before they propagate. It 
nonetheless brings a pleasing symmetry to the theory and could be useful in other contexts.

 \subsubsection{Free energy conservation}

We now consider free-energy conservation in a small perpendicular patch. 
Background inhomogeneity introduces source terms that capture the instantaneous exchange of energy between the fluctuations and the slowly varying mean fields.  
A minor notational inconvenience is useful to flag before proceeding: throughout this subsection we use tildes --- e.g., $\tilde W$ and $\tilde{\mathcal Y}$ for energy and exchange terms, respectively --- to denote quantities that have been averaged  over the perpendicular patch, but not yet over the intermediate timescale.
The transport equations given in the main text and derived below (\cref{sub: transport equations}) are  additionally averaged over time with $\avgi{\cdot}_t$, thereby involving   turbulent-averaged quantities (via $\avgi{\cdot}=\avgi{\avgi{\cdot}_\perp}_t$), and corresponding  quantities are then denoted without tildes, e.g., $W=\langle \tilde W\rangle_t$ or  $\mathcal{Y}=\avgi{\tilde{\mathcal Y}}_t$. The distinction matters only insofar as the time average eliminates  fast time derivatives of the energies.

 In the homogeneous RMHD of \citet{Schekochihin2009} there are five \emph{individually} conserved free energies: 
 \begin{subequations}
 \begin{align}
\tilde{W}^{\pm}_{\perp} &= \frac{1}{4}\rho  \avgp{|\zpm|^{2}}, \quad \text{(Alfv\'enic fluctuations)}\\
\tilde{W}^{\pm}_{\|} &= \frac{1}{4}\rho  \avgp{|\zlpm|^{2}},\quad \text{(Slow-mode fluctuations)}\\
\tilde{W}_{s} &=\frac{1}{2} \frac{p}{\gamma(\gamma-1)} \avg{\frac{\d s^{2}}{c_{v}^{2}}}_{\perp}\!\!,\:\: \text{(Entropy modes).}\label{eq: energy definitions entropy}
\end{align}\label{eq: energy definitions homogeneous}\end{subequations}
Each $\tilde{W}$ satisfies $\partial \tilde{W}_{\rm wave}/\partial t + \nabla\cdot (\bh\, v_{\rm wave} \tilde{W}_{\rm wave})=0$, where $v_{\rm wave}=\{\pm\va,\:\pm\vs,\:0\}$ for $\tilde{W}_{\rm wave}=\{\tilde{W}^{\pm}_{\perp}, \tilde{W}^{\pm}_{\|}, \tilde{W}_{s}\}$, respectively (assuming $U=0$). While  
 the choice of constants in \cref{eq: energy definitions entropy} is arbitrary when $\tilde{W}_{s}$ is individually conserved, the form therein ensures \begin{equation}
\tilde{W}^{\rm tot} \equiv \tilde{W}^{+}_{\perp} + \tilde{W}^{-}_{\perp}+ \tilde{W}^{+}_{\|} + \tilde{W}^{-}_{\|} + \tilde{W}_{s}\label{eq: wtot defn}
\end{equation}
is the fluctuating part of the standard conserved MHD energy,
\begin{equation}
{E}^{\rm MHD} = \frac{1}{V}\int dV\left(\frac{1}{2}\rho |\bm{u}|^{2} + \frac{|\bm{B}|^{2}}{8\pi} + \frac{p}{\gamma -1}\right).
\end{equation}
We will explain this connection more completely below (see \S\ref{sub: mean defns}) and for now just remark that $\tilde{W}_{\|}^{+}+\tilde{W}_{\|}^{-} = \rho\, \langle \duprl^{2}\rangle_\perp/2+ \rho\, \langle \dVprlx^{2}\rangle_\perp/2$ and
\begin{equation}
\frac{1}{2}\rho\, \dVprlx^{2} = \frac{\dbprl^{2}}{8\pi} + \frac{p}{2\gamma} \left(\frac{\d p}{p}\right)^{2},\label{eq: dvprlx energy}
\end{equation}
from which  the second term combines  with $\tilde{W}_{s}$ to give  second-order perturbation to the thermal 
energy.

In the general geometry being considered, variation of the background allows perpendicular and compressive fluctuations to exchange energy, as well as 
exchanging energy with the background field. The latter will be worked out below (\cref{sub: transport equations}), where 
it is shown that energy lost/gained by the background on the slow transport timescale is gained/lost by the
fluctuations, ensuring total energy conservation. Here, at second order, this physics manifests as sources in the 
equations for $\partial \tilde{W}^{\pm}_{\perp}/\partial t$, $\partial \tilde{W}^{\pm}_{\|}/\partial t$, and $\partial \tilde{W}_{s}/\partial t$, as well as
for the total fluctuating energy $\partial \tilde{W}^{\rm tot}/\partial t$. Specifically, we dot \cref{eq:perp-pm-app} with $\rho \zpm/2$, multiply \cref{eq: slow mode zlpm}
by $\rho \zlpm/2$, multiply \cref{eq:s-fluct1} by $\rho T/\gamma\,\d s $, 
then  apply the
perpendicular average $\avgp\cdot$. The latter eliminates terms with $\duprp\cdot\nabla$ or $\dbprp\cdot\nabla$ ,
because $\nabla\cdot\duprp=\nabla\cdot\dbprp=0$ and total divergence terms gain an order in $\epsilon$ (likewise gradients of background quantities).

Summing the result yields
\begin{align}
\frac{\partial \tilde{W}^{\rm tot}}{\partial t}  &+ \nabla\cdot\left[\bh \sum_{\rm wave} (U + v_{\rm wave})  \tilde{W}_{\rm wave}\right] \nn\\ &  = \tilde{\mathcal{Y}}^{\perp} + \tilde{\mathcal{Y}}^{\|} + \tilde{D}^{\rm tot},\label{eq: fluct energy conservation}
\end{align}
 where the divergence  captures the parallel energy transport of each energy type at speed $U+v_{\rm wave}$ and
$\tilde{D}^{\rm tot}$ is the total dissipation of fluctuation energy from viscosity, resistivity, and heat fluxes:
\begin{align}
\tilde{D}^{\rm tot} = \,&\rho \avgp{\duprp\cdot\bm{\mathcal{D}}^{\perp}_{u} }+\frac{1}{4\pi} \avgp{\dbprp\cdot\bm{\mathcal{D}}^{\perp}_{B} } + 
\rho \avgp{\duprl \mathcal{D}^{\|}_{u}}\nn\\&+ \rho \avgp{\dVprlx \mathcal{D}^{\|}_{V}} +\frac{\rho c_{v }T}{\gamma}\left\langle \frac{\d s}{c_{v}}\, \mathcal{D}_{s}\right\rangle_{\perp}.\label{eq:Dtot second order}
\end{align}

The perpendicular source 
 term $\tilde{\mathcal{Y}}^{\perp}$ captures the coupling of compressive fluctuations and  
  Alfv\'enic equations induced by perpendicular gradients, curvature, and gravity;  significant algebra yields the  form:
\begin{align}
&\tilde{\mathcal{Y}}^{\perp}
\equiv -\rho U \left<\duprp \duprl - \frac{\dbprp \dbprl}{4\pi \rho} \right>_{\perp}\!\!\!\cdot\big(\gprp{U} + \bkap\big)
 \nn\\
&
\:+\! \frac{B}{4\pi} \avgp{\duprp \dbprl \!-\! \dbprp \duprl}\!\cdot\!\big(\bkap - \gprp{B}\big) 
\!+\!\avgp{\duprp{\d\rho}}\!\cdot\!\geff \nn\\
&
\:- \frac{1}{\gamma} \avgp{ \d p\,\duprp}\cdot\gprp{p}
\!-\! \frac{p}{\gamma(\gamma-1)} \left\langle { \frac{\d s}{c_{v}}\duprp}\!\!\right\rangle_{\perp}\!\!\!\cdot\gprp{s}. \label{eq: energy sources perp}
\end{align}
 The organization and physical significance of the terms here will be clarified below in \cref{eq: Yperp mech defns app,eq: Y perp th}. In brief, each term results from the exchange with  the free energy contained in background gradients; the first three arise from exchange with the kinetic, magnetic, and (effective) gravitational potential
 energy, respectively, while the final two result from exchange with background pressure and entropy (equivalently,   density and temperature) gradients. 

 The parallel source term $\tilde{\mathcal{Y}}^{\|}$ arises from the parallel background gradients, arising from the 
 work done by fluctuations as they  propagate along an inhomogeneous background. Given that such terms only couple 
 Alfv\'enic or compressive fluctuations to themselves, it is helpful to further split $\tilde{\mathcal{Y}}^{\|}$ as 
 $\tilde{\mathcal{Y}}^{\|} = \tilde{\mathcal{Y}}^{\|}_{\rm A} + \tilde{\mathcal{Y}}^{\|}_{C}$, where
 \begin{align}
\tilde{\mathcal{Y}}^{\|}_{\rm A} &\equiv \left(U \gprl{B} - \frac{\Srho}{\rho  }\!\right)\!\tilde{W}^{\perp}_{u} - U \gprl{U} \tilde{W}^{\perp}_{B}- \rho U \avgp{\zm\!\cdot\!\tsmat\!\cdot\!\zp},\label{eq: parallel energy nonconservation alfvenic}\\[0.5em]
\tilde{\mathcal{Y}}^{\parallel}_{C}
&=
-\left(2U\gprl{U}+\frac{\Srho}{\rho }\right)\tilde{W}^{\|}_{u}+U\left(\gprl{T}-\frac{\Srho}{\rho U}\right)\tilde{W}_{s} \nn\\&
+ 2 U\left[\gprl{\vs} - \frac{\vs^{2}}{\cs^{2}}\gprl{s}+\left(\frac{1}{2}-\gamma\frac{\vs^{2}}{\cs^{2}}\right)\frac{\Srho}{\rho U}\right]\tilde{W}^{\|}_{\tilde{V}}+\nn\\& 
+\frac{\vs}{\cs^{2}}\rho\Su\avgp{\duprl\dVprlx}
-\left(E_{\rm th}\gprl{T}-\frac{\rho}{\gamma}\Su\right)\avg{\duprl\frac{\d s}{c_{v}}}_{\perp}\nn\\ &+\frac{\rho U \vs}{\gamma}\left(\frac{\gprl{s}}{\gamma-1} + \frac{\Srho}{\rho U}\right)\avg{\dVprlx\frac{\d s}{c_{v}}}_{\perp}.\label{eq: parallel energy nonconservation comp}
\end{align}
We have  defined $\rho\Su = \Sm - U\Srho$ for compactness, as well as the additional energies: $
\tilde{W}^{\perp}_{u} = \rho\avgp{|\duprp|^{2}}/2$, $\tilde{W}^{\perp}_{B} ={\avgp{|\dbprp|^{2}}}/{8\pi}$,  $\tilde{W}^{\|}_{u} =\rho\avgp{\duprl^{2}}/2$, and $\tilde{W}^{\|}_{\tilde{V}}=\rho\avgp{\dVprlx^{2}}/2$ 
(note that $\tilde{W}^{\perp}_{+}+\tilde{W}^{\perp}_{-} = \tilde{W}^{\perp}_{u}+\tilde{W}^{\perp}_{B} $).

With $\tsmat=0$ and $\Srho=0$,  $\tilde{\mathcal{Y}}^{\|}_{\rm A}$ can be written as in the more physically enlightening form, $\tilde{\mathcal{Y}}^{\|}_{\rm A}=-\tilde{W}^{\perp}_{B}\nabla\cdot \bm{U} + U \gprl{B} \tilde{W}_{r}^{\perp} $, where $\tilde{W}_{r}^{\perp} = \tilde{W}^{\perp}_{u}-\tilde{W}^{\perp}_{B}$ is the Alfv\'enic residual energy \cite{Chandran2015,Perez2021} --- the terms capture the work done by magnetic pressure and Reynolds/Maxwell stresses on the expanding flow, respectively. $\tilde{\mathcal{Y}}^{\parallel}_{C}$ is not easily interpreted when written in this (slow/entropy) variable choice, and is complicated significantly by the mass and momentum sources. The physical origin of the general structure will be discussed in \S\ref{subsub: local energy conservation}.

 \subsubsection{Summary of the second-order system}
 
 The mean equations at second order yield a homogeneous system for the mean first-order fields. By choosing the 
 initial conditions appropriately such that all first-order fields are initially zero, it is thus consistent for them to remain zero, with corrections to the mean fields first appearing at second order (see below).
 
 From the fluctuating part, equations \eqref{eq:par-mom-slow} and \eqref{eq:par-ind-slow}, along with Eq.~\eqref{eq:s-fluct1} or \eqref{eq:density-evol}, 
provide a closed system for the first-order fluctuations when combined with \cref{eq:perp-pm-app} (or, equivalently, Eqs.~\eqref{eq: potential form}).
Although they contain many terms, they involve just 5 scalar fields, and all gradients ($\gprp{\cdot}$, $\gprl{\cdot}$, and $\bkap$) are treated as constant in the perpendicular direction for their solution; they 
thus present no significant barrier to solution in simple homogeneous domains, using methods similar  to those used for solar-wind flux tubes \citep{vanBallegooijen2016,vanBallegooijen2011,Perez2013,Chandran2019} or
local methods in fusion studies  \cite{Kotschenreuther1995,Jenko2000,Garbet2010,Candy2016,Barnes2019,Mandell2024}.

The system satisfies a free-energy conservation law \eqref{eq: fluct energy conservation}, which describes how fluctuation energy can be sourced by background gradients through $\tilde{\mathcal{Y}}$. Such terms can  sometimes drive
instabilities (see App.~\ref{app: instabilities}) or allow propagating waves to ``liberate'' energy from the background.
Anticipating the transport-order results below, we note that a time-average of \cref{eq: fluct energy conservation}
 eliminates the first term, $\partial \tilde{W}^{\rm tot}/\partial t$, because as an averaged quantity, $W^{\rm tot} = \avgi{\tilde{W}^{\rm tot}}_t$ can vary only on the transport time scale and its time derivative is therefore $\mathcal{O}(\epsilon^{5})$. 
 Thus, \eqref{eq: fluct energy conservation} expresses the balance between turbulence driving, through parallel 
 fluxes and the  source terms $\tilde{\mathcal{Y}}$, and small-scale dissipation, through $\tilde{D}^{\rm tot}$. 
This balance is set up rapidly compared to the rate at which the background changes, 
leaving any given patch of the turbulence in the system in a quasi-stationary state that slowly 
adjusts to the changing background.

\subsection{Third order: the transport equations}\label{sub: transport equations}

 Our expansion at third order will yield slow (transport) timescale evolution equations for the background quantities: $\rho$, $p$ (or $T$), $U$, and $\bm{B}$. 
 These are driven by fluxes induced by turbulence-averaged quadratic products of the fluctuations ($\duprp$, $\dbprl$, $\d \rho$ etc.), describing 
 the physical influence of the turbulence on the background geometry and gradients. 
 These evolution equations capture the work and heating by waves on the background plasma. Combined 
 with a model of the turbulence, they thus can constitute a slow-timescale, arbitrary geometry model of how a 
 structured corona is heated by waves and instabilities, and how it accelerates into the solar wind.
 
Given the complexity of a general third-order expansion, these equations are useful only by leveraging the properties of the
turbulent average. We will thus only derive averaged equations at $\mathcal{O}(\epsilon^{3})$. 
Likewise, we rely heavily on the property that the turbulent average of a product of fluctuating quantities inside a total perpendicular divergence or curl yields an $\mathcal{O}(\epsilon^{3})$ contribution, because
the average yields a slowly varying quantity with gradients that are $\mathcal{O}(\epsilon)$ (see \S\ref{para: average and derivatives} above).
Consequently, rather than the MHD equations in the form \crefrange{eq:cont}{eq: entropy app}, 
it is convenient to instead start from the equivalent conservative form:
\begin{align}
&\frac{\partial  \rho}{\partial t}
+ \nabla\cdot(\rho \bm u)
= \Srho,\label{eq: conserved MHD rho}
\\[0.25em]
&\frac{\partial \bm{m}}{\partial t}
+ \nabla\cdot\left[
\rho\,\bm u\bm u
+ \left(p+\frac{B^2}{8\pi}\right) I
- \frac{\bm B\bm B}{4\pi}
- \bm{\Pi}
\right]
\nn\\& \qquad 
= -\,\rho\nabla\Phi_{\rm tot} + \bm{S}_{\rho U}
- 2\rho\,\bm\Omega\times\bm u ,\label{eq: conserved MHD m}
\\[0.25em]
&\frac{\partial \bm B}{\partial t} 
 - \nabla\times\Big(
\bm u\times\bm B
- \eta\,\nabla\times\bm B
\Big)=0,\label{eq: conserved MHD B}
\\[0.25em]
&\frac{\partial \sigma}{\partial t} 
+ \nabla\cdot(\sigma \bm u)
 = 
\frac{-\nabla\cdot \bm q  + D^{\rm mech} + \Sth}{T} + (s-\gamma c_{v})S_{\rho},\label{eq: conserved MHD s}
\end{align}
where $\bm{m}=\rho \bm{u}$ is the momentum, $\sigma = \rho s$ is the mass-weighted entropy (motivated below), and $D^{\rm mech}= -\bm\Pi\!:\!\nabla \bm u 
 +  \eta\left|\nabla\times \bm B\right|^{2}/4\pi$. These can be combined into the  total energy conservation equation:
 \begin{align}
&\frac{\partial E}{\partial t} 
+ \nabla\!\cdot\!\!\Bigg[
\left(E+p+\frac{B^2}{8\pi}\right)\,\bm u
- \frac{\bm B\cdot\bm u}{4\pi}\,\bm B
+ \bm q+ \bm{\Pi}\cdot\bm u
\nn\\ &\qquad\qquad
+ \frac{\eta}{4\pi}\,(\nabla\times\bm B)\times\bm B
\Bigg]
=S^{\rm tot},\label{eq: conserved MHD E}
\end{align}
where $S^{\rm tot}= \Sth+ \bm{u}\cdot\bm{S}_{\rho U} + (\Phi_{\rm tot} +  u^{2}/2)S_{\rho}$ and 
\begin{equation}
E \equiv \frac{1}{2} \rho u^2 + \frac{p}{\gamma-1}
+ \frac{B^2}{8\pi} + \rho\,\Phi_{\rm tot},\label{eq: full mhd energy}
\end{equation}
with (as above) $\Phi_{\rm tot} \equiv \Phi_{\rm grav} +\Phi_{\rm rot}$.
Note that   $\Phi_{\rm rot}=-\left|\bm{\Omega}\times\bm R\right|^{2}/2\sim \epsilon^{2} \Phi_{\rm grav}$ in our ordering with $\Omega\sim \epsilon^{3}$.
A corresponding balance holds for the {mechanical} energy
$E_{\rm mech}\equiv \rho u^{2}/2 +B^{2}/8\pi+\rho\Phi_{\rm tot}$:
\begin{align}
\frac{\partial}{\partial t} &E_{\rm mech}
+\nabla\cdot\left[
\left(E_{\rm mech}+p+\frac{B^2}{8\pi}\right)\bm u
-\frac{\bm B\cdot \bm u}{4\pi}\bm B
\right]
=\nn\\&p\nabla\cdot\bm u 
+\bm{\Pi}:\nabla\bm u
-\frac{\eta}{4\pi}|\nabla\times\bm B|^{2}+S^{\rm tot}-\Sth,
\label{eq:mech-energy}
\end{align}
which combines with the evolution of the thermal energy \eqref{eq:intE} to yield \eqref{eq: conserved MHD E}, 
thus quantifying how
 pressure work and viscous/resistive dissipation transfer energy between the mechanical and thermal reservoirs.
 
From this conservative form \eqref{eq: conserved MHD rho}--\eqref{eq: conserved MHD E}, all terms inside gradients --- viz., everything other than source-like terms on the right-hand sides --- 
need only be expanded to $\epsilon^{2}$ for the averaged equations.

\subsubsection{Mean-conserved \& mean-accumulative variables and the definition of energy}\label{sub: mean defns}
Before proceeding it is useful to clarify a subtle point about the
multiple-time-scale analysis  that has thus far been implicit in our discussion but not addressed in sufficient detail.
Multiple-time-scale analysis eliminates {secular} growth of mean fields on the fast time \(t\) by
absorbing it into the transport‐time \(\tau \) evolution of the lowest-order 
background.  At third order, 
this implies $\avgi{\partial g^{(2)}/\partial t} = 0$, for any variable $g$; but, it \emph{does not} follow that
\(\langle g^{(2)}\rangle=0\) for every variable, as this leads to  internal 
inconsistenties. A simple example
is the magnetic field: choosing
\(\langle\bm B^{(2)}\rangle=0\) forces
\(\langle (|\bm{B}|)^{(2)}\rangle\neq0\), or vice-versa.

The expansion thus forces the definition of two variable classes \cite{Andrews1978,Buehler2009}. The first, which we term \emph{mean-conserved},
satisfy $\avg{g^{(2)}}=0$ because the second-order contribution is absorbed into the definition of the zeroth-order field
on the transport timescale. The second, termed \emph{mean accumulative}, has a nonzero second-order 
piece $\avg{g^{(2)}}\neq 0$, which  arises due to quadratic combinations of  first-order fluctuating variables. 

In principle, the choice of mean-conserved versus mean-accumulative variables 
is arbitrary, if made in an internally consistent way. However, certain  choices are 
obviously physically superior, and the definition of  energy --- naturally a mean accumulative variable, given it should involve a contribution from the fluctuations --- is a function of the choice: different versions shuffle the wave/fluctuation energy  between the mean and fluctuating reservoirs, obscuring
conservation properties \citep{Myers1991,Buehler2009}.
We take
\begin{equation}
\rho,\quad\sigma=\rho s,\quad\bm{B},\quad \text{and}\quad \bm{m}=\rho \bm{u},\label{eq: mean conserved choices}
\end{equation}
as the mean-conserved variables. We show below that this is the unique choice that yields the standard RMHD  perturbed free energy \eqref{eq: energy definitions homogeneous}, which was  defined above from the second-order system. 
In addition, choosing the conserved MHD quantities (i.e., \eqref{eq: conserved MHD rho}--\eqref{eq: conserved MHD s}) to be mean conserved ensures that the lowest-order background in some patch, $g\equiv \avg{g^{(0)}}$ (where $g=\{\rho,\:\bm{B}, \bm{m},\sigma\}$), changes only through addition/removal of said 
quantity through its boundary (or external sources or $D^{\rm mech}$) --- in other words, in a homogeneous system, perturbing the system conservatively does 
not change its total mass, momentum or flux. These are clearly desirable properties, making the 
choice \eqref{eq: mean conserved choices} natural.

All  quantities other than those in \eqref{eq: mean conserved choices} are automatically mean-accumulative. Most important is the 
total energy \eqref{eq: full mhd energy}, whose accumulated mean we now compute. 
Expanding the magnetic and kinetic energy, $B^{2}/8\pi$ and $\rho u^{2}/2 = m^{2}/\rho/2$, to second order yields  \begin{equation}
\frac{1}{8\pi}\avg{(B^{2})^{(2)}} = \frac{1}{8\pi}\avg{\dbprp^{2}+\dbprl^{2}}
\end{equation}
 and  \begin{equation}
\frac{1}{2}\avg{(\rho u^{2})^{(2)}} = \frac12\rho \avg{\duprp^{2}+\duprl^{2}},
\end{equation}
respectively, 
where the  
$\langle {\d\rho\,\duprl}\rangle$ contributions disappear because $\langle {m_{\|}^{(2)}}\rangle=0$. 
Likewise, though less obvious, the pressure and thermal energy $E_{\rm th}=p/(\gamma-1)$, are mean accumulative. To see this, write $p = \rho^{\gamma}\exp(c_{v}^{-1}\sigma/\rho)$,  expand this to second order 
and take the average, thereby eliminating $\rho^{(2)}$ and $\sigma^{(2)}$; using \cref{eq: s p and rho} and $\d\sigma = \rho\,\d s + \sigma \d\rho/\rho$ to write the result in terms of $\d p$ and $\d s$ 
eliminates cross terms (e.g., $\avg{\d\rho\,\d s}$), giving the fluctuating contribution
to the thermal energy \cite{Myers1991}:
\begin{equation}
\frac{\avg{p^{(2)}}}{\gamma-1} = \frac{p}{2\gamma} \left<\Big(\frac{\d p}{p}\Big)^{2}\right> + \frac{p}{2\gamma(\gamma-1)} \left< \Big(\frac{\d s}{c_{v}}\Big)^{2}\right>.\label{eq: second order Eth}
\end{equation}
The second term defines  $W_{s}$ in Eq.~\eqref{eq: energy definitions entropy}, while the first term is the slow-mode's thermal contribution, combining with the $\avgi{\dbprl^{2}}$ contribution to 
yield $\rho \avgi{\dVprlx^{2}}/2$; see \cref{eq: dvprlx energy}. The second-order perturbation of $\rho\Phi_{\rm tot}$ term is just $\rho\Phi_{\rm rot}$ because $\avgi{\rho^{(2)}}=0$ and $\Phi_{{\rm grav}}$ is fixed; thus, other than this
centrifugal contribution, we obtain the total fluctuating free energy \eqref{eq: wtot defn}.
The definition is also consistent with the gyrokinetic free energy in the collisional limit, a consequence of the distribution 
function $f$ being mean-conservative; see appendix D.2 of \citet{Schekochihin2009}.

Finally, we note that because the zeroth-order flow was restricted to be field parallel, \(\bm{U}=U\bh\), any fluctuation- or rotation-induced perpendicular flow  must  appear explicitly as the second-order mean \(\UperpII\equiv \langle {\bm{U}_{\perp}^{(2)}}\rangle\).  All  fluxes will  therefore be evaluated including to this fluctuation-induced flow.  Its value is not obtained from a simple evolution law but from the implicit solvability condition \eqref{eq:Uperp2-evol}, which expresses the coupling of \(\UperpII\) to the global balance of stress divergences, rotation, and perpendicular pressure forces.
We will assume that the sources $\Srho$, $\Sm$, and $\Sth$ have no transport-order part, though this could be straightforwardly included if desired.

\subsubsection{Overview of the procedure}\label{sub: transport overview}
We wish to obtain turbulent-driven fluxes of the physically interesting quantities for solar-coronal modeling --- density, heat, momentum, and magnetic field ---
as well as understanding how free energy moves between fluctuations (e.g., Alfv\'en waves) and the background equilibrium. 
The first of these goals is partially fulfilled 
via the third-order expansion and average of Eqs.~\eqref{eq: conserved MHD rho}--\eqref{eq: conserved MHD s}, 
yielding, for instance, the density flux from \cref{eq: conserved MHD rho}, or the momentum flux and fluctuation-induced forces from \cref{eq: conserved MHD m}.

However, for the heating and heat flux --- arguably the most physically important quantity in the theory --- this straightforward average yields transport equations that still involve the undetermined dissipation rate $D^{\rm mech}$ and pressure-work term $\langle p\nabla\cdot\bm{u}\rangle$.
To make progress, we  compute the evolution of the mean energies in two ways:
first,  by forming the transport equation for $E_{\rm mech}$ and $E_{\rm th}$ directly from the mean-conserved variables ($\rho U$, $\bm{B}$, $\rho$, and $\sigma$), thereby eliminating  contributions from the fluctuations; 
second, by  directly averaging $\partial E_{\rm mech} /\partial t$ (\cref{eq:mech-energy}) and $\partial E_{\rm th}/\partial t$ (\cref{eq:intE}), which are mean accumulative
and thus  retain the fluctuations' contributions. 
 The comparison of these calculations then yields a number of useful results: 
 (i) it allows one to solve for $D^{\rm mech}$ and $\langle p\nabla\cdot\bm{u}\rangle$ thereby closing the
 transport system; (ii) it \emph{proves} that the dissipation of fluctuations is perpendicularly  local, viz., there is 
no perpendicular transport of fluctuation energy at this order; and
(iii) it provides a stringent algebraic consistency check by recovering the second-order fluctuating energy conservation law \eqref{eq: fluct energy conservation}.
 Along the way, the calculation reveals the physical origin
of each of the 
different fluctuating energy source terms in \cref{eq: energy sources perp}. 
Total energy conservation is assured from the outset by \cref{eq: conserved MHD E}.

With this structure in mind, we organize the calculation by computing the $\partial \rho/\partial \tau$, 
$\partial\bm{B}/\partial \tau$, and $\partial(\rho U)/\partial \tau$ evolution in \S\ref{sub: density transport},
\S\ref{sub: magnetic transport}, and 
\S\ref{sub: momentum transport}, respectively. In \S\ref{sub: thermal transport}, their energies are combined to form the mechanical energy evolution,  then compared to the average of \cref{eq:mech-energy} to solve for $D^{\rm mech}$. \S\ref{sub: thermal transport} then derives the mean thermal energy transport from $\partial\sigma/\partial \tau$ and $\partial\rho/\partial \tau$, and, by direct analogy with $E_{\rm mech}$, compares this with the average of \cref{eq:intE} to 
solve for $\langle p\nabla\cdot\bm{u}\rangle$. Combining all of these 
results yields the local energy conservation law and the total heating rate and heat flux.
We summarize and provide a detailed physical interpretation of the various effects in \S\ref{sub: third order summary}.

\subsubsection{Density transport}\label{sub: density transport}

Expanding the continuity law to third order and averaging gives the stationary-frame
density flux:
\begin{equation}
\frac{\partial \rho}{\partial \tau}=-\nabla\cdot(\avg{\delta\rho\,\duprp}
   +\rho\,\UperpII) = -\nabla\cdot(\rho \Vrhot),
\label{eq:rho-transport}
\end{equation}
where we define the density's turbulent advection velocity, $\Vrhot = \Vrho+\UperpII$ with $\rho\Vrho = \avg{\delta\rho\,\duprp}$, for use below.
The flux is purely perpendicular because parallel momentum density --- not $u_{\|}$ itself --- is treated as mean-conserved, enforcing
$\langle {u_{\|}^{(2)}}\rangle
   =-\avg{\duprl \d\rho}/\rho$ such that the
turbulent parallel transport  vanishes.


\subsubsection{Magnetic-field transport}\label{sub: magnetic transport}

The induction equation averaged to transport order reads
\begin{equation}
\frac{\partial \bm{B}}{\partial \tau}
= \nabla \times \avg{\du\times\db} + \nabla\times (\UperpII\times\bm{B}),
\label{eq:induction-avg}
\end{equation}
where we have dropped the average of the resistive dissipation term, by assuming it acts only on the small scales. 
The first term here involves the turbulent electromotive force $\bm{\mathcal{E}}\equiv \avg{\du\times\db}$  commonly discussed in mean-field dynamo theory \cite{Rincon2019}. Because the perpendicular part of $\bm{\mathcal{E}}$ can be written as $\bm{\mathcal{E}}_{\perp}={\bm{V}}\times\bm{B}$
for some $\bm{V}$, its effect can be understood like an effective velocity and therefore 
conserves flux by Alfv\'en's theorem. In contrast, the parallel part $\mathcal{E}_{\|}=\bh\cdot\bm{\mathcal{E}}$ is not of this form, and can act as 
an anomalous resistivity that  enhances the diffusion of the large-scale field (the so-called ``$\beta$ effect'' in dynamo theory). We prove below that $\mathcal{E}_{\|}$ vanishes, a
nontrivial consequence of the RMHD ordering and incompressibility.  

Decomposing \(\du=\duprp+\bh\,\duprl\) and \(\db=\dbprp+\bh\,\dbprl\) gives three types of cross terms, two of which vanish. 

\paragraph{Parallel-parallel products} 
First,  \( \duprl\bh \times \dbprl\bh = {0} \) trivially, so there are no contributions from products of parallel fluctuations.

\paragraph{Perpendicular-perpendicular products} 
The term \( \duprp\times\dbprp \) is more subtle. At this order both fields are perpendicular and divergence-free, so we write them in terms of the  scalar potentials \(\Phi,\Psi\) (\cref{eq: phi psi defn}). Then
\begin{equation}
\duprp\times\dbprp
= \bh \sqrt{\rho}\,\{\Phi,\Psi\}\label{eq: poisson bracket magnetic}
\end{equation}
where 
$\{\Phi,\Psi\}\equiv \bh\cdot(\nabla\Phi\times\nabla\Psi)$, viz.,  the contribution is parallel to \(\bh\) with its magnitude given by the Poisson bracket of the two scalar potentials.
The turbulence average of the Poisson bracket is itself a perpendicular boundary term  (the perpendicular integral of the Poisson bracket 
vanishes in a homogeneous domain) 
implying that \(\avg{\duprp\times\dbprp}= \bh\sqrt{4\pi \rho}\avg{\,\{\Phi,\Psi\}}=O(\epsilon^3)\). Taking the curl adds another order in $\epsilon$ ($\bh\sqrt{4\pi\rho}\avg{\,\{\Phi,\Psi\}}$ is a large-scale quantity), so the  contribution to \eqref{eq:induction-avg}
is \(O(\epsilon^4)\) and can be dropped at transport order. 

Physically, since an $\mathcal{E}_{\|}$ is needed to break field lines, this is the
RMHD version of the statement from \citet{Abel2013} that there that  is no mean-field anomalous resistivity
driven by  fluctuations in gyrokinetics.
In that theory, this physics is contained within the evolution of the safety factor $q$ (their equation (143)), which 
cannot be enhanced by the turbulent fluctuations. 
Here, it appears in a more general form without restricting the geometry to closed flux surfaces. 
In particular, it seems that the perpendicular incompressibility of RMHD, together with the enforced scale separation, forbids the Richardson-type explosive separation of neighbouring field lines.  This appears to break a key assumption behind the stochastic-flux-freezing proofs of \citet{Eyink2011} and therefore precludes the enhanced “reconnection-diffusion’’ resistivity posited in \citet{Lazarian2020}.

\paragraph{Parallel-perpendicular products} 
The remaining pieces
$\avg{\duprl\bh\times\dbprp + \duprp\times\dbprl\bh}$
can be grouped using the potential representation  to yield 
$\avg{\du\times\db} = \Vpsi\times\bm{B}$ with 
\begin{equation}
\Vpsi \;\equiv \left<\duprp\frac{\dbprl}{B} - \frac{\dbprp}{B}\duprl\right>.
\label{eq:Vpsi-def}
\end{equation}
Substituting into \eqref{eq:induction-avg} gives the compact transport-time induction equation
\begin{equation}
\frac{\partial\bm{B}}{\partial \tau} = \nabla\times\big[\,( \UperpII + \Vpsi )\times\bm{B}\,\big] = \nabla\times (\Vpsit\times \bm{B}).
\label{eq:induction-transport}
\end{equation}
Thus,  magnetic flux surfaces are advected by the turbulence at the {effective}
perpendicular speed \(\Vpsit=\Vpsi+\UperpII\) relative to the zeroth-order background.

 \paragraph{Magnetic-energy transport}
As anticipated in   \S\ref{sub: transport overview}, the evolution of the energy of the mean magnetic field can be used to 
prove the locality of the small-scale turbulence. We thus dot \cref{eq:induction-transport} with \(\bm{B}/4\pi\).
 Expanding \(\nabla\times({\Vpsit}\times\bm{B})\) and using the vector identity 
\(\bh\cdot\grad\tilde{\Vpsi}\cdot\bh = -\,\bkap\cdot\tilde{\Vpsi}\) (valid in RMHD ordering with \(\Vpsit\cdot\bh=0\)), one obtains the transport-time
evolution of the energy in the mean magnetic field:
\begin{equation}
\frac{\partial}{\partial \tau} \frac{B^{2}}{8\pi} 
= -\frac{B^{2}}{4\pi}\,{\Vpsit}\cdot\left(\bkap-\frac{\gradperp B}{B}\right)
-\nabla\cdot \left( \frac{B^{2}}{4\pi}{\Vpsit} \right).
\label{eq:Bstrength-transport}
\end{equation}
Aside from the $\UperpII$ contribution, we recognize the first term as the negative of the second term in $\tilde{\mathcal{Y}}^{\perp}$   from the fluctuating energy conservation 
law (see Eqs.~\eqref{eq: fluct energy conservation} and \eqref{eq: energy sources perp}), thus foreshadowing the  general exchange of energy between the background and  fluctuations explored in 
detail below.
One consequence of this form of exchange is that a force-free field, for which $\bkap=\gradperp{B}/B$, 
provides no source of free energy for the fluctuations. The equivalent statement discussed in \citet{Abel2013} is that there is no free-energy injection purely from magnetic field gradients, {viz.}, a $\gradperp{p}$ or $\geff$ is needed in the perpendicular equilibrium \cref{eqapp: perp equil} for the background and fluctuations to exchange energy (for instance, to drive instabilities).
 
\subsubsection{Momentum and kinetic energy transport}\label{sub: momentum transport}

From the momentum equation \eqref{eq: conserved MHD m}
we are interested in the transport of the lowest-order \emph{parallel} mean flow \(U\equiv\bm{U}\!\cdot\!\bh\). Dotting with \(\bh\), noting that \(\bh\!\cdot\! \partial\bh /\partial t=0\), and   bringing \(\bh\) under the divergence yields
\begin{align}
&\frac{\partial ( \rho U)}{\partial t}
=-\grad\!\cdot\!\left[\rho\,(\bh\!\cdot\!\bm{u})\,\bm{u}-\frac{(\bh\!\cdot\!\bm{B})\,\bm{B}}{4\pi}\right]
+\rho\,\bm{u}\!\cdot\!(\nabla\bh)_{\perp}\!\cdot\!\bm{u}
\nn\\& 
-\frac{1}{4\pi}{\bm{B}\!\cdot\!(\nabla\bh)_{\perp}\!\cdot\!\bm{B}}+\bh\!\cdot\!\left(\rho\,\bm{u}\bm{u}-\frac{\bm{B}\bm{B}}{4\pi}\right)\!\cdot\!\bkap +  S_{\rho U}
\nn\\&-\bh\cdot\nabla \left( p+\frac{B^{2}}{8\pi}\right)- \rho\, \bh\cdot\nabla\Phi_{\rm tot} + \bh\cdot\mathcal{D}_{\rho u}.
\label{eq:Upar-pre}
\end{align}
Here we have used the decomposition \eqref{eq: grad bh} of $\nabla\bh$ and removed the Coriolis force  in anticipation of 
the $\epsilon^{3}$ ordering of $\Omega$.

Expanding and averaging \eqref{eq:Upar-pre} to transport order yields 
\begin{align}
& \frac{\partial(\rho U)}{\partial \tau}
=-\grad\!\cdot\!\left[ \rho U  (\VU+\Vrhot)\right]+\rho U\bkap\!\cdot \left( \VU+\Vrhot\right) \nn\\
&\quad-\bh\cdot\nabla p_{\mathrm{tot}}^{(2)} + F^{\perp}_{\rm RM}
 -2\grad\!\cdot(\bh W_{r}^{\|}) - \rho\bh\cdot\nabla\Phi_{\rm rot}.
\label{eq:Upar-transport}
\end{align}
where we have defined by analogy with \(\Vpsi\) and $\Vrho =\avg{\duprp{\d\rho}/\rho}$ the effective perpendicular velocity associated with velocity transport:
\begin{equation}
\VU \equiv\avg{\frac{\duprl\,\duprp}{U}-\frac{\dbprl\,\dbprp}{4\pi \rho\,U}},
\label{eq:VU-def}
\end{equation}
as well as the parallel force from the Reynolds/Maxwell stresses of Alfv\'enic fluctuations due to the expanding/shearing geometry,
\begin{equation}
F^{\perp}_{\rm RM}\equiv- W_{r}^{\perp}\gprl{B}+\rho\avg{\zp\cdot\tsmat\cdot\zm}.\label{eq: RM perp def}
\end{equation}
Here 
\(W_{r}^{\perp}\equiv W_{u}^{\perp}-W_{B}^{\perp}\) and \(W_{r}^{\|}\equiv W_{u}^{\|}-W_{B}^{\|}\) are the  Alfv\'enic and parallel residual energies (with $W^{\|}_{B} = \langle \dbprl^{2}\rangle/8\pi$), and 
\begin{equation}
p^{(2)}_{\rm tot} = \langle {p^{(2)}} \rangle+ W^{\perp}_{B} + W^{\|}_{B} \label{eq: p2tot}
\end{equation}
is the total pressure perturbation with $\langle {p^{(2)}} \rangle $ given in \cref{eq: second order Eth}.
We have neglected dissipation $\mathcal{D}_{\rho u}$, since it  should not drive significant dissipation of the large-scale momentum.

\Cref{eq:Upar-transport} captures transport of the mean momentum due to the fluctuations, noting that the equilibrium forces accelerating the flow 
have already been captured in the first-order equilibrium \eqref{eqapp: parallel equil}.  Most importantly, the theory  reveals 
a turbulence-driven \emph{perpendicular} transport of parallel wind momentum, driven with flux  $\rho U(\VU+\Vrhot)$.
We also see some familiar terms from past solar-wind literature, particularly the role of  gradients in the Alfv\'en-wave pressure from $W^{\perp}_{B} $ in $p^{(2)}_{\rm tot}$, as well 
as the Alfv\'enic 
residual energy $W_{r}^{\perp}$, which has arisen from the magnetic expansion converting part of the perpendicular Reynolds/Maxwell stress into a parallel net force. The new term \(\zp\!\cdot\!\tsmat\!\cdot\!\zm\) has the same origin as the residual energy term (which can be written $\zp\!\cdot\! \zm\,\nabla\cdot\bh$). These effects are generalized by \cref{eq:Upar-transport} to also capture forces and parallel transport from compressive fluctuations, which
are likely unimportant compared to those of Alfv\'enic fluctuations in the solar-wind context.

\paragraph{Kinetic and potential energy}

It is useful to combine the kinetic with the potential 
energy at this stage, because acceleration effects combine with  gravity to yield the effective gravity $\bm{g}_{\rm eff}$ (see Eqs.~\eqref{eqapp: perp equil}-\eqref{eqapp: parallel equil}). 
The evolution of the background kinetic and potential energy is formed  using the requirement that transport-timescale derivatives are only evaluated on mean-conserved quantities, giving
\begin{equation}
\frac{\partial}{\partial \tau} \left( \frac{1}{2}\rho U^{2} + \rho \Phi_{\rm tot}\right) = U \frac{\partial (\rho U)}{\partial \tau} - (U^{2}-{\Phi}^{U}_{{\rm eff}}) \frac{\partial \rho}{\partial \tau},
\end{equation}
where ${\Phi}^{U}_{{\rm eff}}= U^{2}/2 +\Phi_{\rm grav}$ is a total effective potential ($\Phi_{\rm rot}$ is higher order).
Using \cref{eq:rho-transport}, \eqref{eq:Upar-transport}, and the $\geff$ definition \eqref{eq: effective gravity}, then collecting terms into  total divergences plus 
remainders yields
\begin{align}
\frac{\partial}{\partial \tau}&\left(\frac12\rho U^{2} + \rho \Phi_{\rm tot}\right) =-\grad\!\cdot\!\left(\rho U^{2}\VU+\rho{\Phi}^{U}_{{\rm eff}} \Vrhot\right)\label{eq:KEpar-transport}\\
&+\rho U\VU\!\cdot\!\big(\gradperp{U}+U\bkap\big)
-\rho\Vrhot\cdot\geff- \rho U \bh\cdot\nabla\Phi_{\rm rot}\nn\\
&+UF^{\perp}_{\rm RM} -U \bh\cdot\nabla p_{\mathrm{tot}}^{(2)} -\grad\cdot(2\bh  U W_{r}^{\|}) + 2  W_{r}^{\|} \gradl{U}.\nn
\end{align}

As for the magnetic-energy evolution \eqref{eq:Bstrength-transport}, we see that the 
two terms on the second line in \cref{eq:KEpar-transport} are (aside from $\UperpII$ contribution) the negative of the
first and third terms in $\tilde{\mathcal{Y}}^{\perp}$   from the fluctuating energy conservation 
law \eqref{eq: fluct energy conservation}, capturing the exchange of energy between the fluctuating
and mean fields due to perpendicular gradients in velocity and the energy required to move mass across the effective gravitational force. 
Unlike the magnetic field, we now have other source terms due to parallel gradients, parallel fluxes, 
and the work done by the centrifugal force. 
 

\subsubsection{Total mechanical energy and turbulence locality}\label{sub: mechanical energy}

Adding  the kinetic and potential energy \eqref{eq:KEpar-transport} to the magnetic energy \eqref{eq:Bstrength-transport} yields 
the transport of the total mechanical energy
\begin{align}
\frac{\partial}{\partial \tau}\!&\left(\frac12\rho U^{2} + \frac{B^{2}}{8\pi}+\rho \Phi_{\rm tot}\right) = -\mathcal{Y}^{\perp}_{\rm mech}\label{eq:Emech-transport}\\&
-\grad\!\cdot\!\left[\rho U^{2}\VU+\frac{B^{2}}{4\pi}{\Vpsit} +{\Phi}^{U}_{{\rm eff}}\Vrhot\right]- \rho U \bh\cdot\nabla\Phi_{\rm rot} \nn\\
&+UF^{\perp}_{\rm RM} -U \bh\cdot\nabla p_{\mathrm{tot}}^{(2)}-\grad\cdot(2\bh  U W_{r}^{\|}) + 2  W_{r}^{\|} \gradl{U},\nn
\end{align}
where we now collect the terms that exchange between mean and fluctuation reservoirs to define \begin{align}
\mathcal{Y}^{\perp}_{\rm mech}=&\,\mathcal{Y}^{\perp}_{U}+\mathcal{Y}^{\perp}_{B}+\mathcal{Y}^{\perp}_{g}\nn\\
=&-\rho U^{2}\,\VU\!\cdot\!(\bkap+\gprp{U})+\frac{B^{2}}{4\pi}{\Vpsit}\cdot(\,\bkap-\gprp{B}\,)
\nn\\&+\rho\Vrhot\cdot\geff.\label{eq: Yperp mech defns app}
\end{align}
$\mathcal{Y}^{\perp}_{U}$ and $\mathcal{Y}^{\perp}_{B}$ capture the effect of transport-induced movement of kinetic or magnetic energy across their respective gradients, while  $\mathcal{Y}^{\perp}_{g}$ captures energy exchange from density 
movement across the effective gravitational potential. 

As outlined above, we then compare this to the direct transport-scale average of the mechanical-energy evolution \eqref{eq:mech-energy}:
\begin{align}
\frac{\partial}{\partial \tau}\!&\left(\frac12\rho U^{2} + \frac{B^{2}}{8\pi}+\rho \Phi_{\rm tot}\right)= \nn\\&
- \avg{D^{\rm mech}}+\avg{p\nabla\cdot\bm u } +\avg{(S^{\rm tot})^{(2)}}\nn\\
&-\nabla\cdot\left<
\left(E_{\rm mech}+p+\frac{B^2}{8\pi}\right)\bm u
-\frac{\bm B\cdot \bm u}{4\pi}\bm B
\right>,
\label{eq:mech-energy-av}
\end{align}
where \begin{equation}
\avg{(S^{\rm tot})^{(2)}} = \avgi{ u_{\|}^{(2)}}\Sm +\left(\Phi_{\rm rot} -\frac{1}{2}\avgi{u^{2}}\right)\Srho
\end{equation}
is the transport-order expansion of $S^{\rm tot}-\Sth$, with $\rho \avgi{u^{2}}/2 = \rho U \avgi{ u_{\|}^{(2)}} + W_{u}^{\perp} + W_{u}^{\|} $.
Expanding the terms in the divergence on the second line of \eqref{eq:mech-energy-av} to $\epsilon^{2}$, then splitting them into perpendicular and parallel 
parts and using perpendicular pressure balance \eqref{eq:perp-pressure-balance}, one finds 
that perpendicular flux terms are identical to those of \eqref{eq:Emech-transport}, aside from one term, $p\,\UperpII$.
Thus, subtracting \cref{eq:Emech-transport} from \eqref{eq:mech-energy-av} cancels the perpendicular fluxes and time derivatives, yielding
\begin{align}
&\avg{D^{\rm mech}}-\avg{p\nabla\cdot\bm u } = \mathcal{Y}^{\perp}_{\rm mech}+ U \bh\cdot\nabla p_{\mathrm{tot}}^{(2)} \nn\\
&\qquad-\nabla\cdot(p\,\UperpII)- U F^{\perp}_{\rm RM}  - 2  W_{r}^{\|} \gradl{U} \nn\\
&\qquad +\avgi{ u_{\|}^{(2)}}\Sm -\left(\rho U\avgi{ u_{\|}^{(2)}} + W_{u}^{\perp}+ W_{u}^{\|}\right)\frac{\Srho}{\rho}\nn\\
& \qquad- \nabla\cdot \Big[\bh\Big(U W_{u}^{\perp} + 2 U W_{B}^{\perp} - \frac{B}{4\pi} \avg{\duprp\cdot\dbprp} \nn\\ 
&\qquad \qquad \qquad+ UW_{u}^{\|} +UW^{\|}_{B} + \langle{(p u_{\|})^{(2)}}\rangle \Big) \Big],
\label{eq:mech-energy-diff}
\end{align}
where
$$\langle{(p u_{\|})^{(2)}}\rangle =U\langle {p^{(2)}}\rangle+\frac{p}{\gamma}\avg{\frac{\d s}{c_{v}} \duprl } + \frac{\gamma-1}{\gamma} \avg{\d p\,\duprl}$$
and  $ \langle u_{\|}^{(2)}\rangle = -\avg{ \duprl\,\d\rho/\rho}$. The expression \eqref{eq:mech-energy-diff} \emph{proves} that the turbulence acts locally in the perpendicular direction at this order, since there are no perpendicular gradients of turbulence-related quantities (see \citet{Abel2013} for further discussion). We will show below 
how the final parallel divergence term combines with similar thermal contributions to yield the enthalpy flux of fluctuations, 
making \cref{eq:mech-energy-diff} a statement about how the local dissipation of mechanical energy can be driven 
by combinations of perpendicular gradients (via $\mathcal{Y}^{\perp}_{\rm mech}$), work terms from background gradients (the second line),
and the divergence of the  wave enthalpy flux. Note that the $\Phi_{\rm rot}$ terms disappeared because the parallel flux and source in \cref{eq:mech-energy-av} combined with the work in \cref{eq:Emech-transport} to yield $\Phi_{\rm rot}(\nabla\cdot(\rho \bm{U})-\Srho)=0$.

\subsubsection{Thermal energy}\label{sub: thermal transport}

We start by writing the thermal energy evolution in 
terms of mean-conserved variables $\rho$ and $\sigma$:
\begin{equation}
\frac{\partial E_{\rm th}}{\partial \tau} = \frac{p}{(\gamma-1)\rho} \left[\frac{1}{c_{v}}\left(\frac{\partial \sigma}{\partial \tau} - s \frac{\partial \rho}{\partial \tau} \right) + \gamma  \frac{\partial \rho}{\partial \tau} \right].\label{eq: Eth in terms of sigma}
\end{equation}
From \cref{eq: conserved MHD s}, we compute
\begin{align}
\frac{\partial \sigma}{\partial \tau} & = - \nabla\cdot (s\avg{\d\rho\, \duprp} +\rho \avg{\d s\, \duprp} + \sigma \UperpII) \label{eq: sigma transport}\\ 
& - \nabla\cdot (\bh\, \rho \avg{\d s \duprl}) + \frac{\avg{D^{\rm mech}}}{T} - \frac{\nabla\cdot\langle \bm{q}^{(2)}\rangle}{T}+ \avgi{s^{(2)}}\Srho \nn\\
& +\avg{\frac{\d T}{T}\,\frac{\nabla\cdot\d \bm{q}}{T}} +(\rho U \gradl{s} + c_{v}\gamma \Srho) \,T\avgi{(1/T)^{(2)}} \nn
\end{align}
where we used  $\avg{\sigma^{(2)}}=0$ and $\rho \langle u_{\|}^{(2)}\rangle = -\avg{\d\rho \duprl}$ to 
simplify the parallel flux and \cref{eq:entropy-mean} for the heat  and density source. Note also that perturbed contributions to $D^{\rm mech}$ (e.g., from $\Pi:\nabla\bm{U}$) 
are at least $\mathcal{O}(\epsilon^{3})$, so there is no $\avg{\d T \d \mathcal{D}}$ contribution.

Inserting \cref{eq: sigma transport} into \cref{eq: Eth in terms of sigma}, canceling terms involving $s$, then collecting the remaining terms
into  total divergences, we find the time evolution of $E_{\rm th}$:
\begin{align}
\frac{\partial E_{\rm th}}{\partial \tau}   = & -\nabla\cdot\left[ E_{\rm th} \left( \avg{\frac{\d p}{p}\duprp} + \gamma \UperpII  + \bh\avg{\frac{\d s}{c_{v}}\duprl}\right)\right] \nn\\
& + \UperpII \cdot\gradperp{p}  -\mathcal{Y}^{\perp}_{\rm th} -\mathcal{Y}^{\|}_{\rm th}   +\avg{D^{\rm tot}} -\nabla\cdot\langle \bm{q}^{(2)}\rangle ,\label{eq: thermal energy with diss}
\end{align}
where 
\begin{align}
\mathcal{Y}^{\perp}_{\rm th}&\equiv - \frac{p}{\gamma} \avg{\frac{\d p}{p}\duprp\!} \cdot\gprp{p} - \frac{E_{\rm th}}{\gamma } \avg{\frac{\d s}{c_{v}}\duprp\!}\cdot\gprp{s}\nn\\&
=- p \avg{\frac{\d \rho }{\rho}\duprp\!}\cdot\gprp{\rho}-E_{\rm th} \avg{\frac{\d T}{T}\duprp\!} \cdot\gprp{T} \label{eq: Y perp th}
\end{align}
and 
\begin{align}
\mathcal{Y}^{\|}_{\rm th}\equiv& - E_{\rm th}  \avg{\frac{\d s}{c_{v}}\duprl}\gprl{T}+E_{\rm th} \avg{\frac{\d s}{c_{v}}\frac{\d\rho}{\rho}}\frac{\Srho}{\rho}\nn\\&
 -E_{\rm th} \left(U \gprl{s} +\gamma \frac{\Srho}{\rho} \right)\, T\avgi{(1/T)^{(2)}} \label{eq: Y prl th}
\end{align}
capture the exchange between mean thermal gradients and fluctuations in the perpendicular and parallel direction, respectively.  Likewise, 
\begin{equation}
\avg{D^{\rm tot}} \equiv \avg{D^{\rm mech}} + \avg{\frac{\d T}{T}\,\nabla\cdot\d \bm{q}}
\end{equation}
is identified as the total dissipation rate of fluctuations, which matches the intermediate-time average of that from the second-order system \cref{eq:Dtot second order}. Combining \eqref{eq: Y perp th} with \eqref{eq: Yperp mech defns app}, we see that $\tilde{\mathcal{Y}}^{\perp}$ from the second-order calculation \eqref{eq: energy sources perp} is $\avgi{\tilde{\mathcal{Y}}^{\perp}}_t=\mathcal{Y}^{\perp} =\mathcal{Y}^{\perp}_{U}+\mathcal{Y}^{\perp}_{B}+\mathcal{Y}^{\perp}_{g}+\mathcal{Y}^{\perp}_{\rm th}$,
representing how fluctuations are driven (exchange energy) due to gradients of the mean flow, magnetic field, gravitational potential, and  thermal energy (density and temperature, or equivalently, pressure and entropy), respectively. 

\paragraph{Local thermal energy dissipation}
\Cref{eq: thermal energy with diss}  captures the heating of the background via fluctuations, heat transport perpendicular and 
parallel to the field, and the work done by the fluctuations against background gradients. As above, we compute 
$\partial E_{\rm th}/\partial \tau$ from the mean-accumulative form \eqref{eq:intE} directly, to constrain $\avg{D^{\rm tot}} $ and 
 total energy conservation. Averaging \cref{eq:intE}, noting that no transport-order source appears in this form, we get
\begin{align}
\frac{\partial E_{\rm th}}{\partial \tau}  = & -\nabla\cdot\left[ E_{\rm th} \left( \avg{\frac{\d p}{p}\duprp} + \UperpII\right)\right] -\avg{p\nabla\cdot\bm u } \nn\\
& - \nabla\cdot\left(\bh  \frac{\langle {(p u_{\|})^{(2)}}\rangle}{\gamma -1}\right) -  \nabla\cdot\langle \bm{q}^{(2)}\rangle  + \avg{D^{\rm mech}}.\label{eq: thermal energy second}
\end{align}
Subtracting  \eqref{eq: thermal energy with diss} from \eqref{eq: thermal energy second} yields the dissipation rate of fluctuating thermal energy:
\begin{align}
& \avg{\frac{\d T}{T}\,\nabla\cdot\d \bm{q}} +\avg{p\nabla\cdot\bm{u}} = \mathcal{Y}^{\perp}_{\rm th} + \mathcal{Y}^{\|}_{\rm th} + p\nabla\cdot\UperpII \nn\\
& \quad-\nabla\cdot\left[\bh \left( \frac{\langle {(p u_{\|})^{(2)}}\rangle}{\gamma -1} - E_{\rm th} \avg{\frac{\d s}{c_{v}}\duprl}\right)\right]. \label{eq:thermal-energy-diff}
\end{align}
The procedure here has been identical to that leading to \cref{eq:mech-energy-diff}. We likewise  find a similar structure, 
with the thermal energy dissipation driven by parallel fluxes, perpendicular and parallel gradients, and exchange via pressure dilation $\avg{p\nabla\cdot\bm{u}}$.  

\subsubsection{Local energy conservation}\label{subsub: local energy conservation}
Summing \cref{eq:mech-energy-diff} and \cref{eq:thermal-energy-diff}, we note that the two $\langle {(p u_{\|})^{(2)}}\rangle$ terms combine 
to yield the expansion of the total energy flux \eqref{eq: conserved MHD E}, causing all of the parallel divergence terms to combine
into the total enthalpy flux of the fluctuations:
\begin{equation}
\nabla\cdot\left[\bh \sum_{\rm wave} (U + v_{\rm wave})  W_{\rm wave} + \bh\, U p_{\rm tot}^{(2)}\right].
\end{equation}
Here, we have used $\avg{\d p\, \duprl}= -\rho \vs \langle {\duprl\dVprlx}\rangle$ to convert to the slow-wave energies defined above, and the sum runs over $v_{\rm wave}=\{\pm\va,\:\pm\vs,\:0\}$ for $W_{\rm wave}=\{W^{\pm}_{\perp}, W^{\pm}_{\|}, W_{s}\}$; see \cref{eq: energy definitions homogeneous}.
Putting this together, we derive the total intermediate-time-averaged fluctuating energy conservation law:
\begin{align}
&\avg{D^{\rm tot}} = - \nabla\cdot\left[\bh \sum_{\rm wave} (U + v_{\rm wave})  W_{\rm wave} \right] + \mathcal{Y}^{\perp}_{\rm mech} \nn\\ &\:
+ \mathcal{Y}^{\perp}_{\rm th}+ \mathcal{Y}^{\|}_{\rm th}   - \UperpII\cdot \gradperp{p}- p_{\rm tot}^{(2)}\nabla\cdot\bm{U} - U F^{\perp}_{\rm RM}
\nn\\ &\:    - 2 W_{r}^{\|} \gradl{U}-(W_{u}^{\perp}+ W_{u}^{\|} )\frac{\Srho}{\rho} -\rho \avg{\duprl\frac{\d\rho}{\rho}}\Su, \label{eq: local energy conservation}
\end{align}
where $\rho \Su\equiv \Sm - U \Srho$.
This expression generalizes equation (2.12) of \citet{Perez2021} from Alfv\'enic fluctuations on a straight field line to include compressive fluctuations on a completely general background field and external mass/momentum sources. 
It makes explicit how, in quasi-steady state, the local heating rate, $\langle D^{\rm tot} \rangle$, must balance with various distinct physical processes, namely (in the order they enter in \cref{eq: local energy conservation}):
\begin{enumerate}[label=(\roman*), itemsep=2pt, topsep=2pt, parsep=0pt]
\item Changes in the parallel flux of fluctuation energy;
\item Free energy exchange with the background via the relaxation/creation of perpendicular gradients --- either in kinetic and magnetic energy ($\mathcal{Y}^{\perp}_{U} $ and $\mathcal{Y}^{\perp}_{B} $ in $\mathcal{Y}^{\perp}_{\rm mech} $), in density and temperature  ($\mathcal{Y}^{\perp}_{\rm th} $),  or by transporting mass across the effective gravitational field ($\mathcal{Y}^{\perp}_{g}$ in $\mathcal{Y}^{\perp}_{\rm mech} $);
\item Free energy associated with fluctuation-induced changes to the temperature gradient (compressive correlations in $\mathcal{Y}^{\|}_{\rm th}$) and thermal changes when  mass is added/removed ($\Srho$ terms in $\mathcal{Y}^{\|}_{\rm th}$); 
\item Work done by the fluctuations in driving the self-induced mean flow $\UperpII$ ($  \UperpII\cdot \gradperp{p}$);
\item Work done by the fluctuating pressure on the expanding flow ($p_{\rm tot}^{(2)}\nabla\cdot\bm{U} $);
\item Work done by the Reynolds and Maxwell stresses of the perpendicular fluctuations due to 
expansion ($\gprl{B}$) and shearing ($\tsmat$) in $F^{\perp}_{\rm RM}$;
\item Work done by the Reynolds and Maxwell stresses of the parallel fluctuations ($2  W_{r}^{\|} \gradl{U}$).
\end{enumerate}
As promised, the expression \eqref{eq: local energy conservation} can be rearranged to match the 
intermediate-time average of the local energy conservation law \eqref{eq: fluct energy conservation}, which was derived from the second-order 
fluctuating system.  For terms related to perpendicular rearrangements ($\mathcal{Y}^{\perp}$ and $\UperpII$), this can be seen by noting that the $\UperpII\cdot\gradperp{p}$ term combines with the $\UperpII$ in $\Vpsit$
and $\Vrhot$ to dot into the total perpendicular equilibrium force balance \eqref{eqapp: perp equil}; it thus vanishes, recovering \cref{eq: energy sources perp}.
For terms relating to parallel forces, in the Alfv\'enic sector the $W_{B}^{\perp}$ part of $p^{(2)}_{\rm tot}$ combines, using $\nabla\cdot\bm{U} = U(\gprl{U}-\gprl{B})$,  with  $-U W_{r}^{\perp}\,\gprl{B}$ in $UF^{\perp}_{\rm RM}$ \eqref{eq: RM perp def} and $-W_{u}^{\perp}{\Srho}/{\rho}$  to give the first two terms in  \cref{eq: parallel energy nonconservation alfvenic}.
In the compressive sector, \cref{eq: parallel energy nonconservation comp} is recovered using 
$\gprl{\vs} = (\vs/\cs)^{2}\gprl{p}/2 + (\vs/\cs)^{2} \gprl{B} -\gprl{\rho}/2 $,
\begin{equation}
T \avgi{(1/T)^{(2)}} = \frac{1}{\gamma}\avg{\frac{\gamma-1}{2} \frac{\d p^{2}}{p^{2}} +  \frac{\d p}{p} \frac{\d s}{c_{v}} - \frac{1}{2} \frac{\d s^{2}}{c_{v}^{2}}},
\end{equation}
the identity $\gamma\, \d\rho/\rho = \d p/p- \d s/c_{v}$, 
and the conversion  $\dVprlx/\va=(\va/\vs)(\dbprl/B) = - \beta\,(\va/\vs)(\d p/p)/2$ to write $\dbprl$ and $\d p$ terms using $\dVprlx$.
Given that the derivation here was entirely formulated in terms of quadratic averages, never using results from the 
$\mathcal{O}(\epsilon^{2})$ fluctuating system, this provides  a stringent algebraic and methodological check, 
as well as revealing the physical origin of the driving terms in the second-order system.

\subsubsection{Perpendicular flow}\label{sub: perpendicular flow}

Throughout our derivation we have kept the second-order fluctuation-induced perpendicular flow $\UperpII$ general. Its special status 
arises from our original assumption on the lowest-order flow as being parallel $\bm{U}=U\bh$, meaning that $\langle \bm{u}_{\perp}\rangle$ is
not a dynamically evolving field, but rather evolves as a \emph{constraint} for the application of multiple-time-scale analysis, {viz.,} the system 
must self-generate $\UperpII$ to yield a system that is quasi-stationary on the fast timescale. 

We can understand  this constraint by starting with the perpendicular momentum equation \eqref{eq:mom}, which at lowest order is
\begin{equation}
\bm{R}_{\perp}
\;\equiv\;
\va^{2}\bigl(\bkap-\gprp{B}\bigr)
-\frac{\cs^{2}}{\gamma}\gprp{p}
+\geff = 0.
\label{eq:Rperp_def}
\end{equation}
We then expand the conserved form of the momentum equation \eqref{eq: conserved MHD m} to transport order, project perpendicular to $\bh$, and average.  The fast-time derivatives drop out, giving (including all terms up to $\epsilon^{3}$)
\begin{align}
0
=\rho\bm{R}_{\perp}
+\mathcal{P}_{\perp}\!\Big\{&
\grad\!\cdot\Big\langle {\rho\bm{u}\bm{u}-\frac{\bm{B}\bm{B}}{4\pi}}\Big\rangle
+\grad p^{(2)}_{\rm tot}
\nn\\ &-\rho\nabla\Phi_{\rm rot}
-2\rho U \bm{\Omega}\times \bh
\Big\},
\label{eq:perp_mom_constraint_Rperp}
\end{align}
i.e., the transport-order expansion is the statement that the leading-order equilibrium $\bm{R}_{\perp}=0$ must also be maintained once the quadratic stresses and other $\mathcal{O}(\epsilon^{3})$ pieces are included.  In this viewpoint, $\UperpII$ is not an additional degree of freedom; rather, it is the mean perpendicular drift that must be chosen so that the transport update of $(\rho,U,p,\bm{B})$ remains compatible with $\bm{R}_{\perp}=0$ at each transport step.

Practically, this suggests a natural strategy: evolve the mean transport system (for $\rho$, $\rho U$, $E_{\rm th}$, and $\bm{B}$) with $\UperpII$ \emph{chosen}  to ensure that the perpendicular equilibrium  \eqref{eq:Rperp_def} remains satisfied on the updated profiles (this likely requires an iterative solver of some sort).  This is closely analogous to the fusion practice of working in flux coordinates: one solves for an equilibrium magnetic geometry (e.g.\ via Grad--Shafranov in axisymmetry), evolves transport on those surfaces, and only then (if needed) reconstructs the implied motion of the flux surfaces; in our notation, that “surface motion’’ is $\Vpsit$. The transport equations are recast in this form below (see \cref{eq: general flux transport}).

One can also obtain an explicit along-field representation of the same constraint.  Expanding the Reynolds/Maxwell stress yields the following turbulence-averaged quadratic pieces,
\begin{subequations}
\begin{align}
\mathsf{F}^{\perp\perp}\equiv&\avg{\rho\,\duprp\duprp-\frac{\dbprp\dbprp}{4\pi}},\\
\bm{F}^{\|\perp}\equiv&\avg{\rho\,\duprl\duprp-\frac{\dbprl\dbprp}{4\pi}+U\d\rho\,\duprp } ,\\
F^{\|\|}\equiv&\avg{\rho\,\duprl^{2}-\frac{\dbprl^{2}}{4\pi}}
\end{align}
\end{subequations}
After commuting $\mathcal{P}_{\perp}$ through the tensor divergence, the perpendicular constraint \eqref{eq:perp_mom_constraint_Rperp} becomes
\begin{align}
&\rho  U\big(\bh\!\cdot\!\grad+\bh\,\bkap\!\cdot\big)\UperpII + \rho U (\nabla\bh) \cdot\UperpII + \Srho\UperpII \nn\\ 
&=-\grad\!\cdot\mathsf{F}^{\perp\perp}
-\bh\!\left(W_{r}^{\perp}\nabla\cdot\bh+\rho\avg{\zp\!\cdot\!\tsmat\!\cdot\!\zm}\right) -\grad_{\perp}p_{\mathrm{tot}}^{(2)} \nn\\&
\quad -\big(\bh\!\cdot\!\grad+\bh\,\bkap\!\cdot\big)\bm{F}^{\|\perp} -\frac32\nabla\cdot\bh\,\bm{F}^{\|\perp}
-(\tsmat+\mathsf{A})\!\cdot\!\bm{F}^{\|\perp}
\nn\\ & \quad -\bkap\,F^{\|\|}
+ \rho \nabla_{\perp}\Phi_{\rm rot} + 2\rho U \bm{\Omega}\times \bh. 
\label{eq:Uperp2-evol}
\end{align}
The terms involving $\bh$, $\bkap$, $\nabla\cdot\bh$, $\nabla\bh$,  and $\tsmat$ arise from commuting the projector through the divergence, a similar origin to the $F^{\perp}_{\rm RM}$ parallel Maxwell/Reynolds stress introduced for \cref{eq:Upar-transport}. 

It is important to emphasize the interpretation of \eqref{eq:Uperp2-evol}, which does \emph{not} remove the need to enforce the perpendicular force-balance constraint to compute $\UperpII$: it is simply a convenient along-$\bh$ form of the same condition.  Consequently, any $\UperpII$ that maintains $\bm{R}_{\perp}=0$  along the entire field line will automatically satisfy \eqref{eq:Uperp2-evol}, because \eqref{eq:Uperp2-evol} was explictly derived by rewriting that constraint.  What \eqref{eq:Uperp2-evol} {does}  provide is a potentially useful shortcut for solution: one may determine $\UperpII$ at a single $\ell$ then integrate \eqref{eq:Uperp2-evol} to reconstruct its parallel structure along the field line, rather than solving for $\UperpII$ from $\bm{R}_{\perp}(\ell)$ at each $\ell$. It also helps illustrate various physical effects, for instance how the Coriolis force (the final term  $2\rho  \bm{\Omega}\times \bm{U}$) will drive the nascent beginnings of the Parker spiral in $\UperpII$, as described by \citet{Weber1967}; that said, once this grows to larger angles in super-Alfv\'enic regions the ordering $\UperpII\sim \epsilon^{2}$ will break down 
and the theory ceases to apply.

Finally, note the special behaviour as $U\to 0$: in that limit the explicit $\UperpII$ terms on the left-hand side vanish, reflecting the fact that the perpendicular momentum constraint reduces to a balance between turbulent quantities (perpendicular stresses and pressure forces), while $\UperpII$ becomes underdetermined by \eqref{eq:Uperp2-evol} alone.  This is not a contradiction: it is the statement that when the mean parallel advection is absent, the constraint fixes an instantaneous force balance but does not uniquely determine the “label velocity’’ of the large-scale  geometry.  Closely related degeneracies (and their resolution only at higher order, or via global/ambipolarity constraints) are familiar in low-flow gyrokinetic transport theory \cite{Parra2011,Calvo2012}.

\subsubsection{Summary of the third-order system}\label{sub: third order summary}

The averaged third-order system has yielded a set of evolution equations for the slow-timescale
evolution of density \eqref{eq:rho-transport}, magnetic field \eqref{eq:induction-transport}, momentum  \eqref{eq:Upar-transport}, and thermal 
energy \eqref{eq: thermal energy with diss}. Each of these are split into a turbulence-driven flux --- the part inside a total derivative -- and, for the momentum and heat, additional forces/sources. With a closure or simulations of how the fluctuations depend on the large-scale gradients, as captured by the second-order system, these equations could be evolved in time to yield the slow evolution of the mean background due to injected waves or self-generated instabilities.  Subject to a few caveats discussed below in App.~\ref{app:parker_like_wind} they yield 
equations for a Parker-like wind with additional effects from wave pressure, wave heating, and cross-field turbulent transport
of particles, field, momentum and heat. Total energy conservation on the transport time scale is ensured from the outset by our method; the mechanical and thermal parts sum to yield the third-order expansion of the total 
energy  \cref{eq: conserved MHD E}, which --- as a total divergence --- has no sources other than the 
external ones. 
Thus, although the various terms involving compressive products that appear  are likely to be very small and physically unimportant in most situations (e.g., $2W_{r}^{\|}\gradl{U}$), perhaps 
warranting their neglect in any practical implementation, their inclusion is crucial for providing  
overall consistency to the theory.

The magnetic-field transport equation \eqref{eq:induction-transport} revealed that $\bm{B}$ is transported perpendicularly in the plasma frame at  net velocity $\Vpsi$, implying magnetic surfaces  move at speed $\Vpsit \equiv \Vpsi + \UperpII $ in the zeroth-order stationary frame. 
In keeping with the convention of gyrokinetic theories \cite{Callen2010,Barnes2010,Abel2013}  it is natural to   choose to measure fluxes with respect to $\Vpsit$. Physically, this is transparent --- fluxes 
of density, heat, and momentum are ``cross-field'' fluxes ---  it also has the advantage of removing the explicit appearance of $\UperpII$ in perpendicular fluxes, which may be difficult to
determine in some geometries. In \citet{Abel2013}, where the magnetic field is used to define the radial coordinate, 
this choice appears  in the commutation of the
time derivative through the flux-surface average: this corresponds to either the $\Vpsi$ term or the $\partial \psi/\partial t$ contribution in their equations  (42) or (45).

With this choice, we write each of the transport equations in the form 
\begin{equation}
\left.\frac{\rmd}{\rmd\tau} \right|_{\psi} G + \nabla\cdot \bm{\Gamma}_{G} = \mathcal{S}_{G},\label{eq: general flux transport}
\end{equation}
with $G=\{\rho,\rho U, E_{\rm th}\}$. Here $\bm{\Gamma}_{G}$ is the cross-field flux, $\mathcal{S}_{G}$ collects fluctuation-induced work/force terms for momentum and heat, and we have defined the transport time derivative evaluated in the magnetic-field frame
\begin{align}
    \left.\frac{\rmd}{\rmd \tau}\right|_\psi G &= \left(\frac{\partial }{\partial \tau} + \Vpsit \cdot \grad\right) G + G \grad \cdot \Vpsit \nn\\
    &= \frac{\partial G}{\partial \tau}+ \nabla\cdot (G \Vpsit).
    \label{eq:time_derivative_magnetic_field}
\end{align}
This definition of ${\rmd}/{\rmd \tau}|_\psi$ is chosen as the natural form that eliminates $\UperpII$ and reproduces fusion results \cite{Abel2013} (see \cref{sec:flux_coordinates}),  following a control volume whose boundary moves with $\Vpsit$; accordingly  $\bm{\Gamma}_{G}$ is formed by subtracting 
$G \Vpsit$ from the stationary-frame flux \eqref{eq: general flux transport}.  $\mathcal{S}_{G}$ remains unchanged from the stationary frame, although it will be helpful to rearrange  terms between
$\bm{\Gamma}_{G}$ and $\mathcal{S}_{G}$ to yield physically enlightening forms.

There is no transport of the magnetic flux in this frame by construction, although its $B^{2}$ changes due to the compression and relative shear  
of the moving control volume (this is encoded in \cref{eq:Bstrength-transport}).
For the other fields we obtain: for  density,
\begin{equation}
\bm{\Gamma}_\rho
   =
   \rho( \Vrho-\Vpsi),\quad \mathcal{S}_{\rho}=0;
\label{eq:rho-flux-rel}
\end{equation}
for parallel momentum,
\begin{align}
&\bm{\Gamma}_{\rho U}
   =
  \rho U (\VU + \Vrho - \Vpsi) ,\nn\\&
   \mathcal{S}_{\rho U}= W_{r}^{\perp} \nabla\cdot\bh+\rho\avg{\zp\!\cdot\!\tsmat\!\cdot\!\zm}+\rho U\bkap\!\cdot ( \VU+\Vrhot)
\nn\\
&\qquad\:-\bh\cdot\nabla p_{\mathrm{tot}}^{(2)} -2\grad\!\cdot(\bh W_{r}^{\|}) - \rho\bh\cdot\nabla\Phi_{\rm rot};
\label{eq:M-flux-rel}
\end{align}
and for thermal energy, 
\begin{align}
&\bm{\Gamma}_{\rm th}
   =
  E_{\rm th} \left(\Vp - \gamma\Vpsi \right) ,\nn\\&
   \mathcal{S}_{\rm th}= - p\nabla\cdot\Vpsit - \rho U^{2}\VU\cdot(\bkap+\gprp{U})+\rho \geff \cdot(\Vrho-\Vpsi)\nn\\&
 \quad -\!\nabla\!\cdot\!\left[\bh \!\sum_{\rm wave} (U \!+\! v_{\rm wave})  W_{\rm wave}  \!+\! \bh E_{\rm th} \!\avg{\frac{\d s}{c_{v}}\duprl\!}\!\right]\nn\\&
 \quad  -p^{(2)}_{\rm tot}\nabla\cdot\bm{U} - U F^{\perp}_{\rm RM} - 2W_{r}^{\|}\gradl{U}  -  \nabla\cdot\langle \bm{q}^{(2)}\rangle \nn\\
 &\quad- \avg{\duprl\frac{\d\rho}{\rho}}(\Sm-U\Srho)-(W_{u}^{\perp}+ W_{u}^{\|} )\frac{\Srho}{\rho} ,
\label{eq:Eth-flux-rel}
\end{align}
 with the turbulent transport velocities collected here for convenience,
\begin{flalign}
&\Vrho \equiv \avg{\frac{\d\rho}{\rho}\duprp},\: \Vpsi \equiv \avg{\frac{\dbprl}{B} \duprp - \duprl \frac{\dbprp}{B}},\nn\\
&\Vp \equiv \avg{\frac{\d p}{p}\duprp},\:\VU \equiv \avg{\frac{\duprl\duprp}{U} -\frac{\dbprl\dbprp}{4\pi \rho\,U}}.
\label{eq:transport_velocities}
\end{flalign}
In deriving \cref{eq:Eth-flux-rel} we have inserted the fluctuation's dissipation $\avg{ D^{\rm tot}}$ into \cref{eq: thermal energy with diss}, which cancels thermal-driving terms,  then rearranged terms between $\bm{\Gamma}_{\rm th}$ and $\mathcal{S}_{\rm th}$ to eliminate $\gamma \UperpII$ in the divergence of $\bm{\Gamma}_{\rm th}$. 
This form, which has explicitly removed the heating from  fluctuations in favor of the conservation law \eqref{eq: local energy conservation} that arises from the assumed quasi-stationarity, captures explicitly the fluctuation-mediated 
transfer of kinetic, potential, and/or magnetic  energy into heat (the first line of $\mathcal{S}_{\rm th}$). On the other hand,
it hides the relevance of thermal gradients in driving fluctuations through $\mathcal{Y}^{\perp}_{\rm th}$ \eqref{eq: Y perp th} --- any energy deposited into fluctuations is immediately (on the transport timescale) 
dissipated back into heat, thereby driving transport but not heating.
 
The fluctuation-induced forces ($\mathcal{S}_{\rho U}$) and heating ($\mathcal{S}_{\rm th}$) provide a complementary physical picture to the discussion of the factors influencing the local heating rate  below \cref{eq: local energy conservation}. 
In the force $\mathcal{S}_{\rho U}$ we have (in order of appearance in \eqref{eq:M-flux-rel}):
\begin{enumerate}[label=(\roman*), itemsep=2pt, topsep=2pt, parsep=0pt]
\item The force on the mean flow from Reynolds and Maxwell stresses of Alfv\'enic fluctuations, which arises from the expanding ($\nabla\!\cdot\!\bh$) or shearing ($\tsmat$) geometry; 
\item The ``redirection'' of the cross-field momentum transport along curved field lines (likely unimportant);
\item The parallel force from gradients of the perturbed total thermal and magnetic pressure;
\item The Reynolds and Maxwell stresses of parallel fluctuations; and
\item The centrifugal force. 
\end{enumerate}
In the fluctuation-induced heating $\mathcal{S}_{\rm th}$ we have (in order of appearance in \eqref{eq:Eth-flux-rel}):
\begin{enumerate}[label=(\roman*), itemsep=2pt, topsep=2pt, parsep=0pt]
\item Heating from the compression of the plasma in the magnetic-field frame (note that there is some flexibility in this term, which can be changed in form by changing $\bm{\Gamma}_{\rm th}$; we follow \citet{Abel2013} equation (194)); 
\item Heating, enabled by exchange with the fluctuations,  from the relaxation of perpendicular gradients of the mean flow;
\item Heating, enabled by the fluctuations,  from the free energy lost/released when mass is moved across the effective gravitational potential;
\item The dissipation (or growth) of fluctuation free energy as it propagates through the domain in the parallel direction, mediated by an adiabatic exchange of thermal fluctuations with  the background $\avg{(\d s/c_{v})\duprl}$;
\item The work done by the fluctuating pressure $p^{(2)}_{\rm tot}$ on the expanding flow; 
\item The work done by the Reynolds and Maxwell stresses of perpendicular and parallel fluctuations, through the force in the momentum transport ($U F^{\perp}_{\rm RM}$ and $2 W^{\|}_{r}\gradl{U}$); 
\item Corrections to the standard thermal heat flux due to second-order fields; and
\item The fluctuation's contribution to the work done by the external force ($\Sm$) or mass addition ($\Srho$).
\end{enumerate}
As described in the main text, the source of the fluctuations that yield non-zero $\Vpsi$, $\VU$, $\Vrho$, Alfv\'enic energies, or parallel correlations 
can be either from externally driven waves propagating through the domain, or from self-generated instabilities. The latter, a subset of which are 
explored below in App.~\ref{app: instabilities},
has been the focus of fusion-related studies, while the former seems likely to be more important in understanding coronal heating.

\section{Solving multiscale RMHD as a 3-D heliospheric model}
\label{app:parker_like_wind}

Here we provide further commentary on the mathematics and interpretation of multiscale RMHD, towards the purpose of explaining how  a global model of the heliosphere could be constructed and  evolved in time as described in \cref{sub: sec 2 summary}.

\subsubsection{How a Parker-like wind appears in the multiscale system.}
The multiscale RMHD expansion contains a standard field-aligned ``Parker wind'' already in the  equilibrium at $\mathcal{O}(\epsilon)$.  In particular, the parallel momentum balance \eqref{eq: parallel equil}, together with mass (and flux) conservation along the field, $\gradl(\rho U/B)=0$ ($\gprl{\rho}+\gprl{U}-\gprl{B}=0$), form a closed 1D system for $(\rho,U,p)$ along a given field line. With the standard transport-time ordering, these relations are best viewed as  instantaneous field-aligned constraints that reconstruct $(\rho,U,p)$ along each tube using ordinary differential equations from lower boundary conditions.

More concretely, following \citet{Parker1958}, we can assume isothermality with  $\cs^{2}= p/\rho={\rm const.}$, $\gamma=1$, and $\Srho=0$, meaning $\gprl{p}=\gprl{\rho}=\gprl{B}-\gprl{U}$. The parallel projection of the $\mathcal{O}(\epsilon)$ momentum equation \eqref{eq: parallel equil} then becomes 
\begin{equation}
(U^{2}-\cs^{2})\frac{1}{U} \frac{\partial U}{\partial\ell} = \cs^{2}\nabla\cdot\bh-\gradl\Phi_{\rm grav} ,
\label{eq: parker_like_basic}
\end{equation}
which, if the field is locally radial ($\gradl = \partial/\partial\ell=\partial/\partial R$), is the standard Parker equation generalized to an arbitrary magnetic expansion that is quantified by $\nabla\cdot\bh = -\gprl{B}$ (note that $\nabla\cdot\bh=2/R$ for a purely radial field). This illustrates how the general field-line geometry does not change the basic wind physics.

\paragraph{Interpretation of the transport system.}
As presented in App.~\ref{app: derivation}, the parallel and perpendicular dynamics in the transport system must be treated in fundamentally different ways.  On the slow time $\tau$, the  equations evolve a field-line-labeled perpendicular structure in 2D by updating flux-tube-integrated budgets (mass, momentum, and thermal energy). In contrast, the  field-aligned profiles along each flux tube are slaved constraints: at each transport step they are reconstructed instantaneously from the parallel equilibrium and boundary conditions, rather than being advanced as local-in-$\ell$ initial-value equations.  This is the same organizing principle used in multiscale gyrokinetics \cite{Barnes2010,Abel2013}, where the procedure would be to compute perpendicular mass/momentum/heat fluxes, 
evolve the flux-surface-averaged fields on the transport time, then reconstruct the 2D equilibrium via a Grad--Shafranov solve.  In our open-field setting, the  reconstruction would also involve the 1D Parker-like boundary-value problem along $\bh$ (\cref{eq: parker_like_basic}), and the underlying scale-separation assumption is that information can propagate along the tube (at speeds $\sim U$ plus the relevant wave speeds) rapidly compared to the time over which perpendicular transport significantly modifies the field-line-labeled state.  

Within this viewpoint, the second-order mean perpendicular flow $\UperpII$ plays a central  role 
in explaining the different roles of the parallel and perpendicular equilibria at $\mathcal{O}(\epsilon)$: $\UperpII$ is not prescribed a priori, but is instead determined by the requirement that the averaged system remain slowly evolving, which requires it satisfy the perpendicular momentum balance at each transport step (see \cref{sub: perpendicular flow} for further discussion).  A straightforward approach to achieve this, taken by fusion transport solvers, is to frame the transport system in terms of cross-field, as opposed to stationary-frame,  fluxes \eqref{eq: cross field fluxes}, then use the equilibrium to infer $\UperpII$ if it is needed directly.

\subsubsection{A difficulty: the parallel equilibrium constraints}
An attentive reader will nevertheless have noticed a potentially serious  issue: in the strict multiscale ordering \eqref{eq:ordering} the leading-order mean solution is adiabatic ($\gprl{s}=0$) unless external sources are imposed, with the wave/turbulent dissipation that is crucial for driving faster wind entering the thermal-energy evolution only on the transport time (schematically $\partial E_{\rm th}/\partial \tau$ in \eqref{eq: thermal energy}). Likewise, wave-pressure forces are absent in the momentum equation. A strict interpretation of the multiscale system 
would spread any heat/momentum added across the full length of the tube in question, disallowing localized forces or heating effects ---
 the  corresponding solutions will  disagree strongly with observations.

This mismatch is related to both the 2+1D interpretation discussed above and the fluctuation ordering: standard RMHD assumes $z^{+}/\cs\ll1$, whereas a wave-driven wind generally involves a $\zp$ whose energy is comparable to the enthalpy change required to accelerate the wind over an expansion length ($W^{+}_{\perp}\sim E_{\rm th}$, or $z^{+}\sim \cs$).  Indeed, in the solar corona and wind, it is often the case that $z^{+}\sim \cs$ while still maintaining $\d\rho/\rho\ll1$  (i.e. large transverse motions without strong compressibility).  In this regime, it is not realistic to treat  wave heating  as a small correction to an otherwise adiabatic equilibrium. 

\paragraph{Robustness of the turbulent dynamics.}

Importantly, this mismatch is primarily a statement about the \emph{mean} ordering --- {i.e.}, about the cumulative heating and acceleration produced by dissipation over large distances --- rather than about the utility of the \emph{fluctuation} equations and the local phenomenologies developed above.  This can be understood more formally by considering an alternative low-$\beta$,  imbalanced ordering, as developed in App.~\ref{app: lowbeta_transonic}. There, we consider a transonic outward wave with $z^{+}\sim \cs\ll\va$, while still requiring $\d\rho/\rho\ll1$, and show that the resulting fluctuation equations are precisely the appropriate low-$\beta$ subsidiary limits of the system derived from the standard ordering \eqref{eq: ordering}. In this sense, the asymptotic procedures commute at the level of the fluctuation equations: starting from full MHD with transonic, low-$\beta$, highly imbalanced fluctuations gives the same reduced fluctuation dynamics as taking the low-$\beta$ limit of the present RMHD system.

This logic is, moreover, closely aligned with standard practice in the solar-wind literature. In many applications, RMHD-like or incompressible Alfv\'enic dynamics are coupled directly to Parker-like backgrounds modified by wave pressure and dissipation, without insisting on the formal condition $z^{+}\ll\cs$ \citep[e.g.,][]{Velli1989,vanBallegooijen2011,Chandran2019,Jacques1978,Chandran2011,Chandran2025}. The practical requirement is instead that the fluctuations remain predominantly transverse and only weakly compressive, so that they do not produce shocks or large density variations. From this perspective, App.~\ref{app: lowbeta_transonic} simply provides a controlled asymptotic justification for a modeling logic that is already widely used.

However, what does change in the imbalanced low-$\beta$ ordering is the partitioning of terms in the mean equations. In particular, wave-pressure, thermal-pressure, and transport/heating terms move between equilibrium and transport orders, so a strict transport derivation no longer has the same clean separation as in the standard multiscale system (see App.~\ref{subapp: equil and transport}). This repartitioning  suggests that the strict ordering of terms between the parallel equilibrium (at $\mathcal{O}(\epsilon)$) and transport (at $\mathcal{O}(\epsilon^{3})$) is not robust, and might be --- with care --- profitably reorganized.

\subsubsection{Resummed equilibrium}
We now exploit this separation --- robust small-scale dynamics with a cumulatively important mean feedback --- to propose 
a \emph{resummed equilibrium} method that allows wave-driven Parker-like solutions within the standard multiscale RMHD ordering \eqref{eq:ordering}: rather than taking the adiabatic equilibrium \eqref{eq: equilibrium} as the lowest-order background, we re-define the leading-order mean state to be a 1D, field-aligned wind that already includes a selection of  heating terms and  wave-pressure forces.
Operationally, this is achieved by retaining the mean system, \cref{eq:first-cont-mean,eqapp: parallel equil,eq:entropy-mean}, but with  chosen  terms
from the transport equations repartitioned from $\mathcal{O}(\epsilon^{3})$ into the arbitrary sources $\Srho$, $\Sth$, and $S_{\rho U}$ at $\mathcal{O}(\epsilon)$ (similar 
to how the low-$\beta$ ordering of App.~\ref{app: lowbeta_transonic} moves terms between equilibrium and transport orders). This 
thus ``resums'' the parallel equilibrium.
The idea is justified by the fact that the fluctuation equations, and therefore the turbulent dynamics, do not depend explicitly on $\Srho$, $\Sth$, or $\Su$ themselves; their influence therein
comes only through their effect on changing the background parallel gradients ($\gprl{s}$, $\gprl{U}$ etc.). 
Equivalently, one may regard the repartitioning as a partial resummation of the perturbation series for the mean profiles, in which terms that are formally higher order but accumulate coherently along $\bh$, are absorbed into the $\mathcal{O}(\epsilon)$ sources to define an improved leading-order equilibrium. 
From the perspective of conservation laws, this repartitioning is essentially irrelevant: parallel variation in $E_{\rm th}$ or $\rho U$ that would have been fixed by the lowest-order equations is now not, but the total integrated exchange between 
background and fluctuations, or via parallel fluxes in the case of waves, remains the same.


\paragraph{Practical solution strategies.}
In practice, the resummed framework can be used in (at least) two closely related ways.  In both cases, one specifies an (initial) large-scale geometry of $\bm{B}$ and other mean quantities, boundary conditions for the injected outward-wave amplitude at the coronal base, and a form for the quadratic correlators/fluxes as functionals of the local background gradients and wave energies. The latter could be a phenomenological closure such as that outlined   in \cref{subsec: preliminaries}, or, more accurately, simulations of the second-order  RMHD equations along flux-tubes.  

(i) \emph{Fixed-point (steady) resummation.}
One seeks a steady, field-aligned background in which the mean profiles and the wave dissipation computed from the closure are mutually consistent.
Operationally, this amounts to iterating between solving the 1D mean equations along $\bh$ for $(\rho,U,p)$ together with the outward-wave evolution, and updating the resummed sources ($\Srho$, $S_{\rho U}=\rho \Su+U\Srho$, and/or $\Sth$) from the closure until convergence.  Conceptually, this is simply the familiar flux-tube problem of a wind driven by wave pressure and wave heating (now augmented by  additional curvature/stratification/transport-mediated channels), as in standard wave-driven wind models and codes \citep[e.g.,][]{Cranmer2007,Chandran2011}.  In this form, the method is most naturally applied when the perpendicular structure is prescribed (or varies only parametrically), so that each field line may be treated as an approximately independent 1D problem.

(ii) \emph{Running resummation (time-dependent).}
A particularly simple alternative is to {evolve} the slow-time system directly, continuously evaluating the fluctuation/closure terms using the \emph{current} background gradients, rather than enforcing the formally lowest-order adiabatic parallel equilibrium at each step.  In mathematical terms, the equations solved are simply the \emph{sum} of the $\mathcal{O}(\epsilon)$ parallel-equilibrium constraints and the $\mathcal{O}(\epsilon^{3})$ transport corrections (which involves the slow time derivatives), thereby yielding a single resummed parallel structure \footnote{A reader may reasonably ask why we should preference the parallel equilibrium equation in this way, as opposed to 
the perpendicular one. The answer is that one could, if desired, also sum the perpendicular force balance at $\mathcal{O}(\epsilon)$ and $\mathcal{O}(\epsilon^{3})$. The result is simply \cref{eq:perp_mom_constraint_Rperp}, which captures 
both the mean perpendicular forces and mean turbulent stresses (Reynolds/Maxwell stresses and turbulent pressure) ---
if this were solved as a single equation, rather than separately enforced at first and third orders, the equilibrium would 
also include such turbulent stresses. However, such corrections, unlike the parallel ones, are likely not of significant importance, and care would be needed to ensure energy conservation and that ensure fast timescales are not inadvertently introduced into the system. } Equivalently, one simply considers the mass, parallel momentum and heat sources ($\Srho$, ${S}_{\rho U} $, and $\Sth$, respectively) to 
be the negative of the relevant $\mathcal{O}(\epsilon^{3})$ transport equation, safe with the knowledge that the fluctuation equations remain valid for arbitrary sources. 

The resulting system is then
 \begin{align}
\frac{\partial \rho}{\partial \tau}
&=
-\nabla_{\perp}\!\cdot\!\left(\rho\Vrhot\right)
-\nabla\!\cdot\!\left(\bh \rho U \right),\label{eq:rho-transport-resummed}
\\[0.3em]
\frac{\partial \bm B}{\partial \tau}
&=
\nabla\times\left(\Vpsit\times\bm B\right),
\label{eq:B-transport-resummed}
\\[0.3em]
\frac{\partial(\rho U)}{\partial \tau}
&= -\left(\nabla_{\perp}-\bkap\right)\!\cdot\!
\left[\rho U\left(\VU+\Vrhot\right)\right] - \nabla\!\cdot\!\left(\bh\rho U^{2}\right)\nn\\
&\!\!\!\!\!\!\!- \gradl(p+p_{\rm tot}^{(2)})
+F^{\perp}_{\rm RM}-2\nabla\!\cdot\!\left(\bh W_{r}^{\|}\right)-\rho\gradl\Phi_{\rm tot}
\label{eq:Upar-transport-resummed}
\\[0.3em]
\frac{\partial E_{\rm th}}{\partial \tau}
&=
-\nabla_{\perp}\!\cdot\!\left(E_{\rm th}\Vpt\right)-c_{v}^{-1}E_{\rm th}U \,\gradl{s}
+\Sth -\nabla\!\cdot\!\avgi{\bm q}\nn\\
&
\!\!\!\!\!\!\!-\nabla\!\cdot\!\left[\bh \sum_{\rm wave}(U+v_{\rm wave})W_{\rm wave}+E_{\rm th}\bh\avg{\frac{\d s}{c_{v}}\duprl}\right]
\nn\\
&
\!\!\!\!\!+\mathcal{Y}^{\perp}_{U}
+\mathcal{Y}^{\perp}_{B}
+\mathcal{Y}^{\perp}_{g}
-p_{\rm tot}^{(2)}\nabla\cdot\bm{U}
\nn\\
&
\!\!\!\!\!\!\!-UF^{\perp}_{\rm RM}
-2 W_{r}^{\|}\gradl{U},
\label{eq:Eth-transport-resummed}
\end{align}
where in the momentum equation \eqref{eq:Upar-transport-resummed}, we have 
summed the  parallel equilibrium \eqref{eq: parallel equil} and $U\Srho = - U\nabla\cdot(\bh\rho U)$
to obtain an equation for the momentum (as opposed to the velocity). 
We have set $\Srho=\Su=0$,
 given the lack of external forces and mass sources in the coronal context, but the heat source $\Sth$ is included due to its physical importance for capturing radiative cooling. 
 The system \eqref{eq:rho-transport-resummed}--\eqref{eq:Eth-transport-resummed} additionally requires the  perpendicular equilibrium \eqref{eq: perp equil}, which  determines $\UperpII$ as a constraint (see discussion in \cref{sub: perpendicular flow}).

This ``running'' procedure, which thus simply amounts  to combining all parallel-structure-related mean equations before their solution, is attractive because it avoids any special bookkeeping: the departure of the parallel mean state from the adiabatic background is produced self-consistently by the same wave-pressure and heating terms that appear in the transport equations, and the steady solution (if one exists) will emerge as the fixed point of the evolution. Indeed, 
we can see \cref{eq:Upar-transport-resummed} as a Parker-wind-like equation (similar to \eqref{eq: parker_like_basic}) that 
includes the wave pressure \cite{Jacques1978}. In addition, the system contains all of the new effects discussed in \cref{sec: phenomenology}, including from perpendicular transport of mass, momentum, and heat, and heating via the relaxation of perpendicular gradients (the $\mathcal{Y}^{\perp}$ terms). The system  therefore provides a  direct route to obtaining realistic wave-driven profiles within the multiscale framework for a true 3-D heliospheric model. Further 
discussion of related subtleties, particularly how the ideas connect with explicitly transsonic orderings, is provided in App.~\ref{subapp: equil and transport}.


\paragraph{Limitations.}
We emphasize that the  strategy described above is best viewed as a practical modeling device rather than a  controlled  procedure: it amounts to a partial resummation in which formally higher-order terms that accumulate coherently in the parallel structure are promoted to define an improved leading-order wind.  Making this logic fully systematic would require a more formal treatment of this reorganization; e.g., a modified ordering, as considered in App.~\ref{app: lowbeta_transonic}, or renormalization-group methods for differential equations (see \cite{Chen1994,Veysey2007} and references therein).  The resulting resummed wind need not be unique and may exhibit multiple branches or instabilities (e.g.,  multiple transonic solutions or  unstable equilibria), so numerical strategies should be designed to diagnose non-uniqueness and select the physically relevant branch.  Moreover, given that the method does not preserve the strict ordering, care is required to ensure the intended scale separation is preserved: any terms promoted into the parallel ``equilibrium'' via $\Sth$, $\Sm$, and/or $\Srho$ must be  slowly varying in time and space --- there exists the risk of contaminating the reduced equilibrium problem with fast-time variability that lies outside the ordering assumptions.
A final related danger, which is shared by the standard ordering of App.~\ref{app: derivation} (i.e., it is not related to the resummation procedure),  is that the ordering formally assumes that the profile is communicated along  the full length of a flux tube on timescales shorter than the transport time; depending on the heliospheric altitude to which the the system is solved, this may not be well satisfied.

\section{A subsidiary low-$\beta$ ordering for transonic, highly imbalanced fluctuations}
\label{app: lowbeta_transonic}

The ordering of the fluctuations used in the main text and App.~\ref{app: derivation} assumes $\delta z^{+}/\cs \ll 1$. While this could be well satisfied in closed and structured boundary regions --- where $\beta\sim1$ due to the weak field meaning that Alfv\'enic fluctuations are subsonic if $|\dbprp|/B<1$ (see \cref{fig: model}; \cite{Gary2001}) --- in many regions of the corona the dominant outwards propagating waves are expected to be transonic ($z^{+}\sim \cs$). In the main text, we ignored this issue by arguing that the Alfv\'enic fluctuation dynamics should not depend sensitively on $\cs$ itself: these fluctuations are nearly incompressible, suggesting that the reduced equations should remain useful even outside their formal regime of validity, particularly in near-Sun low-$\beta$ coronal-hole-like regions where $|\dbprp|/B$ remains small. Similar  assumptions have been regularly made implicitly in past works that use incompressible or reduced MHD to model the solar wind \cite{Oughton2011,Bruno2013,Meyrand2025}.

The purpose of this appendix is to move beyond these qualitative expectations and \emph{prove} that this is indeed the case for one choice of ordering for which $z^{+}\sim\cs$. More precisely, we show that starting from the full compressible MHD equations with a transonic, highly imbalanced outward wave that satisfies  $z^{+}\sim \cs$ and $\cs/\va \ll 1$ leads, at lowest order, to the same fluctuation system obtained by taking the low-$\beta$ limit of the equations derived in \cref{app: derivation}. In this sense, the two asymptotic procedures commute: no new leading-order fluctuation terms appear in a formally valid expansion that are not contained in the standard RMHD equations and relevant in our slaved phenomenology. This result provides a formal justification for applying the fluctuation equations from the main text across both structured subsonic regions and low-$\beta$ transonic regions of the corona and solar wind.

Our focus here will be on the fluctuation equations, relegating a brief discussion of equilibrium and transport to \S\ref{subapp: equil and transport} because the mixed ordering yields  a system that does not cleanly separate into different orders (although the terms remain very similar to those in the main text).  This is sufficient for the practical purpose of the main text, namely to justify the use of a single fluctuation model across regions with rather different values of $\beta$ and $\delta z^{+}/\cs$ (see \cref{sub: sec 2 summary} and App.~\ref{app:parker_like_wind}). The ordering presented below is not unique; it is simply a compact proof of principle, and further work along these lines would prove valuable.

\subsection{Imbalanced, transonic fluctuation equations}
\paragraph{Starting equations}
In order to build imbalance directly into the ordering, we begin from the compressible MHD equations written in terms of the total Els\"asser fields,
\begin{equation}
\bm Z^{\pm} \equiv \bm u \mp \bm V,\qquad
\bm V \equiv \frac{\bm B}{\sqrt{4\pi\rho}},
\end{equation}
for which the continuity, momentum, induction, and entropy equations (\crefrange{eq:cont}{eq: entropy app}) become
\begin{align}
\frac{\partial \rho}{\partial t} + \nabla\cdot(\rho \bm u) &= 0,\label{eq:appC_cont}\\
\frac{\partial \bm Z^{\pm}}{\partial t}
+ \bm Z^{\mp}\cdot\nabla \bm Z^{\pm}
&+ \bm V\left(\nabla\cdot\bm V \mp \frac{1}{2}\nabla\cdot\bm u\right)
\nn\\
&= -\,\frac{1}{\rho}\nabla\!\left(p+\frac{B^{2}}{8\pi}\right)+ \bm g,\label{eq:appC_Zpm}\\
\left(\frac{\partial}{\partial t}+ \bm u\cdot\nabla\right)\frac{s}{c_{v}}&=0.
\label{eq:appC_entropy}
\end{align}
As in App.~\ref{app: derivation}, unadorned symbols here represent full, unexpanded fields, but below will represent in the lowest-order background. We omit explicit dissipation and sources for simplicity, because sources do not significantly affect the fluctuation equations and dissipation terms can be assumed implicitly throughout. Rotation is likewise ignored. The gravitational acceleration is written $\bm g = -\nabla \Phi_{\rm grav}$, with characteristic size
$g \sim {v_{\rm esc}^{2}}/({2R})$.

Note that in most equations below we  drop terms that do not involve fluctuations (i.e., those that are equal to their turbulent average), because such terms 
do not contribute to the fluctuating equations. It transpires that the procedure is nearly identical to the standard RMHD procedure from App.~\ref{subapp: second order}, albeit formulated 
in terms of $\bm{Z}^{\pm}$ instead, and with many more terms dropped on ordering grounds.

\paragraph{Ordering}

This appendix uses a nested two-parameter expansion. The first, $\imord\ll1$, is effectively the low-$\beta$ parameter, with $\imord^{2}\sim\beta$, and simultaneously orders the amplitude and anisotropy of the dominant outward wave $\zo$. The second, $\epsilon\ll\imord$, orders the smaller subsidiary fluctuations and the weak background gradients. Because $\zo$  is an exact local solution in a straight field, the carrier-wave problem is treated to all orders in $\imord$ alone, while the present calculation retains only the leading corrections in $\epsilon$, namely the $\epsilon$ and $\imord\epsilon$ terms, the latter being the order at which the first nontrivial fluctuation equations appear.

More precisely, we take  $\zo$ to satisfy
\begin{equation}
\frac{|\bm{u}^{(0)}|}{\va}\sim\frac{|\bm{B}^{(0)}|}{B}\sim \frac{|\zo|}{\va}\sim \imord.
\end{equation}
By setting $\bm{u}^{(0)} = -\bm{V}^{(0)} = \zo/2$, there is no $\mathcal{O}(\imord)$ correction to $\bm{Z}^{-}$. We further order 
\begin{equation}
\frac{\cs}{\va}\sim \frac{U}{\va}\sim \imord,\quad
\frac{k_{\parallel}}{k_{\perp}}\sim \imord,\quad
\frac{\omega}{k_{\perp}\va}\sim \imord,
\label{eq:appC_ordering1}
\end{equation}
so that the dominant outward fluctuation is transonic, $z^{+}\sim \cs$, and ${v_{\rm esc}}\sim {\va}$ (a weaker gravity ordering, ${v_{\rm esc}}\sim {\cs}$, can 
be used instead, but reproduces a subsidiary limit of our final system). The background inhomogeneity and subsidiary fluctuations are then ordered by $\epsilon$ through
\begin{equation}
\gprl{B}\sim \gprl{\rho}\sim \gprl{U}\sim
\gprp{B}\sim \gprp{\rho}\sim \gprp{U}\sim \epsilon k_{\perp},
\label{eq:appC_ordering2}
\end{equation}
and
\begin{equation}
\frac{\dzpprp}{\va}\sim 
\frac{\dzmprp}{\va}\sim 
\frac{\dzpprl}{\va}\sim 
\frac{\dzmprl}{\va}\sim
\frac{\d\rho}{\rho}\sim 
\frac{\dbprl}{B}\sim \epsilon.
\label{eq:appC_dz_ordering1}
\end{equation}
(We use the different notation $\dzp$ and $\dzm$ to distinguish these corrections from the Els\"asser variables used in the main text.) The pressure, $\d p/p$, can be ordered as either $\epsilon/\imord$ or $\epsilon/\imord^{2}$, depending on the strength of its driving at the outer scale (see below).

We choose to define the Els\"asser perturbations with respect to the background density $\rho$, such that $\zo = \bm{u}^{(0)}- \bm{V}^{(0)}=\bm{u}^{(0)}- \bm{B}^{(0)}/\sqrt{4\pi\rho}$. Keeping  arbitrary orders in $\imord$ and to $\mathcal{O}(\epsilon)$ (as needed below), the  total Els\"asser fields are then
\begin{align}
\bm Z^{+} &= (U-\va)\bh + \zo + \dzp +\frac{1}{2}\va\bh\frac{\d\rho}{\rho} ,\label{eq:appC_Zplus_exp}\\
\bm Z^{-} &= (U+\va)\bh + \dzm-\frac{1}{2}\va\bh\frac{\d\rho}{\rho} .\label{eq:appC_Zminus_exp}
\end{align}
Likewise,  $\d\zpm_{\perp} = \duprp\mp \dbprp/\sqrt{4\pi\rho}$ and $\d z^{\pm}_{\|} = \duprl\mp \dbprl/\sqrt{4\pi\rho}$ are 
 defined with respect to the background density, as in the main text.

\paragraph{The carrier wave \texorpdfstring{$\zo$}{Z0}}
The field $\zo$ is  defined to locally satisfy
\begin{equation}
\left[\frac{\partial }{\partial t} + (U+\va)\,\bh\cdot\nabla\right]\zo = 0
\label{eq:appC_z0eq}
\end{equation}
at a given local point in the domain. All WKB, reflection, and gradient-induced corrections are absorbed into $\dzp$ at higher order. This is therefore a local leading-order definition, rather than a global solution on an inhomogeneous background.

$\zo$ also includes a $\imord^{2}$ parallel correction required to make it spherically polarized, so that the $\mathcal{O}(\imord^{2})$ changes to the  magnetic-field strength  it induces do not vary on the $k_{\perp}$ or $k_{\|}$ scale over which $\zo$ itself varies.  Mathematically, we take
\begin{equation}
\nabla | \bm B + \bm B^{(0)}|^{2} \sim \epsilon,
\label{eq:appC_spherical}
\end{equation}
which effectively means that the magnetic pressure of the wave adds to the equilibrium, not the fluctuation equations (see \S\ref{subapp: equil and transport}).
There are novel fluctuation-scale  field-strength contributions such as $\dbprp\cdot\bm{B}^{(0)}$, but it will transpire that these do not contribute in the final equations.

\paragraph{Modified derivatives}
It is convenient to define the usual advection operator built from the background flow and the carrier wave (cf.~\cref{eq: total time deriv}),
\begin{equation}
\frac{\rmd}{\rmd t}
\;\equiv\;
\frac{\partial }{\partial t} + \bm{U}\cdot\nabla + \frac{1}{2}\,\zo\cdot\nabla,
\label{eq:appC_dt}
\end{equation}
and the nonlinear parallel derivative
\begin{equation}
\bhdg
\;\equiv\;
\bh\cdot\nabla - \frac{1}{2}\,\frac{\zo}{\va}\cdot\nabla,
\label{eq:appC_bhdg}
\end{equation}
These derivatives involve the carrier wave only; the fluctuation corrections $\dzp$ and $\dzm$ will not appear in our final equations, as for the 
slaved phenomenology in the main text.

\paragraph{Lowest-order constraints}
Before expanding the Els\"asser equations, it is useful to record the lowest-order consequences of the ordering. First, using  $\nabla\cdot \bm{B}^{(0)}=0$ and 
$\bm{u}^{(0)} = -\bm{V}^{(0)} = \zo/2$, one finds
\begin{equation}
\nabla\cdot\zo=2\nabla\cdot\bm{u}^{(0)} = -2\nabla\cdot\bm{V}^{(0)} = -\frac{1}{2} \zo\cdot\gprp{\rho}.\label{eq: Z0 divergence}
\end{equation}
Second, continuity \eqref{eq:appC_cont} at $\mathcal{O}(\epsilon)$ gives
$\nabla\cdot \duprp = 0$,
implying that  Alfv\'enic corrections remain divergence free at this order.
Third, fluctuating perpendicular momentum balance at $\mathcal{O}(\epsilon)$ implies
\begin{equation}
\nabla_{\perp}\!\left(\d p + \frac{B\,\dbprl}{4\pi}\right)=0,
\label{eq:appC_pbal}
\end{equation}
which is the usual pressure-balance relation. Because $p\sim \rho \cs^{2}\sim \imord^{2}\rho \va^{2}$, \eqref{eq:appC_pbal} requires
\begin{equation}
\frac{\d p}{p}\sim \frac{\epsilon}{\imord^{2}}
\label{eq:appC_porder}
\end{equation}
if $\dbprl/B\sim \epsilon$ is to be retained as an independent fluctuation. As we discuss below, while this scaling may appear restrictive, requiring $\epsilon< \imord^{2}$, it is stronger than is required for the rest of the fluctuation system: if \eqref{eq:appC_porder} is not satisfied, then $\dbprl$ ceases to satisfy the passive equation derived below, but the remaining reduced equations still hold so long as the pressure required to maintain $\nabla\cdot\duprp=0$ remains small compared to the background pressure. This weaker requirement is simply $\epsilon \ll \imord$, which was already assumed at the outset.

\paragraph{Ordered Els\"asser advection terms}
We now expand the nonlinear advection terms in \eqref{eq:appC_Zpm} using \eqref{eq:appC_Zplus_exp}--\eqref{eq:appC_Zminus_exp}, together with the same perpendicular/parallel decompositions used in \cref{eq:PP-splits}. Retaining terms through $\mathcal{O}(\imord\epsilon)$ and $\mathcal{O}(\imord^{3})$ and splitting the carrier field into its perpendicular and parallel parts,
$\zo = \zo_{\perp} + \bh Z^{(0)}_{\parallel}$,
with $Z^{(0)}_{\parallel}/\va \sim \imord^{2}$ from spherical polarisation, one finds
\begin{subequations}\label{eq:appC_ZgradZ_split}
\begin{align}
\big(\bm Z^{-}\!\cdot\!\nabla \bm Z^{+}\big)_{\perp}
&=(U+\va)\,\bh\!\cdot\!\nabla \zo
+\va\,\bh\,\bkap\!\cdot\!\zo
\nn\\
&\quad
+\va\,\bh\!\cdot\!\nabla \dzpprp
+\dzmprp\!\cdot\!\nabla \zo_{\perp},
\label{eq:appC_ZmgradZp_perp}
\\
\big(\bm Z^{+}\!\cdot\!\nabla \bm Z^{-}\big)_{\perp}
&=-\va\,\bh\!\cdot\!\nabla \dzmprp
+\zo_{\perp}\!\cdot\!\nabla \dzmprp
\nn\\
&\quad
+\va\,\zo_{\perp}\!\cdot\!\nabla \bh,
\label{eq:appC_ZpgradZm_perp}
\\
\big(\bm Z^{-}\!\cdot\!\nabla \bm Z^{+}\big)_{\parallel}
&=(U+\va)\,\bh\!\cdot\!\nabla Z^{(0)}_{\parallel}
-\va\,\bkap\!\cdot\!\zo
\nn\\
&\quad
+\va\,\bh\!\cdot\!\nabla \dzpprl
+\frac{\va^{2}}{2}\,\bh\!\cdot\!\nabla \frac{\d\rho}{\rho},
\label{eq:appC_ZmgradZp_prl}
\\
\big(\bm Z^{+}\!\cdot\!\nabla \bm Z^{-}\big)_{\parallel}
&=-\va\,\bh\!\cdot\!\nabla \dzmprl
+\zo_{\perp}\!\cdot\!\nabla \dzmprl
\label{eq:appC_ZpgradZm_prl}\\
&\!\!\!\!\!\!\!\!\!\!\!\!+\va\,\zo\!\cdot\!\gprp{\va}
+\frac{\va^{2}}{2}\,\bh\!\cdot\!\nabla \frac{\d\rho}{\rho}
-\frac{\va}{2}\,\zo_{\perp}\!\cdot\!\nabla \frac{\d\rho}{\rho}.\nn
\end{align}
\end{subequations}
Here only the leading perpendicular part of $\zo\!\cdot\!\nabla$ is retained; terms involving the formally smaller parallel part of $\zo$ are consistently neglected at this order.

\paragraph{Divergence terms}
The remaining terms in \eqref{eq:appC_Zpm} involve 
$\bm V(\nabla\!\cdot\!\bm V \mp \nabla\!\cdot\!\bm u/2)$. 
For the perpendicular part, it is sufficient to note that $\zo\sim \imord$, while the only contribution to 
$\nabla\!\cdot\!\bm V \mp \nabla\!\cdot\!\bm u/2$ at order $\epsilon$ is the mean piece 
$\nabla\!\cdot(\va\bh)=-\va\,\gprl{\rho}/2$. Thus, to $\mathcal{O}(\imord\epsilon)$,
\begin{equation}
\left[\bm V\!\left(\nabla\!\cdot\!\bm V \mp \frac{1}{2}\nabla\!\cdot\!\bm u\right)\right]_{\perp}
=
\frac{\va}{4}\,\zo\,\gprl{\rho}.
\label{eq:appC_divterm_perp}
\end{equation}

For the parallel part, only the leading field-parallel piece of $\bm V$ is needed, so we consider
$\va\bh(\nabla\!\cdot\!\bm V \mp \nabla\!\cdot\!\bm u/2)$. Using \cref{eq: Z0 divergence}, one finds
\begin{align}
\nabla\!\cdot\!\bm V \mp \frac{1}{2}\nabla\!\cdot\!\bm u
&=
\frac{1}{4}\!\left(1\pm\frac{1}{2}\right)\zo\!\cdot\!\gprp{\rho}
+\frac{1}{4}\,\zo\!\cdot\!\nabla \frac{\d\rho}{\rho}
\nn\\
&\quad
\mp \frac{1}{2}\big(\nabla\!\cdot\!\d\bm u\big)^{(\imord\epsilon)}
-\frac{\va}{2}\,\bh\!\cdot\!\nabla \frac{\d\rho}{\rho},
\label{eq:appC_divterm_prl}
\end{align}
where $(\nabla\!\cdot\!\d\bm u)^{(\imord\epsilon)}$ is the next-order $\mathcal{O}(\imord\epsilon)$ contribution to the divergence of $\d\bm{u}$.

\paragraph{Pressure term}
At $\mathcal{O}(\imord\epsilon)$, the pressure term in \eqref{eq:appC_Zpm} contains several contributions, including gradients of the form $\nabla(\bm B^{(0)}\!\cdot\!\dbprp)$ and a next-order pressure fluctuation $\d p^{(2)}$. However, these do not need to be evaluated explicitly in the Alfv\'enic equations, because they contribute only through the perpendicular gradient required to maintain
$\nabla\cdot\dzpprp = \nabla\cdot\dzmprp = 0.$
Accordingly, we absorb all such pieces into a single incompressible pressure, denoted $\tilde p$, exactly as in the main derivation. We also recall that, because the carrier wave is assumed to be spherically polarized, all terms involving only $\bm B^{(0)}$ and its parallel corrections contribute only background-sized gradients and therefore belong to the equilibrium/transport at this order.

\paragraph{Perpendicular Elsasser equations}
Equations \eqref{eq:appC_ZmgradZp_perp}, \eqref{eq:appC_ZpgradZm_perp}, and \eqref{eq:appC_Zpm} yield the evolution of the Alfv\'enic corrections,
\begin{align}
\frac{\partial \dzpprp}{\partial t}
&+ \va\,\bh\!\cdot\!\nabla \dzpprp + \va \bh\,\bkap\cdot\zo
+\frac{\va}{4}\,\zo\,\gprl{\rho}
\nn\\ &= -\dzmprp\!\cdot\!\nabla \zo_{\perp}-\,\frac{1}{\rho}\nabla_{\perp}\tilde p
\label{eq:appC_dzp_perp}
\\
\frac{\partial \dzmprp}{\partial t}
&- \va\,\bh\!\cdot\!\nabla \dzmprp
+\va\,\zo_{\perp}\!\cdot\!\nabla \bh+\frac{\va}{4}\,\zo\,\gprl{\rho}
\nn\\&= -\zo_{\perp}\!\cdot\!\nabla \dzmprp-\,\frac{1}{\rho}\nabla_{\perp}\tilde p
\label{eq:appC_dzm_perp}
\end{align}
Adding \eqref{eq:appC_dzp_perp} to the carrier-wave equation \eqref{eq:appC_z0eq} then gives an equation for the total outward fluctuation $\zo+\dzpprp$.

Equations \eqref{eq:appC_dzp_perp} and \eqref{eq:appC_dzm_perp} are precisely the same as the subsidiary $\zp\gg\zm$, low-$\beta$, low-$\ma$ limit of the perpendicular fluctuation equations derived in the main expansion, \cref{eq:perp-pm-app}. The compressive feedback terms have been ordered out because $\ma$ and $\beta$ are small and the  parallel scale of the fluctuations has been taken to be shorter than that of the background gradients. Likewise, in a cylindrically symmetric flux tube where $(\nabla\bh)_{\perp} =  (\nabla\cdot\bh)\mathsf I_\perp/2$ and $\bkap=0$,  they reduce to the standard straight-flux-tube system \eqref{eq: straight tube} from \citet{Chandran2009,Chandran2019} in the highly imbalanced limit, as expected. In this sense, the perpendicular dynamics of the transonic low-$\beta$ subsidiary ordering commutes with the standard reduced-MHD expansion.

\paragraph{Entropy and magnetic-pressure fluctuations}
With the stronger pressure ordering \eqref{eq:appC_porder}, the entropy fluctuation satisfies
\begin{equation}
\frac{\d s}{c_v}
=
\frac{\d p}{p}
-\gamma\,\frac{\d\rho}{\rho}
=
\frac{\d p}{p} + \mathcal{O}(\epsilon)
\label{eq:appC_entropy_scaling}
\end{equation}
because $\d p/p \sim \epsilon/\imord^2$ while $\d\rho/\rho \sim \epsilon$. Since the entropy equation \eqref{eq:appC_entropy} is linear in $s$ and background gradients are smaller, its dominant evolution is particularly simple:
\begin{equation}
\frac{\rmd}{\rmd t}\,\frac{\d s}{c_v}=0.
\label{eq:appC_entropy_passive}
\end{equation}
Using the pressure-balance relation \eqref{eq:appC_pbal}, this is equivalently
\begin{equation}
\frac{\rmd}{\rmd t}\frac{\dbprl}{B}=0,
\label{eq:appC_dbprl_passive}
\end{equation}
so magnetic-pressure fluctuations, if present at this order, are passively advected by the carrier-wave flow.

\paragraph{Compressive subsystem}
Expanding the continuity equation \eqref{eq:appC_cont} and using \eqref{eq: Z0 divergence} gives
\begin{equation}
\frac{\rmd}{\rmd t}\frac{\d\rho}{\rho}
+\big(\nabla\!\cdot\!\bm u\big)^{(\imord\epsilon)}
=0.
\label{eq:appC_rho_raw}
\end{equation}
We then take the sum and difference of  \cref{eq:appC_ZmgradZp_prl,eq:appC_ZpgradZm_prl}, along with \eqref{eq:appC_divterm_prl}, to form the evolution of 
$\duprl = (\dzpprl + \dzmprl)/2$ and $\d V_{\|} = (\dzmprl - \dzpprl)/2$,
noting first that $\partial\bm{Z}^{\pm}/\partial t$ picks up a contribution $- (\va/2) (\partial/\partial t) (\d\rho/\rho)$ because $\zo$ and $\d\zpm$ are defined with respect to the background density. 

For $\duprl$, one obtains, after splitting the $\zo_{\perp}\cdot\nabla\dzmprl$ term in \eqref{eq:appC_ZpgradZm_prl} and noting that $\bhdg \d V_{\|} = \va \bhdg (\dbprl/B)$ to the relevant order,
\begin{equation}
\frac{\rmd \duprl}{\rmd t}
- \va^{2}\bhdg \frac{\dbprl}{B}
=
-\frac{\va}{2}\,\zo\!\cdot\!\big(\gprp{B}-\bkap\big).
\label{eq:appC_upar}
\end{equation}
This is exactly the low-$\beta$, low-$\ma$ limit of the RMHD parallel-momentum equation \eqref{eq:par-mom}, with the carrier wave providing the compressive forcing. As in the main text, the combination $\bkap-\gprp{B}$ vanishes for a force-free equilibrium, but becomes nonzero in the presence of gravity, so \eqref{eq:appC_upar} contains a physically meaningful source even in the low-$\beta$ limit (the only change to the equations under the  alternate ordering $v_{\rm esc}\sim \cs$ is that this term disappears).

Subtracting the $\dzpprl$ and $\dzmprl$ equations,  including also the $\partial\d\rho/\partial t$ contribution from $\partial \bm Z^{\pm}/\partial t$ noted above, gives
\begin{align}
\frac{\rmd \d V_{\|}}{\rmd t}
&-\va\,\bhdg \duprl
+\frac{\va}{2}\,\zo\!\cdot\!\left(\bkap+\gprp{\va}-\frac{1}{2}\gprp{\rho}\right)
\nn\\
&\quad
+\frac{\va}{2}\big(\nabla\!\cdot\!\d\bm u\big)^{(\imord\epsilon)}
-\frac{\va}{2}\frac{\rmd }{\rmd t} \frac{\d\rho}{\rho}=0.
\label{eq:appC_dVpar}
\end{align}
 Equation \eqref{eq:appC_dbprl_passive}, with $\d V_{\|}=\dbprl/\sqrt{4\pi\rho}$, then implies $\rmd \d V_{\|}/\rmd t=0$ at this order, whereby subtracting $\va/2\times$\eqref{eq:appC_rho_raw} and rearranging yields
\begin{equation}
\frac{\rmd}{\rmd t}\frac{\d\rho}{\rho}
+\bhdg \duprl
=
\frac{1}{2}\,\zo\!\cdot\!\big(\gprp{B}+\bkap-\gprp{\rho}\big).
\label{eq:appC_rho}
\end{equation}
This is again exactly the low-$\beta$, low-$\ma$ limit of the standard density equation \eqref{eq:density-evol}, which remains nontrivial 
in this limit and will drive, for example, density transport. 

\paragraph{Interpretation and limitations}
The five fluctuation equations
\cref{eq:appC_dzp_perp,eq:appC_dzm_perp,eq:appC_dbprl_passive,eq:appC_upar,eq:appC_rho}, together with the $\zo$ equation \eqref{eq:appC_z0eq},
are precisely the relevant $\beta\sim \imord^{2}\ll1$, $\ma\sim \imord\ll1$, large-$z^{+}$ subsidiary limits of the standard fluctuation system derived in \cref{app: derivation}; they
 thus recover the dominant terms in our slaved phenomenology for imbalanced transonic turbulence.  In particular, the passive advection of $\dbprl/B$ is not a new assumption but simply the low-$\beta$ limit of the RMHD $\dbprl$ equation \eqref{eq:par-ind-comb}: in the present ordering, its source terms are suppressed by $\beta\sim \imord^{2}$ and so drop out at lowest order. By contrast, the $\duprl$ and $\d\rho/\rho$ equations remain nontrivial, retaining the couplings to the Alfv\'enic fluctuations that govern density transport, parallel flows, and low-$\beta$, low-$\ma$ compressive dynamics. Thus, although slow-wave-like propagation is demoted in this limit (they were likewise dropped in our slaved phenomenology), the rest of the compressive subsystem has survived unchanged.

The key formal result is therefore that the asymptotic procedures commute at the order considered here: starting from the full compressible MHD equations and expanding for a low-$\beta$, low-$\ma$, highly imbalanced, transonic outward wave yields the same fluctuation equations as taking the appropriate  limit of the RMHD system derived above. This is not obvious a priori. It could have failed, for example, if the transonic ordering had destroyed the perpendicular incompressibility of the Alfv\'enic corrections $\d\zpm_{\perp}$, or if additional feedback terms appeared in the compressive equations from the parallel (spherically polarized) part of $\zo$.

Let us also comment on the interpretation of the stronger pressure ordering \eqref{eq:appC_porder}, which  does not mean that magnetic-pressure fluctuations or pressure perturbations must in practice be as large as $\dbprl/B\sim \epsilon$ and $\d p/p\sim \epsilon/\imord^{2}$. Rather, it shows that \emph{if} fluctuations of this size are retained, then they satisfy the passive equation \eqref{eq:appC_dbprl_passive} in this subsidiary ordering. If instead $\d p/p$ is smaller, then $\d p$ and $\dbprl$ simply become the $\mathcal{O}(\imord\epsilon)$ incompressible pressure required to maintain $\nabla\!\cdot\!\duprp=0$, while the remaining fluctuation equations still follow so long as the weaker condition $\epsilon\ll\imord$ is satisfied so that $\d p/p\ll1$. 
Given $\duprl$ and $\d\rho$ are each driven to their assumed order by the background gradients, while $\dbprl$ is not, the most natural situation is indeed for $\dbprl$ to be $\mathcal{O}(\imord\epsilon)$ and therefore not passively advected.

Finally, the present subsidiary expansion is intended only as a simple proof of principle, not as the most general possible ordering. Other consistent choices can retain additional terms. For example, taking $\gprp{U}\sim \epsilon/\imord$ brings the velocity-shear couplings into \cref{eq:appC_upar,eq:appC_rho} while remaining fully self-consistent; equivalently, such terms also appear without assuming $\ma\sim \imord$. This argues that the velocity-shear-based ACR heating discussed in \cref{subsec: stream_dissipation} should be robust, at least for sub-Alfv\'enic fluctuations. By contrast, taking $\gprp{\rho}$ parametrically larger tends to mix perpendicular-to-parallel coupling terms, destroying the ordering of $\d\rho/\rho$   in the present scheme (this can be seen directly from \cref{eq:appC_rho}). Moreover, the commutation described above \emph{does not}  persist to  higher orders: at $\mathcal{O}(\imord^{2}\epsilon)$, for example, the parallel part of $\zo$ feeds curvature couplings back into the $\dzp$ equation, in a manner analogous to the $\bkap\,\dbprl$ term in \cref{eq: Alfvenic fluctuation equations main}, together with other corrections of the same order. This is entirely expected, but suggests that more extreme orderings --- for example those allowing even larger outward amplitudes or larger perpendicular gradients --- would be interesting to explore in future work to understand such effects.

\subsection{Equilibrium and transport in the transonic ordering}\label{subapp: equil and transport}

A complete transport-order derivation is much less clean in the transonic ordering than in the standard expansion. Because $U\sim \imord \va$ and $p\sim \imord^{2}B^{2}$, the separation between equilibrium and transport is no longer sharply defined: the mean thermal-pressure force, the wave-pressure force associated with $|\bm B+\bm B^{(0)}|^{2}$, and the quadratic fluctuation stresses all enter over a nearby range of orders, including in the perpendicular balance. Likewise, there is no unique slow ordering for all mean fields, because transport-time derivatives can naturally appear at different orders for different quantities. For this reason, along with the system's  corresponding lack of a clear
energy-conservation structure, we do not attempt here a full transport derivation analogous to App.~\ref{sub: transport equations}.

What does remain robust are the turbulent fluxes. As in the standard theory, a turbulence average over a region whose size is comparable to the background-gradient scale adds one power of $\epsilon$ to any averaged divergence, so the transport equations are naturally controlled by quadratic-fluctuation correlators at $\mathcal{O}(\epsilon^{2}\imord)$. For example, the density flux $\avgi{\rho \bm u}$ gives the lowest-order contribution $\rho\,\avgi{\zo\,\d\rho/\rho}/2$, giving  exactly the same slaved-closure structure used in the main text. The magnetic transport is likewise governed by the same physics, yielding the  result that there is no parallel EMF  at lowest relevant order, and thus only advective rather than diffusive magnetic transport:   $\bm u\times\bm B \propto \bm Z^{+}\times\bm Z^{-}$ so that the leading parallel piece is $\avgi{\zo\times\dzmprp}$; because $\dzmprp$ remains perpendicular and divergence free, the same argument made in App.~\ref{sub: magnetic transport} holds.  More generally, for a pure outward Alfv\'enic carrier, the leading $\imord^{2}$ pieces of $\rho \bm u\bm u-\bm B\bm B/4\pi$ cancel, so the surviving cross-field transport requires the $\mathcal{O}(\epsilon)$ corrections, just as in our 
slaved closures --- there are no new dominant transport contributions.

The practical outcome is that the same slaved phenomenological heating and transport closures used in the main text should remain appropriate in this regime, even though their asymptotic placement within a transport hierarchy is less clean. In particular, the turbulent Reynolds--Maxwell stresses, density fluxes, magnetic-flux transport, and heat fluxes retain the same local correlation structure, so the leading perpendicular transport and heating channels are unchanged from the standard ordering. What changes is the status of the mean evolution equations: because \(W_\perp^+\sim p\sim \rho U^2\) in a transonic, low-\(\beta\), strongly wave-driven wind, wave-pressure forces and wave-energy fluxes can enter at the same order as the equilibrium force balance, while the slow-time derivatives of some mean quantities are pushed to higher order. The leading mean equations are therefore more naturally interpreted as quasi-steady, wave-supported balance conditions, with time-dependent transport appearing only after an additional slow ordering is specified. This complicates the construction of a clean, energy-conserving transport theory analogous to App.~\ref{sub: transport equations}.

In this sense, the present subsidiary expansion is naturally connected to the resummed-equilibrium viewpoint discussed in App.~\ref{app:parker_like_wind}: the fluctuation equations and their closure objects remain robust, but equilibrium, wave-pressure, wave-action, and heating terms may move between different formal orders, or drop out, depending on the local asymptotic regime. The resummed approach is then to retain those terms that can be comparable somewhere in the domain, rather than assigning a single formal order globally. Where the low-\(\beta\) or subsonic limits are well satisfied, the formally smaller terms then become numerically negligible automatically; where the wind is instead close to a quasi-steady, wave-supported state, the slow-time derivatives may be the small quantities, leaving the retained wave-pressure and heating terms to enter the leading balance.

\section{Ideal MHD instabilities from the generalized RMHD system}\label{app: instabilities}

A useful feature of the multiscale RMHD system derived in App.~\ref{app: derivation} is that, in the same spirit  as local gyrokinetics, arbitrary background curvature and perpendicular gradients enter the fluctuation equations only as coefficients. One may therefore analyze a wide range of standard ideal-MHD instabilities with little additional effort by linearizing the reduced system about a given local equilibrium, then specializing to appropriate choices of $\bkap$, $\gprp{B}$, $\gprp{p}$, $\gprp{\rho}$, $\gprp{U}$, and $\bm g_{\rm eff}$. As well as providing an algebraically simple  method that can be
readily generalized to kinetic instabilities also, this could help for general understanding and analysis of instabilities, facilitating their categorization into classes driven by similar physics. In this appendix, we do not attempt such an exhaustive analysis, but simply  record the  local linear system and  use it to recover a number of familiar limits. 

We start from the RMHD equations in the potential form \eqref{eq: potential form}, \eqref{eq:par-ind-comb}--\eqref{eq:par-mom}, together with $\duprp=\bh\times\grad\Phi$ and ${\dbprp}/{\sqrt{4\pi\rho}}=\bh\times\grad\Psi$, as this results in the most natural and physically intuitive formulation of the linear system. We neglect dissipation, and, for simplicity,  set all background parallel gradients to zero,
\begin{equation}
\gprl{B}=\gprl{U}=\gprl{\rho}=\gprl{p}=\gprl{s}=g^{\|}_{\rm eff}=0,
\end{equation}
so instabilities can be easily  Fourier analyzed in the parallel direction. If these parallel gradients are retained, the coefficients vary along the field line and the local algebraic dispersion relation is replaced by a field-line-following eigenvalue problem, as in ideal-MHD ballooning theory \citep[e.g.,][]{Connor1978}. We will also assume, purely for the sake of simplicity, that all background perpendicular variation lies in the same plane. The setup captures most situations that are homogeneous along the local magnetic field, including  curvature- and stratification-driven couplings relevant to buoyancy, Parker, interchange,  kink, and magnetorotational-like modes. It excludes  reflection effects from $\gprl{\va}$ or wind expansion, as well as the parallel variation of $B$ needed for ballooning-like instabilities in closed-field-line configurations \cite{Freidberg2014}. The standard Kelvin-Helmholtz instability, with $k_{\|}\sim k_{\perp}$, is not captured by the system, due to the original assumption of anisotropy. With appropriate generalizations to include dissipation, the same local framework could also capture resistive tearing modes \cite{Furth1963,Schekochihin2020} and double-diffusive  instabilities \cite{Turner1974,Hughes1995}.

Mathematically, these assumptions naturally lead to a Cartesian frame with $\bh=\hat{\bm z}$ and all perpendicular equilibrium gradients, together with the curvature vector, in the $x$ direction:
\begin{equation}
\bkap=\kappa\,\hat{\bm x},\qquad
\gprp{G}=\gprps{G}\,\hat{\bm x},\qquad
\geff=g^{\perp}_{\rm eff}\,\hat{\bm x},
\end{equation}
with perturbations
\begin{equation}
\propto \exp\!\left(i k_x x+i k_y y+i k_\| z-i\omega t\right).
\end{equation}
Although the equilibrium varies only in $x$, the destabilizing couplings are proportional to $k_y$, so one must retain both components of $\kp$.

The linear system is then
\begin{align}
&-i\tilde{\omega} \Phi
= ik_\|\va \Psi
-2i k_y \frac{\kappa}{k_\perp^2}
\left(U\duprl-\va^2\frac{\dbprl}{B}\right)
\nn\\
&\qquad\qquad
+i k_y \frac{g_{\rm eff}^{\perp}}{k_\perp^2}\frac{\d\rho}{\rho},
\nn
\\
&-i\tilde{\omega} \Psi
= ik_\|\va \Phi,
\nn
\\
&-i\tilde{\omega} \left(1+\frac{\va^2}{\cs^2}\right)\frac{\dbprl}{B}
= ik_\|\duprl
-i k_y \frac{U}{\va}\mathcal{C}^{+}_{U}\Psi
\nn\\& \qquad\qquad +i k_y \left(\mathcal{C}^{+}_{B}-\frac{1}{\gamma}\gprps{p}\right)\Phi,
\nn
\\
&-i\tilde{\omega} \left(1+\frac{\cs^2}{\va^2}\right)\frac{\d\rho}{\rho}
= -ik_\|\duprl
+i k_y \frac{U}{\va}\mathcal{C}^{+}_{U}\Psi
\nn\\
&\qquad\qquad
-i k_y \left[\mathcal{C}^{+}_{B}
+\frac{1}{\gamma}\frac{\cs^2}{\va^2}\gprps{p}
-\left(1+\frac{\cs^2}{\va^2}\right)\gprps{\rho}\right]\Phi,
\nn
\\
&-i\tilde{\omega} \duprl
= i k_\|\va^{2}\frac{\dbprl}{B}
+i k_y \va\,\mathcal{C}^{-}_{B}\Psi
-i k_y U\,\mathcal{C}^{-}_{U}\Phi.
\label{eq:app-instab-upar-k}
\end{align}
where $\tilde{\omega} \equiv\omega - k_{\|}U$ and  we have defined for convenience $k_\perp^2=k_x^2+k_y^2$ and $\mathcal{C}^{\pm}_{U} \equiv \kappa\pm\gprps{U}$,
$\mathcal{C}^{\pm}_{B} \equiv \kappa\pm\gprps{B}$.
Familiar instability criteria then follow after imposing  equilibrium force balance \eqref{eq: perp equil},
\begin{equation}
\va^{2}\mathcal{C}^{-}_{B}- \frac{{c}_{s}^{2}}{\gamma}\gprps{p}  +g^{\perp}_{\rm eff}=0,\quad g^{\perp}_{\rm eff} = -U^{2}\kappa+g, \label{eq: perp equil linear}
\end{equation}
where $g= -\nabla_{\perp}\Phi_{\rm grav}$,
and solving for the eigenvalues $\omega$. There is one trivial solution, $\tilde{\omega} = 0$, corresponding to a perturbed force balance in the frame moving at speed $U$ along the mean magnetic field. The remaining modes are contained within the quartic
\begin{align}
&\left(\tilde{\omega}^2 - k_\|^2\vs^2\right)\left(\tilde{\omega}^2 - \frac{k_y^2}{k_\perp^2}N_{\rm eff}^2 - k_\|^2\va^2\right)\label{eq: full_disp_rel}\\
&\quad- 4\frac{k_y^2}{k_\perp^2}\tilde{\omega}k_\|\vs^2\kappa U\left(2\kappa - \frac{g^{\perp}_{\rm eff}}{\cs^2}\right)\nn\\
&\quad= \frac{k_y^2}{k_\perp^2}k_\|^2\vs^2\left[4\kappa^2\left(U^2 + \vs^2\right) - \frac{\vs^2}{\cs^2}g^{\perp}_{\rm eff}\left(4\kappa - \frac{g^{\perp}_{\rm eff}}{\cs^2}\right)\right],\nn
\end{align}
where
\begin{align}
    N_{\rm eff}^2 &= -2\kappa\left[\vs^2\left(\frac{1}{\gamma}\gprps{p} - \mathcal{C}^{+}_{B}\right) - U^2\mathcal{C}^{-}_{U}\right]\nn\\
    &- g^{\perp}_{\rm eff}\left[\frac{\vs^2}{\cs^2}\left(\mathcal{C}^{+}_{B} + \frac{1}{\gamma}\frac{\cs^2}{\va^2}\gprps{p}\right) - \gprps{\rho}\right] \label{eq: full_BV_freq}
\end{align}
is the effective Brunt--V\"ais\"al\"a frequency, with $N_{\rm eff}^2 < 0$ corresponding to growing interchange-like modes and $N_{\rm eff}^2 > 0$ corresponding to buoyant oscillations. 
In the limit
\begin{align*}
    &\tilde{\omega} \sim k_\|\vs \sim k_\|\va \gg N_{\rm eff},
    \qquad k_\| \gg \kappa,\\
    &N_{\rm eff}^{2}\sim \kappa g^{\perp}_{\rm eff},\qquad
    \va \sim \vs \sim \cs \gg U \sim \sqrt{\frac{g^{\perp}_{\rm eff}}{\kappa}},
\end{align*}
we recover
\begin{align*}
    \left(\tilde{\omega}^2 - k_\|^2\vs^2\right)\left(\tilde{\omega}^2 - k_\|^{2}\va^2\right) = 0,
\end{align*}
corresponding to slow and Alfv\'en waves in the frame moving with the mean flow.

In what follows, we present some common physical limits of \eqref{eq: full_disp_rel}, all of which drop the term linear in $\tilde{\omega}$, simplifying the analysis. In the first two limits this term vanishes identically, while in the third one (the magnetorotational instability) it is subdominant in the assumed \(U\gg v_A\) ordering. 

\subsection{Plane-parallel stratification: buoyancy and Parker limits.}\label{appsec: parker instability}
A useful first specialization is a plane-parallel equilibrium, for which
\begin{equation}
U=0,\qquad \kappa=0,\qquad \gprps{U}=0.
\end{equation}
The perpendicular force balance \eqref{eq: perp equil linear} then reduces to the local form of magnetohydrostatic balance in a stratified atmosphere,
$g^{\perp}_{\rm eff}=g
= \va^{2}\gprps{B} + {\cs^{2}}\gprps{p}/{\gamma},$
which  is the starting point for both Brunt--V\"ais\"al\"a and Parker-type buoyancy modes \cite{Parker1966,Goedbloed2004}. In this limit, the effective Brunt--V\"ais\"al\"a frequency \eqref{eq: full_BV_freq} reduces to the usual expression for a magnetized, stratified atmosphere in \citet[equation 7.206]{Goedbloed2004}
\begin{align*}
    -N_{\rm eff}^2 = \frac{g}{\va^2+\cs^2}\left[g - \left(\va^2+\cs^2\right)\gprps\rho\right].
\end{align*}

\paragraph{Brunt--V\"ais\"al\"a with magnetic compression}
The simplest buoyancy limit is obtained by taking a uniform magnetic field, $\gprps{B}=0$,
and setting $k_\|=0$, so that field-line bending and magnetic tension do not enter. Then $\Psi$, $\duprl$, and $\dbprl$ decouple as zero-frequency branches, leaving a $2\times2$ buoyancy system for $\Phi$ and $\d\rho/\rho$. Its dispersion relation is
\begin{align}
\omega^{2}
&= \frac{k_y^2}{k_\perp^2}N_{\rm eff}^2\nn\\
&=
-\frac{k_y^2}{k_\perp^2}\,
\frac{g}{\va^2+\cs^2}
\left[
g
-(\va^2+\cs^2)\gprps{\rho}
\right].
\label{eq: magnetized-BV-2}
\end{align}
giving instability for 
$g
\left[
g
-(\va^2+\cs^2)\gprps{\rho}
\right] >0.$
In the hydrodynamic limit $\beta\propto\cs^2/\va^2\to\infty$, \eqref{eq: magnetized-BV-2} reduces to the usual Brunt--V\"ais\"al\"a form,
\begin{equation}
\omega^2 = -
\frac{k_y^2}{k_\perp^2}\,
{g}\nabla\ln \frac{p^{1/\gamma}}{\rho}
\label{eq: hydro-BV}
\end{equation}
(note that $g$ is defined in the positive direction, leading to 
the sign difference compared to standard conventions).
At finite $\beta$ there is an additional stabilizing contribution because  even a uniform magnetic field is compressed by the fluid's displacement, increasing the restoring force relative to the pure hydrodynamic case \cite{Newcomb1961}.

\paragraph{General Parker system.}
Retaining $\gprps{B}\neq0$, one obtains the local magnetic-buoyancy problem in its general plane-parallel form. Simplifying \eqref{eq: full_disp_rel} yields
\begin{align}
&\left(\omega^2 - k_\|^2\vs^2\right)\!\!\left(\omega^2 - \frac{k_y^2}{k_\perp^2}N_{\rm eff}^2 - k_\|^2\va^2\right) = \frac{k_y^2}{k_\perp^2}k_\|^2\vs^4\frac{g^2}{\cs^4}.
\label{eq: parker-quartic-dim}
\end{align}
This matches with the standard local magnetic-buoyancy dispersion relation given in \citet[equation 7.211]{Goedbloed2004} after expanding and translating notation. 

Equation \eqref{eq: parker-quartic-dim} contains both the tension-free $k_\|=0$ buoyancy branch and its finite-$k_\|$ generalization. The former gives the most direct Parker/interchange-type destabilization, whereas finite $k_\|$ introduces magnetic tension and leads to quasi-interchange/undular behavior \citep{Newcomb1961,Acheson1979,Goedbloed2004}. Which branch dominates depends on the relative sizes of the stratification terms and $k_\|\va$ \citep[see][]{Goedbloed2004}; for example, under the ordering
\begin{align*}
    k_\|\vs \sim k_\|\va \sim N_{\rm eff} \ll \omega \ll \vs\frac{g}{\cs^2}
\end{align*}
\eqref{eq: parker-quartic-dim}  gives
\begin{align}
    \omega^2 = \pm \frac{k_y}{k_\perp}k_\|\vs^2\frac{g}{\cs^2},
\end{align}
leading to  finite-$k_{\|}$ quasi-interchange instability on the negative branch.

\subsection{Curved, shearless field: interchange and kink-like modes.}\label{appsec: interchange instability}
A second useful specialization is a static, curved field with no gravity,
\begin{equation}
U=0,\qquad g^{\perp}_{\rm eff}=0,
\end{equation}
intended to describe a local patch of a confined $z$-pinch plasma at radius $R$. We take the field-line curvature to be constant and inward,
\begin{equation}
\bkap=-R^{-1}\hat{\bm x},
\end{equation}
and identify the parallel wavenumber with the usual cylindrical mode number,
$k_\|={m}/{R}=-m\kappa$ .
Perpendicular equilibrium then gives
\begin{equation}
\gprps{B}=-\frac{1}{R}-\frac{\cs^2}{\gamma\va^2}\gprps{p},
\label{eq: zpinch-equil}
\end{equation}
which is simply the local form of radial force balance in a shearless pinch. The effective Brunt--V\"ais\"al\"a frequency reduces to
\begin{align}
    N_{\rm eff}^2 = \frac{2\cs^2}{\gamma R^2}\left[R\gprps{p} + \frac{2\gamma\vs^2}{\cs^2}\right].
\end{align}

The $m=0$ and $m\neq0$ limits correspond to two familiar classes of instability. The first is the pure interchange, which involves no variation along the field. The second contains the helical $m=1$ internal-kink branch, together with higher-$m$ generalizations. This  is the cylindrical counterpart of the slab interchange/quasi-interchange structure discussed above, better studied due to their relevance to fusion confinement \cite{Freidberg2014}. 

\paragraph{Pure interchange.}
Setting $m=0$ in the local system leaves a particularly simple dispersion relation (discarding zero-frequency branches)
\begin{equation}
\omega^2=
\frac{2k_y^2}{k_\perp^2}\frac{\cs^2}{\gamma R^2}
\left(
R\gprps{p}
+\frac{2\gamma\vs^2}{\cs^2}
\right).
\end{equation}
Instability requires
\begin{equation}
R\gprps{p}
+\frac{2\gamma\vs^2}{\cs^2} = R\gprps{p}
+\frac{4\gamma}{2+\gamma\beta}<0,
\label{eq: zpinch-interchange-criterion}
\end{equation}
where we have substituted
$\gamma\beta/2 = {\cs^2}/{\va^2}$ into the definition $\vs^2 = \cs^2/(1+\cs^2/\va^2)$
to match a common  form of the standard Kadomtsev interchange criterion  for a shearless $z$-pinch \cite{Kadomtsev1966}. 

\paragraph{Finite $m$: quasi-interchanges and the internal kink.}
For $m\neq0$, the same local reduction gives the quartic
\begin{align}
&\left(\omega^2 - \frac{m^2\vs^2}{R^2}\right)\!\!\left(\omega^2 - \frac{k_y^2}{k_\perp^2}N_{\rm eff}^2 - \frac{m^2\va^2}{R^2}\right)= \frac{4k_y^2}{k_\perp^2}\frac{m^2\vs^4}{R^4}.
\label{eq: zpinch-kink-quartic}
\end{align}
This expression is the same as the localized-mode limit of the constant-pitch cylindrical dispersion relation given by Eq.~(9.137) of \citet{Goedbloed2004} (specifically, associating their $j_{mn}^2/a^2\to k_\perp^2$ for the short-radial-wavelength limit). There are obvious similarities to the  Parker quartic \eqref{eq: parker-quartic-dim}, using the replacements $k_{\|}\rightarrow m/R$ and $g\rightarrow c_{s}^{2}/R$.  
For \(m\neq0\), stability of the local quartic requires the constant term of the quadratic in \(\omega^2\) to be positive. Setting \(k_y^2/k_\perp^2=1\) for the fastest growing mode, this condition becomes
\(
R\gprps{p}>-{m^2}/{\beta},
\)
or, using the  equilibrium,
\begin{equation}
B^{-2}\frac{d(RB^2)}{dR}<m^2-1
\end{equation}
for stability \cite{Freidberg2014}. Like for the plane atmosphere, the ($m=1$) quasi-interchange can dominate the $m=0$ interchange at higher $\beta$.

The shearless assumption is important here. Once magnetic shear is restored, as in screw pinches or tokamaks, the local field-following equations still fit naturally within the present framework, but one no longer has a simple algebraic dispersion relation: instead, the problem becomes an ODE/eigenvalue problem along the field, with singular structure near resonant surfaces. Those more general cases are treated extensively in \citet{Goedbloed2004} and \citet{Freidberg2014}, and are not pursued here.

Tayler-like instabilities for stellar-interior applications provide another natural extension of the present analysis. In stellar interiors, predominantly toroidal fields are generically unstable to non-axisymmetric $m=1$ modes but modified by rotation, buoyancy, shear, and other effects. The same local operator used above should provide a convenient starting point for studying such hybrid instabilities within an algebraically simple  framework \cite{Pitts1985,Spruit1999,Kirillov2014}.

\subsection{Magnetorotational instability}\label{sec: magnetorotational}
A third useful specialization, yielding an instability of a different character,  is a rotating, curved-field equilibrium appropriate to a local patch of an accretion disc. We take the field to lie in the azimuthal-vertical plane, with dominant gradients and curvature in the radial direction $R$. The dominant force balance is between external gravity and centrifugal acceleration, so that
\begin{equation}
{g}^{\perp}_{\rm eff}= -U^{2}\kappa - \frac{\partial}{\partial R} \Phi_{\rm grav} =0,
\end{equation}
while the geometric coefficients, using $U\propto R^{-1/2}$ and $\gprps{p}=\gprps{\rho}=0$ for  the simplest  limit, are
\begin{equation}
\kappa=-\frac{1}{R},\qquad
\gprps{B}=-\frac{1}{R},\qquad
\gprps{U}=-\frac{1}{2R}.
\end{equation}
 This describes a helical (vertical--azimuthal) curved  field with Keplerian flow shear, as in standard local MRI analyses  \cite{Balbus1991}.  

To recover the standard MRI, we work in the flow-following frame and take the limit $U\gg \va$, with
$\omega\sim k_\| \va\sim \Omega\equiv {U}/{R}$,
so that the reduced equations retain the Coriolis/epicyclic terms while discarding smaller corrections of order $\va/U$. In this limit, and then additionally taking $\cs^2/\va^2\to\infty$ (high $\beta$) as assumed for standard treatments of the MRI \cite{Balbus1998}, the density fluctuation decouples from the system \eqref{eq:app-instab-upar-k}, and the remaining four fields satisfy
\begin{align}
-i\omega \Phi &= i k_\|\va \Psi + \frac{2i}{k_{y}}\Omega\,\duprl,
\nn\\
-i\omega \Psi &= i k_\|\va \Phi,
\nn\\
-i\omega \duprl &= i k_\|\va^2\frac{\dbprl}{B} + \frac{i k_{y}\Omega}{2}\,\Phi,
\nn\\
-i\omega \frac{\dbprl}{B} &= i k_\|\duprl + \frac{3ik_{y}\Omega}{2\va}\,\Psi,
\label{eq: MRI-4x4}
\end{align}
while the  effective Brunt--V\"ais\"al\"a frequency is simply
    $N_{\rm eff}^2 = \Omega^2.$
Here, we have also moved into the fluid frame by replacing $\tilde{\omega}$ with $\omega$ and set $k_{x}=0$ for simplicity, in line with standard treatments. \Cref{eq: MRI-4x4} is the standard local linear MRI system, equivalent to the usual form computed from the local shearing-sheet \cite{Balbus1998}. Note that  the same procedure, but without assuming $\beta\ll1$ and keeping nonlinear terms, yields the rotating RMHD (RRMHD) system of \cite{Kawazura2022}; a comparison of RRMHD to the full MHD system, demonstrating its accurate results, is provided therein and in \citet{Kawazura2024}.

Solving for $\omega$ in \eqref{eq: MRI-4x4}, or equivalently simplifying \eqref{eq: full_disp_rel}, gives
\begin{equation}
\left(\omega^2 - k_\|^2\va^2\right)\left(\omega^2 - \Omega^2 - k_\|^2\va^2\right) = 4\Omega^2k_\|^2\va^2,\label{eq: mri dr}
\end{equation}
with unstable solutions
\begin{equation}
\omega^2=
\frac{1}{2}\left[
\Omega^2+2k_\|^2\va^2
-\Omega\sqrt{\Omega^2+16k_\|^2\va^2}
\right].
\end{equation}
This is the usual MRI dispersion relation for a Keplerian flow, with maximum growth rate $\omega=3i\Omega/4$ at $k_{\|}\va/\Omega=\sqrt{15}/4$ \citep{Balbus1991}.  Note that this relation is often listed in terms of $k_{Z}v_{\rm AZ}$, where $Z$ denotes the vertical direction in the disk, which is simply $k_{\|}\va$ under the common assumption of  axisymmetric linear perturbations. 

The main point for the present paper is not this standard limit itself, but that it emerges straightforwardly from the same reduced operator used above for other modes that are generally considered in a very different context, namely Parker and interchange. The same framework therefore also provides an immediate starting point for more general disk  instabilities, including disks with strong pressure support and non-Keplerian rotation, strong-field modifications of the MRI  \cite{Pessah2006,Das2018,Begelman2023}, and mixed Parker--MRI behaviour in vertically stratified disk atmospheres/coronae \citep[e.g.,][]{Johansen2008,Squire2025}.

\begin{section}{Comparison to multiscale gyrokinetics for magnetic-confinement fusion}\label{app: gyrokinetics}
    In this appendix, we demonstrate that the governing equations derived in App.~\ref{app: derivation} can be recovered, in certain limits, via a subsidiary expansion of the multiscale gyrokinetic system,  originally developed for studying turbulence and transport in the context of magnetic-confinement fusion \cite{Frieman1982,Sugama1998,Abel2013}. This not only serves as a consistency check of the expressions derived here, but also allows us to clarify how features and effects that arise 
    in the closed-flux-surface geometry assumed for fusion applications manifests in the coordinate-free setting used here. It is likewise our hope that the vast conceptual simplification provided by a fluid over a kinetic model could 
    prove useful for qualitatively understanding  aspects of fusion turbulent transport. 
    
    The remainder of this appendix is organised as follows. After briefly introducing the gyrokinetic system of equations and notation (\cref{sec:df_gyrokinetics}), we outline the subsidiary ordering that will form the basis for our asymptotic expansion thereof (\cref{sec:large_scale_ordering}). We then show that both the generalized RMHD equations (\cref{sec:rmhd_equations}) and the associated set of transport equations (\cref{sec:transport_equations}) can be recovered directly from within this subsidiary expansion. Readers already familiar with gyrokinetic theory may wish to skip ahead to \cref{sec:large_scale_ordering}, working backwards where further clarification is required. 

    \subsection{Local $\df$ gyrokinetics}
    \label{sec:df_gyrokinetics}
    Many studies of plasma turbulence in magnetic-confinement fusion are conducted within the framework of local $\df$ gyrokinetics, which describes the evolution of highly-anisotropic, small amplitude fluctuations in the presence of an equilibrium magnetic field. Here, the word `local' refers to the fact that there is an assumed scale separation between the characteristic scales of the turbulence and those of the equilibrium, an entirely analogous setup to that considered in this present work. The fluctuations are assumed to obey the standard gyrokinetic ordering \cite{Abel2013}:
    \begin{align}
        \frac{\omega}{\Omegas} \sim \frac{\nussp}{\Omegas} \sim \frac{\kpar}{\kperp} \sim \frac{\qs\dphipot}{\Ts} \sim \frac{|\d\bm{B}|}{B} \sim  \frac{\dfs}{\fs} \sim \frac{\rhos}{L} \ll 1,
	\label{eq:gyrokinetic_ordering}
    \end{align}
    where $\omega$ is a characteristic rate of change of the fluctuations, $L\sim (\gprps{\cdot})^{-1}$ is the typical length scale over which the plasma equilibrium varies and, for each species $s$, we define its charge $\qs$, mass $\ms$, thermal speed $\vths$, equilibrium temperature \(\Ts = \ms \vths^2/2\), gyrofrequency \(\Omegas = \qs e  B / \ms\), and gyroradius \(\rhos=\vths / |\Omegas|\). Additionally, $\nu_{\s \s'}$ is the collision frequency between species $\s$ and $\s'$, and $\dphipot$ is the fluctuating electrostatic potential. The distribution function $\fs$ can then be expressed as the sum of some equilibrium $\Fs$ and fluctuations $\dfs$, viz., 
    \begin{align}
        \fs = \Fs + \dfs,
        \label{eq:fs_split}
    \end{align}
    where the latter is further split into its gyroangle dependent and independent parts as
    \begin{align}
            \dfs = -\frac{\qs \dphipot(\vr, t)}{\Ts} \Fs(\vRs, \energy) + \hs(\vRs, \energy, \mus, t).
            \label{eq:dfs_hs_decomposition}
    \end{align}
    Here, $\vRs = \vr - \ub \times \vv / \Omegas$ is the guiding center position, $\energy$ is the particle energy, $\mus = \ms \vperp^2/2B$ is the particle magnetic moment, and $\vv$ is the particle velocity. The equilibrium distribution function
    \begin{equation}\begin{aligned}
        \Fs = \frac{\dens}{\left(\sqrt{\pi} \vths \right)^{3/2}} e^{-\energy / \Ts},
    \end{aligned}\end{equation}
    is a Maxwellian of density $\dens$ and temperature $\Ts$. The gyroangle-independent piece of the fluctuating distribution function $\hs$ evolves according to the gyrokinetic equation:
    \begin{align}
        &\frac{\partial}{\partial t}\left( \hs - \frac{\qs \avgRs{\chipot}}{\Ts} \Fs\right) + ( \vpar \ub + \vds + \avgRs{\vchi}) \cdot \grad \hs \nonumber \\
        & \quad \quad \quad \quad  + \avgRs{\vchi} \cdot \grad \Fs = \sum_{\s'} \avgRs{C_{\s\s'}},
        \label{eq:gyrokinetic_equation}
    \end{align} 
    in which $\vchi = (c/B) \ub \times \grad \chipot$ is the drift due to the gyrokinetic potential $\chipot = \phipot - \vv\cdot\vdA$, with $\vdA$ the fluctuating magnetic vector potential, \mbox{$\vds = (\ub/2\Omegas) \times (2 \vpar^2 \curvature + \vperp^2 \grad \ln B)$} are the magnetic drifts, and $\avgRsinline{\dots}$ is the gyroaverage at constant gyrocentre position. The $\avgRs{C_{\s\s'}}$ on the right-hand side is the operator encoding the effect of collisions between species $\s$ and $\s'$ on $\hs$. The electromagnetic fields appearing in the gyrokinetic equation \eqref{eq:gyrokinetic_equation} are determined by Amp\`ere's law split into its parallel and perpendicular parts,
    \begin{align}
    	\grad_\perp^2 \dApar &= - \frac{4\pi}{c} \sum_\s \qs \intv \vpar \avgr{\hs},
    	\label{eq:parallel_amperes_law} \\
    	\grad_\perp^2 \dBpar& = - \frac{4\pi}{B} \grad_\perp \grad_\perp : \sum_\s \ms \intv \avgr{\vvperp \vvperp \hs},
    	\label{eq:perpendicular_amperes_law}
    \end{align}
    and the quasineutrality condition
    \begin{equation}
    	0 = \sum_{\s} \qs \dns = \sum_\s \qs \left[ -\frac{\qs \phipot}{\Ts} \dens + \intv\!\avgr{\hs}\right].
    	\label{eq:quasineutrality}
    \end{equation}
    \Cref{eq:perpendicular_amperes_law} expresses perpendicular pressure balance. Here, and throughout, $\avgr{\dots}$ denotes the standard gyroaverage at constant particle position. 

    It is important to note that unlike in the remainder of this work, equilibrium plasma quantities such as the density $\dens$ and temperature $\Ts$ here only have gradients in the direction perpendicular to magnetic flux surfaces, i.e., they are functions of only the magnetic flux $\psi$. This means that we will only be able to recover a subset of the equations derived in \cref{app: derivation}, {because gradients of plasma quantities along the field ($\gprl{\rho}$, $\gprl{p}$ etc.) are formally small as a consequence of the gyrokinetic ordering \eqref{eq:gyrokinetic_ordering} combined with the absence of (leading-order) sources and non-sourced boundaries in standard fusion applications}. We  also neglected all equilibrium flows because the toroidal flow often included {in this context} \cite{Abel2013} has projections both parallel and perpendicular to the equilibrium magnetic field, unlike the purely field-aligned flow assumed in the rest of this paper. We therefore neglect these terms, with which we cannot eventually compare. As such, we expect to obtain agreement with the expressions from \cref{app: derivation} only in the limit where the field-aligned flow $U$, gravitational potential $\Phi_{\rm grav}$, and parallel gradients of the plasma equilibrium $\gprl{\rho}$, $\gprl{p}$ (and their derived quantities) are all assumed to vanish, viz.,
    \begin{align}
        U = 0, \quad \Phi_{\rm grav} = 0, \quad \gprl{\rho} = \gprl{p} = 0,
        \label{eq:rmhd_to_gk_mapping}
    \end{align}
    though we do retain parallel gradients of the equilibrium magnetic field $B$.      

    \subsection{Large-scale ordering}
    \label{sec:large_scale_ordering}
    While there are obvious similarities between the gyrokinetic and multiscale RMHD orderings  (\eqref{eq:gyrokinetic_ordering} and \eqref{eq:ordering}), additional assumptions are required in order to simplify the gyrokinetic system of equations. We consider fluctuations on scales much larger than the Larmor radii of all species, viz., 
    \begin{align}
        \kperp \rhos \ll 1,
        \label{eq:subsidiary_ordering}
    \end{align}
    with frequencies comparable to both the Alfv\'en frequency and the parallel-streaming rate of species $\s$
    \begin{align}
        \omega \sim  \kpar \va \sim \kpar \vths,
        \label{eq:subsidiary_ordering_freqs}
    \end{align}
    while the density $\dns$, velocity $\dupars$, and temperature $\dTs$ fluctuations of each species $\s$ are ordered analogously to \cref{eq:ordering}:
    \begin{align}
        \frac{\dns}{\dens} \sim\frac{\dTs}{\Ts} \sim \frac{\dupars}{v_A} \sim \frac{|\vduperp|}{v_A} \sim \frac{|\vdBperp|}{B} \sim \frac{\dBpar}{B} \sim  \frac{\rhos}{L}.
        \label{eq:subsidiary_ordering_fields}
    \end{align}
    The resultant subsidiary expansion in $\kperp \rhos$ is performed with all other dimensionless parameters held fixed, including the electron-ion mass ratio $\me/\mi$. Note that all quantities will only need to be determined to leading order in $\kperp \rhos$, and so we will not use any specific notation to distinguish orders in $\kperp \rhos$. 
    
    An immediate consequence of combining \eqref{eq:subsidiary_ordering} and \eqref{eq:subsidiary_ordering_fields} is that the electrostatic and parallel-magnetic-vector potential fluctuations are an order larger than the thermal fluctuations, viz.,
    \begin{align}
        \frac{\qs \dphipot}{\Ts} \sim \frac{\dApar}{\rhos B} \sim \frac{1}{\kperp \rhos} \frac{\dBpar}{B} \sim \frac{1}{\kperp \rhos} \frac{\dns}{\dens}.
        \label{eq:electromagnetic_field_size}
    \end{align}
    We will make extensive use of \eqref{eq:electromagnetic_field_size} in what follows. Furthermore, given that the MHD equations are derived in the limit of large collisionality, we also assume from the outset that the non-adiabatic distribution function of all species is a perturbed Maxwellian 
    \begin{align}
        \hs = \left[\frac{\dns}{\dens} + \frac{\qs \dphipot}{\Ts}  + \frac{\ms \dupars}{\Ts} + \frac{\dTs}{\Ts} \left( \frac{\energy}{\Ts} - \frac{3}{2} \right)\right] \Fs,
        \label{eq:perturbed_maxwellian}
    \end{align}
    with all species having the same perturbed parallel velocity $\dupars = \dupar$, perturbed temperature $\dTs = \dT$, and equilibrium temperature $\Ts = T$. We will retain species indices on $\dupars$, $\dTs$, and $\Ts$ for clarity, but will make use of the fact that these quantities are the same for all species where appropriate.

    While the choice of \eqref{eq:perturbed_maxwellian} can simply be regarded as an ansatz in order to simplify the derivation presented here, it can be made exact in the following way. Expanding all quantities to leading-order in \eqref{eq:subsidiary_ordering}, one obtains kinetic dynamical equations for  large-scale dynamics --- so-called ``kinetic reduced MHD'' (KRMHD), but including variation of the equilibrium quantities unlike standard derivations \cite{Schekochihin2009}. These equations, however, depend both on higher-order moments of the distribution function (through, e.g., the perturbed perpendicular and parallel heat fluxes) as well as its anisotropy (e.g., through the difference between the parallel and perpendicular pressures). To allow for a direct comparison with the equations derived in App.~\ref{app: derivation}, one must eliminate these terms by taking the collisional limit $\omega/\nussp \rightarrow 0$, enforcing the perturbed Maxwellian solution \eqref{eq:perturbed_maxwellian} (note that this subsidiary ordering in collisionality cannot be done simultaneously with the large-scale one \eqref{eq:subsidiary_ordering} as this leads to different equations that are not relevant to the present study). Given that the purpose of this appendix is to illustrate the connection to the gyrokinetic limit, we choose to simplify the  derivation by adopting the ansatz \eqref{eq:perturbed_maxwellian} from the outset, acknowledging these subtleties in taking the collisional limit.

    \subsection{Recovering the generalized RMHD equations}
    \label{sec:rmhd_equations}
    We now derive evolution equations for the perturbed quantities appearing in \eqref{eq:perturbed_maxwellian}, from which we can then recover the equations of generalized RMHD derived in \cref{subapp: second order}. 

    \subsubsection{Alfv\'enic fluctuations}
    \label{sec:alfvenic_fluctuations_gk}
    To obtain the perpendicular momentum equation \eqref{eq: potential form phi}, it will be useful to construct the so-called `gyrokinetic vorticity equation': multiplying \eqref{eq:gyrokinetic_equation} by the species charge $\qs$, integrating over all velocities, and summing over all species, one can use quasineutrality \eqref{eq:quasineutrality} and properties of the gyroaverage \cite{Howes2008} to find 
    \begin{align}
        & \frac{\partial}{\partial t} \sum_{\vkperp} e^{i \vkperp \cdot \vr} \sum_\s \left[ \frac{\qs^2 \dens }{\Ts} \left(1- \rmGamma_{0\s} \right) \dphipotkperp -  \qs \dens \rmGamma_{1\s} \frac{\dBparkperp}{B}\right] \nonumber \\
        & + \sum_\s \qs \int \rmd^3 \vv \avgr{(\vpar \ub + \vds + \avgRs{\vchi}) \cdot \grad \hs } \nonumber \\
        & + \sum_\s \qs \int \rmd^3 \vv \avgr{\avgRs{\vchi} \cdot \grad \Fs} \nonumber \\
        & = \sum_{\s ,\s'} \int \rmd^3 \vv \avgr{\avgRs{C_{\s\s'}[\hs]}}.
        \label{eq:gyrokinetic_vorticity}
    \end{align}
    In \eqref{eq:gyrokinetic_vorticity}, $\dphipotkperp$ and $\dBparkperp$ are the relevant Fourier transforms in the plane perpendicular to the equilibrium magnetic field, while
    \begin{align}
    	\rmGamma_{0\s}(\besselargint) &= \rmI_0(\besselargint) e^{-\besselargint}, \quad \rmGamma_{1\s}(\besselargint) = \left[\rmI_0(\besselargint) - \rmI_1(\besselargint)\right] e^{-\besselargint}, 
        \label{eq:gammas}
    \end{align}
    with $\besselargint = (\kperp \rhos)^2/2$, are functions of the perpendicular wavenumber $\kperp = |\vkperp|$ that capture finite-Larmor-radius (FLR) effects; $\rmI_1$ and $\rmI_2$ are modified bessel functions of the first kind \cite{Abramowitz1970}. We note that a careful handling of these FLR effects is only required to recover \cref{eq: potential form phi}; for all other equations, the gyroaverages appearing in  \eqref{eq:gyrokinetic_equation} can be treated as unity operations, a fact that we will make extensive use of in what follows.  

  Expanding each of the terms in \eqref{eq:gyrokinetic_vorticity} to leading-order in \eqref{eq:subsidiary_ordering} with asymptotic expansions $\rmGamma_{0\s} = 1 - \besselargint + \dots$ and $\rmGamma_{1\s} = 1 - (3/2) \besselargint + \dots$, the time-derivative  becomes:
    \begin{align}
        & \frac{\partial}{\partial t} \sum_{\vkperp} e^{i \vkperp \cdot \vr} \sum_\s \left[ \frac{\qs^2 \dens }{\Ts} \left(1- \rmGamma_{0\s} \right) \dphipotkperp -  \qs \dens \rmGamma_{1\s} \frac{\dBparkperp}{B}\right] \nonumber \\
        & = - \frac{\partial}{\partial t} \left( \sum_\s \frac{\qs^2 \dens}{2 \Ts} \rhos^2 \gperp^2 \dphipot \right) + \dots.
        \label{eq:vorticity_time_derivative}
    \end{align}
    The leading-order contribution to the $\dBpar$ term has vanished via the equilibrium quasineutrality constraint
    \begin{align}
        \sum_\s \qs \dens = 0, \quad \sum_\s \qs \grad \dens = 0
        \label{eq:equilibrium_quasineutrality}
    \end{align}
    while the next-order contribution is also negligible in comparison to the remaining term in \eqref{eq:vorticity_time_derivative} by \eqref{eq:electromagnetic_field_size}. The first two terms on the second line of \eqref{eq:gyrokinetic_vorticity} follow straightforwardly from performing the velocity integration given \eqref{eq:perturbed_maxwellian}, using parallel Amp\`ere's law \eqref{eq:parallel_amperes_law}, and making use of the first expression in \eqref{eq:equilibrium_quasineutrality}, while the collisional term on the right-hand side vanishes at leading order (collisions conserve particle number). The term involving the derivative of the equilibrium distribution function $\Fs$ can be evaluated using
    \begin{align}
        \grad \Fs = \left[\frac{\grad \dens}{\dens} + \frac{\grad \Ts }{\Ts} \left(\frac{\energy}{\Ts} - \frac{3}{2} \right)\right]\Fs,
    \end{align}
    alongside the second expression in \eqref{eq:equilibrium_quasineutrality}. 
    
    The nonlinear term involving the advection of $\hs$ by $\vchi$ requires explicitly evaluating the gyroaverage acting on the gyrokinetic potential, the integrand of which can be written explicitly in terms of the Poisson bracket (see \cref{eq: poisson bracket magnetic}):
    \begin{flalign}
        &\avgr{\avgRs{\vchi} \cdot \grad \hs} \nonumber \\
        &= \frac{c}{B}\left< \left\{\sum_{\vkperp} e^{i \vkperp \cdot \vRs } \left[\rmJ_0(\besselarg) \left(\dphipotkperp - \frac{\vpar \dAparkperp}{c} \right) \right. \right. \right. \nonumber \\
        & \quad\quad\quad\quad + \left. \left. \left. \frac{2 \rmJ_1 (\besselarg)}{\besselarg} \frac{\Ts}{\qs} \frac{\vperp^2}{\vths^2} \frac{\dBparkperp}{B}\right] , \hs(\vRs)\right\} \right>_{\vr}
        \label{eq:expanded_vchi}
    \end{flalign}
    where $\rmJ_0$, $\rmJ_1$ are the zeroth- and first-order Bessel functions of the first kind \cite{Abramowitz1970} with $\besselarg = \kperp \vperp/\Omegas$. Writing $\vRs = \vr - \vrhos$, with $\vrhos = \ub \times \vv / \Omegas$, the first term expands as:
    \begin{align}
        &\avgr{\pbra{\sum_{\vkperp} e^{\rmi\vkperp \cdot \vRs} \rmJ_0(b_\s) \phipotkperp}{\hs (\vRs)}} \nonumber \\
        & = \avgr{\pbra{\left(1 + \frac{1}{4} \frac{\vperp^2}{\vths^2} \rho_\s^2 \gperp^2 + \dots \right) \phipot(\vRs) }{\hs (\vRs)}} \nonumber \\
        & = \bigg< \pbra{\phipot(\vr)}{\hs(\vRs)} - \vrhos \cdot \pbra{\grad \phipot(\vr)}{\hs (\vr)} \nonumber \\
        & \quad +  \vrhos \vrhos : \pbra{\grad \phipot(\vr)}{\grad \hs (\vr)} + \frac{1}{2} \vrhos \vrhos : \pbra{\grad \grad \phipot(\vr)}{\hs (\vr)} \nonumber \\
        &\quad + \left.\pbra{\frac{1}{4} \frac{\vperp^2}{\vths^2} \rho_\s^2 \gperp^2  \phipot(\vr) }{\hs (\vr)} \right>_{\vr} + \dots \nonumber \\
        & = \pbra{\phipot(\vr)}{\avgr{\hs (\vRs)}} \nonumber \\
        & \quad + \frac{1}{2} \frac{\vperp^2}{\vths^2} \rhos^2 \left(\boldsymbol{\rmI}- \ub\ub\right) : \pbra{\grad \phipot(\vr)}{\grad \hs (\vr)} \nonumber \\
        & \quad + \pbra{\frac{1}{2} \frac{\vperp^2}{\vths^2} \rho_\s^2 \gperp^2  \phipot(\vr) }{\hs (\vr)}.
        \label{eq:expanded_vchi_phi_term}
    \end{align}
    In going from the first line to the second, we have expanded the Bessel function $\rmJ_0$; from the second to the third, we have expanded both $\dphipot(\vRs) = \dphipot(\vr - \vrhos)$ and $\hs(\vRs) = \hs(\vr - \vrhos)$; and from the third to the fourth, we have used the identities
       $ \avgr{\vrhos} = 0$ and $\avgr{\vrhos \vrhos} = \tfrac{1}{2} ({\vperp^2}/{\vths^2})\rhos^2\left(\boldsymbol{\rmI} -\ub\ub\right)$
 to evaluate the remaining gyroaverages. 
 Multiplying \eqref{eq:expanded_vchi_phi_term} by $\qs c/B$, integrating over all velocities, and summing over all species, the first term vanishes identically by quasineutrality \eqref{eq:quasineutrality}. The remaining terms give
    \begin{align}
        &\sum_\s \qs \int \rmd^3 \vv  \avgr{\frac{c}{B} \left(\ub \times \grad \avgRs{\dphipot} \right) \cdot \grad \hs } \nonumber \\
        & = - \frac{c}{B} \left(\ub \times \grad \dphipot \right) \cdot \grad \left( \sum_\s \frac{\qs^2 \dens}{2 \Ts} \rhos^2 \gperp^2 \dphipot  \right),
        \label{eq:vorticity_phi_nonlinearity}
    \end{align}
    where we have neglected the all terms involving the density and temperature perturbations as these are small in $\kperp \rhos \ll 1$ (recall \eqref{eq:electromagnetic_field_size}). Following an entirely analogous procedure, one can show that the leading-order contribution to the second term in \eqref{eq:expanded_vchi} is
    \begin{align}
        &\sum_\s \qs \int \rmd^3 \vv  \avgr{\frac{c}{B} \left(\ub \times \grad \avgRs{- \frac{\vpar \dApar}{c}} \right) \cdot \grad \hs } \nonumber \\
        & = \frac{c}{B} \left(\ub \times \grad \dApar \right) \cdot \grad \left( \frac{c}{4\pi} \gperp^2 \dApar \right),
        \label{eq:vorticity_apar_nonlinearity}
    \end{align}
    where we have made use of parallel Amp\`ere's law in evaluating some of the velocity integrals. Finally, one can show that the leading-order contribution from the third term in \eqref{eq:expanded_vchi} is negligible  because the resulting terms are two orders higher in $\kperp \rhos$. 

    Assembling all of these contributions, the leading-order gyrokinetic vorticity equation becomes
    \begin{align}
        & \frac{\rmd}{\rmd t} \left( \sum_\s \frac{\qs^2 \dens}{2 \Ts} \rhos^2 \gperp^2 \dphipot \right) + \bhdg  \left(\frac{c}{4 \pi} \gperp^2 \dApar\right) \nonumber \\
        & \quad- \gprl{B} \frac{c}{4\pi } \gperp^2 \dApar - \frac{c}{B} \left(\ub \times \grad \frac{\dBpar}{B} \right)\cdot \grad p \nonumber\\
        & \quad- \frac{c}{B} \ub \times \left( \bkap + \gprp{B}\right) \cdot \grad \dps[] = 0 ,
        \label{eq:vorticity_equation_final}
    \end{align}
    where we have made use of \eqref{eq: total time deriv} with
    \begin{align}
        \vduperp = \frac{c}{B} \ub \times \grad \dphipot, \quad \vdBperp = - \ub \times \grad \dApar, 
        \label{eq:ub_perturbations}
    \end{align}
    and defined
    \begin{align}
        p = \sum_\s \dens \Ts, \quad \dps[] = \sum_\s \left( \dns \Ts + \dens \dTs \right),
        \label{eq:pressure_definitions}
    \end{align}
    The mass-density and its perturbation are defined analogously:
    \begin{align}
        \rho = \sum_\s \ms \dens, \quad \drho = \sum_\s \ms \dns.
        \label{eq:mass_density_definitions}
    \end{align}
    (and not to be confused with the gyroradius $\rhos$).
    Finally, noting that the perturbed-Maxwellian solution \eqref{eq:perturbed_maxwellian} combined with the parallel component of Amp\`ere's law \eqref{eq:parallel_amperes_law} implies that perturbations are in perpendicular pressure balance to leading order, viz., 
    \begin{align}
        \frac{\dBpar}{B} = - \frac{4\pi}{B} \sum_\s \dps = - \frac{4\pi}{B} \dps[],
        \label{eq:pressure_balance}
    \end{align}
    and introducing the Alfv\'enic potentials, equivalent to those defined in  \eqref{eq: phi psi defn},
    \begin{align}
        \Phi = \frac{c}{B} \dphipot, \quad \Psi = - \frac{v_A}{B} \dApar, 
        \label{eq:alfvenic_potentials}
    \end{align}
    \eqref{eq:vorticity_equation_final} becomes
    \begin{align}
        \frac{\rmd}{\rmd t}\nabla_{\perp}^{2}\Phi&  = \va \bh_{T}\!\cdot\!\nabla \nabla_{\perp}^{2}\Psi - \va \gprl{B}  \nabla_{\perp}^{2}\Psi,\nn\\
         & - 2\va^{2} \,\bh\times \bkap\cdot\nabla\frac{\dbprl}{B},
         \label{eq:potential_form_phi_gk}
    \end{align}
    where we have also made use of the perpendicular equilibrium constraint $B^2(\bkap-\gprp{B})/4\pi = p \gprp{p}$ \eqref{eqapp: perp equil} to rewrite the final two terms in \eqref{eq:vorticity_equation_final}.
   As promised, \cref{eq:potential_form_phi_gk} is precisely \cref{eq: potential form phi} for $U=\bm g_{\rm eff}^{\perp} = \mathcal{D}^{\perp}_{\Phi} = 0$ and   $\gprl{\rho}=0$ (such that $\gprl{B}=\gprl{\va}$). {Note that there are no collisional/viscous terms in this equation as the assumption of large collisions has rendered them vanishingly small.}
    
    To recover the induction equation \eqref{eq: potential form psi}, multiply the gyrokinetic equation \eqref{eq:gyrokinetic_equation} by the species mass $\ms$, take parallel-velocity moment, and use the ansatz \eqref{eq:perturbed_maxwellian}, after which one obtains:
    \begin{align}
        \frac{\rmd }{\rmd t}(\ms \dens \dupars) + \bhdg \dps + \frac{\vdBperp}{B} \cdot \grad \ps \nonumber \\
        = - \qs \dens \left(\frac{1}{c} \frac{\rmd \dApar}{\rmd t} +  \ub \cdot \grad \dphipot \right).
        \label{eq:parallel_velocity_with_epar}
    \end{align}
    The right-hand side of \eqref{eq:parallel_velocity_with_epar} is an order larger than the left-hand side (see \eqref{eq:electromagnetic_field_size}) and so must vanish identically, viz., 
    \begin{align}
        \frac{1}{c} \frac{\rmd \dApar}{\rmd t} +  \ub \cdot \grad \dphipot = 0.
        \label{eq:epar_constraint}
    \end{align}
    In terms of the Alfv\`enic potentials \eqref{eq:alfvenic_potentials}, \eqref{eq:epar_constraint} straightforwardly becomes
    \begin{align}
          \frac{\rmd}{\rmd t}\Psi =& \va \bh\cdot\nabla\Phi  + \va\gprl{B} \Phi.
          \label{eq:potential_form_psi_gk}
    \end{align}
    This is \eqref{eq: potential form psi} for $U = \mathcal{D}^{\perp}_{\Psi} = 0$. We have thus been able to recover both of the equations from App.~\ref{sec:alfvenic_fluctuations} describing the evolution of the Alfv\'enic perturbations in the appropriate limit \eqref{eq:rmhd_to_gk_mapping}.  

    \subsubsection{Compressive fluctuations}
    \label{sec:compressive_fluctuations}
    To recover the parallel momentum equation \eqref{eq:par-mom}, {we sum \eqref{eq:parallel_velocity_with_epar} over species. The right-hand side vanishes via quasineutrality \eqref{eq:quasineutrality}, and we use the} definitions \eqref{eq:pressure_definitions} and \eqref{eq:mass_density_definitions} to obtain
    \begin{align}
        \frac{\rmd}{\rmd t} \left( \rho\, \dupar \right) + \bhdg \dps[] + \frac{\vdBperp}{B} \cdot \grad p = 0. 
    \end{align}
    Rewriting the second term using \eqref{eq:pressure_balance} as
    \begin{flalign}
        \bhdg \dps[] = - \frac{B^2}{4 \pi} \left[ \bhdg \frac{\dBpar}{B} + 2 \gprl{B} \frac{\dBpar}{B} \right],
    \end{flalign}
    and using \eqref{eqapp: perp equil} to rewrite the third in terms of magnetic-field gradients, we find:
    \begin{align}
       & \frac{\rmd \dupar}{\rmd t} - \va^2 \bhdg \frac{\dBpar}{B} \nonumber \\
        &\quad= \va^2 \frac{\vdBperp}{B}\cdot \left( \gprp{B} - \curvature \right) + 2\va^2 \gprl{B} \frac{\dBpar}{B}.
        \label{eq:parallel_momentum}
    \end{align}
    This is \eqref{eq:par-mom} with $U = g^\|_{\rm eff} = \mathcal{D}^{\|}_{u} = 0$. 

    Next, we multiply the gyrokinetic equation \eqref{eq:gyrokinetic_equation} by the mass $\ms$, integrate it over all velocities, sum the result over species, and use \eqref{eq:pressure_definitions}-\eqref{eq:pressure_balance} to obtain at leading order (cf.~\cref{eq:par-ind-raw}),
    \begin{align}
        &\frac{\rmd}{\rmd t} \left(\frac{\drho}{\rho} - \frac{\dBpar}{B} \right) +  \bhdg \dupar     \nn     \\
        &\quad = \gprl{B} \dupar + \vduperp \cdot \left( \curvature +\gprp{B} - \gprp{\rho} \right), \label{eq:density_moment_rearranged} 
    \end{align}
    where we have also made use of \eqref{eq:quasineutrality} and \eqref{eq:electromagnetic_field_size}, as well as  dropping  parallel gradients of the plasma equilibrium.
    To obtain an equation for only the mass-density perturbations $\drho$, the time derivative of  $\dBpar$ needs to be eliminated from \eqref{eq:density_moment_rearranged}: we multiply the gyrokinetic equation by $(2/3) \energy$, integrate over all velocities, and follow an analogous procedure to that which led to \eqref{eq:density_moment_rearranged} to give
    \begin{align}
        & \frac{\rmd}{\rmd t} \left( \frac{\dps[]}{p} - \gamma \frac{\dBpar}{B} \right) + \gamma \bhdg \dupar      \label{eq:pressure_moment_rearranged} \\
        & \quad= \gamma \gprl{B} \dupar - \vduperp \cdot \left( \gprp{p} - \gamma \bkap - \gamma \gprp{B} \right). \nonumber
    \end{align}
    We have written the result in terms of the adiabatic index $\gamma$, which here takes a value of $5/3$. Using  the equilibrium constraint \eqref{eqapp: perp equil} and  fluctuating perpendicular pressure balance \eqref{eq:pressure_balance}, \eqref{eq:pressure_moment_rearranged}  becomes
    \begin{align}
        &\left(1  + \frac{\va^2}{\cs^2}\right)  \frac{\rmd }{\rmd t} \frac{\dBpar}{B} - \bhdg \dupar \nonumber \\
        &\quad = - \gprl{B} \dupar - \vduperp \cdot \left(\curvature + \gprp{B} - \frac{1}{\gamma} \gprp{p} \right). \label{eq:dbpar_equation}
    \end{align}
    This is \eqref{eq:par-ind-comb} for $U = \gprl{p} = \tilde{\mathcal{D}}^{\|}_{B} = 0$. Finally, \eqref{eq:dbpar_equation} can be combined with \eqref{eq:density_moment_rearranged} to give 
    \begin{align}
        &\left(1 + \frac{\cs^2}{\va^2} \right) \frac{\rmd}{\rmd t} \frac{\drho}{\rho} + \bhdg \dupar = \gprl{B} \dupar\label{eq:drho_equation} \\
        & \quad + \vduperp \left[\curvature + \gprp{B} + \frac{1}{\gamma} \frac{\cs^2}{\va^2} \gprp{p} - \left( 1 + \frac{\cs^2}{\va^2} \right) \gprp{\rho} \right]. \nonumber
    \end{align}
    This is \eqref{eq:density-evol} with $U = \tilde{\mathcal{D}}^{\|}_{\rho} = 0$, as well as all parallel gradients of the plasma equilibrium set to zero. 

    We have thus shown that an appropriate subset of the generalized RMHD equations that were derived in App.~\ref{subapp: second order} can also be obtained via a direct expansion of the gyrokinetic system of equations under the assumption of a Maxwellian distribution function.

    \subsection{Recovering the transport equations}
    \label{sec:transport_equations}
    We now show that the form of the transport equations derived in \cref{sub: third order summary} can be recovered from the standard form of the transport equations adopted in transport theories of gyrokinetics \cite{Callen2010,Barnes2010,Abel2013}. We first briefly introduce the flux-surface average in \cref{sec:flux_coordinates} (in particular, its relation to our average and flux-following time derivative), before considering the density and thermal energy transport in sections \cref{sec:density_transport} and \cref{sec:thermal_energy_transport}, respectively. We will not consider momentum transport here since we are neglecting equilibrium flows for the purposes of this comparison.

    \subsubsection{The flux-surface average}
    \label{sec:flux_coordinates}
    As mentioned at the end of \cref{sec:df_gyrokinetics}, equilibrium quantities such as the density $\dens$ and temperature $\Ts$ are functions of only the flux-surface coordinate $\psi$, often taken to be either the poloidal or toroidal magnetic flux in the toroidal geometries relevant to magnetic-confinement fusion contexts. This means that the relevant transport of these quantities is flux-surface perpendicular and in the flux-surface-following frame, viz., the frame that moves at the velocity $\Vpsit$ (see \cref{eq: general flux transport} and preceding text). Fusion gyrokinetic theories deal with this by defining the flux-surface average \cite{Hinton1976,Dhaeseleer1991,Abel2013}, 
    \begin{align}
    	\fsa{\dots} & = \lim_{\Delta \psi \rightarrow 0} \left[\int_{\Delta V(\psi)} \rmd^3 \vr \left(\dots\right) \Bigg/ \int_{\Delta V(\psi)} \rmd^3 \vr\right], 
    	\label{eq:fsa_definition}
    \end{align}
    where $\Delta V(\psi) = V(\psi + \Delta \psi) - V(\psi)$ and $V(\psi)$ is the volume of the flux-surface labelled by $\psi$. The flux-surface average is thus a \textit{volume average} over an infinitesimal spatial region rather than a surface average, and so measures the transport of quantities across a given volume labeled by $\psi$. Evaluating the flux-surface average of the divergence of a perpendicular flux $\bm{\Gamma}_G$, the general transport divergences in  App.~\ref{app: derivation} become (see \citet{Abel2013})
    \begin{align}
    \fsa{\grad \cdot \bm{\Gamma}_G} = \frac{1}{V'} \frac{\partial}{\partial \psi} \left(V' \fsa{\bm{\Gamma}_G \cdot \grad \psi} \right)
       \label{eq:fsa_divergence}
    \end{align}
    where $V' = \partial V/\partial \psi$. Similarly, the magnetic-surface following time-derivative time derivative $\rmd /\rmd \tau|_\psi$ defined in \cref{eq:time_derivative_magnetic_field} satisfies
 \begin{equation}
      \fsa{ \left.\frac{\rmd}{\rmd \tau}\right|_\psi G}= \frac{1}{V'} \left. \frac{\partial}{\partial \tau} \right|_\psi \left(V' G \right), \label{eq:fsa_time_derivative}
 \end{equation}
where the $V'$ factors account for the $\nabla\cdot \Vpsit$ factor in \eqref{eq:time_derivative_magnetic_field}.
    \subsubsection{Density transport}
    \label{sec:density_transport}
    The equilibrium density $\dens$ can be shown to evolve according to [cf. (166) of \citet{Abel2013}]:
    \begin{align}
        \frac{1}{V'} \left. \frac{\partial}{\partial \tau} \right|_\psi \left(V' \dens \right) + \frac{1}{V'} \frac{\partial}{\partial \psi} \left(V' \fsa{\Gamma_\s} \right) = 0
        \label{eq:density_transport_mcf}
    \end{align}
    in which  
    the particle flux is given by
    \begin{align}
        \Gamma_\s = \turbavg{\int \rmd^3 \vec{v} \avgr{\hs \vchi} \cdot \grad \psi},
        \label{eq:particle_flux_mcf}
    \end{align}
    where $\turbavg{\dots}$ is the turbulent average of \cref{sub: notation}. Note that we have only included the turbulent contribution to the particle flux on the right-hand side of \eqref{eq:particle_flux_mcf}; {both the classical and neoclassical contributions will vanish as a consequence of the assumption of strong collisions.} 
    
    Using the ansatz \eqref{eq:perturbed_maxwellian} for $\hs$ and evaluating the resultant velocity integral in \eqref{eq:particle_flux_mcf} to leading-order in \eqref{eq:subsidiary_ordering}, one obtains
    \begin{align}
        \Gamma_\s = \turbavg{\left[ \dns \vduperp - \dens \left(\frac{\dBpar}{B} \vduperp - \dupars \frac{\vdBperp}{B}\right) \right]\cdot \grad \psi},
        \label{eq:particle_flux_mcf_integral}
    \end{align}
    where we have made use of \eqref{eq:electromagnetic_field_size} and \eqref{eq:pressure_balance}, and integrated by parts using the perpendicular spatial integral contained in the turbulent average. Then, multiplying \eqref{eq:density_transport_mcf} by the mass $\ms$, summing over all species, and using \eqref{eq:mass_density_definitions}, we find:
    \begin{align}
        \frac{1}{V'} \left. \frac{\partial}{\partial \tau} \right|_\psi \left(V' \rho \right) + \frac{1}{V'} \frac{\partial}{\partial \psi} \left(V' \fsa{\bm{\Gamma}_\rho \cdot \grad \psi} \right) = 0 
        \label{eq:particle_flux_mcf_final}
    \end{align}
    or equivalently (noting the identities \eqref{eq:fsa_divergence} and \eqref{eq:fsa_time_derivative}),
    \begin{align}
        \fsa{ \left.\frac{\rmd}{\rmd \tau}\right|_\psi \rho + \grad \cdot \bm{\Gamma}_{ \rho}} = 0,
    \end{align}
    where 
    $\bm{\Gamma}_\rho$ is given by the first expression in \eqref{eq:rho-flux-rel}.
    This is precisely the flux-surface-averaged version of \eqref{eq: general flux transport} for $G = \rho$, as promised. 

    \subsubsection{Thermal energy transport}
    \label{sec:thermal_energy_transport}
    The equilibrium pressure of species $\s$ evolves according to (cf. equation (194) of \citet{Abel2013}, noting that their $\Vpsi$ is our $\Vpsit$):
    \begin{align}
        & \frac{3}{2} \frac{1}{V'} \left. \frac{\partial}{\partial \tau} \right|_\psi \left(V' \dens \Ts \right) + \frac{1}{V'} \frac{\partial}{\partial \psi} \left(V' \fsa{Q_\s} \right) \nonumber \\
        &= - \dens \Ts \fsa{\grad \cdot \Vpsit}   + \turbheating,
        \label{eq:thermal_transport_mcf}
    \end{align}
    where the heat (energy) flux, 
    \begin{align}
        Q_\s = \turbavg{\int \rmd^3 \vec{v} \: \energy \avgr{\hs \vchi} \cdot \grad \psi},
        \label{eq:heat_flux_mcf}
    \end{align}
     only includes the turbulent contribution (as with the particle flux \eqref{eq:particle_flux_mcf}). The first term on the right-hand side of \eqref{eq:thermal_transport_mcf} is  the compressional heating due to the motion of flux surfaces, 
     and $\turbheating$ is turbulent heating due to energy exchange with the fluctuations,
    \begin{align}
       & \turbheating  = - \fsa{\turbavg{\int \rmd^3 \vec{v} \: \frac{\Ts \hs}{\Fs} \sum_{\s'} C_{\s\s'}[\hs]}} \label{eq:turbulent_heating} \\
                     & \quad+  \Ts \left(\frac{\rmd  \ln \dens}{\rmd \psi} - \frac{3}{2} \frac{\rmd \ln \Ts}{\rmd \psi} \right) \fsa{\Gamma_\s} + \frac{\rmd \ln \Ts}{\rmd \psi} \fsa{Q_\s}. \nn
    \end{align}
    The first term represents the turbulent dissipation of fluctuations on collisions, while the latter terms are the energy injection into the fluctuations due to the equilibrium density and temperature gradients. 

    Substituting the ansatz \eqref{eq:perturbed_maxwellian} for $\hs$ into the heat flux \eqref{eq:heat_flux_mcf} and evaluating the resultant velocity integral to leading order in \eqref{eq:subsidiary_ordering}, one obtains:
    \begin{align}
        Q_\s = \frac{1}{\gamma-1} \turbavg{\left[ \dps \vduperp - \gamma \left(\frac{\dBpar}{B} \vduperp - \dupars \frac{\vdBperp}{B}\right) \right] \cdot \grad \psi},
        \label{eq:heat_flux_mcf_integral}
    \end{align}
    where we have once again introduced the adiabatic index $\gamma = 5/3$. Similarly, the turbulent heating \eqref{eq:turbulent_heating} can be written as:
    \begin{align}
        & \turbheating  = - \fsa{\turbavg{\int \rmd^3 \vec{v} \: \frac{\Ts \hs}{\Fs} \sum_{\s'} C_{\s\s'}[\hs]}} \nonumber \\
        & \quad - \fsa{\turbavg{\Vpsi \cdot \grad (\dens \Ts)}} +  \fsa{\ps \turbavg{ \frac{\dns}{\dens} \vduperp} \cdot \grad \ln \dens} \nonumber \\
        & \quad + \fsa{\frac{\ps}{\gamma-1} \turbavg{ \frac{\dTs}{\Ts} \vduperp} \cdot \grad \ln \Ts}.\label{eq:turbulent_heating_integral}
    \end{align}
    Note that while \cref{eq:perturbed_maxwellian} would formally cause the first term in \eqref{eq:turbulent_heating} to formally vanish, it has been retained in \eqref{eq:turbulent_heating_integral} to represent the dissipation (viscous and heat-flux) terms included in our RMHD expansion, 
    which (like the collision operator $C_{ss'}$) must ultimately be responsible for converting fluctuating energy into heat.

    In the absence of equilibrium flows the turbulent heating vanishes when summed over all species \cite{Abel2013}, viz.,
    \begin{align}
        \sum_\s \turbheating = 0.
        \label{eq:turbulent_heating_zero}
    \end{align}
    This is the kinetic equivalent of the earlier statement \eqref{eq:thermal-energy-diff} that the thermal heating rate $\mathcal{Y}^{\perp}_{\rm th}$ balances the rate at which the fluctuations are dissipated --- any energy injected into the fluctuations must ultimately be returned to the equilibrium through turbulent heating. Indeed, associating the first term on the right-hand side of \eqref{eq:turbulent_heating_integral} with the total dissipation $\avgi{D^{\rm tot}}$ and taking $\dns/\dens$ to be independent of species, \cref{eq:turbulent_heating_zero} becomes \cref{eq: local energy conservation} in the absence of parallel terms: $\mathcal{Y}^\perp_{\rm mech} = p \gprp{p}\cdot \Vpsit$ due to the equilibrium, which partially cancels $p\UperpII\cdot \gprp{p}$ to give the negative of the first term on the second line of \eqref{eq:turbulent_heating_integral}; the remaining terms are the (negative) thermal heating rate $-\mathcal{Y}^{\perp}_{\rm th}$, representing the exchange of energy between the equilibrium and fluctuations due to the gradients in the former.

    Then, summing \eqref{eq:thermal_transport_mcf} over species and using \eqref{eq:fsa_divergence}, \eqref{eq:heat_flux_mcf_integral}, and \eqref{eq:turbulent_heating_zero}, it follows that
    \begin{align}
        \fsa{ \left.\frac{\rmd}{\rmd \tau}\right|_\psi E_{\text{th}} + \grad \cdot \bm{\Gamma}_{\text{th}} } = - p \fsa{\grad \cdot \Vpsit},
    \end{align}
    where $\bm{\Gamma}_{\text{th}}$ is given by the first expression in \eqref{eq:Eth-flux-rel}, and the right-hand side is the second expression in \eqref{eq:Eth-flux-rel} after taking $U = \bm g_{\rm eff}^{\perp} = 0$ and noting that the flux-surface average eliminates the parallel divergence of the wave-energy flux.
    The simplicity of the source in this limit should be understood as resulting from the lack of other
    energy reservoirs with which to exchange energy: dissipation of large-scale 
    thermal gradients via turbulence cannot, ultimately, heat the plasma (though any fluctuations that driven and dissipated will still cause heat transport), 
    meaning that the only true heating is via energy exchange with the background magnetic field. In this field-following frame, the latter manifests as a compressional heating via $\grad \cdot \Vpsit$.
    
\end{section}

%

\end{document}